\documentclass{aa}
\usepackage{graphicx}
\usepackage{txfonts}
\usepackage{placeins}
\usepackage{color}
\usepackage{adjustbox}
\usepackage[draft]{hyperref}
\usepackage[nolist]{acronym}
\newacro{imf}[IMF]{initial mass function}
\newacro{agb}[AGB]{asymptotic giant branch}
\newacro{rgb}[RGB]{red giant branch}
\newacro{hb}[HB]{horizontal branch}
\newacro{lte}[LTE]{local thermodynamic equilibrium}
\newacro{nlte}[NLTE]{non-LTE}
\newacro{ssp}[SSP]{simple stellar population}
\newacro{gc}[GC]{globular cluster}
\newacro{hrd}[HRD]{Hertzsprung--Russell diagram}
\newacro{cmd}[CMD]{colour--magnitude diagram}
\newacro{pc}[PC]{principal component}

\defcitealias{Yong2005}{Y2005}
\defcitealias{Ramirez2011}{RA2011}
\defcitealias{Worley2009}{W2009}
\defcitealias{vanderSwaelmen2013}{vdS2013}
\defcitealias{Castelli2004}{CH2004}
\defcitealias{Grevesse1998}{GS1998}
\defcitealias{Asplund2009}{A2009}
\defcitealias{Roediger2014}{R2014}
\defcitealias{Thygesen2014}{T2014}
\defcitealias{Larsen2017}{L2017}
\defcitealias{McWilliam2008}{MB2008}
\defcitealias{Sakari2013}{S2013}
\defcitealias{Colucci2014}{C2014}
\defcitealias{Sakari2016}{S2016}
\defcitealias{Venn2004}{V2004}
\defcitealias{Ishigaki2013}{I2013}
\defcitealias{Carretta2009}{C2009}

\makeatletter
\renewcommand*\aa@pageof{, page \thepage{} of \pageref*{LastPage}}

\begin{document}
\title{The chemical composition of globular clusters in the Local Group\thanks{The Tables in Appendices~\ref{app:specwin}-\ref{app:iam} are available in electronic form at the CDS via anonymous ftp to cdsarc.u-strasbg.fr (130.79.128.5) or via http://cdsweb.u-strasbg.fr/cgi-bin/qcat?J/A+A/}}
% \subtitle{}

\author{
   S. S. Larsen
   \inst{1}
   \and
   P. Eitner
   \inst{2,3}
   \and 
   E. Magg
   \inst{3}
   \and
   M. Bergemann
   \inst{3,4}
   \and
   C.\ A.\ S.\ Moltzer
   \inst{1}   
   \and
    J.\ P.\ Brodie
   \inst{5,7}
   \and
     A. J. Romanowsky
   \inst{6,7}   
   \and
    J. Strader
   \inst{8}
}

\institute{
  Department of Astrophysics/IMAPP,
              Radboud University, PO Box 9010, 6500 GL Nijmegen, The Netherlands\\
              \email{s.larsen@astro.ru.nl}
  \and
  Ruprecht-Karls-Universit{\"a}t, Grabengasse 1, 69117 Heidelberg, Germany
  \and
  Max-Planck-Institute for Astronomy, K{\"o}nigstuhl 17, 69117 Heidelberg, Germany
  \and
  Niels Bohr International Academy, Niels Bohr Institute, Blegdamsvej 17, DK-2100 Copenhagen {\O}, Denmark
  \and
  Centre for Astrophysics and Supercomputing, Swinburne University of Technology, Hawthorn, VIC 3122, Australia
  \and
  Department of Physics \& Astronomy, One Washington Square, San Jos{\'e} State University, San Jose, CA 95192, USA
  \and
  University of California Observatories, 1156 High Street,
  Santa Cruz, CA 95064, USA
  \and
  Department of Physics and Astronomy, Michigan State University, East Lansing, Michigan 48824, USA
}

\date{Submitted 17 September 2021; Accepted 23 Nov 2021}

\abstract{We present detailed chemical abundance measurements for 45 globular clusters (GCs) associated with galaxies in (and, in one case, beyond) the Local Group. The measurements are based on new high-resolution integrated-light spectra of GCs in the galaxies NGC~185, NGC~205, M31,  M33, and NGC~2403, combined with reanalysis of previously published observations of GCs in the Fornax dSph, WLM, NGC~147, NGC~6822, and the Milky Way.
The GCs cover the range $-2.8 < \mathrm{[Fe/H]} < -0.1$ and we
determined abundances for Fe, Na, Mg, Si, Ca, Sc, Ti, Cr, Mn, Ni, Cu, Zn, Zr, Ba, and Eu.
Corrections for non local thermodynamic equilibrium effects are included for Na, Mg, Ca, Ti, Mn, Fe, Ni, and Ba, building on a recently developed procedure. 
For several of the galaxies, our measurements provide the first quantitative constraints on the detailed composition of their metal-poor stellar populations. 
Overall, the GCs in different galaxies exhibit remarkably uniform abundance patterns of the $\alpha$, iron-peak, and neutron-capture elements, with a dispersion of less than 0.1~dex in [$\alpha$/Fe] for the full sample. There is a hint that GCs in dwarf galaxies are slightly less $\alpha$-enhanced (by $\sim0.04$~dex on average) than those in larger galaxies. 
One GC in M33 (HM33-B) resembles the most metal-rich GCs in the Fornax dSph (Fornax~4) and NGC~6822 (SC7) by having $\alpha$-element abundances closer to scaled-solar values, possibly hinting at an accretion origin. A principal components analysis shows that the $\alpha$-element abundances strongly correlate  with those of Na, Sc, Ni, and Zn. Several GCs with $\mathrm{[Fe/H]}<-1.5$ are deficient in Mg compared to other $\alpha$-elements.  We find no GCs with strongly enhanced $r$-process abundances as reported for metal-poor stars in some ultra-faint dwarfs and the Magellanic Clouds. The similarity of the abundance patterns for metal-poor GCs in different environments points to similar early enrichment histories and only allow for minor variations in the initial mass function.
}


\maketitle

\section{Introduction}

The relative weakness of `metallic' lines in the integrated spectra of \acp{gc}, which
in some cases implies metallicities of less than 1\% of the solar value, was noted long ago \citep{Mayall1946,Morgan1956,Kinman1959}. Combined with the realisation that the more metal-poor \acp{gc} tend to be less concentrated towards the Galactic plane, this was an early harbinger of the first quantitative scenarios for the formation of the Milky Way \citep{Eggen1962,Searle1978}. In modern theories of galaxy formation, which are closely linked to the $\Lambda$-cold-dark-matter cosmological paradigm, 
present-day galaxies comprise a combination of stars that formed `in situ' within the main progenitor halo and an
`ex situ' component that was built up through a sequence of mergers and accretion events   \citep{Navarro1994,Cooper2010,Genel2014,Schaye2015}. Simulations indicate that the in situ component typically dominates in low-mass galaxies and in the central regions of relatively massive (Milky-Way-like) galaxies, while ex situ stars become increasingly dominant at larger radii, especially in massive galaxies \citep{Pillepich2015,Cook2016,Davison2021}.

Direct evidence of these galaxy assembly processes abounds, not only in the obvious form of on-going major mergers, but also through identification of disrupted Milky Way satellite galaxies such as Sagittarius, Gaia-Enceladus, and others via analysis of the kinematics and chemistry of stars and \acp{gc} \citep{Ibata1994,Helmi2018,Belokurov2018,Bergemann2018,Forbes2020,Kruijssen2020,Woody2021}.  
The large number of substructures in the halo of M31 likewise attest to an active accretion history \citep{Ibata2014,McConnachie2018}, again with a close correspondence between features traced by halo field stars and \acp{gc} \citep{Mackey2019a}.  
The wealth of detailed phase-space information that is now available from the Gaia mission has helped paint a rich and detailed picture of the accretion history of the Milky Way \citep{Malhan2018,Brown2021}, especially in combination with chemical abundance information from ground-based spectroscopic surveys \citep{Mackereth2019,Cordoni2021,Buder2021}.
However, because each galaxy has its own unique hierarchical assembly history, it is essential to establish to what extent lessons learned from detailed studies of the Milky Way can be generalised to other galaxies. 

The chemical abundance patterns of stellar populations in galaxies contain valuable information about the assembly- and star formation histories.
The various chemical elements are produced on different time scales by different mechanisms, and their relative abundances are therefore sensitive to the time scales of chemical enrichment and the relative importance of the various nucleosynthetic mechanisms. The ratio of $\alpha$-capture elements to iron is a well-known indicator of the relative contributions from  core-collapse (Type II) supernovae (SNe) on short timescales and Type Ia SNe with longer-lived progenitors \citep{Tinsley1979,Matteucci1986}. The elements beyond the iron peak are mostly produced by neutron-capture processes in \ac{agb} stars ($s$-process), neutron star mergers, or various types of exotic SNe ($r$-process) \citep{Burbidge1957,Kobayashi2020}. Within these broad categories, individual elements do not vary strictly in lockstep, as most elements are not produced by just a single mechanism. Among the $\alpha$-elements, O and Mg are, at least in the Milky Way, the purest tracers of Type II SN nucleosynthesis, whereas Si and especially Ca and Ti also have significant contributions from Type Ia SNe \citep{Kobayashi2020}. However, Mg abundances can  also be modified by hot hydrogen burning in \ac{agb} stars or massive stars, which may be responsible for the anomalous Mg abundances observed in some \ac{gc} member stars \citep{Gratton2012,Bastian2018}. The iron-peak elements (e.g.\ Cr, Mn, Fe, and Ni) are thought to be produced mainly in Type Ia SNe, but they also have contributions from core collapse SNe \citep{Kobayashi2020} and the production of Mn in particular is sensitive to SN Ia explosion physics and progenitor properties \citep{McWilliam2003,Kirby2019,Eitner2020,Sanders2021}. 

In metal-poor Milky Way halo stars and \acp{gc}, the abundances of the $\alpha$-elements are typically enhanced by about a factor of two compared to scaled-solar composition 
\citep{Cohen1978,Pilachowski1980,Sneden1979,Luck1981}. Similarly $\alpha$-enhanced abundance patterns have been found for \acp{gc} and stars in the inner part of the M31 halo \citep{Beasley2005,Colucci2014,Sakari2016,Escala2019,Escala2020a}. In accordance with the above discussion, this suggests enrichment on time scales that were short relative to the delay before significant Type Ia SN enrichment set in \citep{McWilliam1997,Gilmore1998}.
However, the full picture is now known to be much more complex. At intermediate 
metallicities ($-1.7 \la \mathrm{[Fe/H]} \la -0.5$), stars in the Milky Way halo display at least two distinct sequences in the $[\alpha/\mathrm{Fe}]$ vs.\ [Fe/H] plane, of which the
$\alpha$-rich sequence is thought to be associated with the in situ component, while the
less $\alpha$-enhanced stars appear to be linked to the Gaia-Enceladus accretion event. The latter stars are also characterised by an enhancement of $r$-process elements relative to the $\alpha$-elements \citep{Nissen2010,Helmi2018,Matsuno2021,Matsuno2021a}.
The abundance patterns of the Gaia-Enceladus stars are reminiscent of those observed in nearby extant dwarf galaxies and likely reflect differences in the star formation histories relative to the more $\alpha$-enhanced Galactic halo stars, with chemical enrichment proceeding at a slower pace in the dwarf galaxies \citep{Shetrone2001,Venn2004,Tolstoy2009,McWilliam2013,Lemasle2014}.
The [$\alpha$/Fe] patterns of stars in the outer parts of the M31 halo (beyond $\sim$40 kpc) also tend to resemble those of stars in M31 dwarf satellites more closely than stars nearer the centre \citep{Gilbert2020}, again indicative of a link between the dwarf satellites and the outer halo. 
These examples illustrate the role that chemical abundances can play in tracing hierarchical assembly histories of galaxies.

Detailed chemical abundance analysis of individual stars associated with old stellar populations is only feasible in the Milky Way and its nearest neighbouring galaxies with current astronomical facilities. The integrated light of entire galaxies can be observed to much greater distances, and spectroscopy of early-type galaxies has shown that they are typically dominated by relatively metal-rich, old stellar populations with increasingly enhanced $\alpha$-element abundances for higher masses and velocity dispersions \citep{Worthey1992,Kuntschner2000,Trager2000,Thomas2005,Conroy2014,Kriek2019,Parikh2019}.
However, disentangling the mix of stellar populations with different ages and compositions that contribute to the integrated light is challenging, although some constraints on star formation histories and age-metallicity relations can be obtained from spectral inversion techniques \citep{Peterken2020,Greener2021}.
\acp{gc} occupy an intermediate step between detailed studies of individual stars in nearby galaxies and the integrated light of more distant galaxies. They tend to be preferentially associated with the metal-poor, old  components of galaxies, which usually contribute only a minor fraction of the integrated light, and they are therefore particularly useful tracers of these components. Apart from the Milky Way, association of \acp{gc} with substructure has been demonstrated in external galaxies such as M31 \citep{Mackey2019a} and M87 \citep{Romanowsky2012}.

Measurements of spectroscopic line indices on medium-resolution spectra of \acp{gc} is a well established technique for determining their ages and metallicities, and even obtaining some information about detailed abundances such as $[\alpha/\mathrm{Fe}]$ ratios and nitrogen-enrichment \citep{Brodie2006,Schiavon2013}. Based on such analyses, \acp{gc} around other galaxies tend to have similar old ages ($\sim10$ Gyr) and $\alpha$-enhanced composition to their Galactic counterparts \citep{Larsen2002e,Beasley2008,Puzia2005,Strader2005,Cenarro2007}. However, it is not yet entirely clear just how similar the abundances of \acp{gc} in different environments are. 
For \acp{gc} in the Local Group dwarf galaxies NGC~147, NGC~185, and NGC~205, \citet{Sharina2006} found $\alpha$-element abundances consistent with scaled-solar values, and \citet{Puzia2006} found strongly $\alpha$-enhanced ($[\alpha/\mathrm{Fe}] > +0.5$) abundances for relatively metal-rich ($\mathrm{[Fe/H]} > -1$) \acp{gc} in a sample of early-type galaxies.
In contrast, \citet{Woodley2010} found \acp{gc} in the nearest giant elliptical, NGC~5128, to be only moderately $\alpha$-enhanced with on average $[\alpha/\mathrm{Fe}] = +0.14$, whereas $[\alpha/\mathrm{Fe}]$ values for NGC~5128 \acp{gc} more similar to, or even slightly higher than those in Milky Way \acp{gc}, have been reported from detailed modelling of integrated-light spectra \citep{Colucci2013,Hernandez2018}.  Differences between the abundance patterns in different types of galaxies could have important consequences for constraining their early chemical evolution, and could provide a basis for identification of different progenitor systems via `chemical tagging' \citep{Freeman2002,Sakari2014,Sakari2015,Horta2020,Minelli2021}.

Over the past decade, techniques to measure chemical abundances of individual elements from detailed modelling of integrated-light spectra, either from analysis of individual lines or from spectral fitting, have matured and have been applied to \acp{gc} in several studies. \citet[][hereafter MB2008]{McWilliam2008} showed that abundances consistent with those measured for individual stars could be obtained from an integrated-light spectrum of the Galactic \ac{gc} NGC~104 (47~Tuc). This type of analysis has since been further developed, tested, and applied in several studies \citep{Colucci2009,Colucci2017,Larsen2012a,Larsen2014,Larsen2017,Sakari2013,Sakari2015,Sakari2016,Conroy2018,Renno2020}, and the abundances determined from integrated light generally agree with those obtained from individual stars within $\sim0.1$~dex. So far, these integrated-light studies have mostly adopted the standard simplifying assumptions of 1-D, static model atmospheres and \ac{lte} in the analysis. 
Corrections for \ac{nlte} effects are now becoming increasingly commonplace in abundance analyses of individual stars, and can in some cases lead to substantial differences. 
For example, \citet{Bergemann2017a,Bergemann2017} showed that the detailed [Mg/Fe] ratios in the low-$\alpha$ Galactic stars are sensitive to 3-D and \ac{nlte} effects, although a distinction between Mg-rich and Mg-poor stars remains also in $\langle 3-\mathrm{D} \rangle$ \ac{nlte} analysis \citep{Bergemann2017}.
Application of \ac{nlte} corrections to integrated-light measurements is complicated by the fact that the corrections vary depending on the physical parameters (effective temperature $T_\mathrm{eff}$, surface gravity $\log g$, composition) of stars in different parts of the \ac{hrd}.
\ac{nlte} corrections for integrated-light spectra were computed by \citet{Eitner2019} for Mg, Mn, and Ba, and were applied to observations of \acp{gc} by \citet{Eitner2020}. Especially for Mn, the application of \ac{nlte} corrections significantly modified the results, largely eliminating the trend of decreasing $\mathrm{[Mn/Fe]}$ towards low metallicities seen in \ac{lte} analysis. As a consequence, the preferred model for Galactic chemical evolution changed from one in which Mn is produced in Type Ia SNe with sub-Chandrasekhar mass progenitors to one in which the progenitors have masses near the Chandrasekhar mass.
Hence there is a clear need to further develop techniques for applying \ac{nlte} corrections to integrated-light abundance measurements. 

In this paper we present a homogeneous analysis of integrated-light, high-resolution spectra of 45 \acp{gc}, mostly associated with Local Group galaxies but also including a cluster in the Sc-type galaxy NGC~2403. The galaxies span a range of morphological types, including all three Local Group spirals as well as several dwarf spheroidal and irregular galaxies. 
From detailed modelling of the \ac{gc} spectra we measure the abundances of a large number of elements,
including light- and $\alpha$-elements (Na, Mg, Si, Ca, and Ti), iron-group elements (Sc, Cr, Mn, and Ni), and heavy elements (Cu, Zn, Zr, Ba, and Eu). 
A major update compared to previous papers is the inclusion of \ac{nlte} corrections for several elements, building on the work of \citet{Eitner2019,Eitner2020}. For the dwarf galaxies, our sample includes most of the old Local Group \acp{gc} that are bright enough for integrated-light spectra of sufficient signal-to-noise ratio (S/N, preferably better than about 100 per \AA) to be obtained in a few hours of integration time, which translates to a magnitude limit of about $V=18$. For the larger galaxies, in particular the Milky Way and M31 with their rich \ac{gc} systems, our sample only includes a small subset of the total \ac{gc} populations. Nevertheless, the current sample is large enough that we can gain some insight into the degree of similarity between the chemical abundance patterns of \acp{gc} in different galaxies. As outlined above, this work is complementary to studies of field stars, as the \acp{gc} tend to preferentially trace the more metal-poor populations and their brightness makes it possible to constrain individual element abundances in more detail. In addition to this primary aim of comparing \acp{gc} within the Local Group, it is our hope that the data presented here will also serve as a useful reference for comparison with future work beyond the Local Group.

\section{Data}

\begin{figure}
\centering
\includegraphics[width=\columnwidth]{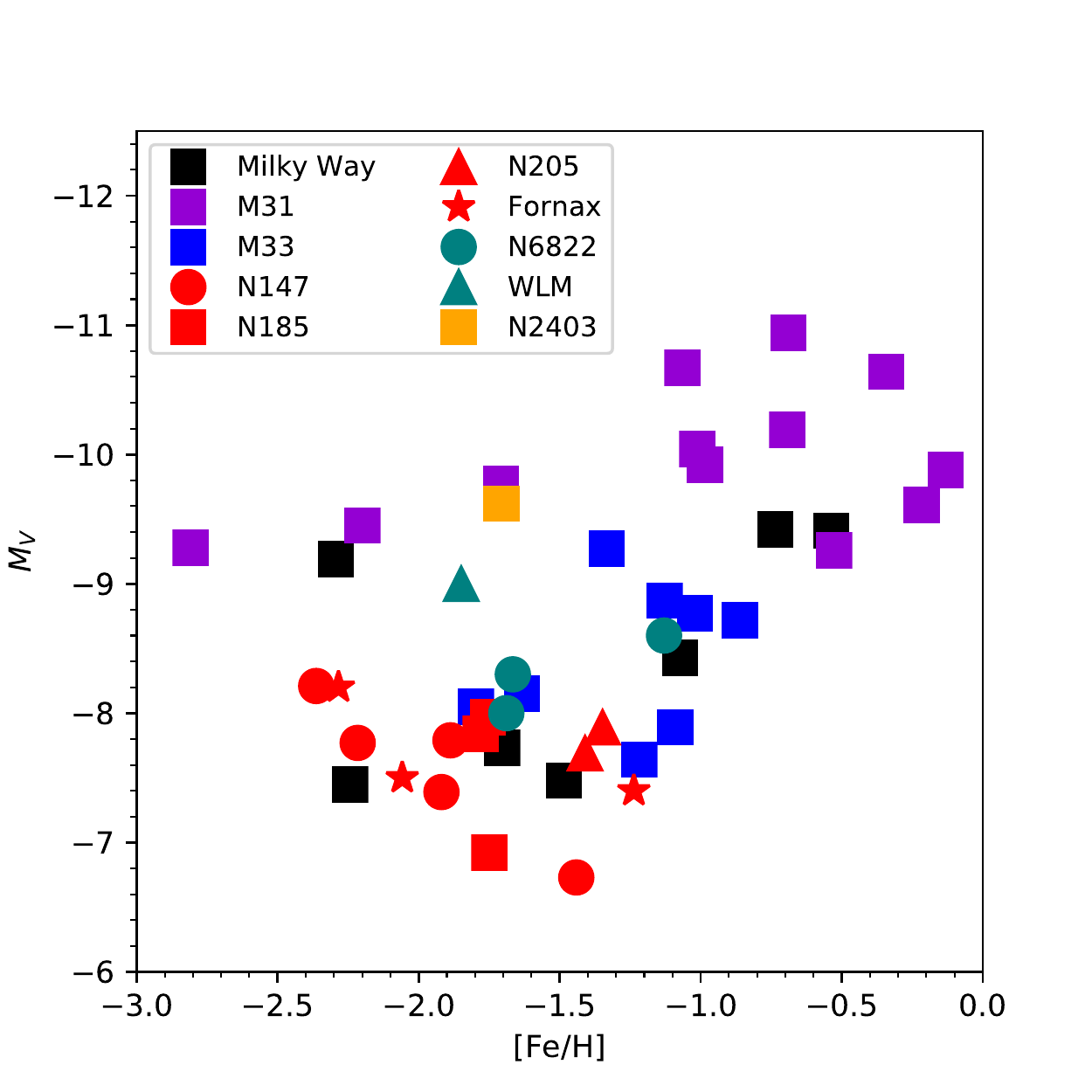}
\caption{\label{fig:femv}Absolute visual magnitude ($M_V$) as a function of metallicity for the observed GCs. Symbol colours and shapes identify the host galaxy as indicated in the legend.}
\end{figure}

The observations are summarised in Table~\ref{tab:obs}. Northern targets were observed with the HIRES spectrograph \citep{Vogt1994} on the Keck~I telescope and for the southern targets we used UVES \citep{Dekker2000} on the ESO Very Large Telescope.
In some cases, abundance analyses based on these observations have been published previously, with references that also provide more information about the reduction of these datasets given in the table. However, many aspects of our analysis technique have been updated (see below) and we here present a full reanalysis of all datasets. As such, the analysis in this paper supersedes the previous work, although the differences with respect to previous results are generally relatively minor (Sect.~\ref{sec:individual_remarks}). 

The July 2015 UVES observations of Milky Way \acp{gc} presented in \citet[][hereafter L2017]{Larsen2017} were combined with new observations of the same \acp{gc} obtained in August 2019. The July 2015 observations were obtained with the UVES red arm and the standard Cross Disperser \#3, centred at 520 nm, while the August 2019 observations used the DIC2 dichroic and the CD\#2 and CD\#4 cross-dispersers in the blue and red arm, respectively. Together the two epochs cover the full spectral range 3300~\AA --9500~\AA\ at a spectral resolving power of $\mathcal{R}\equiv\lambda/\Delta \lambda\sim40\, 000$, where $\Delta \lambda$ is the full width at half maximum of a resolution element. For both epochs of UVES observations the integrated light was sampled using a drift-scan technique whereby the UVES slit was scanned multiple times across the half-light diameter of each \ac{gc}. In most cases, the same scanning patterns and exposure times were used for the two epochs.
The August 2019 data were reduced in the same way as the July 2015 data, using the UVES pipeline running within the ESOREX environment to extract the calibrated 2-D spectra. 
Separate sky exposures, bracketing the science exposures, were used to determine the sky level, which was then subtracted from the science exposures. Finally, the 2-D spectra were collapsed to 1-D spectra which were used in the analysis. For further details we refer to \citetalias{Larsen2017}. The same drift-scan technique was used for the integrated-light spectra of the \acp{gc} in the Fornax dSph, for which more details are given in \citet{Larsen2012a}.

Owing to their larger distances, the remaining \acp{gc} are sufficiently compact that their integrated light could be well sampled without having to rely on the relatively complex slit scanning procedure. The WLM \ac{gc} was observed with UVES in a single setting \citep{Larsen2014}, while the rest of the data were obtained with HIRES.
Most of the M31 \ac{gc} spectra come from two archival datasets, U017Hr (Oct 1-3, 2007, P.I.\ G.\ Smith) and U118Hb (Oct 18-19, 2007, P.I.\ K.\ Gregg), and four of the M33 \ac{gc} spectra are older data \citep{Larsen2002b}. Observations from the programme U017Hr were previously included in the study of mass-to-light ratios for M31 \acp{gc} by \citet{Strader2009}. The remaining HIRES data were obtained for this project as part of dedicated observing programmes. 

Most of the HIRES observations were obtained after the instrument upgrade in 2004 \citep{Butler2017}. Apart from small gaps between the three detectors, the spectral coverage is continuous for wavelengths up to about 6300~\AA. Above this limit the ends of the echelle orders fall off the edges of the detectors, leading to gaps in the wavelength coverage. 
The four M33 \ac{gc} spectra from Oct 1998 were taken prior to the instrument upgrade, when HIRES only had a single detector. For these observations the total spectral range is therefore smaller and the ends of echelle orders already start falling off the ends of the detectors at wavelengths longer than $\sim4500$~\AA . The location of the echellogram on the HIRES detectors can be adjusted by tilting the echelle grating and cross disperser, and not all datasets used the same settings. The exact wavelength coverage and location of the gaps in spectral coverage therefore vary from one dataset to another. The HIRES observations were typically obtained with the C5 decker which has a $7\arcsec\times1\farcs148$ slit and provides a resolving power of $\mathcal{R}=37\, 000$. The 1998 and 2007 observations used somewhat narrower slits of $0\farcs725$ and $0\farcs86$, respectively, with correspondingly higher spectral resolving powers ($\mathcal{R}$ being approximately proportional to the inverse slit width). 

The HIRES data were reduced with the MAKEE (MAuna Kea Echelle Extraction) package\footnote{Available at \url{http://www.astro.caltech.edu/~tb/makee/}} written by T.\ Barlow. MAKEE automatically performs all reduction steps, from bias subtraction and flat-fielding of the raw exposures, to tracing of the spectral orders, optimal extraction, wavelength calibration, and resampling of the spectra to a linear wavelength scale.  
The details of the MAKEE reduction, such as constraints on the spectral extraction and background determination regions, are defined in a configuration file, where in most cases we used the standard configuration file as provided with MAKEE. 
The individual spectra of a given \ac{gc} typically had identical exposure times and similar S/N and combined spectra were obtained as a straight (unweighted) sum of the extracted and calibrated 1-D spectra of each \ac{gc}. In the few cases that involved exposures of unequal duration, a more elaborate weighting scheme might in principle have produced combined spectra of slightly higher S/N. However, in practice the gain would be small: in the extreme case of read-noise limited data for which one exposure is many times longer than the other, the difference in S/N between an error-weighted average of the two exposures and a straight sum would amount to a factor of $\sqrt{2}$, and for the typical exposures used here the difference is less than $\sim5$\%. More details about the reduction of the HIRES spectra can be found in \citet{Larsen2018}.

For each combined \ac{gc} spectrum, the S/N per \AA, averaged over a 50~\AA\ interval near 5000~\AA , is listed in Table~\ref{tab:obs}. The S/N was estimated from the dispersion of the individual combined pixels at each wavelength sampling point. Because the linearisation of the wavelength scale involves interpolation between neighbouring pixel values, the dispersion of the individual values may underestimate the true uncertainties by up to a factor of $\sqrt{2}$, and the final S/N values were therefore reduced by this factor. Nevertheless, the S/N values in Table~\ref{tab:obs} should be considered approximate.

In Fig.~\ref{fig:femv} we plot the absolute visual magnitudes ($M_V$) versus the iron abundances obtained from our analysis for the \acp{gc}. Distances and foreground extinctions mostly come from the references for the $V$ magnitudes in Table~\ref{tab:obs}, except for M31 where a distance modulus of $(m-M)_0=24.47$ was assumed \citep{Stanek1998}, and for M33 where we adopted the distance modulus, $(m-M)_0=24.62$, and reddening ($E(B-V)=0.19$) from \citet{Gieren2013}. 
The clusters span a range between $\mathrm{[Fe/H]}=-2.8$ and $-0.1$, with the more metal-rich \acp{gc} preferentially being associated with the Milky Way and M31 and the more metal-poor ones preferentially with the dwarf galaxies in our sample. We also note that the M31 \acp{gc} in our sample are among the brightest in that galaxy, and are generally brighter than those associated with the dwarf galaxies. 

\section{Analysis}

The basic analysis framework remains similar to that described in several previous papers \citep{Larsen2012a,Larsen2017,Larsen2018}. Briefly stated, we proceed by computing \ac{ssp} model spectra at high spectral resolving power while adjusting the abundances of individual elements until the best fits to the observed spectra are obtained. An outline of the main steps follows below (Sect.~\ref{sec:outline}). 
Compared to previous analyses, some of the main updates for this work include the use of \texttt{ATLAS12} model atmospheres with compositions that self-consistently match those derived from the spectra (Sect.~\ref{sec:atmospheres}), a redefinition of the spectral windows used to fit the abundances of some elements (Sect.~\ref{sec:specwin}), an extensive revision of the line list (Sect.~\ref{sec:linelist}), and a modified prescription for assigning microturbulence velocities to individual stars (Sect.~\ref{sec:sspmod}).
For the first time, we also include \ac{nlte} corrections for our integrated-light abundance measurements for several elements  (Sect.~\ref{sec:nlte}). 

\subsection{General outline of the procedure}
\label{sec:outline}

We assume that \acp{gc} can be modelled as \acp{ssp}, that is, as consisting of stars with a single age and chemical composition. This is clearly an oversimplification, given that a number of (especially light) elements are known to exhibit significant abundance spreads within \acp{gc} \citep{Bastian2018,Gratton2019}. However, attempting to constrain such abundance spreads from integrated-light measurements is beyond the scope of this work \citep[but see][]{Larsen2018a} and what we measure here is thus an average abundance for each element. Populations of stars with different abundances contribute to this average with weights that depend on the response of individual spectral features to the abundance variations \citep{Larsen2017}.

To compute an integrated-light model spectrum, we must first specify the distribution of stars in the \ac{hrd}. In general, this information may come from a theoretical isochrone or from an empirical \ac{cmd}, or some combination of the two. 
In practice, the \ac{hrd} is provided as a set of approximately 100 bins (`HRD-boxes'), each representing a group of stars with specific physical parameters (effective temperature $T_\mathrm{eff}$, surface gravity $\log g$, and radius $R$). The chemical composition is assumed to be the same for all \ac{hrd}-boxes. 
Model atmospheres and synthetic spectra are then computed for each \ac{hrd}-box and the synthesised surface fluxes are scaled by the surface areas of the stars to provide luminosities, which are finally co-added with weights corresponding to the numbers of stars associated with each \ac{hrd}-box. The result is an integrated-light \ac{ssp} model spectrum for the assumed chemical composition and \ac{hrd} parameters, calculated at a spectral resolving power that is sufficiently high to sample the line profiles (typically $\mathcal{R} = 500\, 000$). 
The \ac{ssp} model spectrum is then convolved with a Gaussian kernel to account for instrumental resolution and  velocity broadening in the cluster, it is scaled to the (radial velocity corrected) observed spectrum using a spline or polynomial fitting function to match the continuum levels, and the $\chi^2$ is computed for the model--data difference. The input abundances are then adjusted and the procedure is repeated until the best fit is obtained. In principle, our implementation of this technique allows for any arbitrary number of element abundances to be fitted simultaneously, but in practice we usually fit for one element at a time using spectral windows tailored specifically to the features of interest for each element. Errors are estimated by varying the abundances until the $\chi^2$ value has increased by one, compared to the best-fit value. 

The procedure is implemented as a \texttt{Python 3} package that we have named \texttt{ISPy3} (Integrated-light Spectroscopy with Python 3). The Python code is publically available via Github \citep{ispy3}.
The model atmosphere and spectral synthesis calculations are done via calls to external codes, with the currently supported options being either the Kurucz \texttt{ATLAS9}/\texttt{ATLAS12} and \texttt{SYNTHE} codes  \citep{Kurucz1970,Kurucz1981,Kurucz2005} or \texttt{MARCS} model atmospheres in combination with \texttt{Turbospectrum}  \citep{Gustafsson2008,Alvarez1998,Plez2012}.

\subsection{Model atmospheres}
\label{sec:atmospheres}

In previous papers we have relied mostly on the Linux versions of the \texttt{ATLAS9} and \texttt{SYNTHE} codes \citep{Sbordone2004} for the model atmospheres and spectral synthesis while employing spherically symmetric \texttt{MARCS} models and \texttt{Turbospectrum} \citep{Alvarez1998} for the modelling of the coolest giants. Each combination has pros and cons: the \texttt{ATLAS9}/\texttt{ATLAS12} codes are publically available and can therefore be used to compute models for any desired combination of stellar parameters (effective temperature $T_\mathrm{eff}$, surface gravity $\log g$, and chemical composition), but \texttt{ATLAS} models are limited to plane parallel geometry. The \texttt{MARCS} grid includes models with spherical geometry, but models must be interpolated for physical parameters not included in the pre-computed grid available from the \texttt{MARCS} website\footnote{\url{https://marcs.astro.uu.se}}.

The \texttt{ATLAS} models come in two flavours.
In \texttt{ATLAS9}, the line opacity is modelled via pre-computed opacity distribution functions (ODFs) and models are thus restricted to the abundance patterns used when computing the ODFs. Recomputing the ODFs for different abundance patterns is a time consuming process and becomes impractical if models with many different abundance patterns are needed. The \texttt{ATLAS12} code uses the opacity sampling technique to compute models for arbitrary abundance patterns, but at a much higher computational cost per individual model. It should be noted that, even in \texttt{ATLAS9}, the detailed abundance patterns specified when computing a model do affect the continuum opacity, especially for elements that are important electron donors (such as Na, Mg, Si, and Ca) and therefore have a significant effect on the H$^{-}$ opacity and the resulting atmospheric structure. If the spectral synthesis is subsequently done for abundance patterns that do not match those used when computing the atmosphere models, inconsistencies can arise.

For the analysis presented here we used \texttt{ATLAS12} for stars hotter than $T_\mathrm{eff}=4000$~K to compute model atmospheres with abundance patterns matching those determined from the spectroscopic analysis. As the abundance patterns are not known a priori, this required an iterative approach whereby we started with an initial guess for the input abundances (for the \acp{gc}, typically a 0.3~dex enhancement of the $\alpha$-elements relative to scaled-solar composition), then fitted for the abundances, and recomputed the model atmospheres.  Since the spectral synthesis is, after all, only moderately sensitive to the exact abundances assumed when computing the model atmospheres, this procedure usually required only 2 or 3 iterations. 
For the cooler stars, both dwarfs and giants, we continue to rely on \texttt{MARCS} and \texttt{Turbospectrum}. 
The motivation for this is two-fold: at low surface gravities, departures from plane-parallel geometry become increasingly important, and at high surface gravities the \texttt{ATLAS} models with low temperatures occasionally fail to converge properly, particularly at low metallicities. \texttt{ISPy3} uses the programme \texttt{interpol\_modeles} \citep{Masseron2006} to
interpolate between models for the $T_\mathrm{eff}$, $\log g$, and [Fe/H] values included in the \texttt{MARCS} grid. We used the `standard composition' grid for which the models are computed for an $\alpha$-element enhancement of $[\alpha/\mathrm{Fe}]=+0.4$ at metallicities $\mathrm{[Fe/H]}\leq-1$, gradually decreasing to scaled-solar composition at $\mathrm{[Fe/H]}=0$.
For the spectral synthesis we used the same atomic and molecular line lists for \texttt{SYNTHE} and \texttt{Turbospectrum} (see Sect.~\ref{sec:linelist}), except for TiO for which \texttt{SYNTHE} uses the line list by \citet{Schwenke1998} while \texttt{Turbospectrum} uses the line list by \citet{Plez1998}.

\subsection{Spectral windows}
\label{sec:specwin}

To aid us in updating the line list and (re-)defining the windows used for the spectral fitting,  
we used the \citet{Wallace2000} spectrum of Arcturus (spectral type K1.5 III; \citealt{Keenan1989})
and the 2005 version of the \citet{Kurucz1984} spectrum of the Sun. The Arcturus spectrum has a S/N of about 1000, sampled at 0.06~\AA\ resolution near 5000~\AA\ (corresponding to a $\mathrm{S/N}\sim13000$ per \AA), while the solar spectrum has an even higher S/N of $>2000$ per 0.05~\AA\ sampling interval \citep{Furenlid1988}. In both cases, this is far higher than the S/N of any of our \ac{gc} spectra.
Because this part of the analysis was done at an early stage of the project, we used \texttt{ATLAS9} to compute a model atmosphere for each star, assuming an effective temperature of $T_\mathrm{eff}=4286$~K, surface gravity  $\log g = 1.66$, and an initial metallicity $\mathrm{[Fe/H]}=-0.6$ for Arcturus \citep{Worley2009,Ramirez2011} and $T_\mathrm{eff}=5777$~K, 
$\log g = 4.44$, and $\mathrm{[Fe/H]}=0$ for the Sun \citep{Cox2000}.
For each model atmosphere, we used the \texttt{WIDTH9} code \citep{Castelli2005,Kurucz2005} to calculate equivalent widths for all atomic lines in the most recent version of the line list at the Kurucz website\footnote{\url{http://kurucz.harvard.edu/}} (dated 8 Oct 2017). 
Synthetic spectra were computed with \texttt{SYNTHE}.

For iron we defined 40 new spectral windows. These windows were defined primarily via a visual inspection of the Arcturus spectrum alongside the corresponding \texttt{SYNTHE} model spectrum, using the list of equivalent widths to label the stronger lines. In order to be useful for measuring iron abundances also at low metallicities, we made sure that each spectral window contained lines with a range of equivalent widths, also including relatively strong lines with equivalent widths  $\ga100$~m\AA . 
At wavelengths $<4400$~\AA\ the spectra of late-type stars become strongly affected by CH molecular absorption bands and by increased line blending in general, and we therefore concentrated on the spectral range $\lambda > 4400$~\AA .  Together, our 40 iron windows cover about 45\% of the wavelength range 4570~\AA\ - 6185~\AA\ but include about 60\% of the Fe lines stronger than 100~m\AA .  
We also defined ten new windows for Ca, 14 windows for Ti, and 18 windows for Cr that replace the broader windows used to fit for the abundances of these elements in previous papers. While the velocity broadening of \ac{gc} spectra implies that all lines are affected by blending at some level, we made an effort to define these new windows in such a way that relatively clean lines were prioritised. A full listing of the window definitions can be found in Table~\ref{tab:windows}. 

We added several chemical elements not measured in the previous analyses, in some cases taking advantage of the fact that many of the spectra used here extend well beyond the 6200~\AA\ limit of older analyses. 
For Si, we included six windows in the range 5660~\AA\ - 7430~\AA\ and for Ba we added the \ion{Ba}{ii} line at 6497~\AA . When possible, we also included Zn, Zr, and Eu among the elements measured. 
Our Eu measurements are based on the \ion{Eu}{ii} lines at 4435~\AA\ and at 6645~\AA, but not all observed spectra include both lines. Some of our spectra include the [\ion{O}{i}] line at 6300~\AA\ but the line is very weak even in the spectra of metal-rich \acp{gc} like 47~Tuc and it is often contaminated by telluric O$_2$ and H$_2$O absorption and/or residuals from the corresponding [\ion{O}{i}] night sky line. We therefore did not attempt to measure oxygen.

\subsection{The line list}
\label{sec:linelist}

Previous papers based on the analysis technique used here employed the atomic line list of \citet[][hereafter CH2004]{Castelli2004}, with a few minor modifications, as input for the spectral synthesis. That line list is itself a modified version of an older version of the Kurucz line list (see \citetalias{Castelli2004} for details). 
However, it was clear from a comparison with high-resolution spectra of Arcturus and the Sun that not all lines are well reproduced in model spectra computed with the \citetalias{Castelli2004} list \citep{Larsen2012a}. 
This is a common occurrence when using standard line lists to model observed spectra in detail, owing to the fact that atomic data remain uncertain for many transitions that are detectable even in the spectra of solar-type stars \citep{Jofre2019}. 
One (partial) solution is to derive `astrophysical' oscillator strengths ($\log gf$ values) by requiring that the lines in a model spectrum match those in observations \citep{Shetrone2015,Boeche2016,Laverick2019}. Some limitations of this approach are that the line data are then tied to a chosen abundance scale, to the physics of underlying stellar atmosphere models, and to the details of the analysis method, such as inclusion (or not) of \ac{nlte} effects, and that blended lines can be difficult to treat. 
For this work we opted to critically evaluate the input line list, while still relying as much as possible on existing sources for the atomic data.

Having defined the spectral windows, we proceeded to adjust the input line list via a visual inspection of the fits to the solar and Arcturus spectra. We used the 8 Oct 2017 Kurucz list of atomic transitions as a starting point, and whenever a poor match between the observed and synthetic spectra was found, the $\log gf$ value in the Kurucz list was compared with the values in the \citetalias{Castelli2004} and VALD \citep{Piskunov1995,Kupka1999} lists to see if these gave a better fit. In a few cases, the NIST database \citep{NIST} was also consulted. While it was frequently possible to obtain clear improvements to the fits in this way, no single compilation of line data was found to be satisfactory for all lines. In some cases, the data in all three lists were found to be unsatisfactory, and we resorted to adjusting the $\log gf$ values by hand or removing lines altogether. In total, the $\log gf$ values for some 735 atomic lines were modified (counting lines with hyperfine structure only once), with the \citetalias{Castelli2004} values being preferred for 274 lines, the VALD values for 105 lines, and VALD and \citetalias{Castelli2004} listing identical (preferred) values for 84 lines. For most of the remaining 272 modified entries, the $\log gf$ values were manually adjusted. 
When updating the line data, an additional criterion was to minimise the scatter between abundance determinations for a given element in different windows. 
The median absolute change in the $\log gf$ values was 0.46~dex and for 15\% of the lines the change was greater than 1~dex.  About 4\% of the modified lines were changed by more than 2~dex. The lines we have adjusted represent only a very small fraction of all lines in the Kurucz list, which contains more than $300\, 000$ atomic transitions between 4200~\AA\ and 6200~\AA\ and another $>130\, 000$ between 6200~\AA\ and 7500~\AA. However, most of these are far too weak to be detectable in our spectra. While many of the modified lines are not among those actually measured in a specific spectral window, they may still influence the fits through blending or by biasing the overall scaling of the continuum levels, and we therefore tried to get good fits for as many lines as possible.

The Kurucz line list includes hyperfine splitting for many species
(\ion{Na}{i}, \ion{Al}{i}, \ion{Al}{ii}, \ion{K}{i}, \ion{Sc}{i}, \ion{Sc}{ii}, \ion{V}{i}, \ion{Mn}{i}, \ion{Mn}{ii}, 
\ion{Co}{i}, \ion{Ni}{ii}, \ion{Cu}{i}, \ion{Y}{i}, \ion{Y}{ii}, \ion{Nb}{i}, \ion{Nb}{ii}, \ion{Ba}{i}, \ion{Ba}{ii},
\ion{La}{ii}, and \ion{Eu}{ii}). Since the relative strengths of the hyperfine components are specified in a separate column in the data file, it was straight forward to adjust the oscillator strengths for all components of a given line.

For the lines of \ion{Mg}{i}, we mostly adopted NIST $\log gf$ values,
while for lines of \ion{Si}{i}, \ion{Ti}{i}, and \ion{Fe}{i} the values from the \citetalias{Castelli2004} list were frequently found to give the best results. 
The \ion{Mg}{i} lines at 4351.906~\AA\ and 4354.528~\AA\ are affected by blending with CH molecular lines and are not generally used in our analysis, but we have verified that the results are not very sensitive to inclusion or not of these lines. 
For \ion{Ca}{i}, the most consistent results were typically obtained when using the $\log gf$ values in the VALD database, which come mostly from \citet{Smith1981}. However, for some \ion{Ca}{i} lines we kept the $\log gf$ values in the Kurucz list, some of which date back to \citet{Wiese1969}. Damping coefficients describing line broadening caused by elastic collisions between ions and hydrogen were adopted from \citet{Barklem2000} for some of the stronger lines (\ion{Mg}{i} $b$, many of the \ion{Ca}{i} lines, and the \ion{Ba}{ii} lines). 
For a few lines, mostly from \ion{Sc}{ii} and \ion{Zr}{i}, the wavelengths in the Kurucz line list were found to be off by small amounts (20-40 m\AA) and we adopted wavelengths from VALD or \citetalias{Castelli2004} to match the positions of these lines in the spectra of Arcturus and the Sun. 

At first, the \ion{Zr}{i} lines were found to be systematically too strong in the model spectra computed with \texttt{SYNTHE}. We were unable to attribute this to problems with the oscillator strengths, and found that models computed with \texttt{Turbospectrum} matched the Arcturus spectrum well for these lines. The difference was traced to different ionisation potentials for \ion{Zr}{i} used in the two codes. In \texttt{SYNTHE}, an ionisation potential of 6.840 eV was hard-coded for this species \citep[from][]{Drawin1965}, whereas \texttt{Turbospectrum} instead uses a value of 6.634 eV which agrees with more recent determinations \citep{Liu2019}. We updated the \ion{Zr}{i} ionisation potential and the partition function for \ion{Zr}{i} in \texttt{SYNTHE}, using the same polynomial fitting functions as in \texttt{Turbospectrum} \citep{Irwin1981} for the partition function. With these modifications, the \ion{Zr}{i} lines in the \texttt{SYNTHE} spectrum were found to match those computed with \texttt{Turbospectrum}. 
We also compared the ionisation potentials for other ions, and found any remaining differences between the values used in  \texttt{Turbospectrum} and \texttt{SYNTHE} to be negligible. 

In some cases it was not possible to get a good fit despite our best efforts - typically because a line was present in the observed spectra but not in the synthetic ones, or in cases where complex blends made it difficult or impossible to unambiguously determine the correct $\log gf$ values for the individual lines. In such cases, the affected spectral regions were marked and masked out in the analysis. 

\begin{figure}
\centering
\includegraphics[width=\columnwidth]{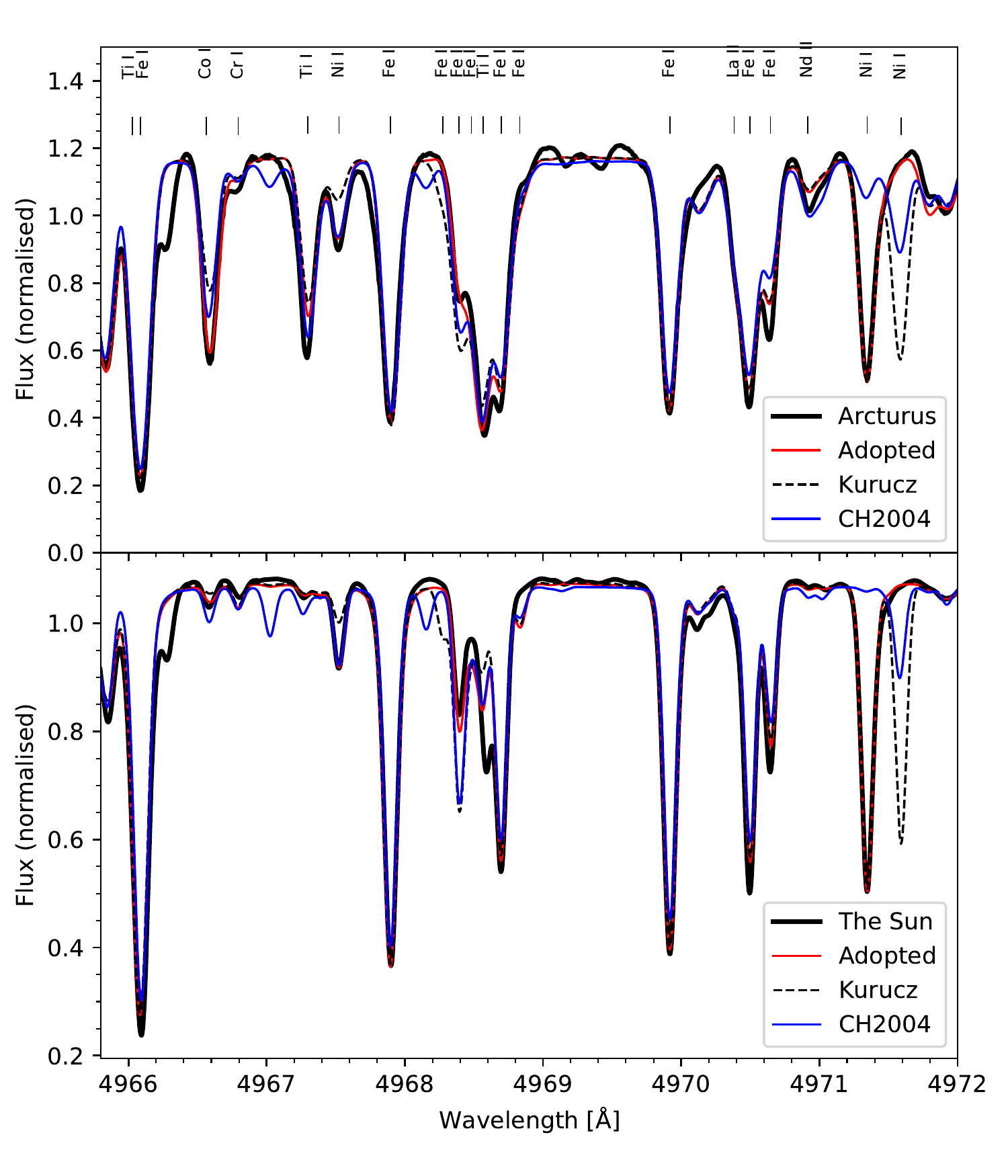}
\caption{\label{fig:Fe496e}Fits to the spectra of Arcturus (top) and the Sun (bottom). Model spectra were computed using the standard Kurucz line list, the \citet{Castelli2004} list, and our adopted version of the Kurucz list, as indicated in the legend.}
\end{figure}

For molecular lines we mostly used the data available at the Kurucz website.  This includes data for the following molecules: H$_2$, NH, OH, NaH, MgH, AlH, SiH, CaH, TiH, CrH, FeH, C$_2$, CN, CO, AlO, SiO, CaO, TiO, VO, and H$_2$O. For the CN line list, an error was detected in the conversion of the $f$-values in the original data \citep{Brooke2014} to the $\log gf$ values in the line list at the Kurucz website. We therefore updated the CN line data accordingly. For CH we used the line list from \citet{Masseron2014}, which is available from the website of B.\ Plez\footnote{\url{https://www.lupm.in2p3.fr/users/plez/}}.  
Nevertheless, these updates are of relatively minor consequence for this work, since the CN and CH lines mostly affect the spectra at wavelengths $<4400$~\AA. 
In principle, the CN and CH features in the wavelength range 4100~\AA --4400~\AA\ can be used to constrain the
abundances of C and N, although the N abundances are better constrained by the stronger CN band near 3800~\AA --3900~\AA, especially at lower metallicities \citep{Cohen2002,Graves2008,Lardo2012,Schiavon2013,Martocchia2021}. However, the interpretation of C and N abundances in integrated-light spectra is complicated by stellar evolutionary effects (mixing) along the \ac{rgb} \citep{Gratton2000,Martell2008}, and we do not here quote abundances of these elements. 

To illustrate the procedure by which the line list was customised, 
Fig.~\ref{fig:Fe496e} shows model fits to a small region of the spectra of Arcturus and the Sun. The fits are shown for the Kurucz line list and the \citetalias{Castelli2004} list, as well as for our final adopted version of the Kurucz list. These fits are fairly typical and reveal significant differences between the observed and synthetic spectra for both the \citetalias{Castelli2004} and Kurucz lists. 
Within this region, spanning only a few \AA , we modified the oscillator strengths for several lines, of which we discuss a few representative examples. 
The most striking mismatch is for the \ion{Ni}{i} line at 4971.591~\AA, which is much too strong when using the data from the Kurucz list.
For this line the Kurucz list has $\log gf = -0.753$, which is the same value listed in the VALD database. In the \citetalias{Castelli2004} list the line has $\log gf = -1.566$, but this still makes the line much too strong in the synthetic spectra. This line was removed altogether from the line list. 
The neighbouring \ion{Ni}{i} line at 4971.34~\AA\ is well matched when using the Kurucz list, but much too weak in the spectrum computed with the \citetalias{Castelli2004} list. 
A more typical example is the \ion{Ni}{i} line at 4967.524~\AA, which is too weak when using the Kurucz list. For this line, the Kurucz list has a lower $\log gf$ value ($-1.989$) than VALD and Castelli ($\log gf = -1.570$), and using the latter value clearly gives a better fit. 
Sorting out the blend between 4968~\AA\ and 4969~\AA\ was more complicated. The \ion{Fe}{i} line at 4968.276~\AA\ was found to be too strong in both the Kurucz and VALD lists ($\log gf = -2.043$), but using the much lower $\log gf$ value from  \citetalias{Castelli2004} ($\log gf = -3.653$) gave a satisfactory fit.  The \ion{Fe}{i} line at 4968.392~\AA\ remained too strong in all three line lists, and the $\log gf$ value was manually adjusted downward from the value in the Kurucz list ($-1.409$) to $\log gf = -1.8$. For the \ion{Ti}{i} line at 4968.567~\AA, the Castelli value ($\log gf = -0.44$) was used instead of the value common to the Kurucz and VALD lists ($\log gf = -0.64$). Nevertheless, the fit remains somewhat unsatisfactory for this rather complex blend, and situations like this could not always be completely resolved. 
The weak feature near 4971.0~\AA\ is a blend that includes the \ion{Nd}{ii} line marked in the figure (with an equivalent width of about 10 m\AA\ in the Arcturus spectrum), as well as various other features (\ion{Co}{i}, \ion{V}{i}). The \citetalias{Castelli2004} data appear to give a better fit to the Arcturus spectrum, but overpredict the strength of the lines in the solar spectrum, and in this case we did not modify the Kurucz data.

The final line list was converted from the format used by \texttt{SYNTHE} to that used by \texttt{Turbospectrum} to ensure consistent modelling with the two codes. This was mostly a straight forward procedure, amounting mainly to unit conversions and some reformatting. We left out more than twice ionised species, which are not supported by \texttt{Turbospectrum} and are, in any case, of no relevance in the cool stars modelled with \texttt{Turbospectrum}. We also excluded a few transitions with excitation potentials greater than 100 eV which caused problems for the \texttt{Turbospectrum} format and would introduce no observable lines in any spectra of relevance here.  

\subsection{Single star validation: The Sun and Arcturus}

As a first verification of the analysis procedure, abundances were determined for the Sun and Arcturus and compared with literature results for these well-studied stars. We discuss each in turn below.

\subsubsection{The Sun}
\label{sec:sun}

\begin{table*}
\caption{Solar analysis.}
\label{tab:sun}
\centering
\begin{tabular}{l c c r c c c}
\hline\hline
 Elem. & \multicolumn{2}{c}{12+$\log \epsilon$} & $N$ & LTE/NLTE & GS1998 & A2009 \\
 & $v_t=0.85$~km~s$^{-1}$ & $v_t=1.0$~km~s$^{-1}$ \\
\hline
%Fe & 7.489 (0.068) & 7.464 (0.068) & 40 & NLTE & $7.50\pm0.05$ & $7.50\pm0.04$\\
%Na & 6.226 (0.025) & 6.215 (0.018) &  2 & NLTE & $6.33\pm0.03$ & $6.24\pm0.04$\\
%Mg & 7.542 (0.053) & 7.543 (0.050) &  4 & NLTE & $7.58\pm0.05$ & $7.60\pm0.04$\\
%Si & 7.554 (0.064) & 7.548 (0.063) &  6 &  LTE & $7.55\pm0.05$ & $7.51\pm0.03$\\
%Ca & 6.388 (0.063) & 6.358 (0.062) &  9 & NLTE & $6.36\pm0.02$ & $6.34\pm0.04$\\
%Ti & 5.051 (0.059) & 5.024 (0.062) & 14 & NLTE & $5.02\pm0.06$ & $4.95\pm0.05$\\
%Sc & 3.072 (0.122) & 3.051 (0.103) &  5 &  LTE & $3.17\pm0.10$ & $3.15\pm0.04$\\
%Cr & 5.606 (0.082) & 5.585 (0.077) & 18 &  LTE & $5.67\pm0.03$ & $5.64\pm0.04$\\
%Mn & 5.541 (0.015) & 5.503 (0.002) &  2 & NLTE & $5.39\pm0.03$ & $5.43\pm0.04$\\
%Ni & 6.277 (0.176) & 6.242 (0.173) & 14 & NLTE & $6.25\pm0.04$ & $6.22\pm0.04$\\
%Cu & 4.215 (0.004) & 4.193 (0.014) &  2 &  LTE & $4.21\pm0.04$ & $4.19\pm0.04$\\
%Zn & 4.591 (0.001) & 4.525 (0.001) &  2 &  LTE & $4.60\pm0.08$ & $4.56\pm0.05$\\
%Zr & 2.415 (0.000) & 2.418 (0.000) &  1 &  LTE & $2.60\pm0.02$ & $2.58\pm0.04$\\
%Ba & 2.123 (0.078) & 2.068 (0.074) &  5 & NLTE & $2.13\pm0.05$ & $2.18\pm0.09$\\
%Eu & 0.186 (0.000) & 0.215 (0.000) &  1 &  LTE & $0.51\pm0.08$ & $0.52\pm0.04$\\
Fe & 7.500 (0.072) & 7.475 (0.073) & 40 & NLTE & $7.50\pm0.05$ & $7.50\pm0.04$\\
Na & 6.240 (0.029) & 6.229 (0.022) &  2 & NLTE & $6.33\pm0.03$ & $6.24\pm0.04$\\
Mg & 7.552 (0.060) & 7.553 (0.057) &  4 & NLTE & $7.58\pm0.05$ & $7.60\pm0.04$\\
Si & 7.554 (0.064) & 7.548 (0.063) &  6 &  LTE & $7.55\pm0.05$ & $7.51\pm0.03$\\
Ca & 6.381 (0.064) & 6.350 (0.064) &  9 & NLTE & $6.36\pm0.02$ & $6.34\pm0.04$\\
Ti & 5.043 (0.072) & 5.017 (0.076) & 14 & NLTE & $5.02\pm0.06$ & $4.95\pm0.05$\\
Sc & 3.072 (0.122) & 3.051 (0.103) &  5 &  LTE & $3.17\pm0.10$ & $3.15\pm0.04$\\
Cr & 5.606 (0.082) & 5.585 (0.077) & 18 &  LTE & $5.67\pm0.03$ & $5.64\pm0.04$\\
Mn & 5.451 (0.009) & 5.414 (0.008) &  2 & NLTE & $5.39\pm0.03$ & $5.43\pm0.04$\\
Ni & 6.204 (0.186) & 6.169 (0.183) & 14 & NLTE & $6.25\pm0.04$ & $6.22\pm0.04$\\
Cu & 4.215 (0.004) & 4.193 (0.014) &  2 &  LTE & $4.21\pm0.04$ & $4.19\pm0.04$\\
Zn & 4.591 (0.001) & 4.525 (0.001) &  2 &  LTE & $4.60\pm0.08$ & $4.56\pm0.05$\\
Zr & 2.415 (0.000) & 2.418 (0.000) &  1 &  LTE & $2.60\pm0.02$ & $2.58\pm0.04$\\
Ba & 2.119 (0.086) & 2.064 (0.081) &  5 & NLTE & $2.13\pm0.05$ & $2.18\pm0.09$\\
Eu & 0.186 (0.000) & 0.215 (0.000) &  1 &  LTE & $0.51\pm0.08$ & $0.52\pm0.04$\\
\hline
\end{tabular}
\tablefoot{Solar abundances were measured using \texttt{ATLAS12}/\texttt{SYNTHE}, assuming $T_\mathrm{eff}=5777$~K and $\log g = 4.44$ \citep{Cox2000}. Numbers in parentheses indicate the rms scatter of the individual measurements. 
}
\end{table*}

The solar spectrum was analysed using an \texttt{ATLAS12} atmosphere and \texttt{SYNTHE}. One remaining parameter to fix is the microturbulence velocity, $v_t$, assumed for the spectral synthesis \citep[e.g.][]{Jefferies1968}. For solar-type stars, typical microturbulence velocities used for classical abundance analysis are $v_t\simeq1$~km~s$^{-1}$.  Recent examples include $v_t=0.85$~km~s$^{-1}$ \citep{Valenti2005,Yong2005,Brewer2015},   0.93~km~s$^{-1}$ \citep{Fulbright2006}, 
0.75~km~s$^{-1}$ \citep{Pavlenko2012}, and $1.10$~km~s$^{-1}$ \citep{Laverick2019}. From a comparison with 3D models, \citet{Dutra-Ferreira2016} found microturbulence velocities of $\sim1$~km~s$^{-1}$ for dwarf stars with their Equation~(2) yielding $v_t = 0.97$~km~s$^{-1}$ for the solar $T_\mathrm{eff}$ and $\log g$. We analysed the solar spectrum using two values of the microturbulence, $v_t=0.85$~km~s$^{-1}$ and $v_t=1.0$~km~s$^{-1}$.

Our average abundance measurements for the Sun are listed in Table~\ref{tab:sun} together with the standard solar compositions of \citet[][hereafter GS1998]{Grevesse1998} and \citet[][A2009]{Asplund2009}. The abundances are normalised to a logarithmic hydrogen abundance of $12+\log \epsilon(\mathrm{H}) = 12$. The numbers in parentheses are the rms dispersions of the individual measurements for each element and $N$ are the numbers of measurements (spectral windows) per element. We list the \ac{nlte} abundances for elements for which these are available and \ac{lte} abundances otherwise. 
Due to the extremely high signal-to-noise ratio of the Kurucz solar spectrum, the formal errors on the fits are negligibly small and the scatter between the abundance measurements in different spectral windows for a given element  is almost entirely due to systematics.  We therefore computed the abundances in the table as a straight average of the individual measurements. 
For most elements, the scatter is less than about 0.1~dex, although a somewhat larger scatter is found for nickel.
As expected, using the larger $v_t$ value reduces the effect of line saturation, which in turn leads to a slight decrease (0.02--0.04~dex) in the abundances of most elements. 

We briefly comment on a few individual elements. Iron is often used as a proxy for metallicity, and when presenting our results for the \acp{gc} we generally follow the usual convention and give the abundances of other elements as $\mathrm{[X/Fe]}$. \citetalias{Grevesse1998} and \citetalias{Asplund2009} both list the same iron abundance of $12+\log \epsilon(\mathrm{Fe}) = 7.50$ for the Sun with uncertainties of 0.04--0.05~dex. Our \ac{nlte} measurement of the solar iron abundance for $v_t=0.85$~km~s$^{-1}$ matches this value exactly. For $v_t=1.0$~km~s$^{-1}$ we find a slightly lower value of $12+\log \epsilon(\mathrm{Fe}) = 7.48$, which is still well within the uncertainties on the reference values. 
For two elements, Zr and Eu, the differences with respect to \citetalias{Grevesse1998} and \citetalias{Asplund2009} are relatively large (0.2-0.3~dex), although \ac{nlte} corrections are not included for these elements. However, our choices of spectral features are not optimised for measuring these elements in the solar spectrum. The blue \ion{Eu}{ii} line at 4435.6~\AA, with an equivalent width of about 25~m\AA, is blended with a much stronger \ion{Ca}{i} line (equivalent width $\sim170$~m\AA) at 4435.7~\AA, and the derived Eu abundance is therefore sensitive to uncertainties in the Ca abundance and to the details of the spectral synthesis, such as the inclusion of velocity-dependent van der Waals broadening constants for the Ca line \citep{Anstee1995}. 
From varying the Ca abundance, we found that a decrease of just 0.06~dex in $\log\epsilon$(Ca) would increase $\log \epsilon$(Eu) to match the reference values. 
The red \ion{Eu}{ii} line, centred at 6645.10~\AA, is quite weak in the solar spectrum (4~m\AA) and is
blended with an \ion{Al}{i} line at 6645.14~\AA\ that has an equivalent width of 13~m\AA . We could not get a reliable estimate of the Eu abundance from this line for the Sun. 
In cool giants the relative strengths of the lines in these blends change in favour of the Eu lines, but it is nevertheless clear that the measurements of Eu must be considered somewhat uncertain. 

The \ion{Zr}{i} lines in the window 6124~\AA --6147~\AA\ are also very weak in the solar spectrum (the equivalent widths are $<$  3 m\AA), and some are blended with stronger lines. In particular, the \ion{Zr}{i} line at 6124.9~\AA\ is blended with a much stronger \ion{Si}{i} line at 6125.0~\AA . There are other lines that would be more suitable for measuring Zr in the solar spectrum, but these are mostly located in the blue ($\lambda<4400$ nm) and are less useful for analysis of \ac{gc} spectra due to blending with molecular and atomic features. 

Excluding Eu and Zr, our measurements agree well with the reference scales overall. For the abundances that include \ac{nlte} corrections, the mean offsets with respect to \citetalias{Grevesse1998} are $\langle \log_{10} \epsilon - \log_{10} \epsilon_\mathrm{GS98} \rangle = -0.01$~dex (for $v_t=0.85$~km~s$^{-1}$) and $-0.04$~dex ($v_t=1.0$~km~s$^{-1}$). Comparing with \citetalias{Asplund2009}, the offsets are instead
$\langle \log_{10} \epsilon - \log_{10} \epsilon_\mathrm{A09} \rangle = +0.00$~dex (for $v_t=0.85$~km~s$^{-1}$) and $-0.02$~dex ($v_t=1.0$~km~s$^{-1}$). 
If we additionally include the elements measured in \ac{lte}, the mean offsets are $-0.02$~dex and $-0.05$~dex for \citetalias{Grevesse1998} for the two $v_t$ values, respectively, and
$0.00$~dex and $-0.03$~dex for \citetalias{Asplund2009}. There is thus no strong preference for either $v_t$ value, but we adopt $v_t=0.85$~km~s$^{-1}$ as the preferred value here as it reproduces the iron reference abundances more closely. We note, however, that the $v_t=1.0$~km~s$^{-1}$ measurements tend to give a slightly smaller rms scatter for most elements. 

At any rate, it is unsurprising that our analysis does not exactly reproduce either of the two standard abundance scales for every element. Our analysis technique is not optimised for the solar spectrum and the \citetalias{Asplund2009} abundance scale, in particular, is based on a much more sophisticated analysis that employs 3D \ac{nlte} hydrodynamical model calculations for many elements. 

In the remainder of this paper, we quote abundances relative to the scale of \citetalias{Grevesse1998} for consistency with previous papers based on our technique. Readers who prefer to convert our measurements to the scale of \citetalias{Asplund2009}, or to adopt a differential comparison with respect to our solar abundance measurements, can do so using the information in Table~\ref{tab:sun}. The caveat should, however, be kept in mind that the spectral windows typically have different weights in the analysis of the \ac{gc} spectra (depending on the uncertainty on each measurement; Sect.~\ref{sec:47tuc}) compared to the uniform weights used for our solar analysis in Table~\ref{tab:sun}. 

\subsubsection{Arcturus}
\label{sec:arcturus}

\begin{table*}
\caption{Arcturus LTE analysis.}
\label{tab:arcturus}
\centering
\begin{adjustbox}{width=1\textwidth}
\begin{tabular}{l c c c c c c c c c c}
\hline\hline
 & \multicolumn{2}{c}{ATLAS12/SYNTHE} & A9/S & A12/T & M/T & $\Delta_\mathrm{NLTE}$ & \multicolumn{4}{c}{Literature studies} \\
 & $v_t=1.5$~km~s$^{-1}$ & $v_t=1.74$~km~s$^{-1}$ & & & & & Y2005  & W2009 & RA2011 & vdS2013 \\
\hline
$\mathrm{[Fe/H]}$ & $-0.580$ (0.104, 40) & $-0.675$ & $-0.578$ & $-0.591$ & $-0.619$ & $+0.009$ & $-0.56$ (0.13, 50) &  $-0.59$ (0.12, 40) & $-0.52$ & $-0.71$ \\
$\mathrm{[Na/Fe]}$ & $+0.191$ (0.070, 2) & $+0.212$  & $+0.201$ & $+0.209$ & $+0.221$ & $-0.192$ & $+0.15$ (0.08, 3) &  $+0.15$ (0.04, 2) & $+0.11\pm0.03$ & $+0.10\pm0.04$ \\
$\mathrm{[Mg/Fe]}$ & $+0.325$ (0.036, 4) & $+0.387$ & $+0.331$ & $+0.310$ & $+0.337$ & $-0.022$ & $+0.45$ (0.14, 4) &  $+0.34$ (0.15, 8) & $+0.37\pm0.03$ & $+0.33\pm0.06$ \\
$\mathrm{[Si/Fe]}$ & $+0.340$ (0.079, 6) & $+0.382$  & $+0.329$ & $+0.395$ & $+0.398$ & \ldots & $+0.35$ (0.06, 5) &  $+0.24$ (0.14, 10) & $+0.33\pm0.04$ & $+0.31\pm0.04$ \\
$\mathrm{[Ca/Fe]}$ & $+0.177$ (0.073, 10) & $+0.124$  & $+0.193$ & $+0.179$ & $+0.203$ & $-0.057$ & $+0.22$ (0.09, 4) &  $+0.19$ (0.06, 12) & $+0.11\pm0.04$ & $+0.03\pm0.04$ \\
$\mathrm{[Ti/Fe]}$ & $+0.235$  (0.052, 14) & $+0.239$  & $+0.250$ & $+0.244$ & $+0.241$ & $+0.162$ & $+0.26$ (0.03, 4) &  $+0.34$ (0.15, 29) & $+0.24\pm0.04$ & $+0.33\pm0.05$ \\
$\mathrm{[Sc/Fe]}$ & $+0.172$ (0.218, 5) & $+0.147$  & $+0.180$ & $+0.185$ & $+0.171$ & \ldots & \ldots & $+0.24$ (0.01, 2) & $+0.21\pm0.04$ & $+0.25\pm0.04$ \\
$\mathrm{[Cr/Fe]}$ & $-0.074$ (0.129, 18) & $-0.069$  & $-0.064$ & $-0.057$ & $-0.053$ & \ldots & \ldots & \ldots & $-0.05\pm0.04$ & $-0.06\pm0.06$ \\
$\mathrm{[Mn/Fe]}$ & $-0.240$ (0.093, 2) & $-0.269$  & $-0.232$ & $-0.202$ & $-0.209$ & $+0.097$ & $-0.25$ (0.06, 3) & \ldots & \ldots & \ldots \\
$\mathrm{[Ni/Fe]}$ & $-0.078$ (0.230, 14) & $-0.089$  & $-0.079$ & $-0.072$ & $-0.080$ & $+0.015$ & $-0.02$ (0.06, 7) &  \ldots & $+0.06\pm0.03$ & $+0.07\pm0.04$ \\
$\mathrm{[Cu/Fe]}$ & $+0.236$ (0.027, 2) & $+0.169$ & $+0.242$ & $+0.205$ & $+0.183$ & \ldots & \ldots & \ldots & \ldots & $-0.03$ \\
$\mathrm{[Zn/Fe]}$ & $+0.033$ (0.016, 2) & $-0.008$  & $+0.026$ & $+0.055$ & $+0.048$ & \ldots & \ldots & $-0.04$ (0.09, 2) & $+0.22\pm0.06$ & \ldots \\
$\mathrm{[Zr/Fe]}$ & $-0.045$ (-, 1) & $+0.016$  & $-0.017$ & $+0.000$ & $-0.021$ & \ldots & $-0.27$ (0.08, 3) & $+0.03$ (0.08, 10) & \ldots & $-0.07\pm0.03$ \\
$\mathrm{[Ba/Fe]}$ & $-0.015$ (0.057, 5) & $-0.099$  & $-0.009$ & $+0.069$ & $+0.064$ & $-0.018$ & $-0.09$ (1) & $-0.19$ (0.08, 2) & \ldots & $-0.19\pm0.03$ \\
$\mathrm{[Eu/Fe]}$ & $+0.300$ (0.062, 2) & $+0.364$  & $+0.298$ & $+0.304$ & $+0.278$ & \ldots & $+0.29$ (1) & $+0.36$ (0.04, 2) & \ldots & $+0.40\pm0.02$ \\
\hline
\end{tabular}
\end{adjustbox}
\tablefoot{\ac{lte} abundances for Arcturus were measured assuming $T_\mathrm{eff}=4286$~K and $\log g = 1.66$. The results obtained with \texttt{ATLAS12}/\texttt{SYNTHE} are given for two values of the microturbulence. For other model atmospheres and spectral synthesis combinations the results are given for $v_t=1.5$~km~s$^{-1}$. For W2009 we list the results from their analysis using spherically symmetric atmospheres. 
}
\end{table*}

As an \ac{rgb} star, the spectrum of Arcturus resembles that of a \ac{gc} much more closely than does the solar spectrum. As such, the Arcturus spectrum provides a better test of the suitability of our line list (and of our analysis technique in general) for analysis of GC spectra. 
The drawback is that the composition of Arcturus is not as well established as that of the Sun.
Nevertheless,
the distance and diameter of Arcturus are well constrained by parallax and interferometric measurements, and consequently other physical parameters are also well determined. 
The effective temperature and surface gravity adopted here (Sect.~\ref{sec:specwin}) are very similar to those used in other studies \citep[e.g.][]{Fulbright2006,Worley2009}.
For the microturbulence, values quoted in the literature range from $v_t=1.2$~km~s$^{-1}$ (from analysis of infrared lines, \citealt{Kondo2019}) to 1.8--1.9~km~s$^{-1}$ \citep[vdS2013]{vanderSwaelmen2013}, with other studies finding intermediate values of 1.50~km~s$^{-1}$ \citep[W2009]{Worley2009}, 1.56~km~s$^{-1}$ \citep[Y2005]{Yong2005}, 1.67~km~s$^{-1}$ \citep{Fulbright2006}, and $v_t=1.74$~km~s$^{-1}$
\citep[RA2011]{Ramirez2011} 

In Table~\ref{tab:arcturus} we list our abundance measurements for Arcturus obtained with \texttt{ATLAS12}/\texttt{SYNTHE} for two values of the microturbulence, $v_t=1.50$~km~s$^{-1}$ and 1.74~km~s$^{-1}$. To assess the sensitivity of the measurements to the choice of model atmospheres and spectral synthesis codes, we also include results obtained with \texttt{ATLAS9}/\texttt{SYNTHE} (A9/S), with \texttt{ATLAS12}/\texttt{Turbospectrum} (A12/T), and with spherical \texttt{MARCS} models and \texttt{Turbospectrum} (M/T). These latter results are given for a single value of the microturbulence, $v_t=1.50$~km~s$^{-1}$. 
We also include abundance measurements from previous studies for comparison.
The literature results are listed as given in the respective papers with no attempt to homogenise the reference abundance scales, oscillator strengths, line lists, or other parameters that can cause systematic offsets in the results.  We comment on some of these issues below. 
To facilitate easier comparison with the literature results (none of which accounts for \ac{nlte} effects) our measurements in the table are also given as \ac{lte} values.
The comparison of our results for different details of the analysis would be unaffected by the inclusion of \ac{nlte} corrections. However, we give the \ac{nlte} corrections, $\Delta_\mathrm{NLTE}$, for elements where these have been determined (Sect.~\ref{sec:nlte}).

The choices of model atmospheres and spectral synthesis codes have a relatively minor effect on the results. The iron abundances obtained with \texttt{SYNTHE}  (for $v_t=1.50$~km~s$^{-1}$)
and \texttt{Turbospectrum} differ by only 0.011~dex when using the same (\texttt{ATLAS12}) model atmospheres, and replacing the \texttt{ATLAS12} atmospheres with \texttt{ATLAS9} models leads to an even smaller difference (0.002~dex). 
The iron abundances fall, in these cases ($\mathrm{[Fe/H]}=-0.58$ to $\mathrm{[Fe/H]}=-0.59$), well within the range found in the literature. 
When using \texttt{MARCS} atmospheres instead of \texttt{ATLAS12} (both in combination with \texttt{Turbospectrum}), the iron abundance decreases by 0.03~dex but remains within the literature range. Using $v_t=1.74$~km~s$^{-1}$ for the microturbulence leads to a decrease of about 0.1~dex in $\mathrm{[Fe/H]}$. This lower iron abundance ($\mathrm{[Fe/H]}=-0.68$) appears to be somewhat disfavoured by comparison with the literature values, although \citetalias{vanderSwaelmen2013} found iron abundances between $\mathrm{[Fe/H]}=-0.58$ and $-0.71$ depending on the amount of noise they added to the Arcturus spectrum. The values with which we compare in Table~\ref{tab:arcturus} are for their `$\infty$S/N' analysis, which, for most elements, yields fairly similar results to our analysis with $v_t=1.74$~km~s$^{-1}$. In this sense, the relatively high $\mathrm{[Fe/H]}=-0.52$ from \citetalias{Ramirez2011}, who used the same high $v_t$ value, is more discrepant. 

For most elemental abundance ratios, the analyses based on \texttt{SYNTHE} or \texttt{Turbospectrum} in combination with \texttt{ATLAS12} model atmospheres yield very similar results that agree well with the literature values. For Si and Ba, the analyses based on \texttt{Turbospectrum} yield somewhat higher abundances ratios (by 0.06~dex and 0.08~dex, respectively) than those based on \texttt{SYNTHE}. More typically the differences are 0.01--0.02~dex. 
Again, the results are relatively insensitive to the choice of \texttt{MARCS} versus \texttt{ATLAS12} models, while the choice of microturbulence can lead to differences of $\sim0.05$~dex in the derived abundance ratios. 

Of the literature results listed in Table~\ref{tab:arcturus}, two are differential analyses with respect to the Sun \citepalias{Yong2005,Ramirez2011}.
The study by \citetalias{Worley2009} is an \ac{lte} analysis with abundances given relative to the solar abundance scale of \citet[][L2003]{Lodders2003}, which differs only slightly from the \citetalias{Grevesse1998} scale used in our analysis for most elements.
The largest differences between the two scales are found for Ti ($12+\log \epsilon(\mathrm{Ti}) = 4.92$ on the L2003 scale) and Sc ($12+\log \epsilon(\mathrm{Sc}) = 3.07$), that is, the solar abundances are 0.1~dex lower than on the \citetalias{Grevesse1998} scale for both elements. This likely accounts for the offsets between our abundance determinations for these elements and those of \citetalias{Worley2009}. The analysis of \citetalias{vanderSwaelmen2013} also assumes \ac{lte} and abundances are given relative to the \citetalias{Grevesse1998} scale, as in the present work. 
The literature results are all based on the same high-resolution, high S/N spectrum of Arcturus that we are using here. 

For Na, systematics at the level of $0.1$~dex arise from two sources: first, the solar reference abundance according to \citetalias{Asplund2009} is 0.09~dex lower than the \citetalias{Grevesse1998} value, so that our $\mathrm{[Na/Fe]}$ values would increase by the same amount if given relative to the \citetalias{Asplund2009} scale. Second, inclusion of \ac{nlte} corrections would decrease our $\mathrm{[Na/Fe]}$ value for Arcturus by $0.19$~dex. Our analysis is most directly comparable with those of \citetalias{Worley2009} and \citetalias{vanderSwaelmen2013}, compared with which studies our $\mathrm{[Na/Fe]}$ value for Arcturus is 0.05--0.09~dex higher.  Our \ac{lte} analysis of the solar spectrum recovers the \citetalias{Grevesse1998} Na abundance almost exactly ($12+\log \epsilon (\mathrm{Na}) = 6.33$) so that we may also reasonably compare our measurements for Arcturus on the \citetalias{Grevesse1998} scale with the differential analyses by \citetalias{Yong2005} and \citetalias{Ramirez2011}. Again, our values are slightly higher (0.04--0.08~dex). However, in \ac{lte} we also find a slightly lower iron abundance for the Sun, $12 + \log \epsilon(\mathrm{Fe}) = 7.47$ (for $v_t=0.85$~km~s$^{-1}$). An 
adjustment for the 0.03~dex difference relative to the \citetalias{Grevesse1998} scale would lead to a corresponding increase in the differential $\mathrm{[Fe/H]}$ value for Arcturus, and therefore a decrease in the differential $\mathrm{[Na/Fe]}$ value by the same amount, which would bring our \ac{lte} measurement of [Na/Fe] very close to that of \citetalias{Yong2005}, and within 0.05~dex of that of \citetalias{Ramirez2011}.

For Zr and Eu our measurements for Arcturus fall within the range quoted in the literature, although the literature values for $\mathrm{[Zr/Fe]}$ span a range of 0.3~dex. For Cu, the only other measurement is that of \citetalias{vanderSwaelmen2013}, whose $\mathrm{[Cu/Fe]}$ ratio is about 0.26~dex lower than ours.  The Cu abundances obtained from our measurements of the two \ion{Cu}{i} lines (at 5106~\AA\ and 5782~\AA) agree quite well for both the Sun (within 0.01~dex) and Arcturus (0.05~dex). However, the \ion{Cu}{i} line at 5782~\AA\ may be contaminated by the diffuse interstellar band (DIB) near 5780~\AA\ \citep{Herbig1975}, and it is therefore omitted from our analysis of the \ac{gc} spectra. 

Overall, we conclude that our abundance measurements for Arcturus are in satisfactory agreement with literature data. The literature values themselves often differ at the level of $\sim0.1$~dex, and for most elements our measurements fall close to or within the range of literature values. Nevertheless, the fact that differences at the level of $\sim0.1$~dex do exist even for a very well-studied star such as Arcturus should be kept in mind later on when we compare our integrated-light abundance measurements for \acp{gc} with other literature data. 

\subsection{Modelling of simple stellar populations}
\label{sec:sspmod}

We based the modelling of the integrated light of stellar clusters on theoretical DSEP (Dartmouth Stellar Evolution Program) isochrones \citep{Dotter2007}. For our purpose, these have the advantage of being available for various compositions (scaled-solar as well as various levels of $\alpha$-enhancement), for any metallicity in the range $-2.5 < \mathrm{[Fe/H]} < +0.5$ (via a web-based interpolation engine), and for ages between 1 Gyr and 15 Gyr. A limitation of the DSEP isochrones is that they only cover stellar evolutionary phases up to the tip of the \ac{rgb}, and we therefore combined them with empirical \ac{hb} data from the Advanced Camera for Surveys (ACS) survey of Galactic Globular Clusters \citep[ACSGCS; ][]{Sarajedini2007}. The empirical \ac{hb} data were binned into typically about ten \ac{hrd}-boxes, for which temperatures and luminosities were derived from the ACSGCS photometry using colour-$T_\mathrm{eff}$ relations and bolometric corrections from the \citet{Castelli2003} model grid. Surface gravities were computed assuming a mass of 0.8~$M_\odot$ for the \ac{hb} stars.  Weights were assigned to each \ac{hb} \ac{hrd}-box by applying a scaling to the observed numbers of stars, based on the number of \ac{rgb} stars in the range $+1 < M_V < +2$ in the observed and isochrone-based \acp{hrd}. 

After the analysis was nearly complete, new isochrones for $\alpha$-enhanced composition were published on the BaSTI website, potentially eliminating the need to combine the theoretical isochrones with empirical \ac{hb} (and \ac{agb}) data as these phases are included in the BaSTI isochrones \citep{Hidalgo2018,Pietrinferni2021}. We repeated the analysis using the BaSTI isochrones and found the results to be very similar to those based on the DSEP isochrones and empirical \acp{hb} (see Sect.~\ref{sec:47tuc}). We kept the DSEP isochrones as the main basis for our analysis.

To assign microturbulence velocities ($v_t$) to each HRD-box, we assumed that $v_t$ can be expressed as a linear function of the logarithmic surface gravity, $\log g$ \citep{McWilliam2008,Colucci2009,Larsen2012a,Sakari2013}.
We used the Sun and Arcturus as anchor points, assuming $v_t=0.85$~km~s$^{-1}$ and 1.50~km~s$^{-1}$ for these two stars, respectively (Sects.~\ref{sec:sun} and \ref{sec:arcturus}). The two points at 
$(\log g, v_t) = (4.44, 0.85$~km~s$^{-1}$) 
and
(1.66, 1.50~km~s$^{-1}$)  
then define the following relation:
\begin{equation}
v_t = (1.88 - 0.23 \log g)\, \mathrm{km} \, \mathrm{s}^{-1} .
\label{eq:vt}
\end{equation}
For \ac{hb} stars we assume $v_t = 1.8$~km~s$^{-1}$ \citep{Pilachowski1996}.
While a parameterisation of $v_t$ in terms of only $\log g$ is probably an oversimplification and other prescriptions have been proposed (e.g.\ in terms of [Fe/H], $T_\mathrm{eff}$, and $\log g$; \citealt{Mashonkina2017a}), we note that a very similar relation was found by \citet{Roederer2014} for metal-poor stars ($v_t = (1.88 -0.20 \log g) \, \mathrm{km} \, \mathrm{s}^{-1} $). The new relation (\ref{eq:vt}) differs slightly from that used in previous papers, in which the reference points were $(\log g, v_t)$ = (1.0, 2.0~km~s$^{-1}$) and (4.0, 1.0~km~s$^{-1}$), which implies $v_t = 2.33 - \frac{1}{3} \log g$ \citep{Larsen2012a}. For the Sun, this gives the same microturbulence velocity, $v_t = 0.85$~km~s$^{-1}$, while a somewhat larger value results for Arcturus ($v_t = 1.78$~km~s$^{-1}$) and for giants in general. 

To model the contribution to the integrated light from stars at different locations along an isochrone, an assumption must also be made about the mass function (MF). A common choice is the segmented power-law  proposed by \citet{Kroupa2001}, 
\begin{equation}
\frac{\mathrm{d}N}{\mathrm{d}M} \propto (M / 0.5 M_\odot)^{\alpha}
\end{equation}
with $\alpha=-2.3$ for $M>0.5 \, M_\odot$ and $\alpha=-1.3$ for $M<0.5 \, M_\odot$.
However, the MFs in \acp{gc} often have substantially shallower slopes, probably as a consequence of dynamical evolution.  \citet{Sollima2017} found that the MFs of \acp{gc} can, in many cases, be approximated by single power-laws over the mass range $0.2 < M/M_\odot < 0.8$, with slopes varying between $\alpha\approx0$ and $\alpha\approx-1.5$. 

We approximated the MF as a power-law with an intermediate slope, $\mathrm{d}N/\mathrm{d}M \propto M^{-1}$, including stars down to a lower mass limit of $M_\mathrm{min} = 0.15 \, M_\odot$. We note that the choice of MF mainly affects the modelling of the \acp{hrd} below the main sequence turn-off, as the \ac{rgb} spans a narrow mass range. 
The sensitivity of our measurements to different MF assumptions is quantified below (Sect.~\ref{sec:47tuc}). 

\subsection{NLTE corrections}
\label{sec:nlte}

\begin{figure*}
\centering
\includegraphics[width=175mm]{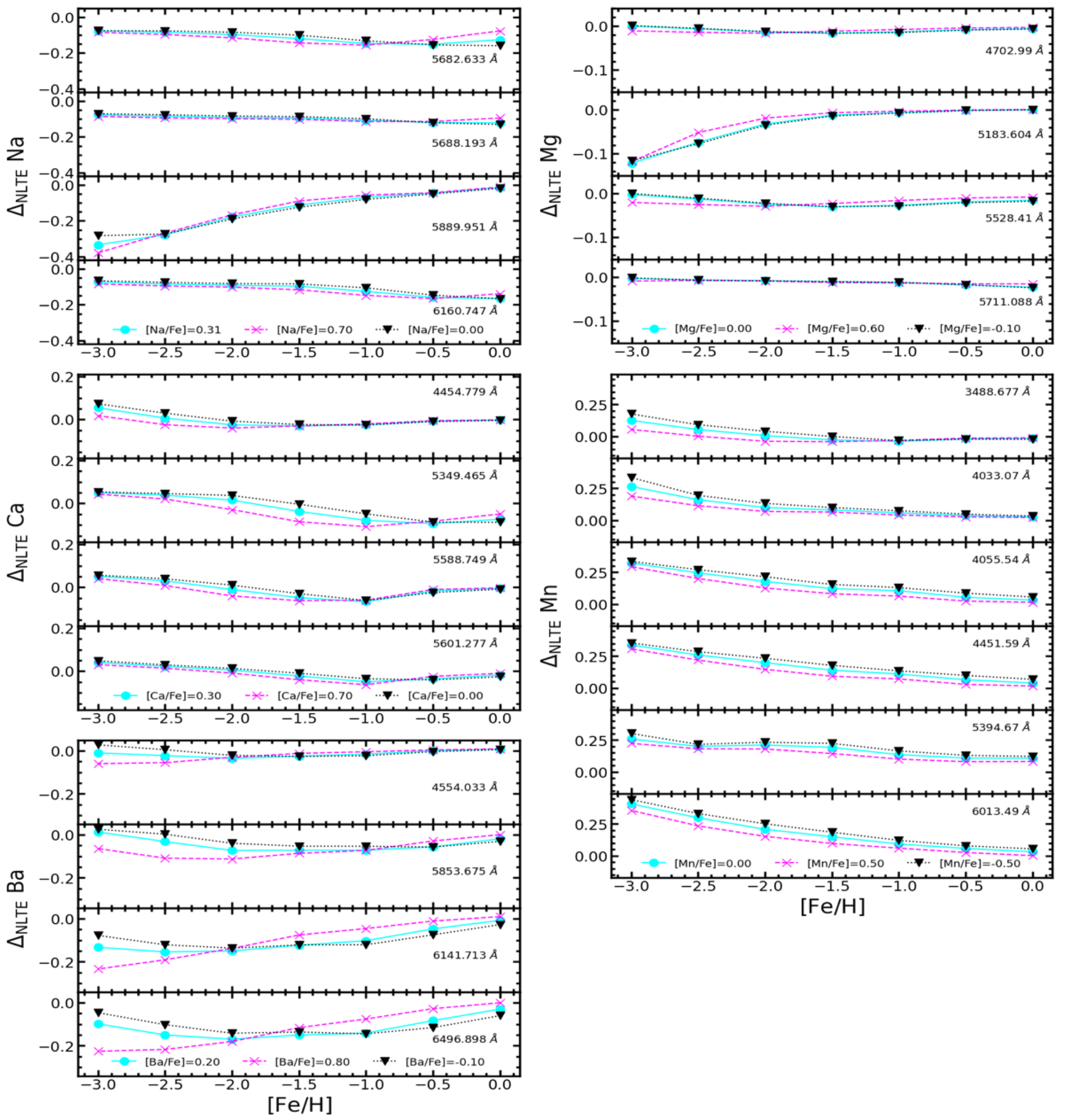}
\caption{\label{fig:nltecorr}Integrated-light \ac{nlte} abundance corrections for Na, Mg, Ca, Mn, and Ba. In each panel, corrections are shown as a function of [Fe/H] for three different abundance ratios, as indicated in the legends.}
\end{figure*}

The spectral modelling with \texttt{ATLAS}/\texttt{SYNTHE} and \texttt{MARCS}/\texttt{Turbospectrum}  operates under the classical approximation of \ac{lte}, in which the atomic energy level populations only depend on the local temperature and electron density via the Saha-Boltzmann equations  \citep{Mihalas1970}. While computationally convenient, the limitations of this approximation have long been recognised and corrections for \ac{nlte} effects are now commonly applied in analyses of individual stars. A procedure for applying \ac{nlte} corrections in the analysis of integrated-light spectra was introduced in \citet{Eitner2019}, who established the basic framework and performed validation tests for Mg, Mn, and Ba. 
Here we apply \ac{nlte} corrections for a larger number of elements (Na, Mg, Ca, Ti, Mn, Fe, Ni, and Ba). We also discuss how to apply the corrections computed for individual lines to the \ac{lte} abundances, which are obtained from spectral fits that typically include several lines within a given spectral window. 

The atomic models will be described in detail in Magg et al. (in prep). In short, the model atoms were taken from \citet{Bergemann2017} for Mg, \citet{Bergemann2019} for Mn, and \citet{Gallagher2020} for Ba.  The models of Fe and Ca are based on \citet{Bergemann2012} and \citet{Mashonkina2007}, respectively, but have been updated with new radiative and collisional data in \citet{Semenova2020}. Our model atom for Ni was presented in \citet{Bergemann2021}, whereas the Ti model is essentially the one adopted from \citet{Bergemann2011}, but updated with new H collisional rates from \citet{Grumer2020}. The model atom of sodium was developed specifically for this study \citep{Moltzer2020}. The model is based on NIST energy levels and bound-bound radiative transitions from the Kurucz\footnote{\url{http://kurucz.harvard.edu/atoms/1100/}} database. In total, the model atom includes 102 energy levels, with 101 levels in \ion{Na}{i} and closed by the ground state of \ion{Na}{II}. Fine structure was retained up to the term 5p $^{2}$P$^{o}$ (energy of 35042.850 cm$^{-1}$). The model also includes 121 bound-bound radiative transitions with oscillator strengths and damping parameters extracted from the Kurucz database, except the van der Waals damping, which was taken from \citet{Barklem2000}
where available. For all other transitions, the standard Uns{\"o}ld value was used.  Photoionisation cross-sections were adopted from the TOPbase\footnote{\url{http://cdsweb.u-strasbg.fr/topbase/topbase.html}} database. The rate coefficients describing bound-bound (excitation) and bound-free (ionisation) transitions due to collisions with electrons  were adopted from \citet{Igenbergs2008}. The values from \citet{Barklem2010,Barklem2017} were used to represent excitation and charge transfer reactions caused by processes in inelastic collisions with hydrogen atoms. The two datasets were merged and tabulated on a denser grid of temperatures to allow a smoother interpolation in \texttt{MULTI1D} \citep{Carlsson1986}. Our Na model is, in this respect, similar to the study by \citet{Lind2011}. For the details about the \ac{nlte} model atoms of all other elements, we refer the reader to the aforementioned papers.

For each element, the \ac{nlte} corrections were calculated using the \texttt{MULTI2.3} statistical equilibrium code \citep{Carlsson1986} and model atmospheres similar to those used in the abundance analysis in this work (Sect.~\ref{sec:atmospheres}). We adopted seven values of the metallicity: $\mathrm{[Fe/H]}=-3, -2.5, \ldots, 0.0$ for several points in the \ac{hrd} \citep{Eitner2019}. As was the case for the spectral fitting, the \acp{hrd} used for the modelling of the integrated-light \ac{nlte} corrections were based on $\alpha$-enhanced DSEP isochrones with an age of 13 Gyr combined with empirical \ac{hb} data and \texttt{ATLAS12} atmospheres, but with a smaller number of HRD-boxes (typically about 25). We then interpolated between these models to find the corrections for each \ac{gc} in our sample. 

For many elements, each spectral window contains multiple lines with different strengths that  contribute with different weights to the abundances derived from the spectral fits. In some cases, the lines within a window correspond to different transitions within the same multiplets, and the level populations are affected in similar ways by \ac{nlte} corrections. Nevertheless, different lines usually have different strengths, and are located on different parts of the curve-of-growth. In general, the average \ac{nlte} correction, $\langle\Delta_\mathrm{NLTE}\rangle$, for the various lines included in the fit can be expressed as a weighted average of the corrections for the individual lines, $\Delta_{\mathrm{NLTE},i}$:
\begin{equation}
    \langle\Delta_\mathrm{NLTE}\rangle = \frac{\sum \omega_i \Delta_{\mathrm{NLTE},i}}{\sum \omega_i}.
    \label{eq:nltew}
\end{equation}
To find the weights $\omega_i$, we assume that the abundance $A$ of an element, measured within a given spectral window that contains multiple lines, is a weighted average of the abundances $A_i$ that would be obtained by measuring each line individually, with weights given by the inverse variances $\sigma_{A_i}^{-2}$. These are then the same weights that apply to the $\Delta_{\mathrm{NLTE},i}$ values. 
Writing the $A_i$ as a function of the equivalent widths $W_i$ of the corresponding lines, the variances can be written as
\begin{equation}
    \sigma_{A_i}^2 = \left(\frac{\partial A_i}{\partial W_i}\right)^2 \sigma_{W_i}^2,
    \label{eq:sigai}
\end{equation}
where $\sigma_{W_i}$ are the uncertainties on the $W_i$. For most lines, the observed line profiles are determined mainly by instrumental and velocity broadening and are thus similar for all lines. We therefore assume that the $\sigma_{W_i}$ are inversely proportional to the S/N of the spectra, $\sigma_{W_i} \propto (\mathrm{S/N})^{-1}$ \citep{Cayrel1988}. 
While we do not actually derive abundances by measuring equivalent widths of individual lines, we assume that the uncertainties on the $A_i$ obtained from spectral fitting still scale with the $\sigma_{W_i}$ as in Eq.~(\ref{eq:sigai}). Assuming further that the S/N is the same at the position of each line, so that the $\sigma_{W_i}$ are the same for all lines, the weights are then given by the squared slopes of the curves-of-growth,
\begin{equation}
 \omega_i = \left( \frac{\partial W_i}{\partial A} \right)^2.
 \label{eq:nltew2}
\end{equation}
For weak lines (on the linear part of the curve-of-growth) this means that the weights scale as the square of the equivalent widths, so that \ac{nlte} corrections can be ignored for lines that are too weak to contribute significantly to the $\chi^2$ of the fit.
Equations~(\ref{eq:nltew}) and (\ref{eq:nltew2}) then allow us to compute the mean corrections $\langle\Delta_\mathrm{NLTE}\rangle$ for each spectral window. We note that the weights $\omega_i$ will, in general, depend on the abundance of the element in question, and therefore must be computed separately for each case.

Figure~\ref{fig:nltecorr} shows representative integrated-light \ac{nlte} corrections for several diagnostic spectral lines of Na, Mg, Ca, Mn, and Ba. The corrections were calculated for several values of the abundance ratios $\mathrm{[Na/Fe]}$ and $\mathrm{[Mg/Fe]}$, because the \ac{nlte} effects are sensitive to the number density of the element and, therefore, the abundance corrections also change slightly. For Na, the differences between the \ac{nlte} and \ac{lte} abundances depend strongly on the atomic properties of the transitions. For some lines, such as the relatively weak sub-ordinate lines at 5682~\AA, 5688~\AA, and 6160 \AA, the \ac{nlte} corrections are negative and reach $-0.2$ dex at solar metallicity. However, in more metal-poor atmospheres, [Fe/H] $\lesssim -1.5$, the differences between \ac{lte} and \ac{nlte} abundances for these lines progressively vanish and do not exceed $-0.05$ dex at [Fe/H]$=-3$. 

For Mg, the \ac{nlte} corrections of all diagnostic lines display a rather smooth behaviour, not exceeding $-0.03$ dex at solar metallicity, but in the regime below [Fe/H]$\approx -1.5$ some lines become more sensitive to \ac{nlte} effects. In particular, the strong Mg $b$ triplet lines in the optical (5167~\AA, 5172~\AA, and 5183 \AA) tend to become even stronger in \ac{nlte} at lower metallicity, however, we do not use these lines in this work.
The profiles of the weaker high-excitation lines at 4572~\AA, 4731~\AA, 4702~\AA, and 5711 \AA~remain either very close to \ac{lte} or are slightly weakened compared to \ac{lte}, which implies that the abundances derived from these lines in \ac{lte} are relative insensitive to \ac{nlte} effects. The \ac{nlte} corrections for the 5528 \AA~line are sensitive to the abundance of Mg at [Fe/H]$=-2$ and below. In the $\alpha$-enhanced regime, [Mg/Fe]$=+0.4$ dex, which is typically seen in Galactic GCs, the line shows very small \ac{nlte} corrections. On the other hand, in $\alpha$-poor conditions, [Mg/Fe]$=-0.1$ the \ac{nlte} correction is slightly negative. 

The \ac{nlte} results for Ca depend on the properties of individual spectral lines. Whereas the overall behaviour is such that the \ac{nlte} line profiles are very similar to \ac{lte} at solar metallicity, [Fe/H] $\approx -1$ represents a transition regime, where the \ac{nlte} corrections change sign and start increasing with decreasing metallicity. In the transition regime, the \ac{nlte} corrections are typically negative for all \ion{Ca}{I} lines in our linelist, and reach $-0.1$ to $-0.2$ dex, depending on the abundance of Ca used in the statistical equilibrium calculations. The \ac{nlte} corrections are typically more negative for lower [Ca/Fe] ratios, and more positive for elevated  [Ca/Fe]. In the most metal-poor systems,  [Fe/H] $\lesssim -2$, the \ac{nlte} corrections to abundances inferred from \ion{Ca}{I} lines reach $\sim 0.1$ dex and they become less sensitive to the Ca abundance in the model atmosphere.

Our results for Mn are very similar to those described in \citet{Eitner2020}. \ion{Mn}{I} is a typical low-ionisation-potential ion with large photo-ionisation cross-sections in the blue and it is subject to over-ionisation in the atmospheres of FGK-type stars. The \ac{nlte} corrections for all \ion{Mn}{I} lines display a very similar behaviour, being close to $+0.05$ in solar-metallicity models, but they linearly increase with decreasing metallicity of the model. The largest \ac{nlte} correction of $\sim +0.4$ dex is attained at [Fe/H]$=-3$, which represents the limit of our model grid. This implies that Mn abundances in \ac{lte} are systematically under-estimated and the bias increases for more metal-poor systems.

The \ion{Ba}{II} lines are qualitatively similar to the \ion{Na}{I} lines in terms of their \ac{nlte} effects, which is not surprising because for both systems the NLTE effects are driven by strong line scattering. The \ac{nlte} corrections are small and slightly negative for the resonance line at 4554~\AA\ and the weaker subordinate line at 5853~\AA. Only in metal-poor models with extreme Ba enhancement ([Ba/Fe] $=0.8$ dex) does the subordinate line show the \ac{nlte} correction of $-0.2$ dex. However, the 6141~\AA\ and the 6496~\AA\ lines show a larger sensitivity to \ac{nlte}, which is reflected in their \ac{nlte} corrections smoothly increasing in amplitude with decreasing metallicity. In the models with [Fe/H] $\lesssim -2$, the corrections reach a plateau at $\Delta_\mathrm{NLTE} \approx -0.2$ dex and then start increasing again.

\subsection{Validation on 47~Tuc}
\label{sec:47tuc}

\begin{table*}
\caption{Analysis of 47~Tuc.}
\label{tab:47tuc}
\centering
{\tiny
\begin{tabular}{l c c c c c c c c c c}
\hline\hline
 & \multicolumn{5}{c}{This analysis}  & L2017$^1$ & MB2008$^2$ & KM2008$^3$ & S2013$^4$ & T2014$^5$ \\
 & \multicolumn{3}{c}{DSEP LTE} & DSEP & BaSTI & \\
 & $\alpha=0$ & $\alpha=-1$ & Kroupa & \multicolumn{2}{c}{NLTE $\alpha=-1$} & & & & \\
\hline
$\mathrm{[Fe/H]}$  & $-0.763$  & $-0.748$ & $-0.726$  & $-0.735$ & $-0.724$ & $-0.863$ & $-0.75$ & $-0.76$ & $-0.81\pm0.02$ & $-0.78\pm0.07$ \\
$\mathrm{[Na/Fe]}$  & $+0.417$  & $+0.393$  & $+0.357$  & $+0.237$ & $+0.237$ & $+0.422$ & $+0.45$ & $+0.21$ & $+0.38\pm0.12$ & $+0.21\pm0.14$ \\
$\mathrm{[Mg/Fe]}$  & $+0.432$  & $+0.404$  & $+0.359$  & $+0.404$ & $+0.392$ & $+0.442$ & $+0.22$ & $+0.46$ & $+0.42\pm0.14$ & $+0.44\pm0.08$ \\
$\mathrm{[Si/Fe]}$  & $+0.392$  & $+0.389$  & $+0.393$  & ($+0.376$) & ($+0.368$) & \ldots & $+0.37$ & $+0.39$ & \ldots & $+0.32\pm0.09$ \\
$\mathrm{[Ca/Fe]}$ & $+0.325$  & $+0.296$ & $+0.237$   & $+0.238$ & $+0.239$ & $+0.412$ & $+0.31$ & $+0.34$ & \ldots & $+0.24\pm0.13$ \\
$\mathrm{[Ti/Fe]}$  & $+0.330$  & $+0.337$  & $+0.341$  & $+0.412$ & $+0.421$ & $+0.370$ & $+0.41$ & $+0.37$ & \ldots & $+0.37\pm0.11$ \\
$\mathrm{[Sc/Fe]}$  & $+0.189$  & $+0.209$  & $+0.248$  & ($+0.197$) & ($+0.207$) & $+0.219$ & $+0.14$ & \ldots & \ldots & $+0.11\pm0.11$ \\
$\mathrm{[Cr/Fe]}$  & $-0.016$  & $-0.017$  &  $-0.019$  & ($-0.029$) & ($-0.036$) & $-0.060$ & $-0.02$ & \ldots & \ldots & $-0.03\pm0.11$ \\
$\mathrm{[Mn/Fe]}$  & $-0.262$  & $-0.256$  & $-0.249$  & $-0.188$ & $-0.205$ & $-0.229$ & $-0.44$ & \ldots & \ldots & $-0.20\pm0.13$ \\
$\mathrm{[Ni/Fe]}$  & $+0.020$  & $+0.028$  & $+0.045$  & $+0.060$ & $+0.068$ & \ldots & $+0.00$ & \ldots & & $-0.12\pm0.04$ \\
$\mathrm{[Cu/Fe]}$ & $-0.046$  & $-0.036$  & $-0.021$  & ($-0.049$) & ($-0.028$) & \ldots & $-0.13$ & \ldots & \ldots & $-0.14\pm0.35$ \\
$\mathrm{[Zn/Fe]}$  & $+0.120$  & $+0.126$  & $+0.143$  & ($+0.113$) & ($+0.165$) & \ldots & \ldots  & \ldots & \ldots & $+0.26\pm0.13$ \\
$\mathrm{[Zr/Fe]}$  & $+0.208$  & $+0.237$  & $+0.288$  & ($+0.224$) & ($+0.165$) & \ldots & $+0.05$ & \ldots & \ldots & $+0.41\pm0.17$ \\
$\mathrm{[Ba/Fe]}$  & $+0.192$  & $+0.205$  & $+0.230$  & $+0.133$ & $+0.159$ & $+0.155$ & $+0.02$ & \ldots & \ldots & $+0.25\pm0.24$ \\
$\mathrm{[Eu/Fe]}$  & $+0.262$  & $+0.238$  &  $+0.232$  & ($+0.225$) & ($+0.258$) & \ldots & $+0.04$ & \ldots & $+0.27\pm0.14$ & $+0.32\pm0.19$ \\
\hline
\end{tabular}
}
\tablefoot{
The \ac{nlte} columns give the abundance ratios, corrected for \ac{nlte} effects, for DSEP and BaSTI isochrones and a MF slope of $\alpha=-1$. For values in parentheses, no explicit \ac{nlte} corrections were computed and the \ac{nlte} values only differ from the \ac{lte} values because of the $0.012$~dex change in the iron abundance.}
\tablebib{
(1)~\citet{Larsen2017};
(2)~\citet{McWilliam2008};
(3)~\citet{Koch2008};  
(4)~\citet{Sakari2013}: $\mathrm{[Fe/H]}$ is $\mathrm{[\ion{Fe}{i}/H]}$;
(5)~\citet{Thygesen2014}: abundances are median and uncertainty on median, except $\mathrm{[Fe/H]} $ for which the mean value is given. 
}
\end{table*}

In \citetalias{Larsen2017} the integrated-light analysis technique was tested by measuring metallicities and chemical abundances for the seven Galactic \acp{gc} that are also included here. In that paper, the sensitivity of the analysis to various model assumptions was also tested, and it is not our intent to repeat those tests here. An extensive discussion of systematic uncertainties in the analysis of integrated-light spectra can also be found in \citet{Sakari2014}. Here, instead, we carry out a more detailed comparison with the well-studied Galactic GC 47~Tuc, for which measurements of a large number of elements for individual stars are available in the literature. 

Table~\ref{tab:47tuc}  lists our integrated-light abundance measurements for 47~Tuc and recent literature data. The measurements of \citetalias{McWilliam2008} and \citet[][S2013]{Sakari2013} are integrated-light measurements, while those of \citet[][KM2008]{Koch2008} and \citet[][T2014]{Thygesen2014} come from individual \ac{rgb} stars. The \citetalias{Thygesen2014} analysis includes \ac{nlte} corrections for Na, Mg, and Ba.
In addition to our default MF ($\alpha=-1$), we list results for a Kroupa MF and for a flat (i.e.\ extremely bottom-light) MF ($\alpha=0$). We also include results obtained from a modelling based on a BaSTI isochrone, as well as the previous integrated-light measurements from \citetalias{Larsen2017}. The abundances in Table~\ref{tab:47tuc} are weighted averages of the values obtained from fits to the individual spectral windows,
\begin{equation}
\langle \mathrm{[X/Fe]} \rangle \equiv \frac{\sum w_i \mathrm{[X/Fe]}_i}{\sum w_i}
\end{equation}
with weights defined as
\begin{equation}
w_i = \frac{1}{\sigma_i^2 + \sigma_0^2}
\label{eq:weights}
\end{equation}
where $\sigma_0=0.05$~dex is a `floor' added in quadrature to account for non-random uncertainties on the abundances derived from the individual fits. For a discussion of the uncertainties on the integrated-light measurements we refer to Sect.~\ref{sec:anal}, but here we note that the window-to-window dispersions for 47~Tuc are mostly fairly similar to those found in the analyses of the Sun and Arcturus. 

The \ac{nlte} columns lists the abundances obtained by applying the \ac{nlte} corrections to the DSEP and BaSTI $\alpha=-1$ analyses. Comparison with Table~\ref{tab:arcturus} shows that the \ac{nlte} corrections for 47~Tuc are fairly similar to those obtained for Arcturus. Values in parentheses are elements for which no \ac{nlte} corrections were computed, but the abundance ratios relative to iron still change by $-0.012$~dex relative to the \ac{lte} values because of the change in $\mathrm{[Fe/H]}$.

A substantial range of iron abundances are quoted in the literature for 47~Tuc, spanning a range of at least 0.2~dex from $\mathrm{[Fe/H]}=-0.83$ \citep{Lapenna2014} to
$\mathrm{[Fe/H]}=-0.62$ \citep{Pritzl2005}. The most recent version of the \citet{Harris1996} catalogue lists $\mathrm{[Fe/H]}=-0.72$. Our new \ac{nlte} measurement, $\mathrm{[Fe/H]}=-0.74$ for the reference MF ($\mathrm{[Fe/H]}=-0.72$ when using the BaSTI isochrone), is thus in better agreement with the literature data than the analysis presented in \citetalias{Larsen2017}. It was found in \citetalias{Larsen2017} that using the empirical \ac{cmd}
instead of theoretical isochrones to model the HRD of 47~Tuc had practically no effect on $\mathrm{[Fe/H]}$, and Table~\ref{tab:47tuc} shows that the effect of changing the MF is relatively minor, too. Most of the difference with respect to \citetalias{Larsen2017} is due to a combination of the updated microturbulence prescription and the use of model atmospheres with self-consistent abundance patterns. While the line list used here also differs from that used in \citetalias{Larsen2017}, this only has a small effect on the iron abundance:  repeating the same analysis as in \citetalias{Larsen2017}, but with our new line list, we found $\mathrm{[Fe/H]}=-0.855$ (\ac{lte}) which differs by less than 0.01~dex from the value in \citetalias{Larsen2017}. 
The uncertainty on the iron abundance due to stochastic sampling of the MF within the slit scan area is about 0.04~dex \citepalias{Larsen2017}. 

A detailed discussion of integrated-light Na and Mg abundance measurements for the Galactic \acp{gc} was given in \citetalias{Larsen2017}. A complication affecting these elements is that their abundances often exhibit large star-to-star variations within individual \acp{gc}. In 47~Tuc, the range in $\mathrm{[Na/Fe]}$ is about 0.5~dex, but there is no significant spread in $\mathrm{[Mg/Fe]}$ \citep[][hereafter C2009]{Carretta2009}.
The \ac{lte} abundances of these elements found from our new analysis differ only slightly from those in \citetalias{Larsen2017}. For Mg, the integrated-light abundance ratio is unaffected by \ac{nlte} corrections, while the Na abundance is 0.16~dex lower in \ac{nlte}. Our $\mathrm{[Mg/Fe]}$ measurement agrees well with those found by other studies, with the exception of \citetalias{McWilliam2008} whose integrated-light  $\mathrm{[Mg/Fe]}$ value is about 0.2 dex lower. 
Our \ac{nlte} $\mathrm{[Na/Fe]}$ value agrees well with the studies of individual stars listed in Table~\ref{tab:47tuc}, of which one \citepalias{Thygesen2014} likewise included \ac{nlte} corrections and the other (KM2008) was a differential analysis with respect to the Sun. The two other literature integrated-light analyses \citepalias{McWilliam2008,Sakari2013} both assumed \ac{lte}, and agree well with our measured \ac{lte} $\mathrm{[Na/Fe]}$ value. 

While not listed in Table~\ref{tab:47tuc}, it is also of interest to compare with the measurements of $\mathrm{[Na/Fe]}$, $\mathrm{[Mg/Fe]}$, and $\mathrm{[Si/Fe]}$ obtained from UVES spectra of 11 individual stars in 47~Tuc by \citetalias{Carretta2009}. They find a significantly higher mean \ac{nlte} Na abundance ratio of $\langle\mathrm{[Na/Fe]}\rangle = +0.53$ than that reported here. They also find a somewhat higher Mg abundance ratio ($\langle\mathrm{[Mg/Fe]}\rangle=+0.52$), while their Si abundance ratio is similar to ours ($\langle\mathrm{[Si/Fe]}\rangle=+0.40$). For Mg, the difference can be attributed to different atomic parameters and solar reference abundances, where their solar Mg abundance (from \citealt{Gratton2003a}) is 0.15~dex lower than that used in our work. For Na, the difference is less easily explained. Part of it (0.12~dex) can again be explained by different solar reference abundances, but this still leaves a difference of about 0.2~dex unaccounted for. The spread in $\mathrm{[Na/Fe]}$ values within GCs in general, including 47~Tuc, contributes to some uncertainty on the mean value computed for a sample of only 11 stars, but seems unlikely to fully explain the 0.2~dex offset. Indeed, for a larger sample (147) of GIRAFFE spectra, \citet{Carretta2009a} quote a minimum $\mathrm{[Na/Fe]}=+0.15$ and a maximum $\mathrm{[Na/Fe]}=+0.74$ for member stars of 47~Tuc, and for this sample we compute a mean value of $\langle\mathrm{[Na/Fe]}\rangle=+0.47$, still considerably higher than our \ac{nlte} measurement and the literature values in Table~\ref{tab:47tuc}. We return to this discrepancy below (Sect.~\ref{sec:abundance_trends}). 

The new spectral windows and the atomic data for Ca (Sect.~\ref{sec:linelist}) differ significantly from those used in our previous work, and our new \ac{lte} $\mathrm{[Ca/Fe]}$ measurement for 47~Tuc is about 0.1~dex lower than in \citetalias{Larsen2017}. 
In the comparisons with Arcturus and the Sun, our $\mathrm{[Ca/Fe]}$ measurements were slightly higher than the literature values, but for 47~Tuc our \ac{lte} measurement ($\mathrm{[Ca/Fe]}=+0.30$ for $\alpha=-1$) falls within the range of values determined by other authors. 
We note that the Ca abundance measurement is sensitive to the MF choice, with a difference of nearly 0.1~dex in $\mathrm{[Ca/Fe]}$ between the fits for a Kroupa and a flat ($\alpha=0$) MF. 
The \ac{nlte} corrections lead to a decrease of 0.06~dex in $\mathrm{[Ca/Fe]}$.
Hence it appears that specific model assumptions can have a relatively large systematic effect, at least $\sim0.1$~dex, on [Ca/Fe].
Other elements for which the spectral windows have changed significantly are Ti and Cr, but for these elements the resulting changes in the abundance ratios are smaller than for Ca and both fall within 0.05~dex of those determined in \citetalias{Larsen2017}. 

We include several elements here that were not previously measured in \citetalias{Larsen2017}: Si, Ni, Cu, Zn, Zr, and Eu. 
Of these, our $\mathrm{[Si/Fe]}$ ratio agrees well with the literature data, while literature data are more scarce for the rest. Our \ac{lte} $\mathrm{[Ni/Fe]}$ measurement falls within 0.03~dex of that found by \citetalias{McWilliam2008}, while it is about 0.15~dex higher than the value found by \citetalias{Thygesen2014}. 
For Zn, our value ($\mathrm{[Zn/Fe]} = +0.13\pm0.05$) is 0.13~dex lower than that found by \citetalias{Thygesen2014} although the latter has a fairly large uncertainty ($0.26\pm0.13$).
\citet{Cerniauskas2018} measured an average Zn abundance of  $\langle\mathrm{[Zn/Fe]}\rangle = +0.11$ for 27 \ac{rgb} stars in 47~Tuc (with an rms variation of 0.09~dex), which agrees well with our measurement. 
Despite the measurement of $\mathrm{[Eu/Fe]}$ being relatively challenging, our $\mathrm{[Eu/Fe]}$ value agrees well with those determined by \citetalias{Thygesen2014} and \citetalias{Sakari2013}, and we get very consistent results from the two \ion{Eu}{ii} lines (see Appendix~\ref{app:iam}). There is a somewhat larger difference of about 0.2~dex with respect to \citetalias{McWilliam2008}. 

For most elements, the results depend only weakly on the detailed HRD modelling assumptions.
The strongest dependency on MF slope occurs for Ca, as mentioned above. For most other elements the effect is small, $\sim0.01$~dex. The same is true for the choice of isochrones, where the most strongly affected abundance ratios are [Zn/Fe], [Zr/Fe], and [Eu/Fe], which vary by $\sim0.05$~dex. Again, more typical variations are $\sim0.01$~dex.

\subsection{Analysis of the globular cluster spectra}
\label{sec:anal}

\begin{table*}
\caption{Summary of model assumptions.}
\label{tab:assump}
\centering
{\small
\begin{tabular}{lccclcc}
\hline\hline
Cluster & $t_\mathrm{iso}$ & $\mathrm{[Fe/H]}_\mathrm{iso}$ & $[\alpha/\mathrm{Fe}]_\mathrm{iso}$ & HB & $\sigma_\mathrm{br}$  & $v_\mathrm{hel}$  \\
& Gyr & & & & km~s$^{-1}$ & km~s$^{-1}$ \\
\hline
NGC~104  & 11 & $-0.8$ & $+0.4$ & NGC~104 & 12.3 & $-17^1$ \\
NGC~362  & 11 & $-1.1$ & $+0.4$ & NGC~362 & 8.5 & $+225^1$ \\
NGC~6254 & 13 & $-1.5$ & $+0.4$ & NGC~6254 & 7.3 & $+75^1$ \\
NGC~6388 & 13 & $-0.6$ & $+0.2$ & NGC~6388 & 18.1 & $+83^1$ \\
NGC~6752 & 13 & $-1.8$ & $+0.4$ & NGC~6752 & 8.7 & $-28^1$ \\
NGC~7078 & 13 & $-2.3$ & $+0.4$ & NGC~7078 & 13.3 & $-105^1$ \\
NGC~7099 & 13 & $-2.3$ & $+0.4$ & NGC~7099 & 6.2 & $-184^1$ \\
M31 006-058 & 13 & $-0.6$ & $+0.4$ & NGC~104 & 12.1 & $-238^2$ \\
M31 012-064 & 13 & $-1.8$ & $+0.4$ & NGC~6093 & 19.2 & $-359^2$ \\
M31 019-072 & 11 & $-0.7$ & $+0.4$ & NGC~104 & 18.0 & $-222^2$ \\
M31 058-119 & 13 & $-1.0$ & $+0.4$ & NGC~362 & 20.0 & $-221^2$ \\
M31 082-144 & 11 & $-0.7$ & $+0.4$ & NGC~104 & 25.8 & $-373^2$ \\
M31 163-217 & 13 & $-0.2$ & $+0.2$ & NGC~6388 & 18.2 & $-163^2$ \\
M31 171-222 & 13 & $-0.2$ & $+0.2$ & NGC~6388 & 15.1 & $-267^2$ \\
M31 174-226 & 13 & $-1.0$ & $+0.4$ & NGC~362 & 14.7 & $-491^2$ \\
M31 225-280 & 11 & $-0.4$ & $+0.4$ & NGC~6388 & 27.6 & $-160^2$ \\
M31 338-076 & 13 & $-1.1$ & $+0.4$ & NGC~362 & 19.9 & $-266^2$ \\
M31 358-219 & 13 & $-2.2$ & $+0.4$ & NGC~7078 & 11.8 & $-315^2$ \\
M31 EXT8 & 13 & $-2.8$ & $+0.4$ & n/a & 13.7 & $-204^3$ \\
M33 H38   & 10 & $-1.1$ & $+0.4$ & NGC~362 & 6.0 & $-241^6$ \\
M33 M9    & 13 & $-1.7$ & $+0.4$ & NGC~6093 & 6.4 & $-249^6$ \\
M33 R12 & 10 & $-0.9$ & $+0.4$ & NGC~104 & 6.9 & $-218^6$ \\
M33 U49 & 10 & $-1.4$ & $+0.4$ & NGC~362 & 7.9 & $-150^6$ \\
M33 R14 & 10 & $-1.1$ & $+0.4$ & NGC~362 & 10.8 & $-214^2$ \\
M33 U77 & 13 & $-1.8$ & $+0.4$ & NGC~6093 & 6.4 & $-222^2$ \\
M33 CBF28 & 10 & $-1.2$ & $+0.4$ & NGC~362 & 8.3 & $-238^2$ \\
M33 HM33B & 13 & $-1.2$ & $+0.4$ & NGC~362 & 4.6 & $-190^2$ \\
NGC~147 Hodge~II & 13 & $-1.5$ & $+0.4$ & NGC~6254 & 4.2 & $-207^6$ \\
NGC~147 Hodge~III & 13 & $-2.4$ & $+0.4$ & NGC~7078 & 7.5 & $-197^6$ \\
NGC~147 PA-1 & 13 & $-2.3$ & $+0.4$ & NGC~7078 & 7.0 & $-221^6$ \\
NGC~147 PA-2 & 13 & $-1.9$ & $+0.4$ & NGC~6779 & 7.0 & $-221^6$ \\
NGC~147 SD7  & 13 & $-1.9$ & $+0.4$ & NGC~6779 & 6.3 & $-197^6$ \\
NGC~185 FJJ-III & 13 & $-1.8$ & $+0.4$ & NGC~6093 & 6.0 & $-243^2$ \\
NGC~185 FJJ-V & 13 & $-1.8$ & $+0.4$ & NGC~6093 & 6.9 & $-173^2$ \\
NGC~185 FJJ-VIII & 13 & $-1.8$ & $+0.4$ & NGC~6093 & 5.7 & $-188^2$ \\
NGC~205 Hubble~I & 13 & $-1.4$ & $+0.4$ & NGC~6254 & 7.6 & $-302^2$ \\
NGC~205 Hubble~II & 13 & $-1.3$ & $+0.4$ & NGC~362 & 8.5 & $-241^2$ \\
NGC~6822 Hubble~VII & 13 & $-1.7$ & $0.0$ & NGC~6093 & 9.6 & $-62^2$ \\
NGC~6822 SC6 & 13 & $-1.7$ & $+0.4$ & NGC~6093 & 9.4 & $-5^6$ \\
NGC~6822 SC7 & 13 & $-1.1$ & $0.0$ & NGC~362 & 9.9 & $-39^6$ \\
WLM GC & 13 & $-1.9$ & $+0.2$ & NGC~6779 & 9.9 & $-109^5$ \\
Fornax 3 & 13 & $-2.3$ & $+0.4$ & NGC~7078 & 8.1 & $+60^4$ \\
Fornax 4 & 13 & $-1.3$ & $0.0$ & NGC~362 & 5.5 & $+47^4$ \\
Fornax 5 & 13 & $-2.1$ & $+0.4$ & NGC~6779 & 6.3 & $+61^4$ \\
NGC~2403 F46 & 13 & $-1.7$ & $+0.4$ & NGC~6093 & 12.2 & $+140^2$ \\
\hline
\end{tabular}
}
\tablefoot{
For each cluster, we list the age ($t_\mathrm{iso}$), metallicity ([Fe/H]$_\mathrm{iso}$), and composition ([$\alpha$/Fe]$_\mathrm{iso}$) of the DSEP isochrone used to model the integrated-light spectrum. The HB column indicates the Galactic GC from which the horizontal branch was adopted, $\sigma_\mathrm{br}$ is the broadening applied to the model spectra, and $v_\mathrm{hel}$ is the heliocentric radial velocity. 
The broadening includes the instrumental resolution, which is typically 2--3 km~s$^{-1}$.
The BaSTI isochrone used for M31~EXT8 already includes the \ac{hb}. 
}
\tablebib{Radial velocities:
(1)~\citet{Larsen2017};
(2)~This work;
(3)~\citet{Larsen2020}; 
(4)~\citet{Larsen2012a};
(5)~Determined for the analysis in \citet{Larsen2014} but not listed in that work;
(6)~\citet{Larsen2014}.
}
\end{table*}

Our full sample of \ac{gc} spectra were analysed using the same procedure as described above for NGC~104.
For uniformity, we based all analyses on theoretical isochrones, even though resolved photometry reaching the main sequence turn-off is available for some clusters (mainly those in the Milky Way and Fornax dSph galaxies).
In Table~\ref{tab:assump} we list the ages and compositions ($\mathrm{[Fe/H]}$ and $[\alpha/\mathrm{Fe}]$) for the isochrones used to model each cluster and the Galactic \acp{gc} from which
\acp{hb} were adopted. Metallicities were chosen to self-consistently match those derived from the \ac{lte} spectral analysis to within a 0.1~dex tolerance. In most cases we assumed ages of 13 Gyr, but for the M31 \acp{gc} we adopted the ages from \citet{Caldwell2011} and somewhat younger ages (10 Gyr) were assumed for several of the M33 \acp{gc} on account of their relatively red \ac{hb} morphologies \citep{Sarajedini2000}.
The analysis in the remainder of this paper is based on DSEP isochrones for all clusters except M31~EXT8, but results based on BaSTI isochrones are included in Appendix~\ref{app:BaSTIfits}.  The DSEP isochrones do not extend to the low metallicity of M31~EXT8 and we based the analysis of this cluster on a BaSTI isochrone, instead of using a MIST isochrone with scaled-solar composition as was done in \citet{Larsen2020}.
Table~\ref{tab:assump} also lists the Gaussian dispersions of the kernels used to smooth each spectrum and the heliocentric radial velocities. The latter are given to the nearest km~s$^{-1}$ and have typical uncertainties of about 1~km~s$^{-1}$.  

\begin{figure}
\centering
\includegraphics[width=\columnwidth]{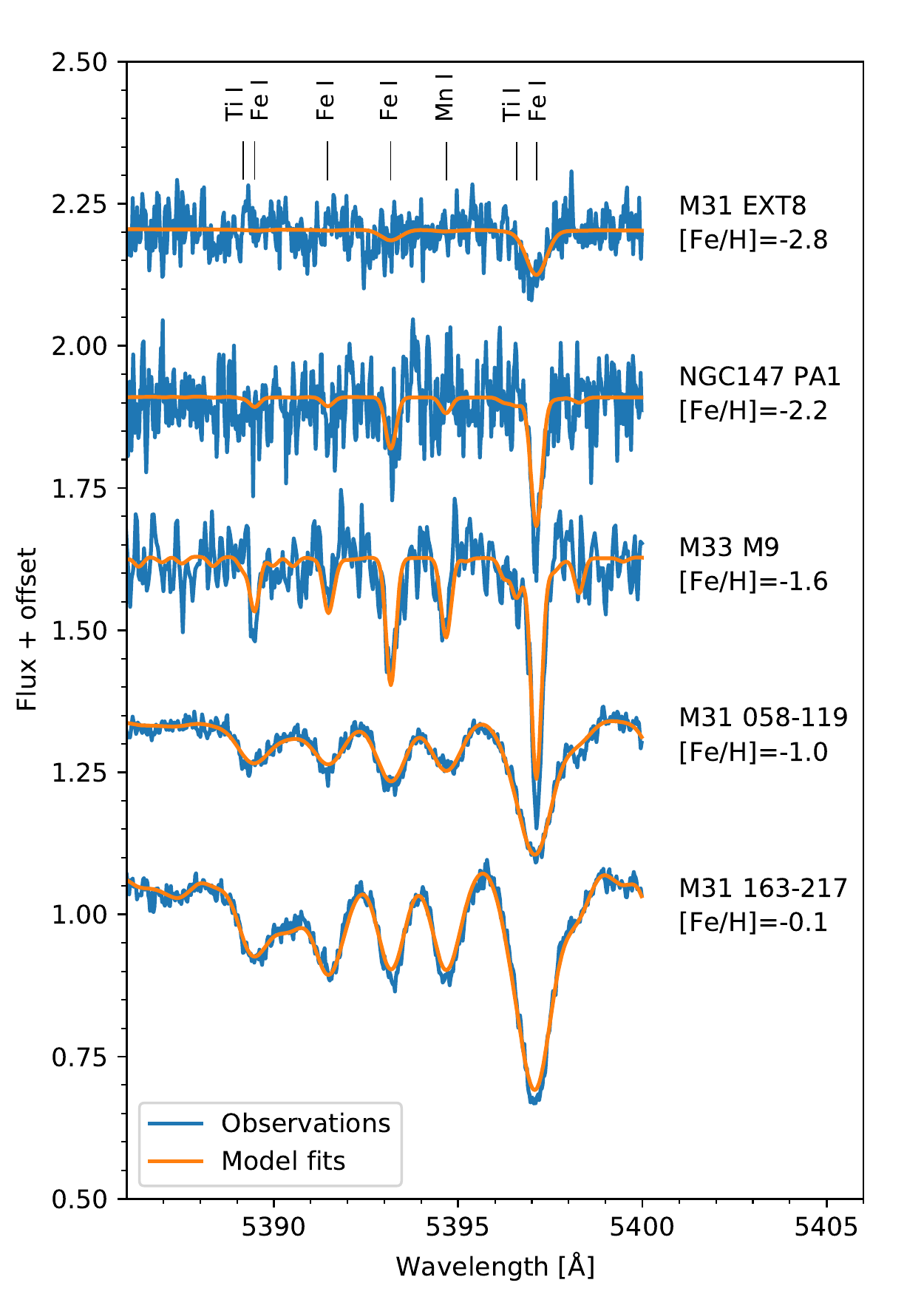}
\caption{\label{fig:spcmp}Integrated-light spectra and model fits for GCs spanning the range of NLTE  metallicities from $\mathrm{[Fe/H]}=-2.8$ (M31~EXT8, top) to $\mathrm{[Fe/H]}=-0.1$ (M31 163-217, bottom).}
\end{figure}

Figure~\ref{fig:spcmp} shows example fits for several GCs, ranging from the metal-poorest to the most metal-rich. Apart from the different metallicities, the figure also illustrates how the appearance of the spectra changes depending on the S/N (Table~\ref{tab:obs}) and the broadening of the spectra. 
For the spectra shown in the figure, the S/N (per \AA) ranges from 81 (M33 M9) to 394 (M31 058-119) while the broadening is between 6.4~km~s$^{-1}$ (for M33 M9) and 18.2~km~s$^{-1}$ (M31 163-217). The broadening is dominated by internal velocity broadening in the clusters but also includes the instrumental broadening of 2--3~km~s$^{-1}$. 

The individual abundance measurements for each spectral window are listed in Appendix~\ref{app:iam}. For each window we list the wavelength range, the \ac{lte} abundance obtained from the spectral fitting ($\mathrm{[X/H]}(\mathrm{LTE}))$, the \ac{nlte} correction $\Delta_\mathrm{NLTE}$, and the uncertainty $\sigma_i$ on the measurement. 
The weighted average abundance measurements are listed in Tables~\ref{tab:results1}-\ref{tab:results3} and include the \ac{nlte} corrections for the elements for which these were computed (the pure \ac{lte} versions are listed in Tables~\ref{tab:results1v4lte}-\ref{tab:results3v4lte}). 
The formal uncertainties on the average measurements ($\sigma_{\langle \mathrm{X}\rangle}$), based on propagation of the measurement errors, were computed as
\begin{equation}
\sigma_{\langle \mathrm{X}\rangle}  = \left(\sum w_i\right)^{-1/2} ,
\end{equation}
with weights $w_i$ defined by Eq.~(\ref{eq:weights}). The total uncertainties on the mean abundance ratios [X/Fe] should, in principle, be estimated as $\sigma_{\langle\mathrm{[X/Fe]}\rangle}^2 = \sigma_{\langle\mathrm{X}\rangle}^2 + \sigma_{\langle\mathrm{Fe}\rangle}^2$, but in practice the (random) uncertainties on the Fe abundances are almost always much smaller than for the other elements.
However, in most cases the scatter of the measurements exceeds the formal uncertainties, so that a propagation of the measurement errors underestimates the true uncertainties. In these cases, a more conservative estimate of the errors comes from the weighted standard deviations,
\begin{equation}
\mathrm{SD}_{X,w} \equiv \left[\frac{N}{N-1} \frac{\sum  \left(\mathrm{[X/Fe]}_i - \langle \mathrm{[X/Fe]} \rangle\right)^2 w_i}{\sum w_i}\right]^{1/2}
\end{equation}
from which the standard errors on the mean can then be estimated as $S_X = \mathrm{SD}_{X,w} / \sqrt{N}$, where $N$ is the number of measurements. We list both estimates of the uncertainties, except when $N=1$ in which case SD$_{X,w}$ is undefined.
In most cases, $S_X > \sigma_{\langle \mathrm{X}\rangle}$, but in a few cases the opposite is true, typically for small $N$ where the estimate of $S_X$ may not be very reliable.

\begin{table}
\caption{Error correlations for NGC~104.}
\label{tab:cerr0104}
\centering
{\small
\begin{tabular}{lrrrrrr}
\hline\hline
  & d(Na) & d(Mg) & d(Si) & d(Ca) & d(Ti)  \\
\hline
$\Delta\mathrm{[Fe/H]}$ &  $0.000$ & $+0.008$ & $+0.004$ & $0.000$ & $+0.008$  \\
$\Delta\mathrm{[Na/Fe]}$ & \ldots & $-0.002$ & $+0.010$ & $+0.006$ & $+0.012$  \\
$\Delta\mathrm{[Mg/Fe]}$ & $+0.002$ & \ldots & $+0.006$ & $+0.001$ & $+0.003$ \\
$\Delta\mathrm{[Si/Fe]}$ & $+0.006$ & $+0.003$ & \ldots & $0.000$ & $-0.005$ \\
$\Delta\mathrm{[Ca/Fe]}$ & $0.000$ & $0.000$ & $-0.001$ & \ldots & $-0.001$  \\
$\Delta\mathrm{[Ti/Fe]}$ & $-0.003$ & $+0.004$ & $+0.002$ & $+0.001$ & \ldots  \\
$\Delta\mathrm{[Sc/Fe]}$ & $+0.001$ & $+0.016$ & $-0.001$ & $-0.002$ & $+0.014$  \\
$\Delta\mathrm{[Cr/Fe]}$ & $-0.001$ & $-0.010$ & $-0.001$ & $0.000$ & $+0.008$  \\
$\Delta\mathrm{[Mn/Fe]}$ & $0.000$ & $-0.007$ & $-0.007$ & $0.000$ & $+0.001$  \\
$\Delta\mathrm{[Ni/Fe]}$ & $+0.001$ & $+0.001$ & $+0.002$ & $0.000$ & $+0.007$  \\
$\Delta\mathrm{[Cu/Fe]}$ & $+0.002$ & $-0.001$ & $+0.003$ & $-0.002$ & $+0.010$  \\
$\Delta\mathrm{[Zn/Fe]}$ & $+0.001$ & $+0.010$ & $+0.003$ & $-0.004$ & $-0.005$  \\
$\Delta\mathrm{[Zr/Fe]}$ & $-0.011$ & $-0.032$ & $+0.011$ & $-0.011$ & $+0.037$  \\
$\Delta\mathrm{[Ba/Fe]}$ & $0.000$ & $+0.015$ & $+0.007$ & $+0.008$ & $+0.008$  \\
$\Delta\mathrm{[Eu/Fe]}$ & $+0.002$ & $+0.022$ & $+0.016$ & $-0.117$ & $+0.025$  \\
\hline
\end{tabular}
}
\tablefoot{For each column, the entries indicate how the corresponding element responds to an increase of 0.1~dex in the abundance of the element listed in the header.
}
\end{table}

\begin{table}
\caption{Error correlations for NGC~7078.}
\label{tab:cerr7078}
\centering
{\small
\begin{tabular}{lrrrrrr}
\hline\hline
  & d(Na) & d(Mg) & d(Si) & d(Ca) & d(Ti) \\
\hline
$\Delta\mathrm{[Fe/H]}$ &  $+0.001$ & $0.000$ & $-0.002$ & $+0.001$ & $+0.002$  \\
$\Delta\mathrm{[Na/Fe]}$ & \ldots & $-0.020$ & $-0.024$ & $-0.004$ & $-0.010$  \\
$\Delta\mathrm{[Mg/Fe]}$ & $0.000$ & \ldots & $+0.001$ & $+0.001$ & $+0.002$  \\
$\Delta\mathrm{[Si/Fe]}$ & $+0.002$ & $0.000$ & \ldots & $-0.002$ & $-0.006$  \\
$\Delta\mathrm{[Ca/Fe]}$ & $0.000$ & $0.000$ & $+0.001$ & \ldots & $-0.003$  \\
$\Delta\mathrm{[Ti/Fe]}$ & $+0.004$ & $+0.002$ & $+0.003$ & $+0.002$ & \ldots  \\
$\Delta\mathrm{[Sc/Fe]}$ & $+0.001$ & $+0.010$ & $+0.004$ & $-0.001$ & $0.000$  \\
$\Delta\mathrm{[Cr/Fe]}$ & $+0.001$ & $0.000$ & $0.000$ & $+0.004$ & $+0.002$  \\
$\Delta\mathrm{[Mn/Fe]}$ & $0.000$ & $-0.002$ & $-0.001$ & $0.000$ & $+0.005$  \\
$\Delta\mathrm{[Ni/Fe]}$ & $+0.001$ & $0.000$ & $+0.004$ & $0.000$ & $+0.006$  \\
$\Delta\mathrm{[Cu/Fe]}$ & \ldots & \ldots & \ldots & \ldots & \ldots  \\
$\Delta\mathrm{[Zn/Fe]}$ & $+0.001$ & $+0.001$ & $+0.003$ & $-0.002$ & $+0.001$  \\
$\Delta\mathrm{[Zr/Fe]}$ & \ldots & \ldots & \ldots & \ldots & \ldots  \\
$\Delta\mathrm{[Ba/Fe]}$ & $-0.001$ & $+0.005$ & $+0.006$ & $+0.009$ & $0.000$  \\
$\Delta\mathrm{[Eu/Fe]}$ & $0.000$ & $+0.003$ & $+0.008$ & $-0.134$ & $+0.005$  \\
\hline
\end{tabular}
}
\tablefoot{For each column, the entries indicate how the corresponding element responds to an increase of 0.1~dex in the abundance of the element listed in the header. 
}
\end{table}

As discussed in Sect.~\ref{sec:sun}, the uncertainties on the $\mathrm{[Eu/Fe]}$ measurements are correlated with those on $\mathrm{[Ca/Fe]}$. More generally, we may expect correlations between the errors on different abundance ratios. In the case of Eu and Ca, the reason is easily identifiable as line blending, so that an increase in $\mathrm{[Ca/Fe]}$ leads to a decrease in $\mathrm{[Eu/Fe]}$. However, more subtle inter-correlations may result from variations in the atmospheric structure, such as may be caused by variations in the abundances of important electron-donor elements (e.g.\ Na, Mg, and Si).
It is not computationally practical to quantify these effects fully for each individual cluster, but in Table~\ref{tab:cerr0104} and Table~\ref{tab:cerr7078} we illustrate the effect of varying the abundances of a subset of the elements by 0.1~dex, taking NGC~104 and NGC~7078 as representative of metal-rich and metal-poor GCs.
The error correlations are quite modest for most elements, the most significant (anti-)correlation indeed being that of Eu vs.\ Ca. In general, these interdependencies are not symmetric; a change in the Ca abundance has a much larger effect on Eu than vice versa, because the Ca measurement is based on many more, stronger lines. 
\begin{figure}
\centering
\includegraphics[width=\columnwidth]{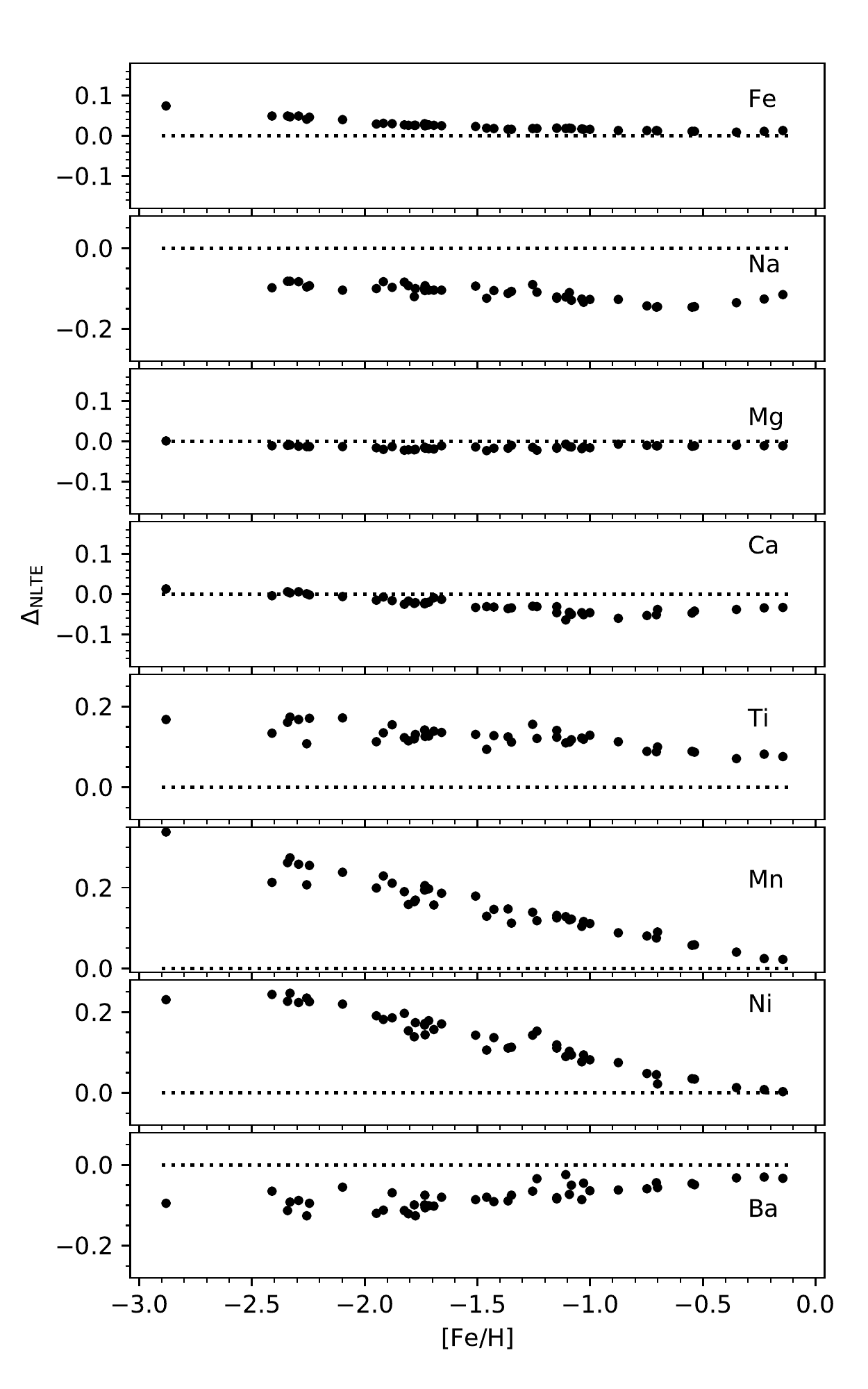}
\caption{\label{fig:dnlte}Average integrated-light NLTE corrections for each element as a function of metallicity.}
\end{figure}

Figure~\ref{fig:dnlte} shows the \ac{nlte} abundance corrections per element as a function of metallicity for each cluster, averaged over all spectral windows. For some elements, \ac{nlte} corrections are not available for all spectral windows and in these cases the missing corrections were estimated as the weighted average of the available corrections.
The \ac{nlte} corrections are plotted for the abundances relative to hydrogen ($\Delta_\mathrm{NLTE}(\mathrm{[X/H]}$) so that the corrections on the abundance ratios with respect to iron are given as
$\Delta_\mathrm{NLTE}(\mathrm{[X/Fe]}) = \Delta_\mathrm{NLTE}(\mathrm{[X/H]}) - \Delta_\mathrm{NLTE}(\mathrm{[Fe/H]})$.
For iron, the corrections are very small at the high metallicity end ($\Delta_\mathrm{NLTE}(\mathrm{[Fe/H]}\approx+0.01$~dex) and gradually increase towards lower metallicities, reaching $\Delta_\mathrm{NLTE}(\mathrm{[Fe/H]})=+0.07$~dex for M31~EXT8 at $\mathrm{[Fe/H]}(\mathrm{LTE})=-2.9$. The corrections for Na are fairly constant at $\Delta_\mathrm{NLTE}(\mathrm{[Na/H]})\approx-0.1$ and most closely resemble the corrections for the 5688.193~\AA\ line, which carries the most weight in the fits. 
For Mg the corrections are very small, although the variations in the corrections for iron introduce a trend of slightly decreasing $\mathrm{[Mg/Fe]}$ \ac{nlte} abundances, compared to the \ac{lte} values, towards the low-metallicity end. For Ca and Ti the trends with wavelength are similar to that for iron, so that the slope of $\mathrm{[Ca/Fe]}$ and $\mathrm{[Ti/Fe]}$ versus $\mathrm{[Fe/H]}$  looks similar in \ac{nlte} and \ac{lte}, although the corrections for Ti are more positive (by $\approx0.15$~dex) overall. As found in previous studies, the corrections for Mn increase strongly towards low metallicities, and a significant trend is seen also for Ni. For Ba the corrections are slightly negative, decreasing by about 0.1~dex from high to low metallicities.

\begin{table}
\caption{Comparison of LTE and NLTE abundances for NGC~104 and NGC~7078.}
\label{tab:ltevsnlte}
\centering
\begin{tabular}{lccc}
\hline\hline
         & $\Delta_\mathrm{NLTE}$ & SD$_{X,w}$ (LTE) & SD$_{X,w}$ (NLTE) \\ \hline
NGC 104 \\
  $\mathrm{[Fe/H]}$  & $+0.013$ & 0.116 & 0.116 \\
  $\mathrm{[Na/Fe]}$ & $-0.156$ & 0.099 & 0.092 \\
  $\mathrm{[Mg/Fe]}$ & $-0.023$ & 0.120 & 0.115 \\
  $\mathrm{[Ca/Fe]}$ & $-0.061$ & 0.093 & 0.085 \\
  $\mathrm{[Ti/Fe]}$ & $+0.076$ & 0.115 & 0.098 \\
  $\mathrm{[Mn/Fe]}$ & $+0.068$ & 0.061 & 0.058 \\
  $\mathrm{[Ni/Fe]}$ & $+0.032$ & 0.207 & 0.205 \\
  $\mathrm{[Ba/Fe]}$ & $-0.072$ & 0.184 & 0.157 \\
NGC 7078 \\
  $\mathrm{[Fe/H]}$  & $+0.049$ & 0.126 & 0.124 \\
  $\mathrm{[Na/Fe]}$ & $-0.131$ & 0.003 & 0.002 \\
  $\mathrm{[Mg/Fe]}$ & $-0.060$ & 0.060 & 0.062 \\
  $\mathrm{[Ca/Fe]}$ & $-0.041$ & 0.072 & 0.086 \\
  $\mathrm{[Ti/Fe]}$ & $+0.108$ & 0.206 & 0.160 \\
  $\mathrm{[Mn/Fe]}$ & $+0.212$ & 0.238 & 0.257 \\
  $\mathrm{[Ni/Fe]}$ & $+0.204$ & 0.219 & 0.244 \\
  $\mathrm{[Ba/Fe]}$ & $-0.162$ & 0.127 & 0.080 \\ \hline
\end{tabular}
\tablefoot{The column labelled $\Delta_\mathrm{NLTE}$ gives the \ac{nlte} corrections, while SD$_{X,w}$ (LTE) and SD$_{X,w}$ (\ac{nlte}) are the weighted standard deviations of the individual \ac{lte} and \ac{nlte} measurements, respectively. 
}

\end{table}

A more detailed comparison of \ac{lte} and \ac{nlte} abundances is given in Table~\ref{tab:ltevsnlte} for NGC~104 and NGC~7078. For each \ac{nlte} element we list the abundance correction $\Delta_\mathrm{NLTE}$ and the weighted standard deviations of the \ac{lte} and \ac{nlte} measurements. The general trends with metallicity noted above are again apparent from this table. The correction for iron is very small for NGC~104 ($+0.01$~dex) and increases to $+0.05$~dex for NGC~7078. In both cases, the dispersions SD$_{\mathrm{Fe},w}$ on [Fe/H] remain the same, which is a consequence of the relatively small dispersion in the corrections themselves ($\sigma_\mathrm{NLTE}(\mathrm{[Fe/H]}) = 0.01$~dex for NGC~104 and $\sigma_\mathrm{NLTE}(\mathrm{[Fe/H]}) =0.039$~dex for NGC~7078) compared to the overall spread in the abundance measurements between the different windows. Except for Na, all corrections are smaller than 0.1~dex for NGC~104, while larger corrections are found for NGC~7078 for several elements. With a few exceptions, the dispersions SD$_{X,w}$ decrease for the \ac{nlte} abundances, which reinforces the need to account for \ac{nlte} effects in accurate studies of chemical abundances of stellar populations.

\begin{figure}
\centering
\includegraphics[width=\columnwidth]{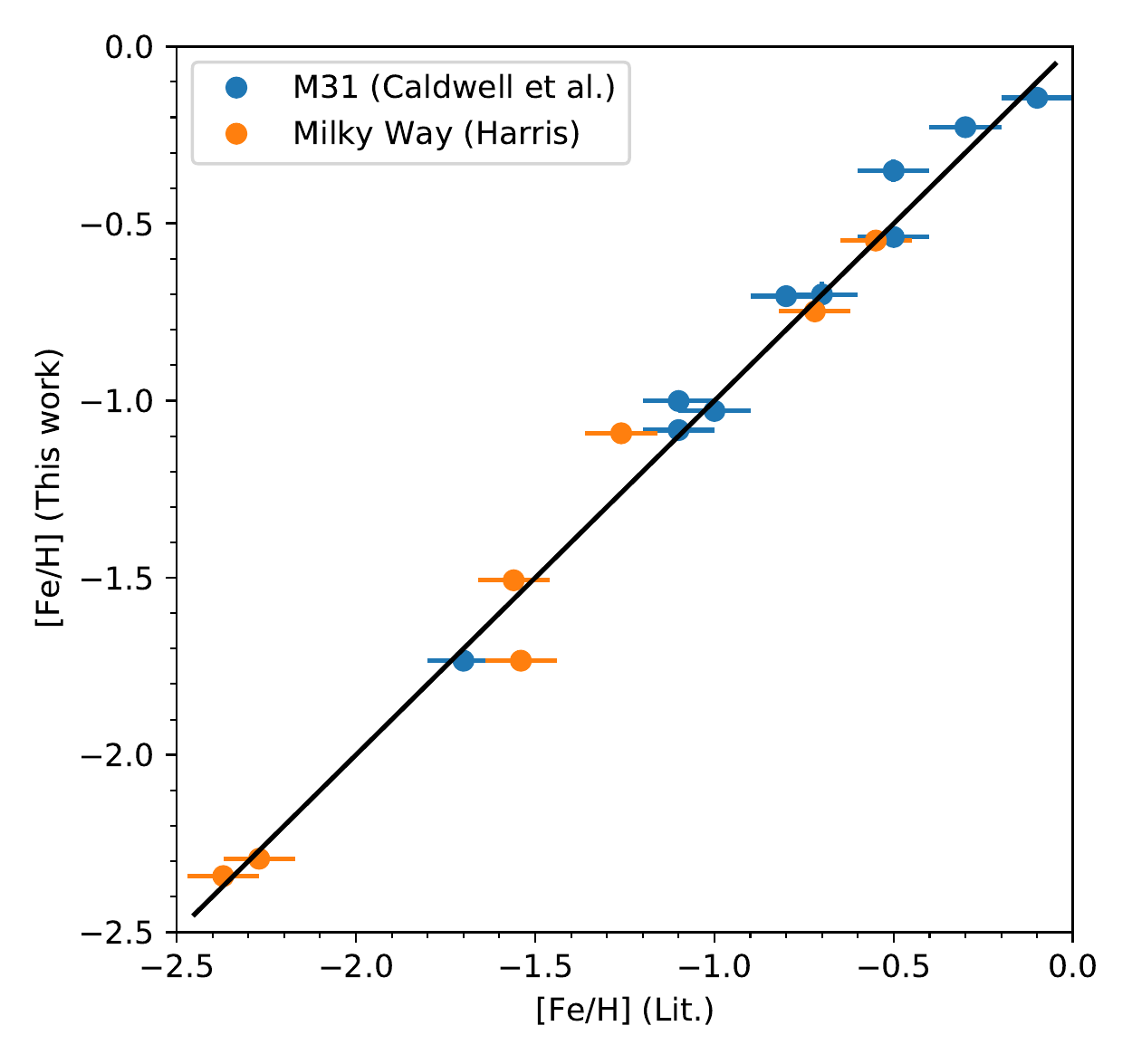}
\caption{\label{fig:cmpfeh}Comparison of our integrated-light \ac{lte} $\mathrm{[Fe/H]}$ values for M31 and Milky Way GCs  with data from \citet{Caldwell2011} and \citet{Harris1996}. The line is the 1:1 relation, not a fit. }
\end{figure}

In Fig.~\ref{fig:cmpfeh} we compare our integrated-light $\mathrm{[Fe/H]}$ measurements with data for GCs in the Milky Way \citep[][2010 revision]{Harris1996} and M31 \citep{Caldwell2011}. The former are based on data compiled from the literature while the latter are based on measurements of Lick indices on Hectospec data. To be consistent with the literature data, which are mostly based on \ac{lte} analyses (ultimately tied to the \citetalias{Carretta2009} and \citealt{Carretta1997} abundance scales), we plot our \ac{lte} measurements in this figure. 
The straight line represents the one-to-one relation and is not a fit to the data. We see that our measurements agree well with the literature data, with a standard deviation around the 1:1 relation of 0.09~dex. The mean metallicity offset between our measurements and those listed in \citet{Harris1996} is
$\langle \Delta \mathrm{[Fe/H]}\rangle_\mathrm{MW} = 0.001$~dex, while for M31 it is
$\langle \Delta \mathrm{[Fe/H]}\rangle_\mathrm{M31} = 0.029$~dex. For the combined sample we find $\langle \Delta \mathrm{[Fe/H]}\rangle_\mathrm{All} = 0.017$~dex.
If we instead compare with our \ac{nlte} abundances, the offset for the combined sample increases to 0.037~dex.
Overall, then, the cluster metallicities derived from our integrated-light analysis agree well with existing data for these well-studied clusters, with no evident systematic trends as a function of metallicity.  

\begin{table}
\caption{Comparison of our integrated-light \ac{nlte} abundance measurements with \ac{nlte} measurements for individual stars \citep[][K2019]{Kovalev2019}.}
\label{tab:kov2019}
\centering
\begin{tabular}{lccc} \hline\hline
& [Fe/H] & [Mg/Fe] & [Ti/Fe] \\
\hline
K2019 \\
NGC~104  & $-0.74\pm0.03$ & $+0.38\pm0.05$ & $+0.30\pm0.07$ \\
NGC~362  & $-1.05\pm0.04$ & $+0.15\pm0.06$ & $+0.29\pm0.06$ \\
NGC~6752 & $-1.48\pm0.06$ & $+0.20\pm0.09$ & $+0.17\pm0.07$ \\
NGC~7078 & $-2.28\pm0.06$ & $+0.22\pm0.19$ & $+0.21\pm0.05$ \\
\multicolumn{3}{l}{This work (from Table~\ref{tab:results1})} \\
NGC~104 & $-0.74\pm0.02$ & $+0.40\pm0.06$ & $+0.41\pm0.03$ \\
NGC~362 & $-1.07\pm0.02$ & $+0.17\pm0.03$ & $+0.42\pm0.03$ \\
NGC~6752 & $-1.70\pm0.01$ & $+0.31\pm0.07$ & $+0.36\pm0.03$ \\
NGC~7078 & $-2.29\pm0.02$ & $+0.15\pm0.04$ & $+0.43\pm0.05$ \\
\hline
\end{tabular}

\end{table}

In addition to the detailed comparison of literature data carried out for NGC~104 in Sect.~\ref{sec:47tuc}, it is of interest to compare our abundance measurements for Galactic \acp{gc} with the \ac{nlte} analysis by \citet[][K2019]{Kovalev2019}. These authors measured abundances for individual stars in four \acp{gc} in common with our sample, namely NGC~104, NGC~362, NGC~6752, and NGC~7078. We list their average \ac{nlte} measurements of [Fe/H], [Mg/Fe], and [Ti/Fe], together with our own \ac{nlte} measurements (from Table~\ref{tab:results1}) in Table~\ref{tab:kov2019}. 
For all clusters except NGC~6752, the $\mathrm{[Fe/H]}$ values are in agreement to within 0.02~dex. 
The case of NGC~6752 has been discussed in \citetalias{Larsen2017}, where it was noted that the relatively low surface brightness and large diameter on the sky make the integrated-light spectrum of this cluster especially susceptible to stochastic fluctuations in the number of red giants. This causes a stochastic uncertainty of 0.1~dex in [Fe/H] for this cluster, which is not included in the uncertainty in Table~\ref{tab:kov2019}. 
The $\mathrm{[Mg/Fe]}$ values agree within the uncertainties, while our $\mathrm{[Ti/Fe]}$ values tend to be somewhat higher (by 0.1--0.2~dex) than those of K2019, which is likely due to us using an updated and more accurate \ac{nlte} model atom of Ti in this work. 

\section{Results}

We next give an overview of previous work on the \ac{gc} systems in the galaxies included in our sample and discuss various aspects of the analysis specific to each galaxy. 

\subsection{Remarks on globular cluster systems of individual galaxies}
\label{sec:individual_remarks}

\paragraph{Fornax}

The \ac{gc} system of the Fornax dSph has an interesting history. 
The brightest cluster, Fornax~3, was discovered long before the Fornax dwarf itself by John Herschel, who described it as `a curious little object and easily mistaken for a star, which, however, it certainly is not' \citep{Herschel1847}. \citet{Shapley1939} added the clusters now generally referred to as Fornax~2 and Fornax~4 and also listed  `a very faint cluster of unidentified character'. This description of Fornax~6 remains quite appropriate today, although spectroscopic observations indicate that it is more metal-rich ($\mathrm{[Fe/H]}=-0.71\pm0.05$), and probably younger, than the other clusters \citep{Pace2021}. Clusters 1 and 5, the outermost known clusters, were discovered by \citet{Hodge1961}. Fornax~1 and 2 are relatively extended, diffuse objects, not ideal for our integrated-light measurements, but from high-dispersion spectroscopy of individual stars these clusters were found to be metal-poor ($\mathrm{[Fe/H]}=-2.5$ and $-2.1$, respectively) and $\alpha$-enhanced \citep{Letarte2006}.

The metallicities measured here for Fornax~3 ($\mathrm{[Fe/H]}=-2.28\pm0.02$) and Fornax~5 ($\mathrm{[Fe/H]}=-2.06\pm0.03$) are very similar to those determined in our previous work, while our new determination of $\mathrm{[Fe/H]}=-1.24$ ($\mathrm{[Fe/H]}=-1.26$ in the \ac{lte} analysis) for Fornax~4 is 0.10--0.15~dex higher  \citep{Larsen2012a,Larsen2018}. 
Within a projected galactocentric radius of 0.6$^\circ$, approximately corresponding to the location of the outermost \acp{gc}, the field star metallicity distribution in Fornax is instead dominated by a component with $\mathrm{[Fe/H]}\simeq-1$ \citep{Battaglia2006,Kirby2011a}.
While this component becomes less prominent beyond 0.6$^\circ$ \citep{Hendricks2014}, the conclusion that the \acp{gc} (except Fornax~6) are significantly more metal-poor than the bulk of the field stars in Fornax thus remains unchanged. 
Fornax~3 and Fornax~5 are (again) found to be moderately $\alpha$-enhanced, and the $\alpha$-element abundances of Fornax~4 remain close to scaled-solar.

Fornax~3 was included in the integrated-light study by \citet{Colucci2017}, who found a metallicity very similar to ours ($\mathrm{[Fe/H]}=-2.27\pm0.05$ from \ion{Fe}{i}). They also reported enhanced $\alpha$-element abundances ($\mathrm{[Mg/Fe]} = +0.21\pm0.22$, $\mathrm{[Ca/Fe]} = +0.24\pm0.10$, [\ion{Ti}{i}/Fe] = $+0.29\pm0.06$, and [\ion{Ti}{ii}/Fe] = $+0.35\pm0.10$) that are consistent with our \ac{lte} values, although they found a somewhat lower Ba abundance ($\mathrm{[Ba/Fe]} = +0.12\pm0.08$) than our measurement. 

While we have assumed an age of 13 Gyr for Fornax~4, some studies have reported a somewhat younger age of $\sim10$~Gyr \citep{Buonanno1999,Hendricks2016,Martocchia2020}. Indeed, a younger age would be consistent with the less $\alpha$-enhanced composition. 
Repeating our analysis for an age of 10~Gyr instead of 13~Gyr makes little difference, with $\mathrm{[Fe/H]}$ increasing very slightly (by 0.006~dex) and the $\alpha$-element abundance ratios changing by 0.01--0.02~dex. 

The composition and nature of Fornax~4 have been subjects of some debate in the literature. \citet{Strader2003} found $\mathrm{[Fe/H]}=-1.5$ and listed literature metallicities ranging between $\mathrm{[Fe/H]}=-2.0$ and $\mathrm{[Fe/H]}=-1.35$. Some of these values are based on measurements of various spectral features that are sensitive not only to Fe, but also to elements such as Ca and Mg. Given that the conversions from line strengths and/or indices to metallicities in some of these studies are calibrated on Galactic \acp{gc} that tend to have $\alpha$-enhanced abundance patterns \citep{Brodie1990,Dubath1992}, the resulting metallicities may be biased by the scaled-solar abundance patterns in Fornax~4. \citet{Hendricks2016} measured $\mathrm{[Fe/H]} = -1.50\pm0.05$ for one likely member star, again somewhat lower than our measurement. Their $\alpha$-element abundances were largely consistent with the roughly scaled-solar composition found from our measurements, but with large uncertainties ($\mathrm{[Si/Fe]} = -0.35\pm0.34$, $\mathrm{[Ca/Fe]} =+0.05\pm0.08$, and $\mathrm{[Ti/Fe]} = -0.27\pm0.23$).

Despite a small radial velocity offset 
(6--9~km~s$^{-1}$; \citealt{Larsen2012a,Hendricks2016})
between Fornax~4 and the field stars, the location of the cluster near the centre of the Fornax dSph has led to suggestions that it may be the nucleus \citep{Hardy2002,Strader2003}. This idea has recently been revisited by \citet{Martocchia2020}, who found a larger colour spread on the \ac{rgb} than could be accounted for by the measurement errors, suggesting a metallicity spread of up to $\approx0.5$~dex. While this should not seriously affect the comparison of integrated-light measurements, a robust comparison with individual stars would require larger samples than the single star of \citet{Hendricks2016}.

\paragraph{Wolf-Lundmark-Melotte} 

The WLM galaxy hosts a single known, but quite luminous ($M_V=-9$), \ac{gc} \citep{Humason1956,Sandage1985,Larsen2014}. A \ac{cmd} from the \textit{Hubble} Space Telescope (HST) shows well-defined giant- and horizontal branches, and confirms the object as an old, metal-poor \ac{gc} \citep{Hodge1999}. 
Our measurement of the iron abundance, $\mathrm{[Fe/H]}=-1.85\pm0.03$, is about 0.11~dex higher than that listed in \citet{Larsen2014}, which is mainly due to the revised microturbulence prescription and the inclusion of \ac{nlte} corrections. However, it is still significantly lower than that listed by \citet{Colucci2011} who found $\mathrm{[Fe/H]}=-1.71\pm0.03$. 
In any case, the \ac{gc} in WLM has a metallicity well below that of the general old field population, which has an average $\mathrm{[Fe/H]}=-1.3$ \citep{Leaman2013}. 
Similar to Fornax, WLM is thus in agreement with the general tendency for \acp{gc} to preferentially be associated with the more metal-poor populations in their parent galaxies \citep{Forte1981,Brodie1991,Harris2007,Lamers2017}.
\citet{Colucci2011} also found $\mathrm{[Ca/Fe]}=+0.25\pm0.05$ and $\mathrm{[Ba/Fe]}=-0.05\pm0.15$, which agrees well with our measurements. 

\paragraph{NGC~147}

This M31 satellite galaxy has ten known \acp{gc}. Six of these were discovered relatively recently, with three each by \citet{Sharina2009} and \citet{Veljanoski2013} who also provide a review of the earlier literature on the \ac{gc} system in NGC~147, going back to \citet{Baade1944a}. 

Our observations of \acp{gc} in NGC~147 were previously presented in \citet{Larsen2018}. 
The metallicities derived here are slightly higher than those found previously ($\sim0.1$ dex), due to the updated assumptions in our modelling procedure which also lead to slight changes in the abundances of individual elements. The overall conclusions remain unchanged, that is, the \acp{gc} are more metal-poor than the bulk of the field stars in NGC~147 and the $\alpha$-element abundances are generally similar to those of Milky Way \acp{gc} and field stars at similar metallicities. For Hodge~III,  $\mathrm{[Mg/Fe]}$  remains lower than the other $\alpha$-element abundance ratios, which now also include $\mathrm{[Si/Fe]}$. As discussed in more detail below (Sect.~\ref{sec:abundance_trends}), the abundances of Mg deviate from those of other $\alpha$-elements in several other \acp{gc} in our sample.

\paragraph{NGC~185}

As in NGC~147, the presence of \acp{gc} in NGC~185 was first noted by \citet{Baade1944a}. Today, a total of eight \acp{gc} are known in NGC~185 \citep{Veljanoski2013}. 
Interestingly, the three \acp{gc} that we have observed in this M31 companion all have nearly identical $\mathrm{[Fe/H]}\simeq-1.75$, and the $\alpha$-elements (Si, Ca, and Ti) are enhanced by similar amounts as in Milky Way \acp{gc} at this metallicity. Larger variations are seen in the abundances of Na and Mg, perhaps related to the presence of multiple populations.

Among the clusters observed here, FJJ-III and FJJ-V were also included in the sample of \citet{DaCosta1988}, whose determinations of the iron abundances are very similar to ours
($\mathrm{[Fe/H]}=-1.7\pm0.15$ for FJJ-III and $\mathrm{[Fe/H]}=-1.8\pm0.15$ for FJJ-V). 
All three \acp{gc} observed here are in common with the study by \citet{Sharina2006}, who found metallicities of
$\mathrm{[Z/H]} = -1.6\pm0.3$ (FJJ-III), $-1.5\pm0.2$ (FJJ-V), and $-1.5\pm0.3$ (FJJ-VIII).
They also measured $\alpha$-element abundance ratios of $[\alpha/\mathrm{Fe}] = +0.1\pm0.3$, $0.0\pm0.3$, and $0.0\pm0.3$, for the three clusters. 
There is a tendency for these [$\alpha/$Fe] ratios to be closer to solar than indicated by our measurements, although the results for individual clusters are largely consistent with our $\alpha$-enhanced values within the relatively large uncertainties on the \citet{Sharina2006} measurements. 
It is less clear how to compare their total metallicities, $\mathrm{[Z/H]}$, with our measurements of $\mathrm{[Fe/H]}$. If we use the relation $\mathrm{[Fe/H]} = \mathrm{[Z/H]} - 0.75 \times \mathrm{[Mg/Fe]}$ \citep{Vazdekis2015} to
convert our $\mathrm{[Fe/H]}$ values to $\mathrm{[Z/H]}$ by adding a rough correction of $+0.2$~dex (assuming $\mathrm{[Mg/Fe]}\approx0.3$) they become similar to the $\mathrm{[Z/H]}$ values obtained by \citet{Sharina2006}, although this procedure may be questionable given the roughly scaled-solar $\alpha$-element abundances found by the latter study.

\paragraph{NGC~205}

The current list of known \acp{gc} in NGC~205 has changed little with respect to the eight objects identified by \citet{Hubble1932}. Hubble~III has a radial velocity that suggests it is most likely a projected M31 \ac{gc} and Hubble~V may be an intermediate-age object. An additional candidate listed by \citet{Sargent1977}, M31C-55, has subsequently been identified as a foreground star \citep{Battistini1987,DaCosta1988,Galleti2004} and indeed has a non-zero proper motion according to Gaia EDR3 (pm(RA,DEC) = ($-0.47\pm0.11$, $-0.96\pm0.10$) mas yr$^{-1}$) which would imply a velocity of about 4000~km~s$^{-1}$ at the distance of NGC~205.

The two \acp{gc} included in our sample, Hubble~I and Hubble~II, both have enhanced $\alpha$-element abundance ratios according to our analysis.  Our metallicity determinations of $\mathrm{[Fe/H]}=-1.41\pm0.03$ (Hubble~I) and $\mathrm{[Fe/H]}=-1.35\pm0.02$ (Hubble~II) agree fairly well with those by \citet{DaCosta1988},  $\mathrm{[Fe/H]}=-1.5\pm0.15$ for both clusters, and with the value of $\mathrm{[Fe/H]}=-1.49\pm0.02$ found for Hubble~I by \citet{Colucci2011}. The latter authors also found both clusters to be $\alpha$-enhanced, although they found a somewhat higher metallicity for Hubble~II ($\mathrm{[Fe/H]}=-1.12\pm0.02$). \citet{Sharina2006} found relatively high metallicities for both clusters
($\mathrm{[Z/H]} = -1.1\pm0.1$ for Hubble~I and $\mathrm{[Z/H]} = -1.2\pm0.1$ for Hubble~II)
and lower $\alpha$-element abundance ratios ($[\alpha/\mathrm{Fe}] = +0.2\pm0.2$ and $[\alpha/\mathrm{Fe}] = 0.0\pm0.2$). 

\paragraph{NGC~6822}

\citet{Hubble1925} listed ten `nebulae' in NGC~6822, of which he tentatively suggested that a few might be stellar clusters. One of these, Hubble~VII, was confirmed as an old, metal-poor \ac{gc} by \citet{Cohen1998}, who found an age of $11^{+4}_{-3}$~Gyr and a metallicity of $\mathrm{[Fe/H]}=-1.95\pm0.15$~dex from spectroscopic line index measurements. A \ac{cmd} obtained from HST data is also consistent with that of an old, relatively metal-poor \ac{gc} \citep{Wyder2000}.
An additional seven \acp{gc}, located outside the main body of NGC~6822, were discovered more recently on ground-based wide-field CCD images \citep{Hwang2011,Huxor2013}.

Our analysis of the clusters SC6 and SC7 was discussed in \citet{Larsen2018}. As for most other clusters that we have reanalysed here, the metallicities have increased by about 0.1~dex (mainly due to the revised microturbulence prescription), but we recover our previous result that the $\alpha$-element abundances for SC7 are close to scaled-solar while SC6 exhibits the $\alpha$-enhanced abundance patterns that are characteristic of other metal-poor \acp{gc} in our sample. 

In addition to SC6 and SC7, we here include the cluster Hubble~VII although the analysis of its spectrum  posed some difficulties. When plotting the abundance measurements as a function of wavelength, the spectral windows with wavelengths $\lambda>5200$~\AA\ tended to yield higher Fe abundances than the bluer windows. Specifically, using only windows with $\lambda<5200$~\AA\ we found $\mathrm{[Fe/H]}=-1.77$ (with an rms of 0.13~dex) while for $\lambda>5200$~\AA\ we found $\mathrm{[Fe/H]}=-1.58$ (rms 0.23~dex). 
Such a trend could potentially be caused by a mismatch between the \ac{hrd} of the actual cluster and that used in the modelling, for example if an incorrect age were assumed. However, the trend persisted even if an age as young as 5~Gyr was assumed: in this case we still found a difference of 0.16~dex between the $\mathrm{[Fe/H]}$ values for $\lambda<5200$~\AA\ and $\lambda>5200$~\AA .

As Hubble~VII is projected onto the main body of NGC~6822, where the background is more complicated, one might suspect poor background subtraction as a possible culprit. We compared our HIRES spectrum with a spectrum obtained with the Keck Cosmic Web Imager (KCWI) and found the two spectra to be very similar (apart from the lower spectral resolution of KCWI). 
Hence, we have not been able to identify a satisfactory explanation for the difficulties with the analysis of Hubble~VII, and the results should therefore be considered more uncertain than for the other clusters in our sample. 
Nevertheless, our analysis agrees fairly well with that of \citet{Colucci2011} who measured $\mathrm{[Fe/H]}=-1.61\pm0.02$, just 0.06~dex higher than our value (or 0.08~dex when compared with our \ac{lte} analysis). In both cases, these are significantly higher iron abundances than the value measured by \citet{Cohen1998}.
\citet{Colucci2011} also found an approximately scaled-solar Ca abundance of $\mathrm{[Ca/Fe]}=+0.01\pm0.07$, which is similar to our measurement of $\mathrm{[Ca/Fe]}=+0.07\pm0.04$ ($+0.06$ in \ac{lte}). Nevertheless, the detailed abundance patterns remain somewhat puzzling, as we find Si and Ti to be enhanced while Mg is depleted ($-0.03$~dex). We also find a slightly higher Ba abundance than \citet{Colucci2011}, $\mathrm{[Ba/Fe]}=+0.35\pm0.08$~dex ($+0.48$ in \ac{lte}) vs.\ their $\mathrm{[Ba/Fe]}=+0.22\pm0.13$. Hubble~VII thus appears worthy of further study. 

\paragraph{M31}

\begin{table*}
\caption{Comparison of our LTE measurements for M31 GCs with literature data.}
\label{tab:m31cmp}
\centering
{\small
\begin{tabular}{llrrrrrr}
\hline\hline
& & 006-058 & 012-064 & 163-217 &171-222 & 225-280 & 358-219 \\ \hline
$\mathrm{[Fe/H]}$ & 
This work & $-0.54$ & $-1.73$ & $-0.15$ & $-0.23$ & $-0.35$ & $-2.25$ \\
& C2014 & $-0.73$ & $-1.61$ & $-0.49$ & $-0.45$ & $-0.66$ & $-2.21$\\
& S2016 & $-0.69$ & $-1.60$ & $-0.42$ & $-0.52$ & $-0.64$ & \ldots \\
$\mathrm{[Na/Fe]}$ & 
This work & $+0.46$ & $+0.24$ & $+0.65$ & $+0.58$ & $+0.54$ & $+0.34$ \\
& C2014 & $+0.52$ & $+0.62$ & $+0.72$ & $+0.73$ & \ldots & \ldots \\
& S2016 & $+0.39$ & \ldots & $+0.57$ & $+0.57$ & \ldots & \ldots \\
$\mathrm{[Mg/Fe]}$ & 
This work & $+0.34$ & $+0.05$ & $+0.22$ & $+0.27$ & $+0.27$ & $+0.18$ \\
& C2014 & $+0.32$ & $-0.08$ & $+0.25$ & $+0.18$ & $+0.45$ & $-0.03$ \\
& S2016 & $+0.43$ & $-0.14$ & $+0.22$ & $+0.37$ & $+0.24$ & \ldots \\
$\mathrm{[Si/Fe]}$ & 
This work & $+0.32$ & $+0.71$ & $+0.29$ & $+0.25$ & $+0.36$ & $+0.42$ \\
& C2014 & $+0.48$ & $+0.35$ & $+0.25$ & $+0.45$ & $+0.49$ & \ldots \\
& S2016 & $+0.37$ & $+0.43$ & $+0.19$ & $+0.27$ & $+0.32$ & \ldots \\
$\mathrm{[Ca/Fe]}$ & 
This work & $+0.25$ & $+0.33$ & $+0.11$ & $+0.11$ & $+0.15$ & $+0.34$ \\
& C2014 & $+0.25$ & $+0.40$ & $+0.28$ & $+0.27$ & $+0.40$ & $+0.27$ \\
& S2016 & $+0.31$ & \ldots & $+0.27$ & $+0.30$ & $+0.34$ & \ldots \\
$\mathrm{[Ti/Fe]}$ & 
This work & $+0.29$ & $+0.18$ & $+0.27$ & $+0.23$ & $+0.38$ & $+0.18$ \\
& C2014 & $+0.22$ & $+0.34$ & $+0.25$ & $-0.06$ & $+0.29$ & $+0.23$ \\
& S2016 & $+0.43$ & \ldots & $+0.22$ & $+0.35$ & $+0.39$ & \ldots \\
\hline
\end{tabular}
}
\tablefoot{For C2014 we give $\mathrm{[Ti/Fe]}$ as the weighted average of \ion{Ti}{i} and \ion{Ti}{ii}.
}
\end{table*}

Since the initial discovery of 140 `nebulous objects \ldots provisionally identified as globular clusters' by \citet{Hubble1932}, the M31 \ac{gc} system has been the subject of numerous studies. We do not attempt to review the literature here, but only note that the spectroscopic observations analysed here were made prior to the identification of about 100 \acp{gc} in the PAndAS survey, mostly located in the outer halo of M31 \citep{Huxor2014}. Many of these are, in any case, relatively extended, and therefore less efficiently observed with a single-slit spectrograph such as HIRES. 

The M31 \acp{gc} in our sample span a \ac{nlte} metallicity range from $\mathrm{[Fe/H]}=-2.8$ (EXT8) to $\mathrm{[Fe/H]}=-0.1$ (163-217). They all show enhanced abundances of Si, Ca, and Ti relative to scaled-solar composition, although EXT8 is extremely Mg-deficient \citep{Larsen2020}. In general, the abundance patterns tend to align with those observed in Milky Way \acp{gc} of similar metallicities. 
From integrated-light spectroscopy, the inner $\sim10\arcmin$ of M31 itself are dominated by old, $\alpha$-enhanced ($[\alpha/\mathrm{Fe}]\simeq0.25$) stars with near-solar metallicities, associated with a bulge or bar-like component \citep{Saglia2009,Saglia2018}. Hence, the enhanced $\alpha$-element abundances of the metal-rich \acp{gc} are similar to those seen in metal-rich stars in the central regions of M31 and may trace the same components at larger galactocentric distances. The three most metal-rich M31 \acp{gc} in our sample are located at projected distances of $7\farcm8$ (171-222), $13\farcm2$ (163-217), and $20\farcm5$ (225-280).

The clusters analysed here are among the brightest \acp{gc} in M31, and many of them have been included in several spectroscopic studies in the past.  In addition to the comparison with the low-dispersion metallicity measurements in Fig.~\ref{fig:cmpfeh}, it is of interest to compare with previous analyses that made use of high-dispersion spectroscopy and analysis techniques similar to ours. In Table~\ref{tab:m31cmp} we compare our \ac{lte} abundance measurements with those published by \citet[][C2014]{Colucci2014}, also using HIRES observations (but from different runs than those used here) and \citet[][S2016]{Sakari2016} (using APOGEE $H$-band observations). We do not include the errors on the measurements in Table~\ref{tab:m31cmp} but they typically amount to $\sim\pm0.1$~dex (see Tables~\ref{tab:results1}-\ref{tab:results3} and the original references). 
 
As noted by \citetalias{Colucci2014}, their measurements of $\mathrm{[Fe/H]}$ show a systematic difference with respect to low-resolution studies in the sense that their [Fe/H] values tend to be lower for metal-rich \acp{gc}.
As discussed above (Sect.~\ref{sec:anal}), we instead find excellent agreement between our iron abundances and those measured by \citet{Caldwell2011} at all metallicities, so that our $\mathrm{[Fe/H]}$ values are systematically higher than those of \citetalias{Colucci2014} and \citetalias{Sakari2016} at the high-metallicity end. At low metallicities (012-064 and 358-219), all three studies agree well on $\mathrm{[Fe/H]}$. 

Despite the systematic differences in the iron abundances, we generally find similar abundance ratios to those measured by \citetalias{Colucci2014} and \citetalias{Sakari2016}. There are some occasional outliers, such as $\mathrm{[Na/Fe]}$ and $\mathrm{[Si/Fe]}$ for 012-064 (where our measurements differ by 0.3-0.4~dex  from those of \citetalias{Colucci2014} and \citetalias{Sakari2016}) and $\mathrm{[Mg/Fe]}$ for 225-280, where the value measured by \citetalias{Colucci2014} is about 0.2~dex higher, but otherwise the three studies mostly agree within $\sim0.1$~dex.

\paragraph{M33}

The characterisation of the M33 \ac{gc} system is, to a large extent, still a work in progress. Part of the difficulty stems from the fact that the \ac{gc} population is relatively sparse (certainly when compared to that of M31), and most of the clusters appear projected onto the highly structured disc of M33. 
This difficulty is compounded by the presence of a large number of intermediate-age clusters which may easily be confused with ancient \acp{gc}, especially if reddened.
Indeed, the prevalence of clusters with blue colours compared to Galactic \acp{gc} was noted already in early studies \citep{Hiltner1960,Kron1960} and was later confirmed by larger samples \citep{Melnick1978,Christian1982}.

Our sample of M33 \acp{gc} mostly consists of clusters with resolved HST photometry from  \citet{Sarajedini2000}, to which we have added CBF~28 \citep{Chandar1999} and the cluster HM33-B from \citet{Huxor2009}. 
We had planned to include several additional \ac{gc} candidates from \citet{Beasley2015} which however turned out to be Galactic foreground stars, as confirmed by their non-zero parallaxes and proper motions listed in the Gaia DR2 catalogue (IDs 1115, 1138, 1322, 1965, 2001, and 2145 in \citealt{Beasley2015}). The observations of the four clusters R14, U77, CBF~28, and HM33-B are published here for the first time, while analysis of data for H38, M9, R12, and U49 was included in \citet{Larsen2018}.

Despite the suggestion that the M33 \acp{gc} are several Gyr younger than their Milky Way counterparts \citep{Sarajedini2000}, most of them have enhanced $\alpha$-element abundances similar to those seen in \acp{gc} in other galaxies. However, it is interesting to note that HM33-B appears chemically more similar to Fornax~4 and NGC~6822-SC7 in some respects (we discuss this further in Sect.~\ref{sec:discussion}), and is also one of a few \acp{gc} located well outside the main body of M33 \citep{Huxor2009}. This is reminiscent of the behaviour of stars in the outer M31 halo, which also tend to have abundances more similar to those seen in dwarf galaxies \citep{Gilbert2020}.

\paragraph{NGC~2403}

At a distance of about 3.2~Mpc \citep{Freedman1988,Radburn-Smith2011}, this low-mass Sc-type spiral (similar to M33) is the only galaxy in our sample that is located outside the Local Group. A list of five cluster candidates was given by \citet{Tammann1968}, of which one had a red colour consistent with it being a \ac{gc}. The cluster included in our sample, F46, was first identified by \citet{Battistini1984}, who listed a relatively blue colour ($B-V=0.47$) and suggested that it might be an intermediate-age object. However, according to SDSS DR12 photometry \citep{Alam2015}, F46 has $(g-z)_0 = 0.92\pm0.02$, which is quite normal for an old \ac{gc} with a metallicity around $\mathrm{[Fe/H]}\simeq-1.5$ \citep{Vanderbeke2013a}. 

The spectrum of F46 is of good quality, despite the relatively modest integration time (Table~\ref{tab:obs}) and most elemental abundances are well constrained. It helps that the cluster is very compact and the spectrum was obtained on a night with good seeing ($0\farcs8$ FWHM), so that slit losses were minimised. 
The cluster is moderately metal-poor ($\mathrm{[Fe/H]} = -1.71\pm0.03$) and the $\alpha$-elements are mostly enhanced, with the exception of $\mathrm{[Mg/Fe]}$ which is \emph{sub}-solar.  In this respect F46 resembles several other clusters with metallicities around $\mathrm{[Fe/H]}\approx-1.7$.

We can use the velocity broadening obtained from the spectral analysis to constrain the virial mass and mass-to-light ratio of F46. Correcting the total broadening in Table~\ref{tab:obs} for an instrumental broadening of 3.4~km~s$^{-1}$ \citep{Larsen2020}, the line-of-sight velocity broadening is $\sigma_\mathrm{1D}=11.7$~km~s$^{-1}$. We used an HST/ACS archival image in the F555W filter (Progr. ID 10402, P.I. R.\ Chandar) to estimate the half-light radius of F46 by measuring the flux in concentric apertures. 
Visually, the cluster can be traced out to a radius of about 50~pixels ($2\farcs5$) in the ACS image, and within this radius the total magnitude is about $m_\mathrm{F555W} = 17.8$, slightly brighter than the value of $V=17.96$ given by \citet{Battistini1984}. The ACS photometry may however be affected by multiple cosmic-ray hits in the single exposure. 
Correcting the measurement in the ACS image for a foreground extinction of $A_V=0.11$~mag (NED) we get $M_V=-9.8$.
Half of the flux is contained within a radius of 6 ACS pixels or $0\farcs30$, corresponding to a half-light radius of $r_h=4.6$~pc at a distance of 3.2~Mpc. Subtracting the half-light radius of the ACS point-spread function \citep[about 1.5 pixels;][]{Bohlin2016} in quadrature decreases $r_h$ by about 0.15~pc. The dynamical mass is then
\begin{equation}
    M_\mathrm{dyn} = 10 \frac{\sigma_\mathrm{1D}^2 r_h}{G} \simeq 1.4\times10^6 \, M_\odot
\end{equation}
and the mass-to-light ratio is $M_\mathrm{dyn}/L_V \simeq 2.0 \, M_\odot/L_{V,\odot}$, which is a typical value for an old \ac{gc} at this metallicity \citep{Strader2011}. For an age of $10^8-10^9$~years, as suggested by \citet{Battistini1984}, the $V$-band mass-to-light ratio would be only 5\%--15\% of that at 10~Gyr \citep[e.g.][]{Bruzual2003}, in strong contrast to the measured value.
We conclude that F46 is indeed most likely a relatively metal-poor, old \ac{gc}.

\begin{figure}
\centering
\includegraphics[width=\columnwidth]{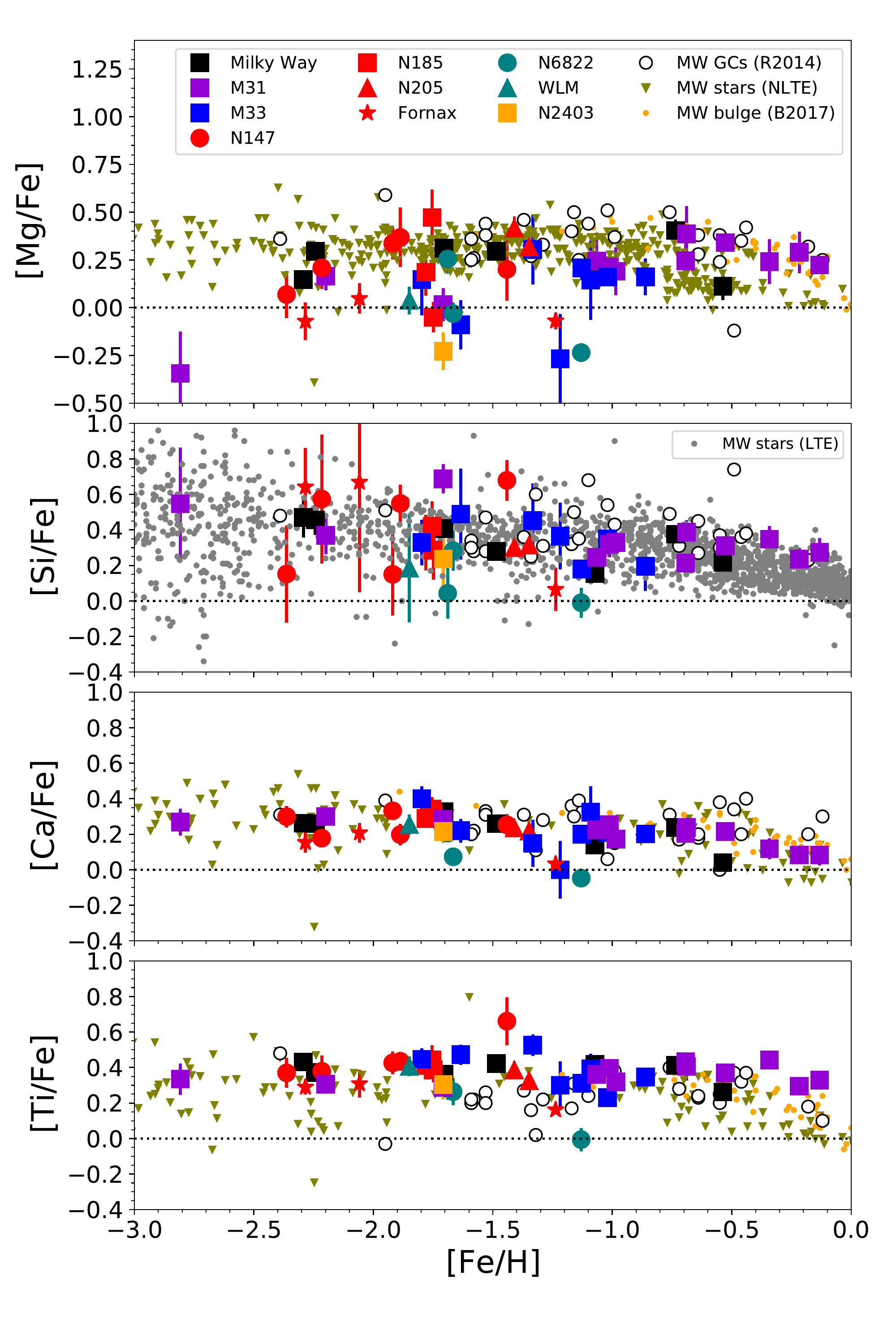}
\caption{\label{fig:alphafe}Integrated-light abundances of $\alpha$-elements relative to iron (Table~\ref{tab:results1}). Large symbols show our measurements, with symbol colours and shapes encoding information about the host galaxies as indicated in the legend.
Error bars indicate $\sigma_{\langle X \rangle}$ or $S_X$, whichever is larger.
Olive coloured small triangles: NLTE measurements for Milky Way field stars \citep{Bergemann2017,Mashonkina2017,Mashonkina2019,Mishenina2017,Zhao2016}. Orange points: Milky Way bulge stars \citep{Bensby2017}.
Grey points in panel with [Si/Fe]: LTE measurements for Milky Way field stars \citep{Suda2008}.
}
\end{figure}

\begin{figure}
\centering
\includegraphics[width=\columnwidth]{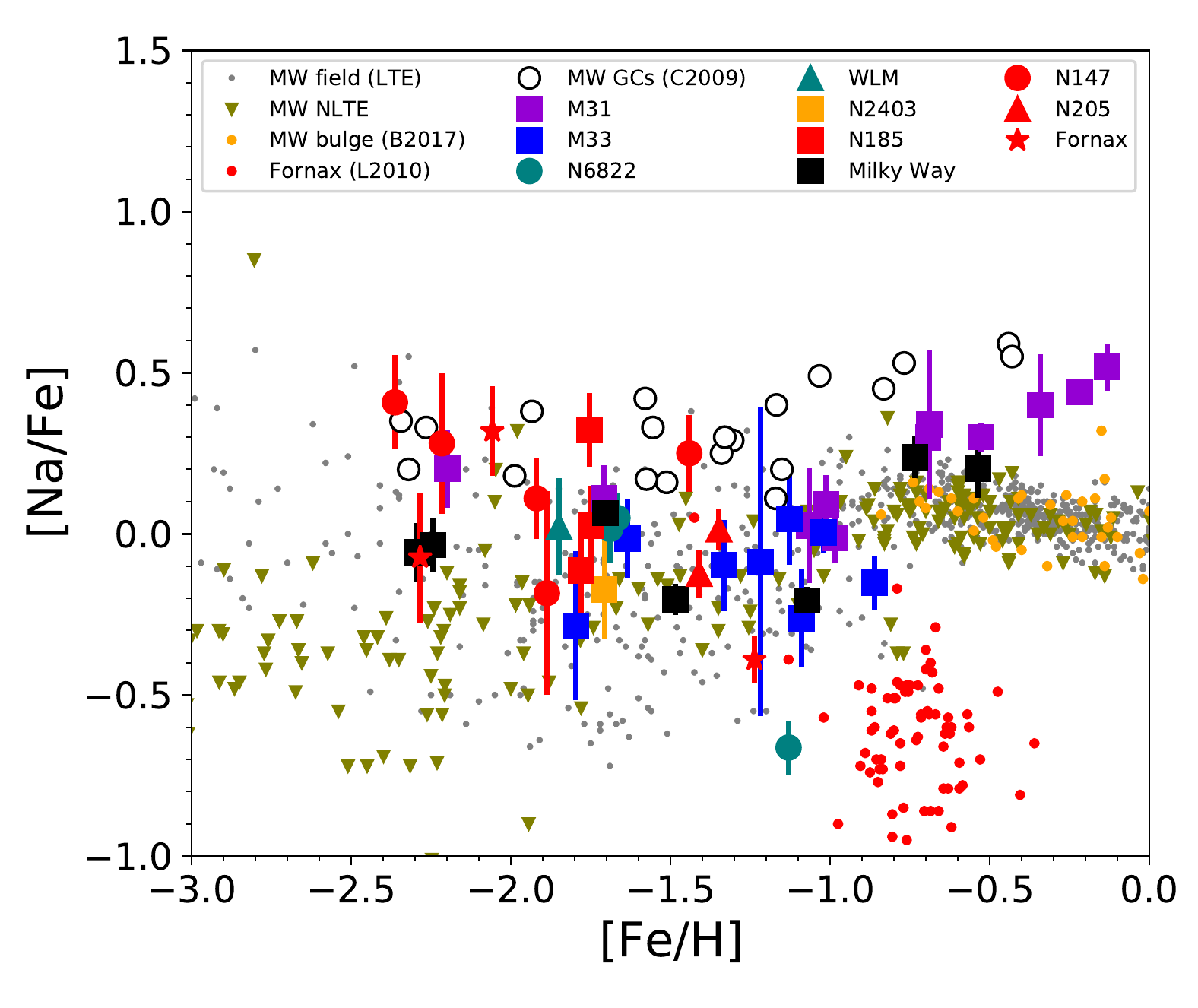}
\caption{\label{fig:nafe}Sodium abundances (NLTE). Symbols for our integrated-light measurements are the same as in Fig.~\ref{fig:alphafe}. References for Milky Way \acp{gc} (open circles), Milky Way field star LTE data (grey points) and NLTE data (olive coloured triangles), and field stars in the Fornax dwarf (small red circles) are given in the text.}
\end{figure}

\subsection{Abundance trends} 
\label{sec:abundance_trends}

\begin{figure}
\centering
\includegraphics[width=\columnwidth]{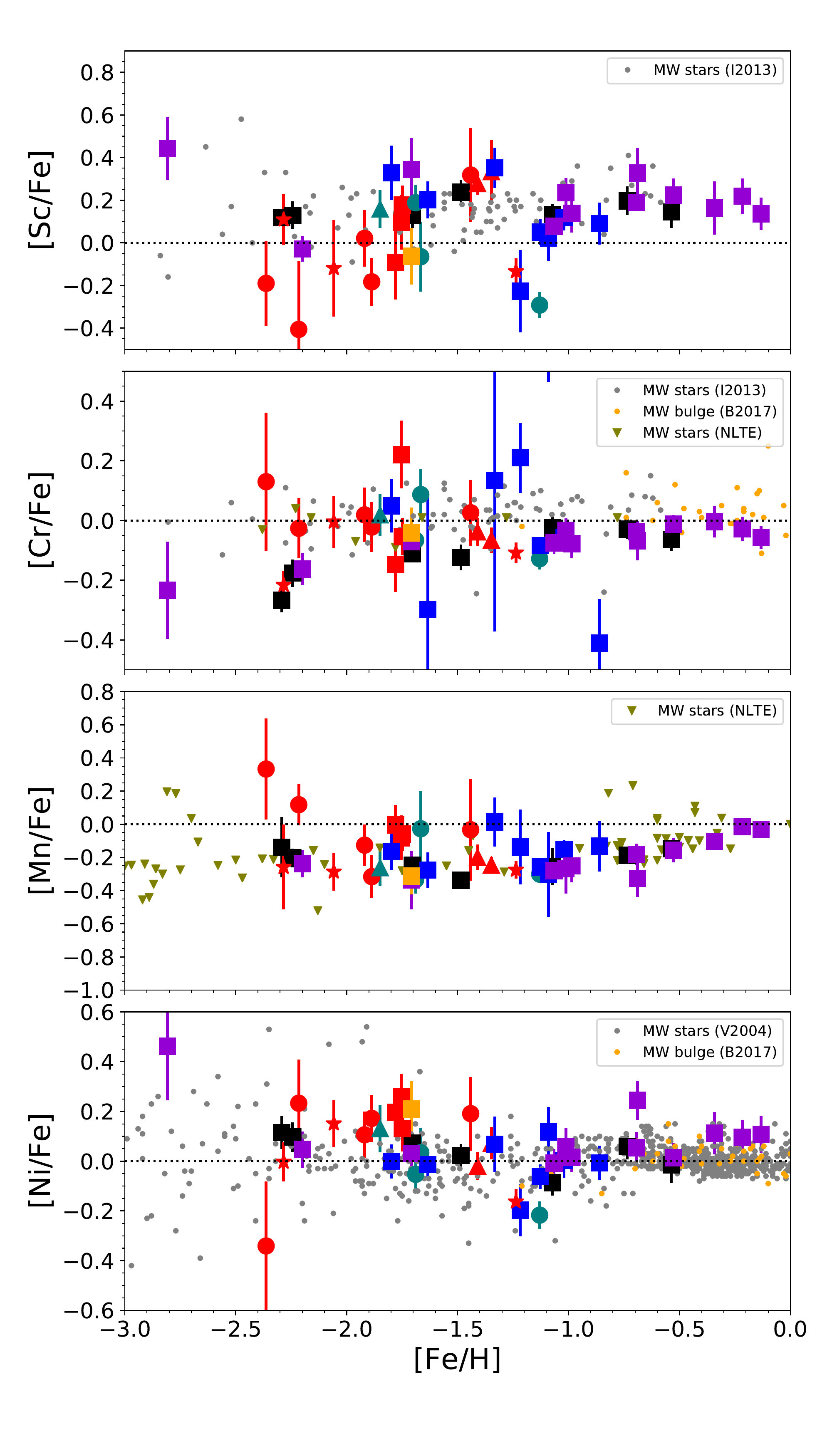}
\caption{\label{fig:fepeak}Abundances of iron-peak elements (Table~\ref{tab:results2}). Symbols for our measurements are the same as in Fig.~\ref{fig:alphafe}. NLTE abundances for Mn in Milky Way field stars are from \citet{Eitner2020}.}
\end{figure}

\begin{figure}
\centering
\includegraphics[width=\columnwidth]{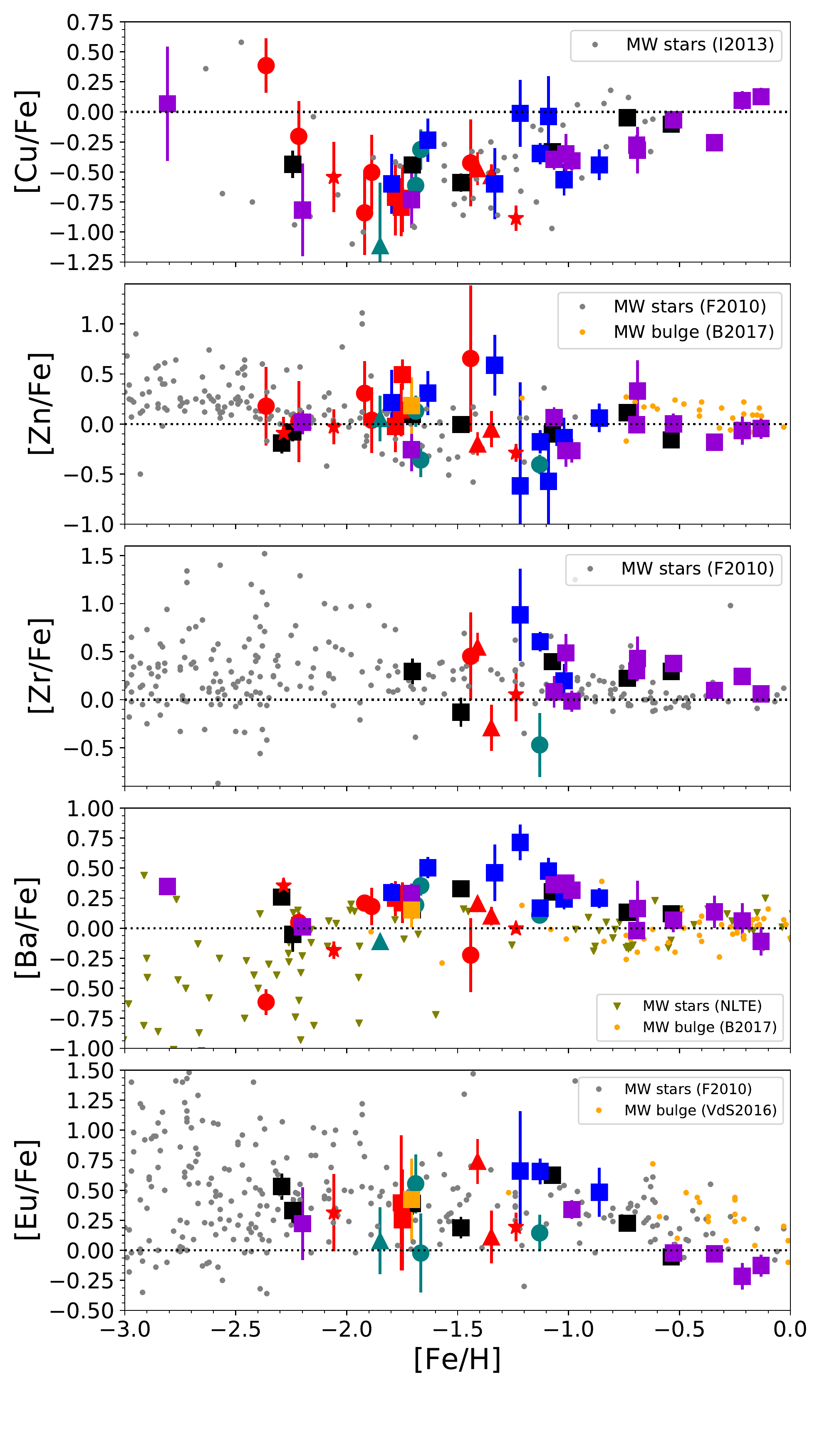}
\caption{\label{fig:heavy}Abundances of heavy elements (Table~\ref{tab:results3}). Symbols for our measurements are the same as in Fig.~\ref{fig:alphafe}.} 
\end{figure}

Having discussed the individual \ac{gc} systems, we next turn to a more general discussion of the abundance trends and comparison with existing measurements of individual stars. 
In Fig.~\ref{fig:alphafe}-\ref{fig:heavy} we plot the measurements  in Tables~\ref{tab:results1}-\ref{tab:results3} as a function of $\mathrm{[Fe/H]}$. In each figure, the  symbols for our integrated-light measurements are colour-coded according to the type of host galaxy, using the same scheme as in Fig.~\ref{fig:femv}. Red symbols indicate \acp{gc} belonging to dwarf spheroidals and ellipticals (NGC~147, NGC~185, NGC~205, and the Fornax dSph) and teal-colour symbols indicate dwarf irregulars (NGC~6822 and WLM). The Milky Way, M31, and M33 \acp{gc} are shown with black, violet, and blue squares, respectively, and the single \ac{gc} in NGC~2403 is represented by an orange square. 

\paragraph{The $\alpha$-elements}

Figure~\ref{fig:alphafe} shows the $\alpha$-element abundance ratios ($\mathrm{[Mg/Fe]}$, $\mathrm{[Si/Fe]}$, $\mathrm{[Ca/Fe]}$, and $\mathrm{[Ti/Fe]}$). 
Although the classification of Ti as an $\alpha$-element may be questionable \citep{Woosley1995}, it usually tends to trace other $\alpha$-elements (such as Mg and Si) fairly closely \citep{Conroy2014,Parikh2019}, and here we group it together with the $\alpha$-elements.
For Mg, Ca, and Ti we include \ac{nlte} measurements for Milky Way field stars \citep{Zhao2016,Mashonkina2017,Mashonkina2019,Mishenina2017} and additionally, for Mg, data for 326 thick-disc stars from \citet{Bergemann2017}. For Si (where our measurements are \ac{lte} values) we show literature data from the Stellar Abundances for Galactic Archeology (SAGA) database \citep{Suda2008}.  The increasing scatter at metallicities below $\mathrm{[Fe/H]}=-2$ for the literature [Si/Fe] data is caused partly by larger measurement uncertainties, although uncertainties are not provided for all stars. We also include measurements for microlensed Milky Way bulge stars from \citet{Bensby2017} and literature data for individual stars in Galactic \acp{gc} \citep[][R2014]{Roediger2014}, the latter indicated with open circles. 

Perhaps the most notable feature of Fig.~\ref{fig:alphafe} is the high degree of uniformity of the abundance patterns for Si, Ca, and Ti. All three elements are enhanced relative to scaled-solar composition with mean abundance ratios of $\langle \mathrm{[Si/Fe]}\rangle = +0.32\pm0.01$, 
$\langle \mathrm{[Ca/Fe]}\rangle = +0.21\pm0.01$, and
$\langle \mathrm{[Ti/Fe]}\rangle = +0.35\pm0.01$. The dispersions of the $\mathrm{[Ca/Fe]}$ and $\mathrm{[Ti/Fe]}$ values are only 0.09~dex and 0.10~dex, respectively, while the $\mathrm{[Si/Fe]}$ values have a larger dispersion (0.17~dex), driven at least in part by the larger uncertainties. Above $\mathrm{[Fe/H]}\approx-0.5$, $\mathrm{[Ca/Fe]}$ shows a declining trend towards higher metallicities while Ti and Si remain elevated throughout the entire metallicity range. We thus confirm the $\alpha$-element like behaviour of Ti. 

Overall, the abundance ratios [Si/Fe], [Ca/Fe], and [Ti/Fe] for our sample are fairly similar to those
seen in the \citetalias{Roediger2014} data and in the Milky Way field stars. The larger differences for Mg are discussed in more detail below. We note that the \ac{nlte} corrections tend to slightly decrease our [Ca/Fe] values and increase the [Ti/Fe] values (Table~\ref{tab:ltevsnlte}). There may indeed be a tendency for our abundances for these elements to be slightly offset in the corresponding directions from the \citetalias{Roediger2014} data, which are generally based on \ac{lte} analyses of individual \ac{gc} member stars.  Slight offsets could, however, also be caused by other effects, such as differences in the atomic data and adopted solar abundance scales (Sect.~\ref{sec:arcturus}). In their compilation, \citetalias{Roediger2014} did not attempt to homogenise the data to account for such differences. 

Superimposed on the general homogeneity of the $\alpha$-element abundance patterns are some higher-order differences. 
Two \acp{gc}, Fornax~4 and NGC~6822-SC7, have consistently low $\alpha$-element abundances. As discussed in \citet{Larsen2018}, this mirrors the tendency for the `knee' in the $\alpha$-element versus [Fe/H] relation to occur at a lower metallicity in dwarf galaxies than in the Milky Way. Our present sample includes an additional cluster with a hint of this behaviour, HM33-B, albeit at lower significance. This cluster has relatively low $\mathrm{[Mg/Fe]}$ and $\mathrm{[Ca/Fe]}$ ratios although the $\mathrm{[Si/Fe]}$ and $\mathrm{[Ti/Fe]}$ ratios are more similar to those of the Galactic stars at this metallicity. The S/N ratio of the HM33-B spectrum is relatively low, and the uncertainties on the abundances thus relatively large, and obtaining a better spectrum of this cluster would be desirable.
At metallicities below those of Fornax~4, NGC~6822-SC7, and HM33-B, $\mathrm{[Fe/H]}<-1.3$, there is a small but systematic difference between the mean $\alpha$-element abundance ratios for the \acp{gc} in spirals and those in dwarf galaxies. For the metal-poor \acp{gc} in spirals we find
$\langle \mathrm{[Si/Fe]}\rangle_\mathrm{MP,spir} = +0.40\pm0.02$, 
$\langle \mathrm{[Ca/Fe]}\rangle_\mathrm{MP,spir} = +0.28\pm0.01$, and
$\langle \mathrm{[Ti/Fe]}\rangle_\mathrm{MP,spir} = +0.38\pm0.01$,
while the corresponding mean values for \acp{gc} in the dwarf galaxies are
$\langle \mathrm{[Si/Fe]}\rangle_\mathrm{MP,dwarf} = +0.35\pm0.03$, 
$\langle \mathrm{[Ca/Fe]}\rangle_\mathrm{MP,dwarf} = +0.23\pm0.01$, and
$\langle \mathrm{[Ti/Fe]}\rangle_\mathrm{MP,dwarf} = +0.36\pm0.01$.
Hence, the \acp{gc} in the spiral galaxies are systematically more $\alpha$-enhanced by about 0.04~dex.

For Mg the situation is more complicated than for the other $\alpha$-elements. The $\mathrm{[Mg/Fe]}$ ratio has the largest dispersion of the $\alpha$-elements ($\sigma_\mathrm{[Mg/Fe]} = 0.19$~dex) and a number of \acp{gc} scatter well below the relation followed by the field stars. 
The tendency for the integrated-light $\mathrm{[Mg/Fe]}$ measurements to fall below those of the other $\alpha$-elements has been noted in several previous studies of extragalactic \acp{gc} \citep{Colucci2009,Colucci2014,Larsen2012a,Larsen2018,Sakari2015} and it has been suggested that it may be caused by internal Mg spreads in the clusters similar to those observed in some Galactic \acp{gc}.  One difficulty with this interpretation is the relatively large fraction of clusters with integrated-light measurements that appear to be affected \citep{Larsen2018}, and another is that the Mg spreads observed in Galactic \acp{gc} appear insufficient produce the observed effect \citep{Pancino2017}. 
In the context of galactic chemical evolution, low [Mg/Fe] values can be produced by a variety of mechanisms, such as gas inflows or extended star formation histories \citep{Buck2020,Buck2021}, but are usually expected to be accompanied by a general deficit of the $\alpha$-elements, unlike the situation for the \acp{gc}.
Hence there is currently no satisfactory explanation for these unusually low Mg abundances.
Apart from Fornax~4, NGC~6822-SC7, and HM33-B, which are also deficient in other $\alpha$-elements, it is interesting to note that the clusters with strongly depleted Mg abundances are mainly found at $\mathrm{[Fe/H]}<-1.5$. Indeed, the most Mg-deficient cluster, EXT8, is also the most metal-poor cluster in our sample. 

\paragraph{Sodium}

Our integrated-light measurements of $\mathrm{[Na/Fe]}$ are shown in Fig.~\ref{fig:nafe} along with 
Galactic field star samples that include \ac{nlte} corrections \citep{Gehren2004,Gehren2006,Mishenina2017,Mashonkina2017,Zhao2016} and the larger \ac{lte} field star samples of \citet[][V2004]{Venn2004} and \citet[][I2013]{Ishigaki2013}. Also included are (\ac{lte}) data for field stars in the Fornax dwarf spheroidal galaxy \citep[][L2010]{Letarte2010} and average $\mathrm{[Na/Fe]}$ values for individual stars in Galactic \acp{gc} \citepalias{Carretta2009}. The latter include \ac{nlte} corrections from \citet{Gratton1999}.

The behaviour of $\mathrm{[Na/Fe]}$ as a function of metallicity and environment is rather complex and the Na abundances in \acp{gc} are, furthermore, affected by intra-cluster abundance spreads associated with multiple populations. The latter most likely account for the tendency for the \ac{gc} $\mathrm{[Na/Fe]}$ values to lie above the bulk of the field star measurements, especially at low and high metallicities. At intermediate metallicities, $-1.5 \la \mathrm{[Fe/H]} \la -1.0$, the offset of the integrated-light measurements with respect to the field star \ac{nlte} literature data is less evident. However, it is worth noting that the trends of $\mathrm{[Na/Fe]}$ vs. $\mathrm{[Fe/H]}$ in the \ac{nlte} corrected field samples are quite different from those seen in the larger samples of \citetalias{Venn2004} and \citetalias{Ishigaki2013}: the latter display a larger scatter in $\mathrm{[Na/Fe]}$ at $\mathrm{[Fe/H]}\la-1$ and reach lower $\mathrm{[Na/Fe]}$ values at intermediate metallicities. Given that our \ac{nlte} corrections for Na tend to be negative, which also tends to be the case for individual stars \citep{Gehren2004,Takeda2003}, a NLTE correction would most likely shift the \citetalias{Venn2004} and \citetalias{Ishigaki2013} data to lower [Na/Fe] values. 
The tendency for [Na/Fe] to reach significantly sub-solar values at low to intermediate metallicities is characteristic of halo populations (\citetalias{Ishigaki2013}; \citealt{Zhao2016}), which are less well represented at intermediate metallicities in the \ac{nlte} comparison samples, but may provide a more direct comparison with the \ac{gc} measurements.

The $\approx+0.3$~dex offset between our integrated-light $\mathrm{[Na/Fe]}$ values and the average values from \citetalias{Carretta2009}, discussed above for 47~Tuc (Sect.~\ref{sec:47tuc}), is also evident in Fig.~\ref{fig:nafe}. If shifted downwards by 0.3~dex, the trend of $\mathrm{[Na/Fe]}$ vs.\ $\mathrm{[Fe/H]}$ seen in the \citetalias{Carretta2009} data would closely match that seen in our integrated-light measurements, including the transition from the scatter around scaled-solar abundance ratios in the range $\mathrm{[Fe/H]}<-1$ to a tighter sequence at super-solar $\mathrm{[Na/Fe]}$ values at higher metallicities.
We recall that a difference of 0.1~dex between our [Na/Fe] values and those measured by \citetalias{Carretta2009} is explained by the different solar abundance scales, but this still leaves 0.2~dex unaccounted for. One possibility could be the use of significantly over-estimated \ac{nlte} corrections in the \citetalias{Carretta2009} analysis. Their work relies on the \ac{nlte} model atom by \citet{Gratton1999}, who employed empirically-calibrated collision rates based on the \citet{Drawin1965} formula. Our analysis, however, relies on state-of-the-art ab initio data for Na$+$H collisions that are available from detailed quantum-mechanical calculations. As \citet{Lind2011} demonstrated previously, the differences in the collisional rates may account for up to $+0.2$ dex difference in \ac{nlte} abundances. In particular,  \citet{Lind2011} also hint a significant difference between their results and those by \citet{Gratton1999}, in the sense that the latter study leads to positive \ac{nlte} corrections, in contrast with predictions with more accurate H collision data (see also \citealt{Asplund2005}). Our \ac{nlte} corrections for Na lines are negative and are in agreement with the \citet{Lind2011} findings, which suggests that the \ac{nlte} Na abundances from \citetalias{Carretta2009} are over-estimated.

While it is tempting to interpret strongly elevated integrated-light $\mathrm{[Na/Fe]}$ values as evidence of multiple populations in the \acp{gc}, this must be tempered by the caveat that the Na abundances of field stars in other galaxies may differ from those observed in the Milky Way. In Fig.~\ref{fig:nafe}, this is evident from the difference between the Na abundances of stars in the Fornax dSph and those in the Milky Way. Similarly sub-solar [Na/Fe] abundance ratios are observed in field stars in other Local Group dwarf galaxies such as Sagittarius \citep{Sbordone2007,Hasselquist2017},
Draco \citep{Cohen2009},
Ursa Minor \citep{Cohen2010} and 
Sculptor \citep{Salgado2019}, while massive early-type galaxies are often found to exhibit super-solar $\mathrm{[Na/Fe]}$ ratios that correlate with velocity dispersion \citep{Conroy2014,Worthey2014,LaBarbera2017}.
In the nuclear regions of M31, Na enhancements as high as $\mathrm{[Na/Fe]}\approx+1.0$~dex have been measured \citep{Conroy2012b}. 
As in previous studies \citep{Colucci2014,Sakari2016},  we find the metal-rich \acp{gc} in M31 to be quite Na-rich, too, albeit not reaching the extreme enhancement found by \citet{Conroy2012b}.

\paragraph{Iron-peak elements: Sc, Cr, Mn, Ni}

Our measurements of the iron-peak elements are shown in Fig.~\ref{fig:fepeak}. Corrections for \ac{nlte} effects are included for Mn and Ni. For field stars we include \ac{nlte} Mn abundances from \citet{Eitner2019}. While our integrated-light measurements do not include \ac{nlte} corrections for Cr, we have included \ac{nlte} data for field dwarfs from \citet{Bergemann2010}. These do not differ strongly from the \ac{lte} data from \citetalias{Ishigaki2013}. 

The iron-peak elements mostly tend to follow the trends observed in Milky Way samples and display only moderate departures from scaled-solar abundance patterns.  
The $\mathrm{[Sc/Fe]}$ ratios tend to be moderately enhanced compared to scaled-solar composition in both the \ac{gc} and field star measurements, with the exception of Fornax~4, NGC~6822-SC7, and HM33-B. A few of the most metal-poor \acp{gc} (mainly associated with NGC~147) also show hints of sub-solar [Sc/Fe] values, but these measurements all have fairly large uncertainties.
Manganese is moderately sub-solar over most of the metallicity range (by about 0.2~dex), but the strongly depleted $\mathrm{[Mn/Fe]}$ ratios typically seen at low metallicities in \ac{lte} analyses of \ion{Mn}{I} lines, which can be as low as $\mathrm{[Mn/Fe]} \approx -1.0$ \citep{Bonifacio2009}, are not present in our \ac{nlte} analysis. Our integrated-light $\mathrm{[Mn/Fe]}$ ratios closely follow those seen in \ac{nlte} analyses of Milky Way field stars, with the \acp{gc} that deviate from the general trend tending to have large uncertainties. 

Nickel is slightly enhanced relative to scaled-solar composition over most of the metallicity range, again with the exception of the clusters Fornax~4, NGC~6822-SC7, and HM33-B, which have clearly sub-solar [Ni/Fe].
The \ac{nlte} corrections for Ni are significant at low metallicities, reaching 0.2~dex at $\mathrm{[Fe/H]}=-2.5$ (see e.g.\ Table~\ref{tab:ltevsnlte}), which causes an offset between the integrated-light \ac{gc} measurements and the \ac{lte} field star data. 
Hence our \ac{gc} analysis does not show the slightly sub-solar [Ni/Fe] values that are typically quoted in the literature for halo stars (\citealt{Gratton1987}; \citetalias{Ishigaki2013}). It should be pointed out, however, that none of the Galactic chemical evolution models reproduces the observational trends across the entire metallicity range. The models by \citet{Kobayashi2019} underpredict the observed [Ni/Fe] values below [Fe/H] $\approx -0.7$, and the same trend is seen in the results by \citet{Palla2021}. It is not clear yet whether a
contribution from  rotating massive stars \citep[e.g.][]{Limongi2018} could explain this discrepancy.

\paragraph{Heavy elements: Cu, Zn, Zr, Ba, Eu}

Figure~\ref{fig:heavy} shows our measurements of the elements beyond the iron-peak, which are thought to be produced mainly via the (slow and rapid) neutron-capture processes. 
Our \ac{lte} measurements for Cu, Zn, Zr, and Eu are plotted together with Milky Way field star data from \citet[][F2010]{Frebel2010}, \citetalias{Ishigaki2013}, \citet[][B2017]{Bensby2017}, and \citet[][VdS2016]{VanderSwaelmen2016}, as indicated in the legends. 

The \ac{gc} data generally align well with the field stars, and there are no obvious differences between the abundance patterns for \acp{gc} in different types of galaxies, apart from the overall lower metallicities of the \acp{gc} in the dwarf galaxies. We note that neither Fornax~4, NGC~6822-SC7, or HM33-B are conspicuous outliers in any of these plots, in contrast to the situation for some $\alpha$- and iron-peak elements.

The top panel in Fig.~\ref{fig:heavy} reveals a strong correlation between $\mathrm{[Cu/Fe]}$ vs.\ $\mathrm{[Fe/H]}$, which has long been observed in field stars (\citealt{Sneden1991a}; \citetalias{Ishigaki2013}). However, recent work suggests that it may be caused mainly by \ac{nlte} effects \citep{Andrievsky2018}.
Zinc is mildly enhanced at the lowest metallicities and approaches scaled-solar abundances at near-solar metallicities, again consistent with the behaviour of the field stars \citep{Bensby2017,DaSilveira2018}. 

The measurements of Zr and Eu are more challenging due to the weakness of the lines and are only available for a subset of the clusters, mainly the more metal-rich ones. 
At $\mathrm{[Fe/H]}\la-1$, the \acp{gc} resemble the field stars in being somewhat enhanced in $\mathrm{[Zr/Fe]}$ on average, although the uncertainties on individual measurements are large.  At higher metallicities, the tendency for Zr to be enhanced is still apparent in the \acp{gc}, while the Galactic (disc) stars approach scaled-solar composition. Measurements of Zr in the bulge are scarce, but \citet{Johnson2012b} tentatively suggested that there may be two sequences, one being characterised by $\mathrm{[Zr/Fe]}\approx+0.25$ and the other by $\mathrm{[Zr/Fe]}\approx-0.10$. The metal-rich \acp{gc} would then appear to align with the Zr-rich sequence. 
For $\mathrm{[Eu/Fe]}$, one can recognise the same pattern observed in field stars, with a roughly constant level of Eu enhancement at lower metallicities ($\mathrm{[Fe/H]} \la -1$) and a decrease at higher metallicities that is also seen in Galactic bulge giants \citep{Johnson2012b,VanderSwaelmen2016}.
However, the decrease is more pronounced for the \acp{gc}, which reach [Eu/Fe] values even below those typical of the disc stars at high metallicities. While the Eu measurements are challenging, we note that the metal-rich Milky Way bulge \ac{gc} NGC~6553 has abundance patterns measured from individual stars that resemble our integrated-light measurements with a roughly solar [Eu/Fe] and enhanced $\alpha$-element abundances \citep{Barbuy1999,Alves-Brito2006}.
We do not recover the very high [Eu/Fe] values for the Fornax \acp{gc} from the analysis in \citet{Larsen2012a}. These were driven in part by the \ion{Eu}{ii} 4205~\AA\ line that is not included in the present analysis. 

The second panel from the bottom of Fig.~\ref{fig:heavy} shows our integrated-light \ac{nlte} measurements for Ba together with \ac{nlte} field star data \citep{Mashonkina2017,Mishenina2017,Zhao2016}. When synthesising the \ion{Ba}{ii} lines, we assumed the $r$-process isotopic ratios of \citet{McWilliam1998}. For $s$-process dominated isotopic mixture, the Ba abundances would be higher by about 0.1~dex, owing to the more dominant contribution from $^{138}$Ba which lacks hyperfine splitting. 
In previous versions of our analysis, the integrated-light $\mathrm{[Ba/Fe]}$ ratios were found to be slightly higher compared to literature data for field stars \citep{Larsen2014,Larsen2018}. A slight hint of this tendency may still be present in Fig.~\ref{fig:heavy}, especially for the M33 \acp{gc}, but the integrated-light \ac{gc} measurements mostly fall within the envelope defined by the field stars.

\subsection{Comparison with individual stars in host galaxies}
\label{sec:icmp}

\begin{figure}
\centering
\includegraphics[width=\columnwidth]{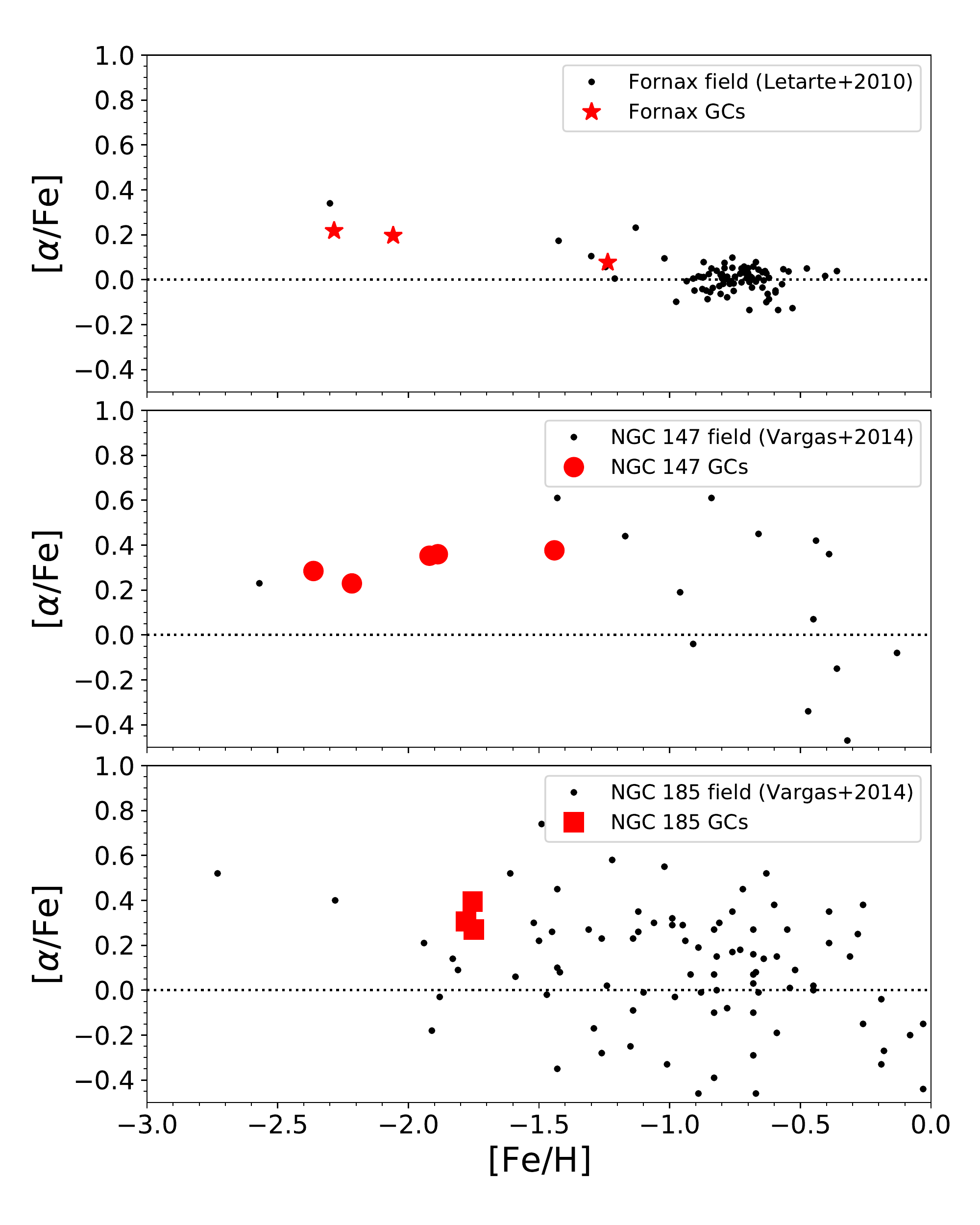}
\caption{\label{fig:alphacmp}Mean $\alpha$-element abundance ratios for GCs and field stars in NGC~147, NGC~185 \citep{Vargas2014}, and Fornax \citep{Letarte2010}.}
\end{figure}

For a few galaxies other than the Milky Way, we can directly compare our integrated-light \ac{gc} measurements with data for field stars for a subset of elements.  In Fig.~\ref{fig:alphacmp} we compare our integrated-light measurements of [$\alpha$/Fe] (the average of [Mg/Fe], [Si/Fe], [Ca/Fe], and [Ti/Fe]) for \acp{gc} in NGC~147, NGC~185, and the Fornax dSph with literature data for individual stars in these systems \citep{Letarte2010,Vargas2014}. This figure illustrates how the \acp{gc} complement data for individual stars: while $\alpha$-element abundances can be measured for individual \ac{rgb} stars in the Andromeda satellites, the scatter is large (mainly due to measurement uncertainties) and the field stars are preferentially more metal-rich than the \acp{gc}. In the Fornax dSph, abundance measurements for individual stars are more accurate, but the distinct difference in the metallicities of field stars and \acp{gc} remains evident. For all three galaxies, the \acp{gc} trace the mean trends seen in the field stars quite well.

\section{Discussion}
\label{sec:discussion}

\begin{figure}
\centering
\includegraphics[width=\columnwidth]{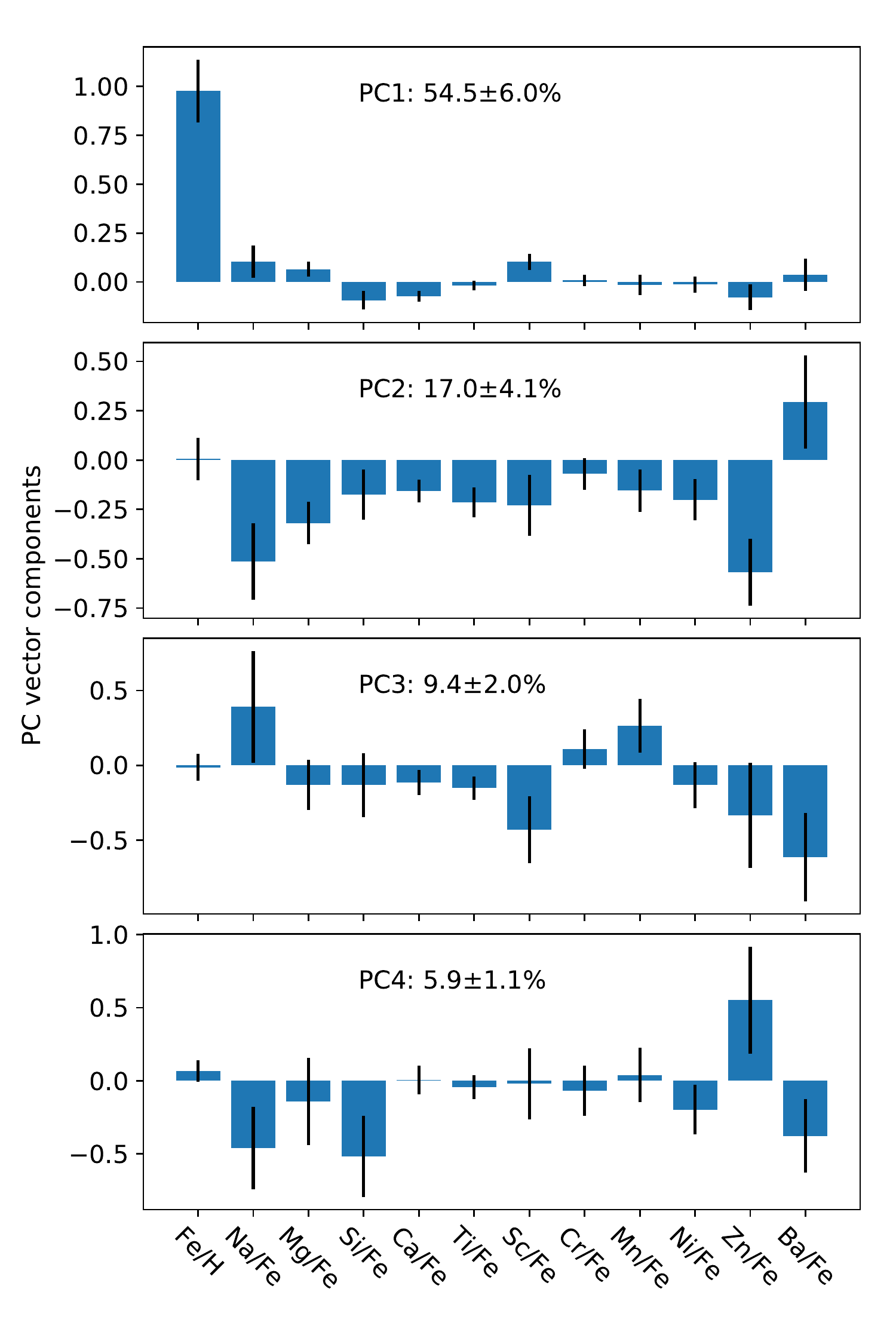}
\caption{\label{fig:pca}Principal components analysis of the abundance measurements, showing the first four principal components.}
\end{figure}

\begin{figure}
\centering
\includegraphics[width=\columnwidth]{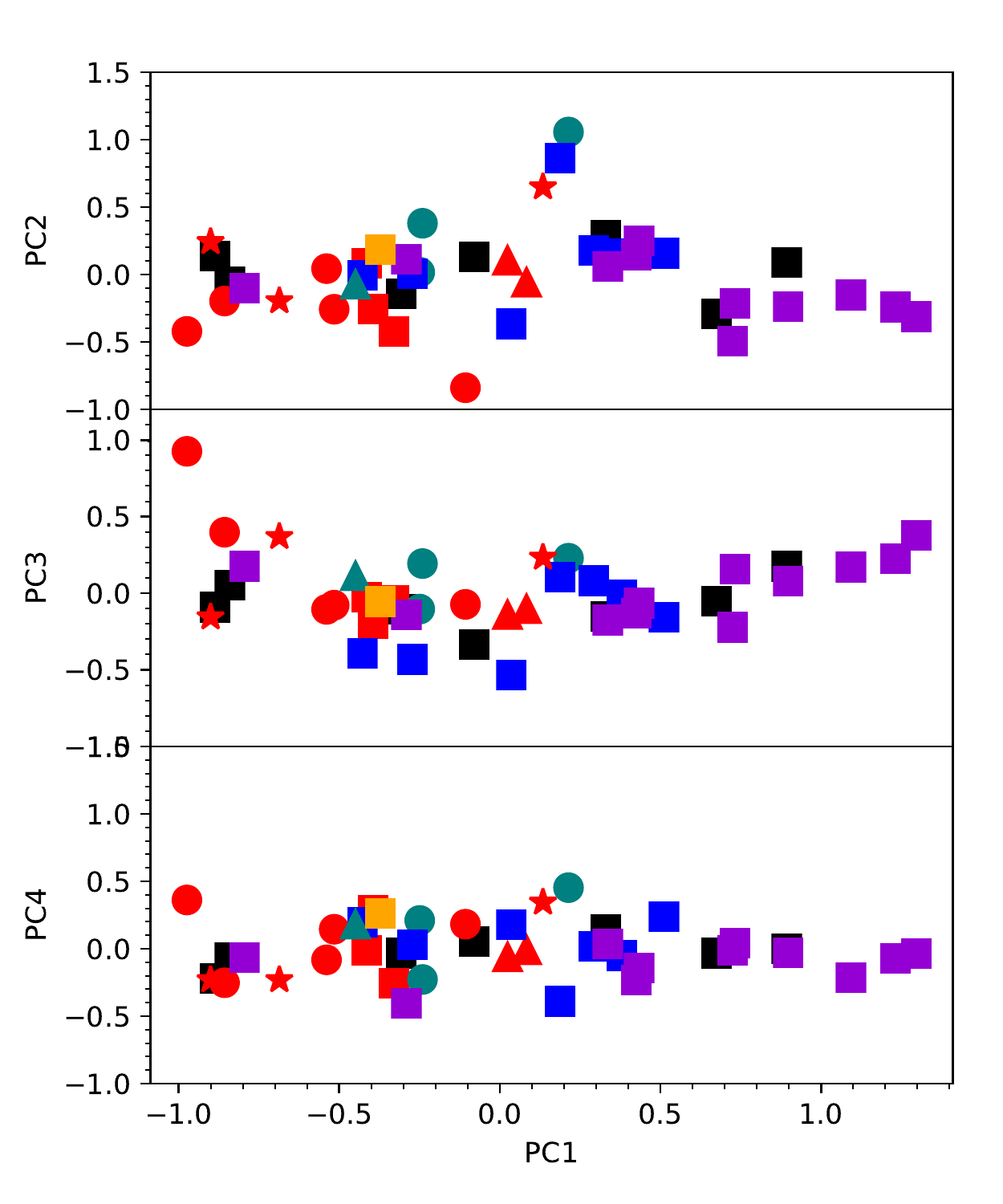}
\caption{\label{fig:pc}Second, third, and fourth PC as a function of the first PC for each cluster. Symbols are the same as in Fig.~\ref{fig:alphafe}.}
\end{figure}

\begin{table}
\caption{Projections of abundance measurement onto principal component vectors.}
\label{tab:pca}
\centering
\small
\begin{tabular}{l r r r r}
\hline\hline
 Cluster & PC1 & PC2 & PC3 & PC4 \\ \hline
NGC 0104        & $ 0.675$ & $-0.290$ & $-0.051$ & $-0.033$  \\
NGC 0362        & $ 0.331$ & $ 0.289$ & $-0.153$ & $ 0.143$  \\
NGC 6254        & $-0.080$ & $ 0.132$ & $-0.333$ & $ 0.053$  \\
NGC 6388        & $ 0.893$ & $ 0.087$ & $ 0.176$ & $ 0.001$  \\
NGC 6752        & $-0.308$ & $-0.143$ & $-0.104$ & $-0.032$  \\
NGC 7078        & $-0.887$ & $ 0.135$ & $-0.092$ & $-0.221$  \\
NGC 7099        & $-0.840$ & $-0.056$ & $ 0.054$ & $-0.066$  \\
N147 HII        & $-0.107$ & $-0.840$ & $-0.073$ & $ 0.182$  \\
N147 HIII       & $-0.974$ & $-0.422$ & $ 0.926$ & $ 0.362$  \\
N147 PA-1       & $-0.857$ & $-0.196$ & $ 0.398$ & $-0.252$  \\
N147 PA-2       & $-0.516$ & $-0.258$ & $-0.078$ & $ 0.145$  \\
N147 SD7        & $-0.540$ & $ 0.044$ & $-0.105$ & $-0.082$  \\
N185 FJJ-III    & $-0.329$ & $-0.423$ & $-0.046$ & $-0.255$  \\
N185 FJJ-V      & $-0.415$ & $ 0.085$ & $-0.027$ & $-0.011$  \\
N185 FJJ-VIII   & $-0.395$ & $-0.255$ & $-0.197$ & $ 0.283$  \\
N205 HubbleI    & $ 0.024$ & $ 0.106$ & $-0.139$ & $-0.059$  \\
N205 HubbleII   & $ 0.083$ & $-0.057$ & $-0.101$ & $-0.007$  \\
N6822 SC6       & $-0.249$ & $ 0.017$ & $-0.103$ & $ 0.210$  \\
N6822 SC7       & $ 0.214$ & $ 1.057$ & $ 0.229$ & $ 0.453$  \\
N6822 HVII      & $-0.241$ & $ 0.380$ & $ 0.194$ & $-0.228$  \\
M33 M9          & $-0.272$ & $ 0.006$ & $-0.432$ & $ 0.029$  \\
M33 R12         & $ 0.511$ & $ 0.158$ & $-0.157$ & $ 0.239$  \\
M33 U49         & $ 0.036$ & $-0.366$ & $-0.534$ & $ 0.179$  \\
M33 R14         & $ 0.380$ & $ 0.152$ & $-0.008$ & $-0.050$  \\
M33 U77         & $-0.428$ & $-0.007$ & $-0.393$ & $ 0.195$  \\
M33 CBF28       & $ 0.293$ & $ 0.178$ & $ 0.082$ & $ 0.017$  \\
M33 HM33B       & $ 0.187$ & $ 0.863$ & $ 0.106$ & $-0.390$  \\
WLM GC          & $-0.449$ & $-0.072$ & $ 0.114$ & $ 0.183$  \\
Fornax 3        & $-0.901$ & $ 0.243$ & $-0.156$ & $-0.231$  \\
Fornax 4        & $ 0.134$ & $ 0.649$ & $ 0.234$ & $ 0.344$  \\
Fornax 5        & $-0.687$ & $-0.196$ & $ 0.369$ & $-0.231$  \\
M31 006-058     & $ 0.898$ & $-0.236$ & $ 0.079$ & $-0.031$  \\
M31 012-064     & $-0.291$ & $ 0.112$ & $-0.140$ & $-0.404$  \\
M31 019-072     & $ 0.732$ & $-0.216$ & $ 0.156$ & $ 0.041$  \\
M31 058-119     & $ 0.434$ & $ 0.250$ & $-0.065$ & $-0.144$  \\
M31 082-114     & $ 0.725$ & $-0.494$ & $-0.223$ & $-0.013$  \\
M31 163-217     & $ 1.297$ & $-0.313$ & $ 0.379$ & $-0.035$  \\
M31 171-222     & $ 1.231$ & $-0.240$ & $ 0.226$ & $-0.069$  \\
M31 174-226     & $ 0.425$ & $ 0.146$ & $-0.131$ & $-0.231$  \\
M31 225-280     & $ 1.093$ & $-0.152$ & $ 0.170$ & $-0.216$  \\
M31 338-076     & $ 0.336$ & $ 0.059$ & $-0.175$ & $ 0.036$  \\
M31 358-219     & $-0.795$ & $-0.101$ & $ 0.176$ & $-0.063$  \\
N2403 F46       & $-0.372$ & $ 0.183$ & $-0.054$ & $ 0.260$  \\
\hline
\end{tabular}
\end{table}

The general homogeneity of the $\alpha$-element abundance patterns among the \acp{gc} in our sample, with at most minor differences between \acp{gc} in dwarf galaxies and in spirals, was noted in Sect.~\ref{sec:abundance_trends}. Indeed, the similarity is not restricted to the $\alpha$-elements, but also extends to most other elements. 
Three clusters, namely Fornax~4, NGC~6822-SC7, and HM33-B, have significantly different abundance patterns than the rest. From inspection of Figs.~\ref{fig:alphafe}-\ref{fig:heavy}, these three clusters have clearly depleted abundances of Na, Mg, Si, Ca, Ti, Sc, Ni, and possibly also Zn and Eu, compared to the other \acp{gc} in our sample. On the other hand, Cr and Mn do not appear to vary in a correlated way with these elements.

A more quantitative way to gain insight into correlations between abundances of different elements is via a Principal Components Analysis (PCA). We used the \texttt{scikit-learn} package \citep{Pedregosa2011} in Python to carry out a PCA for the abundance measurements [Fe/H], [Na/Fe], [Mg/Fe], [Si/Fe], [Ca/Fe], [Ti/Fe],  [Sc/Fe], [Cr/Fe], [Mn/Fe], [Ni/Fe], [Zn/Fe], and [Ba/Fe]. For these elements, measurements are available for all clusters except EXT8 (whose  features are very weak owing to the low metallicity) and M33~H38 (which has a low S/N). Errors on the \ac{pc} vectors were estimated via a Monte-Carlo simulation in which the \acp{pc} were recomputed from 1000 random realisations of the dataset. 
The first four \acp{pc} of the dataset are shown in Fig.~\ref{fig:pca}, ordered according to their contributions to the total variance of the dataset. It is clear, and fairly unsurprising, that the first \ac{pc}, which accounts for more than half of the variance, mainly traces metallicity. The second \ac{pc}, accounting for 17\% of the variance, involves all of the abundance ratios in which the three GCs mentioned above are deficient, also including Zn. This again shows very clearly that variations in the $\alpha$-elements are strongly correlated with variations in [Na/Fe], [Sc/Fe], [Ni/Fe], and [Zn/Fe]. 

Table~\ref{tab:pca} lists, for each cluster, the projection of the abundance measurements onto the first four  \ac{pc} vectors. Hence the first column (PC1) indicates metallicity and the second column (PC2) indicates variations in the $\alpha$-elements and their related elements. We note that PC2 is defined such that a positive increase in this component corresponds to a decrease in the $\alpha$-element abundances. It is again evident that PC2 has the strongest positive contribution in Fornax~4, HM33-B, and NGC~6822-SC7. The PCA clearly confirms that HM33-B shares similar abundance patterns with Fornax~4 and with NGC~6822-SC7.

While the first and second \ac{pc} have clear astrophysical interpretations, this is less clear for the higher-order \acp{pc}. Fig.~\ref{fig:pc} graphically illustrates the data in Table~\ref{tab:pca} by plotting PC2, PC3, and PC4 as a function of PC1. The three $\alpha$-deficient \acp{gc} are clearly identifiable as outliers in the top panel. PC3 shows a U-shaped variation with PC1, implying a similar U-shaped variation of [Na/Fe] with [Fe/H] and an inverse variation of [Ba/Fe]. This tendency for the most metal-poor and metal-rich clusters to have relatively high [Na/Fe] and relatively low [Ba/Fe] ratios is indeed visible in Figs.~\ref{fig:nafe} and \ref{fig:heavy}. PC4 accounts for less than 6\% of the variance, and probably does not trace significant physical variations in the dataset.

The relation between [Na/Fe] and the $\alpha$-elements has previously been discussed in the literature, with conflicting conclusions. Several studies have found significant correlations between [Na/Fe] and [$\alpha$/Fe] similar to that reported here (\citealt{Fulbright2002}; \citetalias{Ishigaki2013}), yet no such correlation was found by \citetalias{Venn2004} in their analysis of a compilation of literature data. 
Evidently the relationship between Na and the $\alpha$-elements is complex, as can be appreciated from the fact that relatively metal-rich field stars in dwarf galaxies reach much lower [Na/Fe] values than Galactic stars of comparable metallicities (e.g.\ Fig.~\ref{fig:nafe}). 

Similar remarks may be made about the Ni-Na correlation, which has received considerable attention in the literature and is thought to be linked to the production of both elements in Type II SN nucleosynthesis \citep{Nissen1997,Cohen2010,Letarte2010,Lemasle2014}. The two abundance ratios, [Ni/Fe] and [Na/Fe], are also correlated in our dataset though PC2, but cannot here be separated in an unambiguous way from the general correlation of both elements with the $\alpha$-element abundances.

The identification of the cluster HM33-B in the above analysis is reminiscent of the idea of chemical tagging, which seeks to establish relations between stars or star clusters based on their chemical composition \citep{Freeman2002}. Given the small variation in the various abundance ratios at low metallicities, it seems clear that prospects for tagging \acp{gc} are most promising for clusters that have metallicities higher than the knee in host galaxies that are sufficiently massive to host \acp{gc}, that is, $\mathrm{[Fe/H]} \ga -1.5$.
Similar conclusions were reached in previous work that explored the chemical association of \acp{gc} with halo substructure in M31 \citep{Sakari2014,Sakari2015} and the Milky Way \citep{Horta2020}.

\subsection{Enrichment time scales and IMF}

As discussed in the introduction, the integrated light of massive early-type galaxies is usually dominated by old stellar populations with enhanced $\alpha$-element abundances.
When broken down by individual elements the picture becomes more complex. The lighter $\alpha$-elements (O, Mg) become increasingly enhanced for higher velocity dispersions (i.e.\ more massive systems), whereas Ca is typically found to be only slightly enhanced relative to scaled-solar composition, or not at all \citep{Conroy2014,Worthey2014}. These differences reflect differences in nucleosynthetic origin, with the light $\alpha$-elements being produced predominantly by hydrostatic burning in massive stars, while Ca and Ti are also produced in explosive nucleosynthesis and have significant contributions from type Ia SNe \citep{Woosley1995,McWilliam2013,Kobayashi2020}. Data for the integrated light of dwarf galaxies are more scarce, but indicate Mg abundances close to scaled-solar values, thus extending the trend of decreasing [$\alpha$/Fe] ratio with decreasing galaxy mass towards lower masses \citep{Gorgas1997,Sen2018}.

Abundance measurements for \acp{gc} provide a complementary insight into the behaviour of the various abundances at lower metallicities compared to the dominant field star component, and hence at an earlier stage of chemical enrichment (Sect.~\ref{sec:icmp}).
In metal-poor \acp{gc}, we find that Ca participates in the general level of $\alpha$-element enhancement, but at higher metallicities it tends to decrease towards scaled-solar values and thus approaches the behaviour seen in integrated galaxy light. 
Most of the \acp{gc} have metallicities below the knee in the relation of [$\alpha$/Fe], and at these relatively low metallicities any differences in the mean abundance patterns between dwarf galaxies and the large Local Group spirals are very minor. 

In the time-delay model for chemical evolution \citep{Tinsley1979,Matteucci1986}, the abundance ratios at early times, for stars with metallicities below the knee, are expected to depend mainly on the Type II SN yields (see Figure~10 in \citealt{Vincenzo2015} for an illustration). The high degree of similarity of the abundance patterns observed for metal-poor \acp{gc} in different environments may therefore not be surprising. However, since the detailed yields do depend on the masses of the Type II SN progenitors (in addition to their metallicity and spin), abundance ratios can, in principle, provide constraints on the shape of the \ac{imf} \citep{Matteucci1990,Nissen1994,McWilliam1997,Wyse1998,Tolstoy2003}.  According to chemical evolution models, the $\alpha$-element abundance patterns of metal-poor stars in the solar neighbourhood are consistent with an early \ac{imf} slope similar to, or slightly steeper, than that of the classical \citet{Salpeter1955} \ac{imf} \citep{Tsujimoto1997,Wyse1998,Hopkins2018}. By way of illustration, \citet{Wyse1992} calculated that a change in \ac{imf} slope from $-2.3$ to $-1.1$ (where the \citet{Salpeter1955} slope is $-2.35$) would lead to an increase in $\mathrm{[O/Fe]}$ by about 0.4~dex for stars in the Galactic bulge. The details may differ at lower metallicities and for other $\alpha$-elements, with the effect on Mg, Si, and Ca expected to be about half that on [O/Fe] \citep{Vincenzo2015}. Very roughly, a difference in mean [$\alpha$/Fe] ratio of about 0.04~dex between dwarf and spiral galaxies (Sect.~\ref{sec:abundance_trends}) might then correspond to a difference in \ac{imf} slope of about 0.2, suggesting that the early \ac{imf} in the dwarf galaxies in our sample did not differ substantially from the \citet{Salpeter1955} \ac{imf}.

The effects of \ac{imf} variations on models for chemical evolution have been explored theoretically for dwarf galaxies \citep{Recchi2014}, including the Sagittarius dSph \citep{Vincenzo2015}, with particular attention to predictions for the Integrated Galactic IMF (IGIMF) theory \citep{Weidner2005}.  In this theory, the \ac{imf} is coupled to the star formation rate (SFR), with lower SFRs leading to a deficit of high-mass stars and a corresponding lowering of the $\alpha$-element abundances. The SFRs at early times are generally not well constrained for the dwarf galaxies in our sample. However, the fact that these galaxies were able to form \acp{gc} suggests relatively high SFRs. The brightest \acp{gc} in each galaxy reach $M_V\sim-8$ or brighter, corresponding to masses greater than $2.7\times10^5 \, M_\odot$ (for $M/L_V = 2 \, M_\odot/L_{V,\odot}$), which translates to SFRs of about $0.8 \, M_\odot \, \mathrm{yr}^{-1}$ or higher in the IGIMF theory \citep{Vincenzo2015}. At such relatively high SFRs, the IGIMF theory predicts only minor differences compared with models that assume classical \acp{imf}. Comparing models with SFRs of 0.5~$M_\odot \, \mathrm{yr}^{-1}$  and 100~$M_\odot \, \mathrm{yr}^{-1}$, \citet{Recchi2014} found a difference of only 0.015~dex in the predicted [$\alpha$/Fe] ratios on the plateau below the knee. 
These arguments are thus consistent with the similarity of the $\alpha$-element abundance ratios for \acp{gc} in dwarf and spiral galaxies, which suggest that only minor differences in the \ac{imf} shape are allowed at low metallicities.

The relatively minor differences in the abundance ratios at low metallicities are in contrast to the more pronounced differences observed at higher metallicities, such as a deficit of hydrostatic elements relative to explosive elements in relatively metal-rich field stars in the Sagittarius dSph \citep{McWilliam2013}. The latter may suggest that the later stages of chemical evolution were characterised by a steeper, more top-light \ac{imf} as predicted in the IGIMF theory for the lower SFRs expected at later times \citep{Vincenzo2015}. 
At lower metallicities, the abundances of stars in Sagittarius are more similar to those observed in the Milky Way halo (and to our \ac{gc} measurements), consistent with a normal \ac{imf} at earlier times \citep{Hansen2018}.
It is again worth recalling that some metal-poor \acp{gc} also show very low Mg abundances, but linking this to \ac{imf} variations is made more difficult by the fact that only some \acp{gc} are affected, and by the possibility that Mg is affected by multiple populations. 

\subsection{Enrichment processes}

\begin{figure}
\centering
\includegraphics[width=\columnwidth]{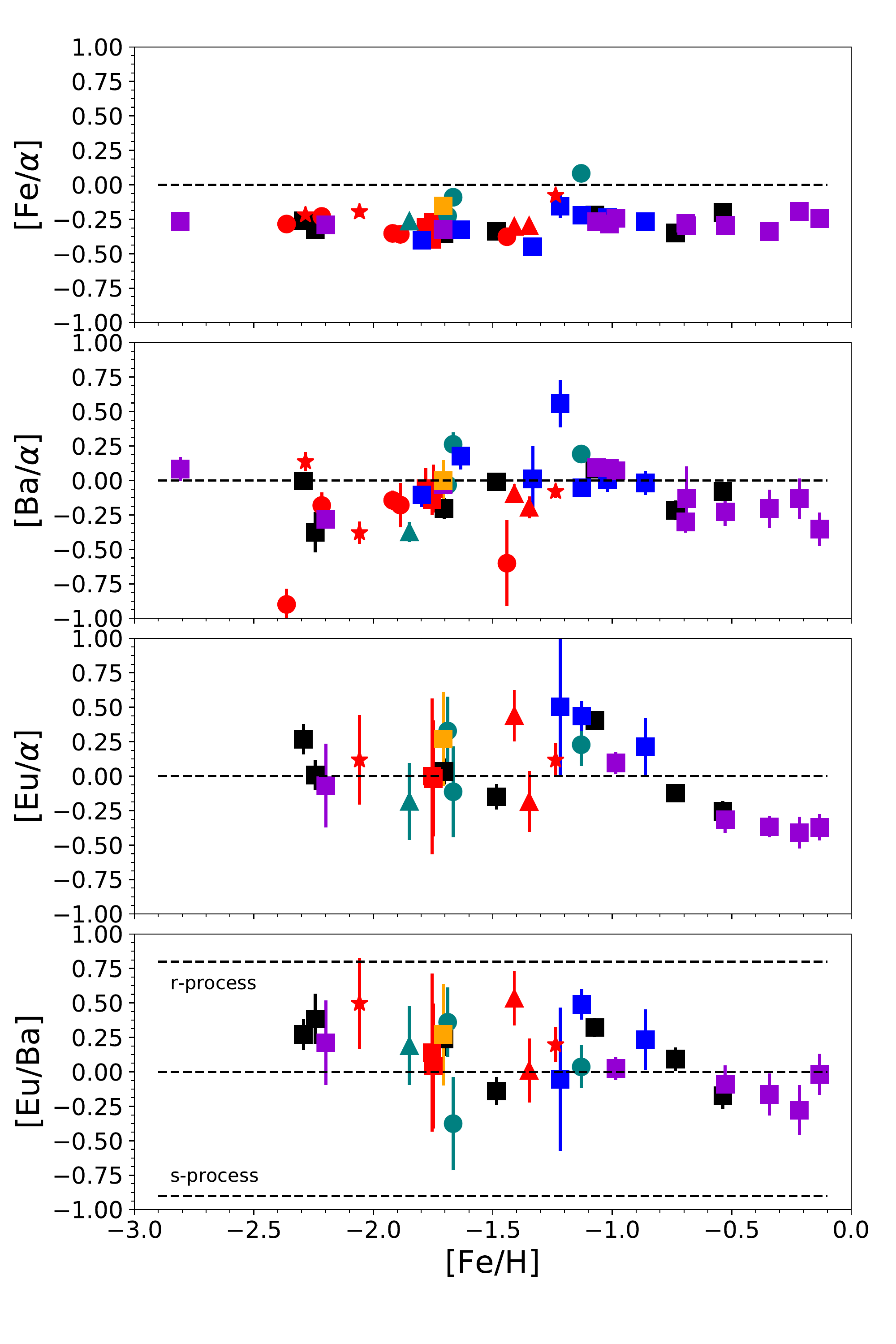}
\caption{\label{fig:heavy2}Abundances of heavy elements with respect to the average of the $\alpha$-elements. Symbols for our measurements as in Fig.~\ref{fig:alphafe}.}
\end{figure}

In Fig.~\ref{fig:heavy2} we compare the abundances of Fe, Ba, and Eu, taken as representative tracers of Type Ia SN, $s$-process, and $r$-process nucleosynthesis, respectively. In the literature it is customary to plot these elements relative to Mg, taking the latter as a tracer of Type II SN nucleosynthesis \citep[e.g.][]{Skuladottir2020}. We have here replaced Mg with an error-weighted average of the $\alpha$-element abundances (Mg, Si, Ca, and Ti) to reduce the uncertainties, and to alleviate the impact of the anomalously low $\mathrm{[Mg/Fe]}$ values seen in some \acp{gc}. As shown in Fig.~\ref{fig:alphafe}, this approach is justified by the fairly similar dependencies of the various $\alpha$-element abundances on metallicity, even though the heavier elements (in particular Ca and Ti) have a more complicated nucleosynthetic history. 

The top panel in Fig.~\ref{fig:heavy2} shows [Fe/$\alpha$] vs.\ $\mathrm{[Fe/H]}$ and is, of course, equivalent to the mean of the panels in Fig.~\ref{fig:alphafe}, with the vertical axis inverted, and leads to the same conclusion: for the most part, the \acp{gc} are preferentially enriched in the $\alpha$-elements compared to scaled-solar composition, consistent with enrichment on time scales shorter than those characteristic of Type Ia SNe. Indeed, Fig.~\ref{fig:heavy2} suggests the alternative, but equivalent and perhaps more intuitive formulation that the metal-poor \acp{gc} are deficient in iron, due to the still `missing' contribution from SN Ia nucleosynthesis. 
As previously discussed, the main exceptions are Fornax~4, NGC~6822-SC7, and HM33-B. While Fig.~\ref{fig:heavy2} shows an average of all four $\alpha$-elements, the scatter is similar to those of [Ca/Fe] and [Ti/Fe] alone, $\sigma_{[\mathrm{Fe}/\alpha]} = 0.09$~dex, suggesting that at least part of the dispersion is real. 

Similar to [Fe/$\alpha$], the Ba and Eu abundance ratios, defined with respect to either iron (Fig.~\ref{fig:heavy}) or the $\alpha$-elements (Fig.~\ref{fig:heavy2}), display no obvious differences depending on the host galaxy. 
In particular, the apparent enhancement in [Ba/Fe] noted for the M33 \acp{gc} (Sect.~\ref{sec:abundance_trends}) becomes less clear relative to the $\alpha$-elements. To be more quantitative, we computed the average [Ba/Fe] for the M33 \acp{gc} and for other \acp{gc} in a similar metallicity range ($-2 < \mathrm{[Fe/H]} < -0.8$). We found $\langle \mathrm{[Ba/Fe]} \rangle_\mathrm{M33} = +0.25\pm0.03$ for the M33 \acp{gc} and $\langle \mathrm{[Ba/Fe]} \rangle_\mathrm{other} = +0.21\pm0.01$ for the other \acp{gc}, which is only a marginally significant difference. 
With respect to the $\alpha$-elements, we instead found $\langle [\mathrm{Ba}/\alpha] \rangle_\mathrm{M33} = -0.01\pm0.03$ and $\langle [\mathrm{Ba}/\alpha] \rangle_\mathrm{other} = 0.00\pm0.01$, which confirms the visual impression from Fig.~\ref{fig:heavy2} that no significant difference is present. 

As noted in the introduction, the n-capture element abundances are often found to differ significantly between field stars in classical dSph galaxies and the Milky Way. However, at low metallicities, these differences are typically less pronounced \citep{Reichert2020}, in agreement with our results for the \acp{gc}. Ultra-faint dwarfs  exhibit a larger scatter in the n-capture element abundances that may reflect stochasticity in the enrichment processes dominated by a few rare events \citep{Ji2016,Marshall2019,Molero2021}. 
It is noteworthy that no \ac{gc} in our sample exhibits the strongly enhanced [Eu/Fe] abundance ratios, up to $\mathrm{[Eu/Fe]}>1$, recently found for metal-poor stars in the Small and Large Magellanic Cloud
\citep[SMC and LMC;][]{Reggiani2021}. These authors suggested that the Eu enhancement in the SMC and LMC stars might result from a more gradual early chemical enrichment history that would have allowed for a more substantial contribution from delayed $r$-process enrichment due to neutron star mergers. By comparison with our sample, the Magellanic Clouds would appear to have been fairly unique in this respect, even when compared with relatively isolated, irregular galaxies such as NGC~6822 and WLM. 
A few \acp{gc} are consistent with a more moderate Eu enhancement of $\mathrm{[Eu/Fe]}\simeq0.6-0.7$, as reported for the Gaia Sausage \citep{Aguado2021} and the Galactic \acp{gc} NGC~1261 and NGC~6934 \citep{Koch-Hansen2021,Marino2021}. Due to the challenging nature of the Eu measurements in the integrated light, independent confirmation of the results for individual clusters would be desirable. However, our measurement of $\mathrm{[Eu/Fe]}=+0.63\pm0.04$ for the Galactic \ac{gc} NGC~362 is fairly similar to the literature value of $\mathrm{[Eu/Fe]}=+0.57\pm0.02$ \citep{Shetrone2000,Pritzl2005}.

The [Eu/Ba] ratio, plotted in the lower panel in Fig.~\ref{fig:heavy2}, is frequently used as an indicator of the importance of $r$- versus $s$-process enrichment. The $r$- and $s$-process reference ratios are  from \citet{McWilliam1997}. Our integrated-light measurements mostly lie between the
solar system and pure $r$-process value, but there is a significant scatter.  Some \acp{gc}, particularly those towards the metal-rich end, even reach sub-solar [Eu/Ba] ratios, implying a more important $s$-process contribution. Overall, these results are similar to those found for Galactic \acp{gc} and field stars in the corresponding metallicity range \citep{McWilliam1997,Gratton2004}. 
While Ba is indeed considered a classical tracer of the $s$-process (according to \citet{Bisterzo2014}, 
about 85\% of the Ba in the solar system is attributed to the $s$-process), Ba production via the $r$-process becomes increasingly important at early times \citep{Pagel1997a,Cescutti2006}. 
Europium is, instead, a fairly pure $r$-process element, with a contribution of only 6\% from the $s$-process in the solar system \citep{Bisterzo2014}.  The identification of \ac{agb} stars as an important $s$-process production site is fairly uncontroversial, but $s$-process elements are also produced in massive stars, so the yields remain not well constrained, as the predictions depend on different parameters in the models, such as rotation, metallicity, and the size of the convective $^{13}$C pocket \citep{Cescutti2006,Prantzos2018, Limongi2018, Kobayashi2020}. The details of the $r$-process are even more uncertain, and it appears increasingly likely that there are at least two $r$-process sources of which one operates on relatively short time scales ($\la10^8$ yr) and is likely linked to massive stars, while the other involves a delay of up to several Gyr and may involve neutron-star mergers
\citep{Skuladottir2020,Molero2021}. Therefore, the variation in the relative abundances of Ba and Eu as a function of time and metallicity may be expected to be rather more complicated than for the [$\alpha$/Fe] ratio. Some of the difficulties encountered when modelling the evolution of n-capture elements have been more fully discussed by \citet{Tautvaisiene2021}.

Internal variations in the abundances of neutron-capture elements have been observed in NGC~7078 (M15) and a few other \acp{gc} \citep{Roederer2011b}, although some cases remain controversial \citep{Cohen2011}. It is reasonable to wonder how this might affect the comparison of n-capture elements in \acp{gc} and field stars. In NGC~7078, which is the clearest and best studied case, the member stars fall into two groups with Ba, La, and Eu abundances differing by 0.2--0.4 dex, but both groups have abundance patterns indicative of $r$-process enrichment. Interestingly, both groups independently show the Na-O anti-correlation that is generally associated with multiple populations in \acp{gc} \citep{Sneden1997,Worley2013}. While the Na spreads in \acp{gc} cause clear systematic differences between field stars and mean \ac{gc} abundances (Fig.~\ref{fig:nafe}), internal spreads in n-capture element abundances appear to be relatively rare, and in any case do not make NGC~7078 a significant outlier in the relevant relations. A satisfactory explanation for the spreads in n-capture elements in \acp{gc} is even more elusive than for the light element abundance variations. The possibility that $r$-process self-enrichment in \acp{gc} may be caused by neutron-star mergers has been discussed in the literature \citep[e.g.][]{Zevin2019}, but it seems clear from Fig.~\ref{fig:heavy} that such processes do not leave any significant signature in the form of a detectable offset in [Eu/Fe] between \acp{gc} and field stars \citep[see also][]{Sakari2013}.

\subsection{The extremely metal-poor GC M31 EXT8}

By far the most metal-poor \ac{gc} in our sample is M31~EXT8. The spectroscopic observations of this cluster were previously discussed in \citet{Larsen2020} and its \ac{cmd} was discussed in \citet{Larsen2021}. 

The metallicity reported here ($\mathrm{[Fe/H]} = -2.81\pm0.04$) is about 0.1~dex higher than that found in \citet{Larsen2020}, mainly due to the application of \ac{nlte} corrections (our \ac{lte} value is $\mathrm{[Fe/H]}=-2.88$; see Appendix~\ref{app:lte}). 
As the \ac{nlte} corrections also lead to an increase in the metallicites of the other clusters (albeit less so at higher metallicities), the differential change with respect to other \acp{gc} in our sample is even smaller. In any case, the slight increase does not change the conclusion that M31~EXT8 is, by a considerable margin, the most metal-poor \ac{gc} currently known. It was already noted in previous analysis that the [Mg/Fe] is peculiarly low, while the other $\alpha$-element abundances are fairly typical of Milky Way field stars at the corresponding metallicity. We note that the [Mg/Fe] measurement in \citet{Larsen2020} was based on the Mg $b$ triplet, while the measurement here is obtained from the weaker \ion{Mg}{i} lines that were also used for other clusters in our sample. Nevertheless, we recover the low [Mg/Fe] value previously found.

At the low metallicity of EXT8, the uncertainties on most abundance ratios are relatively large owing to the weakness of the spectral features. There is a suggestion that [Ni/Fe] is somewhat enhanced, but it is not very unusual compared to the Milky Way data when the $\sim0.2$~dex \ac{nlte} correction is accounted for. 
The [Ba/Fe] value, which has a small uncertainty, is also relatively high, and falls near the upper envelope of $\mathrm{[Ba/Fe]}$ values for Galactic halo field stars (Fig.~\ref{fig:heavy}). However, [Mg/Fe] remains the most significant outlier in abundance space.

\subsection{Outlook: Refining the analysis techniques}

While the analysis presented herein has been updated and improved in a variety of ways, it is certainly not definitive. When comparing with model predictions, it is clearly important to eliminate systematic biases in the analysis as much as possible.  Even when comparing differentially with other datasets, \ac{lte} vs.\ \ac{lte} is not necessarily a like-with-like comparison, owing to the fact that different stellar types and spectral features are affected in different ways by \ac{nlte} effects \citep[e.g.][]{Eitner2019}. 
Here we have applied \ac{nlte} corrections for many lines and elements to integrated-light analysis for the first time, and we find that our \ac{nlte} abundances for the \acp{gc} generally agree well with the corresponding \ac{nlte} values for metal-poor Milky Way field stars. 

As for individual stars, some elements are more strongly affected by \ac{nlte} effects than others. In some cases, application of \ac{nlte} corrections can lead to quantitatively different trends with metallicity and different astrophysical implications. A case in point is [Mn/Fe] which displays a strong increasing trend as a function of metallicity in \ac{lte}, but remains approximately flat in \ac{nlte}.  As discussed in \citet{Eitner2020}, the flat trend of slightly sub-solar [Mn/Fe] ratios can be reproduced in galactic chemical evolution models in which the Type Ia SN contribution to Mn production is mainly due to sub-Chandrasekhar mass progenitors, with a significant contribution from core collapse SNe as well. In contrast, the steeper trend found in \ac{lte} would support Type Ia SN progenitors with near-Chandrasekhar masses. 

A similar case to Mn may be Cu. We have not here included \ac{nlte} corrections for this element, and our integrated-light analysis (Fig.~\ref{fig:heavy}) shows the trend of increasing [Cu/Fe] with metallicity that is also commonly observed in field stars. This is often taken as evidence that Cu behaves as a secondary element, produced by the weak $s$-process in massive stars \citep{Romano2007}. However, \ac{nlte} analyses of Cu tend to find a much shallower trend with metallicity that more closely resembles the behaviour expected for a primary element \citep{Yan2015,Korotin2018,Shi2018}. Hence, Cu is another case for which the details of the analysis have important consequences for the astrophysical interpretation, and it might therefore be worthwhile to compute integrated-light \ac{nlte} corrections for Cu. 
Taken together, the abundances of Mn, Zn, and Cu may provide important constraints on Type Ia SN models, in particular the role of the double-detonation scenario in which detonation of a helium-rich surface layer may lead to increased production of these three elements \citep{Lach2020}.

Of course, correction for \ac{nlte} effects addresses just one of several approximations made in classical abundance analysis.  Another is the assumption that the physical properties of stellar atmospheres are time-independent and depend only on the depth in the atmosphere. In the last decade, significant progress has been made in the computation of 3-dimensional radiation-hydrodynamic models \citep[e.g.][]{Magic2013} and the resulting stellar spectra, from which it is clear that the effects on the derived abundances can be significant at the level of 0.1~dex or more for some elements \citep[][]{Bergemann2019,Semenova2020,Amarsi2019}. 
One appealing aspect of 3-D radiation-hydrodynamic models is that convective motions are explicitly accounted for, which eliminates the mixing length,  micro-, and macro-turbulent velocities as free parameters in the analysis.

\section{Summary and conclusions}

We have presented homogeneous integrated-light abundance measurements for a sample of 45 \acp{gc}, including \ac{nlte} corrections for several elements. The sample includes clusters in most Local Group galaxies that host \ac{gc} systems, although it is mostly limited to the brighter half of the \ac{gc} luminosity function and to relatively compact clusters that are better suited for measurements of their integrated light with single-slit echelle spectrographs. Compared with previously published analyses of a subset of the data, the current analysis has been updated in a number of ways, including a critical revision of the line list, a modified prescription for the assignment of microturbulent velocities, and the use of \texttt{ATLAS12} model atmospheres with self-consistent chemical composition.

Our main results can be summarised as follows:

\begin{itemize}
    \item We have extended the formalism for integrated-light \ac{nlte} corrections developed by \citet{Eitner2019} to account for simultaneous fitting of multiple spectral lines.  
    \item To first order, \acp{gc} in different galaxies have very similar abundance patterns. Metal-poor \acp{gc} ($\mathrm{[Fe/H]}\la-1.5)$ have enhanced $\alpha$-element abundances in all galaxies, and the overall scatter in the $\alpha$-element abundances is less than 0.1~dex. As in previous studies, we find peculiarly low [Mg/Fe] values for a fraction of the metal-poor \acp{gc}.
    \item The similarity of the $\alpha$-element abundances indicates that the \ac{gc} formed in environments with similar \acp{imf}. There is possibly a small offset of about 0.04~dex between the mean $\alpha$-element abundances of \acp{gc} in dwarf galaxies and in the spirals, which we estimate translates to a variation in the \ac{imf} slope of at most 0.2~dex.
   \item From a principal components analysis, the $\alpha$-element abundance variations are found to be correlated with variations in [Na/Fe], [Sc/Fe], [Ni/Fe], and [Zn/Fe]. These correlations are driven mainly by three clusters: Fornax~4, NGC~6822-SC7, and HM33-B. Fornax~4 and NGC~6822-SC7 are the most metal-rich \acp{gc} in their respective host galaxies, and the similarity of the abundance patterns of HM33-B to the two other clusters may suggest an accretion origin for HM33-B. 
   \item The n-capture element abundances are also fairly uniform across our sample. We find no \acp{gc} with strongly enhanced $r$-process signatures, as reported for metal-poor stars in the Magellanic Clouds \citep{Reggiani2021} and in ultra-faint dwarfs such as Reticulum~II \citep{Ji2016,Roederer2016}
  \item In view of the very similar abundance patterns at low metallicities, and in agreement with previous studies, we conclude that chemical tagging of \acp{gc} is likely to work best at metallicities above $\mathrm{[Fe/H]}\approx-1.5$, where differences between the chemical composition of stellar populations in host dwarf galaxies become more apparent. 
  \item We measure a line-of-sight velocity dispersion of about 11.7~km~s$^{-1}$ and a corresponding mass-to-light ratio of $M_\mathrm{dyn}/L_V \simeq 2.0 \, M_\odot/L_{V,\odot}$ for the cluster F46 in NGC~2403, consistent with the value expected for a moderately metal-poor old \ac{gc}.
\end{itemize}

Beyond the results summarised above, the analysis presented here may also serve as a reference for future studies of larger samples of \acp{gc} beyond the Local Group. With the next generation of 30-40 m class telescopes, it will be possible to obtain high S/N spectra for \acp{gc} well beyond the Local Group, including in galaxies with rich \ac{gc} systems (such as the Sombrero galaxy and NGC~5128), of which a substantial fraction will fall in the more metal-rich regime where interesting differences may be expected. 

\begin{acknowledgements}

JS acknowledges support from the Packard Foundation.
AJR was supported by National Science Foundation Grant AST-1616710 and as a Research Corporation for Science Advancement Cottrell Scholar.
MB is supported through the Lise Meitner grant from the Max Planck Society. We acknowledge support by the Collaborative Research centre SFB 881 (projects A5, A10), Heidelberg University, of the Deutsche Forschungsgemeinschaft (DFG, German Research Foundation).  This project has received funding from the European Research Council (ERC) under the European Union’s Horizon 2020 research and innovation programme (Grant agreement No. 949173).

Funding for the Sloan Digital Sky Survey (SDSS) has been provided by the Alfred P. Sloan Foundation, the Participating Institutions, the National Aeronautics and Space Administration, the National Science Foundation, the U.S. Department of Energy, the Japanese Monbukagakusho, and the Max Planck Society. The SDSS Web site is http://www.sdss.org/.

The SDSS is managed by the Astrophysical Research Consortium (ARC) for the Participating Institutions. The Participating Institutions are The University of Chicago, Fermilab, the Institute for Advanced Study, the Japan Participation Group, The Johns Hopkins University, Los Alamos National Laboratory, the Max-Planck-Institute for Astronomy (MPIA), the Max-Planck-Institute for Astrophysics (MPA), New Mexico State University, University of Pittsburgh, Princeton University, the United States Naval Observatory, and the University of Washington.

This research has made use of ESASky, developed by the ESAC Science Data Centre (ESDC) team and maintained alongside other ESA science mission's archives at ESA's European Space Astronomy Centre (ESAC, Madrid, Spain).

Some of the data presented herein were obtained at the W.~M.\ Keck Observatory, which is operated as a scientific partnership among the California Institute of Technology, the University of California and the National Aeronautics and Space Administration. The Observatory was made possible by the generous financial support of the W.~M.\ Keck Foundation. 
The authors wish to recognize and acknowledge the very significant cultural role and reverence that the summit of Maunakea has always had within the indigenous Hawaiian community.  We are most fortunate to have the opportunity to conduct observations from this mountain. 

We thank A.\ Wasserman for assistance with the HIRES observations and J.\ van Eijck for help with selection of lines for \ac{nlte} analysis.  We are grateful to the anonymous referee for carefully reading the manuscript and delivering a timely report which contained several useful suggestions.
\end{acknowledgements}

\bibliographystyle{aa}
\bibliography{refs.bib}

\begin{thebibliography}{352}
\expandafter\ifx\csname natexlab\endcsname\relax\def\natexlab#1{#1}\fi

\bibitem[{Aguado {et~al.}(2021)Aguado, Belokurov, Myeong, Evans, Kobayashi,
  Sbordone, Chanam{\'{e}}, Navarrete, \& Koposov}]{Aguado2021}
Aguado, D.~S., Belokurov, V., Myeong, G.~C., {et~al.} 2021, ApJ, 908, L8

\bibitem[{Alam {et~al.}(2015)Alam, Albareti, Prieto, Anders, Anderson, Andrews,
  Armengaud, Aubourg, Bailey, Bautista, Beaton, Beers, Bender, Berlind,
  Beutler, Bhardwaj, Bird, Bizyaev, Blake, Blanton, Blomqvist, Bochanski,
  Bolton, Bovy, Bradley, Brandt, Brauer, Brinkmann, Brown, Brownstein, Burden,
  Burtin, Busca, Cai, Capozzi, Rosell, Carrera, Chen, Chiappini, Chojnowski,
  Chuang, Clerc, Comparat, Covey, Croft, Cuesta, Cunha, da~Costa, {Da Rio},
  Davenport, Dawson, {De Lee}, Delubac, Deshpande, Dutra-Ferreira, Dwelly,
  Ealet, Ebelke, Edmondson, Eisenstein, Escoffier, Esposito, Fan,
  Fern{\'{a}}ndez-Alvar, Feuillet, Ak, Finley, Finoguenov, Flaherty, Fleming,
  Font-Ribera, Foster, Frinchaboy, Galbraith-Frew,
  Garc{\'{i}}a-Hern{\'{a}}ndez, P{\'{e}}rez, Gaulme, Ge, G{\'{e}}nova-Santos,
  Ghezzi, Gillespie, Girardi, Goddard, Gontcho, Hern{\'{a}}ndez, Grebel, Grieb,
  Grieves, Gunn, Guo, Harding, Hasselquist, Hawley, Hayden, Hearty, Ho, Hogg,
  Holley-Bockelmann, Holtzman, Honscheid, Huehnerhoff, Jiang, Johnson,
  Kinemuchi, Kirkby, Kitaura, Klaene, Kneib, Koenig, Lam, Lan, Lang, Laurent,
  Goff, Leauthaud, Lee, Lee, Licquia, Liu, Long, L{\'{o}}pez-Corredoira,
  Lorenzo-Oliveira, Lucatello, Lundgren, Lupton, Mack, Mahadevan, Maia,
  Majewski, Malanushenko, Malanushenko, Manchado, Manera, Mao, Maraston,
  Marchwinski, Margala, Martell, Martig, Masters, McBride, McGehee, McGreer,
  McMahon, M{\'{e}}nard, Menzel, Merloni, M{\'{e}}sz{\'{a}}ros, Miller,
  Miralda-Escud{\'{e}}, Miyatake, Montero-Dorta, More, Morice-Atkinson,
  Morrison, Muna, Myers, Newman, Neyrinck, Nguyen, Nichol, Nidever, Noterdaeme,
  Nuza, O'Connell, O'Connell, O'Connell, Ogando, Olmstead, Oravetz, Oravetz,
  Osumi, Owen, Padgett, Padmanabhan, Paegert, Palanque-Delabrouille, Pan,
  Parejko, Park, P{\^{a}}ris, Pattarakijwanich, Pellejero-Ibanez, Pepper,
  Percival, P{\'{e}}rez-Fournon, P{\'{e}}rez-R{\`{a}}fols, Petitjean, Pieri,
  Pinsonneault, de~Mello, Prada, Prakash, Price-Whelan, Raddick, Rahman, Reid,
  Rich, Rix, Robin, Rockosi, Rodrigues, Rodr{\'{i}}guez-Rottes, Roe, Ross,
  Ross, Rossi, Ruan, Rubi{\~{n}}o-Mart{\'{i}}n, Rykoff, Salazar-Albornoz,
  Salvato, Samushia, S{\'{a}}nchez, Santiago, Sayres, Schiavon, Schlegel,
  Schmidt, Schneider, Schultheis, Schwope, Sc{\'{o}}ccola, Sellgren, Seo,
  Shane, Shen, Shetrone, Shu, Sivarani, Skrutskie, Slosar, Smith, Sobreira,
  Stassun, Steinmetz, Strauss, Streblyanska, Swanson, Tan, Tayar, Terrien,
  Thakar, Thomas, Thompson, Tinker, Tojeiro, Troup, Vargas-Maga{\~{n}}a,
  Vazquez, Verde, Viel, Vogt, Wake, Wang, Weaver, Weinberg, Weiner, White,
  Wilson, Wisniewski, Wood-Vasey, Y{\`{e}}che, York, Zakamska, Zamora,
  Zasowski, Zehavi, Zhao, Zheng, Zhou, Zhou, Zhu, Zou, Thompson, Tinker,
  Tojeiro, Troup, Vargas-Maga{\~{n}}a, Vazquez, Verde, Viel, Vogt, Wake, Wang,
  Weaver, Weinberg, Weiner, White, Wilson, Wisniewski, Wood-Vasey, Ye`che,
  York, Zakamska, Zamora, Zasowski, Zehavi, Zhao, Zheng, {Zhou (周旭)}, {Zhou
  (周志民)}, {Zou (邹虎)}, \& Zhu}]{Alam2015}
Alam, S., Albareti, F.~D., Prieto, C.~A., {et~al.} 2015, ApJS, 219, 12

\bibitem[{Alvarez \& Plez(1998)}]{Alvarez1998}
Alvarez, R. \& Plez, B. 1998, A\&A, 330, 1109

\bibitem[{Alves-Brito {et~al.}(2006)Alves-Brito, Barbuy, Zoccali, Minniti,
  Ortolani, Hill, Renzini, Pasquini, Bica, Rich, Mel{\'{e}}ndez, \&
  Momany}]{Alves-Brito2006}
Alves-Brito, A., Barbuy, B., Zoccali, M., {et~al.} 2006, A{\&}A, 460, 269

\bibitem[{Amarsi {et~al.}(2019)Amarsi, Nissen, \&
  Sk{\'{u}}lad{\'{o}}ttir}]{Amarsi2019}
Amarsi, A.~M., Nissen, P.~E., \& Sk{\'{u}}lad{\'{o}}ttir, {\'{A}}. 2019,
  A{\&}A, 630, A104

\bibitem[{Andrievsky {et~al.}(2018)Andrievsky, Bonifacio, Caffau, Korotin,
  Spite, Spite, Sbordone, \& Zhukova}]{Andrievsky2018}
Andrievsky, S., Bonifacio, P., Caffau, E., {et~al.} 2018, MNRAS, 473, 3377

\bibitem[{Anstee \& O'Mara(1995)}]{Anstee1995}
Anstee, S.~D. \& O'Mara, B.~J. 1995, MNRAS, 276, 859

\bibitem[{Asplund(2005)}]{Asplund2005}
Asplund, M. 2005, ARA\&A, 43, 481

\bibitem[{Asplund {et~al.}(2009)Asplund, Grevesse, Sauval, \&
  Scott}]{Asplund2009}
Asplund, M., Grevesse, N., Sauval, A.~J., \& Scott, P. 2009, ARA\&A, 47, 481

\bibitem[{Baade(1944)}]{Baade1944a}
Baade, W. 1944, ApJ, 100, 147

\bibitem[{Barbuy {et~al.}(1999)Barbuy, Renzini, Ortolani, Bica, \&
  Guarnieri}]{Barbuy1999}
Barbuy, B., Renzini, A., Ortolani, S., Bica, E., \& Guarnieri, M.~D. 1999,
  A{\&}A, 341, 539

\bibitem[{Barklem {et~al.}(2010)Barklem, Belyaev, Dickinson, \&
  Gadea}]{Barklem2010}
Barklem, P.~S., Belyaev, A.~K., Dickinson, A.~S., \& Gadea, F.~X. 2010, A\&A,
  519, 20

\bibitem[{Barklem {et~al.}(2017)Barklem, Osorio, Fursa, Bray, Zatsarinny,
  Bartschat, \& Jerkstrand}]{Barklem2017}
Barklem, P.~S., Osorio, Y., Fursa, D.~V., {et~al.} 2017, A\&A, 606, 11

\bibitem[{Barklem {et~al.}(2000)Barklem, Piskunov, \& O'Mara}]{Barklem2000}
Barklem, P.~S., Piskunov, N., \& O'Mara, B.~J. 2000, A\&AS, 142, 467

\bibitem[{Bastian \& Lardo(2018)}]{Bastian2018}
Bastian, N. \& Lardo, C. 2018, ARA\&A, 56, 83

\bibitem[{Battaglia {et~al.}(2006)Battaglia, Tolstoy, Helmi, Irwin, Letarte,
  Jablonka, Hill, Venn, Shetrone, Arimoto, Primas, Kaufer, Francois, Szeifert,
  Abel, \& Sadakane}]{Battaglia2006}
Battaglia, G., Tolstoy, E., Helmi, A., {et~al.} 2006, A\&A, 459, 423

\bibitem[{Battistini {et~al.}(1987)Battistini, Bonoli, Braccesi, Federici,
  {Fusi Pecci}, Marano, \& Borngen}]{Battistini1987}
Battistini, P., Bonoli, F., Braccesi, A., {et~al.} 1987, A\&AS, 67, 447

\bibitem[{Battistini {et~al.}(1984)Battistini, Bonoli, Federici, {Fusi Pecci},
  \& Kron}]{Battistini1984}
Battistini, P., Bonoli, F., Federici, L., {Fusi Pecci}, F., \& Kron, R.~G.
  1984, A{\&}A, 130, 162

\bibitem[{Beasley {et~al.}(2008)Beasley, Bridges, Peng, Harris, Harris, Forbes,
  \& Mackie}]{Beasley2008}
Beasley, M.~A., Bridges, T., Peng, E., {et~al.} 2008, MNRAS, 386, 1443

\bibitem[{Beasley {et~al.}(2005)Beasley, Brodie, Strader, Forbes, Proctor,
  Barmby, \& Huchra}]{Beasley2005}
Beasley, M.~A., Brodie, J.~P., Strader, J., {et~al.} 2005, AJ, 129, 1412

\bibitem[{Beasley {et~al.}(2015)Beasley, {San Roman}, Gallart, Sarajedini, \&
  Aparicio}]{Beasley2015}
Beasley, M.~A., {San Roman}, I., Gallart, C., Sarajedini, A., \& Aparicio, A.
  2015, MNRAS, 451, 3400

\bibitem[{Belokurov {et~al.}(2018)Belokurov, Erkal, Evans, Koposov, \&
  Deason}]{Belokurov2018}
Belokurov, V., Erkal, D., Evans, N.~W., Koposov, S.~E., \& Deason, A.~J. 2018,
  MNRAS, 478, 611

\bibitem[{Bensby {et~al.}(2017)Bensby, Feltzing, Gould, Yee, Johnson, Asplund,
  Mel{\'{e}}ndez, Lucatello, Howes, McWilliam, Udalski, Szyma{\'{n}}ski,
  Soszy{\'{n}}ski, Poleski, Wyrzykowski, Ulaczyk, Koz{\l}owski, Pietrukowicz,
  Skowron, Mr{\'{o}}z, Pawlak, Abe, Asakura, Bhattacharya, Bond, Bennett,
  Hirao, Nagakane, Koshimoto, Sumi, Suzuki, \& Tristram}]{Bensby2017}
Bensby, T., Feltzing, S., Gould, A., {et~al.} 2017, A{\&}A, 605, A89

\bibitem[{Bergemann \& Cescutti(2010)}]{Bergemann2010}
Bergemann, M. \& Cescutti, G. 2010, A{\&}A, 522, A9

\bibitem[{Bergemann {et~al.}(2017{\natexlab{a}})Bergemann, Collet, Amarsi,
  Kovalev, Ruchti, \& Magic}]{Bergemann2017a}
Bergemann, M., Collet, R., Amarsi, A.~M., {et~al.} 2017{\natexlab{a}}, ApJ,
  847, 15

\bibitem[{Bergemann {et~al.}(2017{\natexlab{b}})Bergemann, Collet,
  Sch{\"{o}}nrich, Andrae, Kovalev, Ruchti, Hansen, \& Magic}]{Bergemann2017}
Bergemann, M., Collet, R., Sch{\"{o}}nrich, R., {et~al.} 2017{\natexlab{b}},
  ApJ, 847, 16

\bibitem[{Bergemann {et~al.}(2019)Bergemann, Gallagher, Eitner, Hansen, Eitner,
  Bautista, Collet, Yakovleva, Belyaev, Mayriedl, Plez, Carlsson, Carlsson, \&
  Leenaarts}]{Bergemann2019}
Bergemann, M., Gallagher, A.~J., Eitner, P., {et~al.} 2019, A\&A, 631, A80

\bibitem[{Bergemann {et~al.}(2021)Bergemann, Hoppe, Semenova, Carlsson,
  Yakovleva, Voronov, Bautista, Nemer, Belyaev, Leenaarts, Mashonkina, Reiners,
  \& Ellwarth}]{Bergemann2021}
Bergemann, M., Hoppe, R., Semenova, E., {et~al.} 2021, MNRAS

\bibitem[{Bergemann {et~al.}(2011)Bergemann, Lind, Collet, \&
  Asplund}]{Bergemann2011}
Bergemann, M., Lind, K., Collet, R., \& Asplund, M. 2011, JPhCS, 328, 012002

\bibitem[{Bergemann {et~al.}(2012)Bergemann, Lind, Collet, Magic, \&
  Asplund}]{Bergemann2012}
Bergemann, M., Lind, K., Collet, R., Magic, Z., \& Asplund, M. 2012, MNRAS,
  427, 27

\bibitem[{Bergemann {et~al.}(2018)Bergemann, Sesar, Cohen, Serenelli,
  Sheffield, Li, Casagrande, Johnston, Laporte, Price-Whelan, Sch{\"{o}}nrich,
  \& Gould}]{Bergemann2018}
Bergemann, M., Sesar, B., Cohen, J.~G., {et~al.} 2018, Nature, 555, 334

\bibitem[{Bisterzo {et~al.}(2014)Bisterzo, Travaglio, Gallino, Wiescher, \&
  K{\"{a}}ppeler}]{Bisterzo2014}
Bisterzo, S., Travaglio, C., Gallino, R., Wiescher, M., \& K{\"{a}}ppeler, F.
  2014, ApJ, 787, 10

\bibitem[{Boeche \& Grebel(2016)}]{Boeche2016}
Boeche, C. \& Grebel, E.~K. 2016, A\&A, 587, A2

\bibitem[{Bohlin(2016)}]{Bohlin2016}
Bohlin, R.~C. 2016, AJ, 152, 60

\bibitem[{Bonifacio {et~al.}(2009)Bonifacio, Spite, Cayrel, Hill, Spite,
  Francois, Plez, Ludwig, Caffau, Molaro, Depagne, Andersen, Barbuy, Beers,
  Nordstrom, \& Primas}]{Bonifacio2009}
Bonifacio, P., Spite, M., Cayrel, R., {et~al.} 2009, A\&A, 501, 519

\bibitem[{Brewer {et~al.}(2015)Brewer, Fischer, Basu, Valenti, \&
  Piskunov}]{Brewer2015}
Brewer, J.~M., Fischer, D.~A., Basu, S., Valenti, J.~A., \& Piskunov, N. 2015,
  ApJ, 805, 126

\bibitem[{Brodie \& Huchra(1990)}]{Brodie1990}
Brodie, J.~P. \& Huchra, J.~P. 1990, ApJ, 362, 503

\bibitem[{Brodie \& Huchra(1991)}]{Brodie1991}
Brodie, J.~P. \& Huchra, J.~P. 1991, ApJ, 379, 157

\bibitem[{Brodie \& Strader(2006)}]{Brodie2006}
Brodie, J.~P. \& Strader, J. 2006, ARA\&A, 44, 193

\bibitem[{Brooke {et~al.}(2014)Brooke, Ram, Western, Li, Schwenke, \&
  Bernath}]{Brooke2014}
Brooke, J. S.~A., Ram, R.~S., Western, C.~M., {et~al.} 2014, ApJS, 210, 23

\bibitem[{Brown(2021)}]{Brown2021}
Brown, A. G.~A. 2021, ARA\&A, 59, 59

\bibitem[{Brown {et~al.}(2021)Brown, Vallenari, Prusti, de~Bruijne, Babusiaux,
  Biermann, Creevey, Evans, Eyer, Hutton, Jansen, Jordi, Klioner, Lammers,
  Lindegren, Luri, Mignard, Panem, Pourbaix, Randich, Sartoretti, Soubiran,
  Walton, Arenou, Bailer-Jones, Bastian, Cropper, Drimmel, Katz, Lattanzi, van
  Leeuwen, Bakker, Cacciari, Casta{\~{n}}eda, Angeli, Ducourant, Fabricius,
  Fouesneau, Fr{\'{e}}mat, Guerra, Guerrier, Guiraud, Piccolo, Masana,
  Messineo, Mowlavi, Nicolas, Nienartowicz, Pailler, Panuzzo, Riclet, Roux,
  Seabroke, Sordo, Tanga, Th{\'{e}}venin, Gracia-Abril, Portell, Teyssier,
  Altmann, Andrae, Bellas-Velidis, Benson, Berthier, Blomme, Brugaletta,
  Burgess, Busso, Carry, Cellino, Cheek, Clementini, Damerdji, Davidson,
  Delchambre, Dell'Oro, Fern{\'{a}}ndez-Hern{\'{a}}ndez, Galluccio,
  Garc{\'{i}}a-Lario, Garcia-Reinaldos, Gonz{\'{a}}lez-N{\'{u}}{\~{n}}ez,
  Gosset, Haigron, Halbwachs, Hambly, Harrison, Hatzidimitriou, Heiter,
  Hern{\'{a}}ndez, Hestroffer, Hodgkin, Holl, Jan{\ss}en, de~Fombelle, Jordan,
  Krone-Martins, Lanzafame, L{\"{o}}ffler, Lorca, Manteiga, Marchal, Marrese,
  Moitinho, Mora, Muinonen, Osborne, Pancino, Pauwels, Petit, Recio-Blanco,
  Richards, Riello, Rimoldini, Robin, Roegiers, Rybizki, Sarro, Siopis, Smith,
  Sozzetti, Ulla, Utrilla, van Leeuwen, van Reeven, Abbas, Aramburu, Accart,
  Aerts, Aguado, Ajaj, Altavilla, {\'{A}}lvarez, Cid-Fuentes, Alves, Anderson,
  Varela, Antoja, Audard, Baines, Baker, Balaguer-N{\'{u}}{\~{n}}ez, Balbinot,
  Balog, Barache, Barbato, Barros, Barstow, Bartolom{\'{e}}, Bassilana,
  Bauchet, Baudesson-Stella, Becciani, Bellazzini, Bernet, Bertone, Bianchi,
  Blanco-Cuaresma, Boch, Bombrun, Bossini, Bouquillon, Bragaglia, Bramante,
  Breedt, Bressan, Brouillet, Bucciarelli, Burlacu, Busonero, Butkevich, Buzzi,
  Caffau, Cancelliere, C{\'{a}}novas, Cantat-Gaudin, Carballo, Carlucci,
  Carnerero, Carrasco, Casamiquela, Castellani, Castro-Ginard, Sampol, Chaoul,
  Charlot, Chemin, Chiavassa, Cioni, Comoretto, Cooper, Cornez, Cowell, Crifo,
  Crosta, Crowley, Dafonte, Dapergolas, David, David, de~Laverny, Luise, March,
  Ridder, de~Souza, de~Teodoro, de~Torres, del Peloso, del Pozo, Delbo,
  Delgado, Delgado, Delisle, Matteo, Diakite, Diener, Distefano, Dolding,
  Eappachen, Edvardsson, Enke, Esquej, Fabre, Fabrizio, Faigler, Fedorets,
  Fernique, Fienga, Figueras, Fouron, Fragkoudi, Fraile, Franke, Gai, Garabato,
  Garcia-Gutierrez, Garc{\'{i}}a-Torres, Garofalo, Gavras, Gerlach, Geyer,
  Giacobbe, Gilmore, Girona, Giuffrida, Gomel, Gomez, Gonzalez-Santamaria,
  Gonz{\'{a}}lez-Vidal, Granvik, Guti{\'{e}}rrez-S{\'{a}}nchez, Guy, Hauser,
  Haywood, Helmi, Hidalgo, Hilger, H{\l}adczuk, Hobbs, Holland, Huckle,
  Jasniewicz, Jonker, Campillo, Julbe, Karbevska, Kervella, Khanna, Kochoska,
  Kontizas, Kordopatis, Korn, Kostrzewa-Rutkowska, Kruszy{\'{n}}ska, Lambert,
  Lanza, Lasne, Campion, Fustec, Lebreton, Lebzelter, Leccia, Leclerc,
  Lecoeur-Taibi, Liao, Licata, Lindstr{\o}m, Lister, Livanou, Lobel, Pardo,
  Managau, Mann, Marchant, Marconi, Santos, Marinoni, Marocco, Marshall, Polo,
  Mart{\'{i}}n-Fleitas, Masip, Massari, Mastrobuono-Battisti, Mazeh, McMillan,
  Messina, Michalik, Millar, Mints, Molina, Molinaro, Moln{\'{a}}r,
  Montegriffo, Mor, Morbidelli, Morel, Morris, Mulone, Munoz, Muraveva, Murphy,
  Musella, Noval, Ord{\'{e}}novic, Orr{\`{u}}, Osinde, Pagani, Pagano,
  Palaversa, Palicio, Panahi, Pawlak, Esteller, Penttil{\"{a}}, Piersimoni,
  Pineau, Plachy, Plum, Poggio, Poretti, Poujoulet, Pr{\v{s}}a, Pulone, Racero,
  Ragaini, Rainer, Raiteri, Rambaux, Ramos, Ramos-Lerate, Fiorentin, Regibo,
  Reyl{\'{e}}, Ripepi, Riva, Rixon, Robichon, Robin, Roelens, Rohrbasser,
  Romero-G{\'{o}}mez, Rowell, Royer, Rybicki, Sadowski, Sell{\'{e}}s, Sahlmann,
  Salgado, Salguero, Samaras, Gimenez, Sanna, Santove{\~{n}}a, Sarasso,
  Schultheis, Sciacca, Segol, Segovia, S{\'{e}}gransan, Semeux, Shahaf,
  Siddiqui, Siebert, Siltala, Slezak, Smart, Solano, Solitro, Souami, Souchay,
  Spagna, Spoto, Steele, Steidelm{\"{u}}ller, Stephenson, S{\"{u}}veges,
  Szabados, Szegedi-Elek, Taris, Tauran, Taylor, Teixeira, Thuillot, Tonello,
  Torra, Torra, Turon, Unger, Vaillant, van Dillen, Vanel, Vecchiato, Viala,
  Vicente, Voutsinas, Weiler, Wevers, Wyrzykowski, Yoldas, Yvard, Zhao, Zorec,
  Zucker, Zurbach, \& Zwitter}]{Brown2021a}
Brown, A. G.~A., Vallenari, A., Prusti, T., {et~al.} 2021, A\&A, 649, A1

\bibitem[{Bruzual \& Charlot(2003)}]{Bruzual2003}
Bruzual, G. \& Charlot, S. 2003, MNRAS, 344, 1000

\bibitem[{Buck(2020)}]{Buck2020}
Buck, T. 2020, MNRAS, 491, 5435

\bibitem[{Buck {et~al.}(2021)Buck, Rybizki, Buder, Obreja, Macci{\`{o}},
  Pfrommer, Steinmetz, \& Ness}]{Buck2021}
Buck, T., Rybizki, J., Buder, S., {et~al.} 2021, MNRAS, 508, 3365

\bibitem[{Buder {et~al.}(2021)Buder, Sharma, Kos, Amarsi, Nordlander, Lind,
  Martell, Asplund, Bland-Hawthorn, Casey, de~Silva, D'Orazi, Freeman, Hayden,
  Lewis, Lin, Schlesinger, Simpson, Stello, Zucker, Zwitter, Beeson, Buck,
  Casagrande, Clark, {\v{C}}otar, da~Costa, de~Grijs, Feuillet, Horner, Kafle,
  Khanna, Kobayashi, Liu, Montet, Nandakumar, Nataf, Ness, Spina,
  Tepper-Garc{\'{i}}a, Ting, Traven, Vogrin{\v{c}}i{\v{c}}, Wittenmyer, Wyse,
  {\v{Z}}erjal, {\v{Z}}erjal, \& Collaboration}]{Buder2021}
Buder, S., Sharma, S., Kos, J., {et~al.} 2021, MNRAS, 506, 150

\bibitem[{Buonanno {et~al.}(1999)Buonanno, Corsi, Castellani, Marconi, {Fusi
  Pecci}, \& Zinn}]{Buonanno1999}
Buonanno, R., Corsi, C.~E., Castellani, M., {et~al.} 1999, AJ, 118, 1671

\bibitem[{Burbidge {et~al.}(1957)Burbidge, Burbidge, Fowler, \&
  Hoyle}]{Burbidge1957}
Burbidge, E., Burbidge, G., Fowler, W., \& Hoyle, F. 1957, Reviews of Modern
  Physics, 29, 547

\bibitem[{Butler {et~al.}(2017)Butler, Vogt, Laughlin, Burt, Rivera, Tuomi,
  Teske, Arriagada, Diaz, Holden, \& Keiser}]{Butler2017}
Butler, R.~P., Vogt, S.~S., Laughlin, G., {et~al.} 2017, AJ, 153, 208

\bibitem[{Caldwell {et~al.}(2011)Caldwell, Schiavon, Morrison, Rose, \&
  Harding}]{Caldwell2011}
Caldwell, N., Schiavon, R., Morrison, H., Rose, J.~A., \& Harding, P. 2011, AJ,
  141, 61

\bibitem[{Carlsson(1986)}]{Carlsson1986}
Carlsson, M. 1986, Uppsala Astronomical Observatory Reports, 33

\bibitem[{Carretta {et~al.}(2009{\natexlab{a}})Carretta, Bragaglia, Gratton, \&
  Lucatello}]{Carretta2009}
Carretta, E., Bragaglia, A., Gratton, R., \& Lucatello, S. 2009{\natexlab{a}},
  A\&A, 505, 139

\bibitem[{Carretta {et~al.}(2009{\natexlab{b}})Carretta, Bragaglia, Gratton,
  Lucatello, Catanzaro, Leone, Bellazzini, Claudi, D'Orazi, Momany, Ortolani,
  Pancino, Piotto, Recio-Blanco, \& Sabbi}]{Carretta2009a}
Carretta, E., Bragaglia, A., Gratton, R.~G., {et~al.} 2009{\natexlab{b}}, A\&A,
  505, 117

\bibitem[{Carretta \& Gratton(1997)}]{Carretta1997}
Carretta, E. \& Gratton, R.~G. 1997, A\&AS, 121, 95

\bibitem[{Castelli(2005)}]{Castelli2005}
Castelli, F. 2005, Memorie della Societ{\`{a}} Astronomica Italiana Supplement,
  8, 44

\bibitem[{Castelli \& Hubrig(2004)}]{Castelli2004}
Castelli, F. \& Hubrig, S. 2004, A\&A, 425, 263

\bibitem[{Castelli \& Kurucz(2003)}]{Castelli2003}
Castelli, F. \& Kurucz, R.~L. 2003, in Modelling of Stellar Atmospheres, Proc.
  of the 210th Symposium of the IAU, ed. N.~Piskunov, W.~W. Weiss, \& D.~F.
  Gray, Vol. 210, 20

\bibitem[{Cayrel(1988)}]{Cayrel1988}
Cayrel, R. 1988, in IAU Symp. 132, The Impact of Very High S/N Spectroscopy on
  Stellar Physics, ed. G.~{Cayrel de Strobel} \& M.~Spite (Dordrecht: Kluwer
  Academic Publishers), 345

\bibitem[{Cenarro {et~al.}(2007)Cenarro, Beasley, Strader, Brodie, \&
  Forbes}]{Cenarro2007}
Cenarro, A.~J., Beasley, M.~A., Strader, J., Brodie, J.~P., \& Forbes, D.~A.
  2007, AJ, 134, 391

\bibitem[{{\v{C}}erniauskas {et~al.}(2018){\v{C}}erniauskas, Ku{\v{c}}inskas,
  Klevas, Bonifacio, Ludwig, Caffau, \& Steffen}]{Cerniauskas2018}
{\v{C}}erniauskas, A., Ku{\v{c}}inskas, A., Klevas, J., {et~al.} 2018, A\&A,
  616, A142

\bibitem[{Cescutti {et~al.}(2006)Cescutti, Fran{\c{c}}ois, Matteucci, Cayrel,
  \& Spite}]{Cescutti2006}
Cescutti, G., Fran{\c{c}}ois, P., Matteucci, F., Cayrel, R., \& Spite, M. 2006,
  A{\&}A, 448, 557

\bibitem[{Chandar {et~al.}(1999)Chandar, Bianchi, \& Ford}]{Chandar1999}
Chandar, R., Bianchi, L., \& Ford, H.~C. 1999, ApJS, 122, 431

\bibitem[{Christian \& Schommer(1982)}]{Christian1982}
Christian, C.~A. \& Schommer, R.~A. 1982, ApJS, 49, 405

\bibitem[{Cohen(1978)}]{Cohen1978}
Cohen, J.~G. 1978, ApJ, 223, 487

\bibitem[{Cohen(2011)}]{Cohen2011}
Cohen, J.~G. 2011, ApJ, 740, L38

\bibitem[{Cohen \& Blakeslee(1998)}]{Cohen1998}
Cohen, J.~G. \& Blakeslee, J.~P. 1998, AJ, 115, 2356

\bibitem[{Cohen {et~al.}(2002)Cohen, Briley, \& Stetson}]{Cohen2002}
Cohen, J.~G., Briley, M.~M., \& Stetson, P.~B. 2002, AJ, 123, 2525

\bibitem[{Cohen \& Huang(2009)}]{Cohen2009}
Cohen, J.~G. \& Huang, W. 2009, ApJ, 701, 1053

\bibitem[{Cohen \& Huang(2010)}]{Cohen2010}
Cohen, J.~G. \& Huang, W. 2010, ApJ, 719, 931

\bibitem[{Colucci \& Bernstein(2011)}]{Colucci2011}
Colucci, J. \& Bernstein, R. 2011, EAS Publications Series, 48, 275

\bibitem[{Colucci {et~al.}(2009)Colucci, Bernstein, Cameron, McWilliam, \&
  Cohen}]{Colucci2009}
Colucci, J.~E., Bernstein, R.~A., Cameron, S., McWilliam, A., \& Cohen, J.~G.
  2009, ApJ, 704, 385

\bibitem[{Colucci {et~al.}(2014)Colucci, Bernstein, \& Cohen}]{Colucci2014}
Colucci, J.~E., Bernstein, R.~A., \& Cohen, J.~G. 2014, ApJ, 797, 116

\bibitem[{Colucci {et~al.}(2017)Colucci, Bernstein, \& McWilliam}]{Colucci2017}
Colucci, J.~E., Bernstein, R.~A., \& McWilliam, A. 2017, ApJ, 834, 105

\bibitem[{Colucci {et~al.}(2013)Colucci, Duran, Bernstein, \&
  McWilliam}]{Colucci2013}
Colucci, J.~E., Duran, M.~F., Bernstein, R.~A., \& McWilliam, A. 2013, ApJL,
  773, L36

\bibitem[{Conroy {et~al.}(2014)Conroy, Graves, \& van Dokkum}]{Conroy2014}
Conroy, C., Graves, G.~J., \& van Dokkum, P.~G. 2014, ApJ, 780, 33

\bibitem[{Conroy \& van Dokkum(2012)}]{Conroy2012b}
Conroy, C. \& van Dokkum, P.~G. 2012, ApJ, 760, 71

\bibitem[{Conroy {et~al.}(2018)Conroy, Villaume, van Dokkum, \&
  Lind}]{Conroy2018}
Conroy, C., Villaume, A., van Dokkum, P.~G., \& Lind, K. 2018, ApJ, 854, 139

\bibitem[{Cook {et~al.}(2016)Cook, Conroy, Pillepich, Rodriguez-Gomez,
  Hernquist, Cook, Conroy, Pillepich, Rodriguez-Gomez, \& Hernquist}]{Cook2016}
Cook, B.~A., Conroy, C., Pillepich, A., {et~al.} 2016, ApJ, 833, 158

\bibitem[{Cooper {et~al.}(2010)Cooper, Cole, Frenk, White, Helly, Benson, {De
  Lucia}, Helmi, Jenkins, Navarro, Springel, \& Wang}]{Cooper2010}
Cooper, A.~P., Cole, S., Frenk, C.~S., {et~al.} 2010, MNRAS, 406, 744

\bibitem[{Cordoni {et~al.}(2021)Cordoni, {Da Costa}, Yong, Mackey, Marino,
  Monty, Nordlander, Norris, Asplund, Bessell, Casey, Frebel, Lind, Murphy,
  Schmidt, Gao, Xylakis-Dornbusch, Amarsi, \& Milone}]{Cordoni2021}
Cordoni, G., {Da Costa}, G.~S., Yong, D., {et~al.} 2021, MNRAS, 503, 2539

\bibitem[{Cox(2000)}]{Cox2000}
Cox, A.~N. 2000, {Allen's Astrophysical Quantities}, 4th edn. (New York: AIP
  Press)

\bibitem[{{Da Costa} \& Mould(1988)}]{DaCosta1988}
{Da Costa}, G.~S. \& Mould, J.~R. 1988, ApJ, 334, 159

\bibitem[{da~Silveira {et~al.}(2018)da~Silveira, Barbuy, Fria{\c{c}}a, Hill,
  Zoccali, Rafelski, Gonzalez, Minniti, Renzini, \& Ortolani}]{DaSilveira2018}
da~Silveira, C.~R., Barbuy, B., Fria{\c{c}}a, A. C.~S., {et~al.} 2018, A\&A,
  614, A149

\bibitem[{Davison {et~al.}(2021)Davison, Norris, Leaman, Kuntschner, Boecker,
  \& van~de Ven}]{Davison2021}
Davison, T.~A., Norris, M.~A., Leaman, R., {et~al.} 2021, MNRAS, 507, 3089

\bibitem[{Dekker {et~al.}(2000)Dekker, D'Odorico, Kaufer, Delabre, \&
  Kotzlowski}]{Dekker2000}
Dekker, H., D'Odorico, S., Kaufer, A., Delabre, B., \& Kotzlowski, H. 2000,
  Proc. SPIE, 4008, 534

\bibitem[{Dotter {et~al.}(2007)Dotter, Chaboyer, Jevremovi{\'{c}}, Baron,
  Ferguson, Sarajedini, \& Anderson}]{Dotter2007}
Dotter, A., Chaboyer, B., Jevremovi{\'{c}}, D., {et~al.} 2007, AJ, 134, 376

\bibitem[{Drawin \& Felenbok(1965)}]{Drawin1965}
Drawin, H.-W. \& Felenbok, P. 1965, {Data for plasmas in local thermodynamic
  equilibrium} (Paris: Gauthier-Villars)

\bibitem[{Dubath {et~al.}(1992)Dubath, Meylan, \& Mayor}]{Dubath1992}
Dubath, P., Meylan, G., \& Mayor, M. 1992, ApJ, 400, 510

\bibitem[{Dutra-Ferreira {et~al.}(2016)Dutra-Ferreira, Pasquini, Smiljanic,
  de~Mello, \& Steffen}]{Dutra-Ferreira2016}
Dutra-Ferreira, L., Pasquini, L., Smiljanic, R., de~Mello, G. F.~P., \&
  Steffen, M. 2016, A\&A, 585, A75

\bibitem[{Eggen {et~al.}(1962)Eggen, Lynden-Bell, \& Sandage}]{Eggen1962}
Eggen, O.~J., Lynden-Bell, D., \& Sandage, A.~R. 1962, ApJ, 136, 748

\bibitem[{Eitner {et~al.}(2020)Eitner, Bergemann, Hansen, Cescutti, Seitenzahl,
  Larsen, \& Plez}]{Eitner2020}
Eitner, P., Bergemann, M., Hansen, C., {et~al.} 2020, A\&A, 635, A38

\bibitem[{Eitner {et~al.}(2019)Eitner, Bergemann, \& Larsen}]{Eitner2019}
Eitner, P., Bergemann, M., \& Larsen, S.~S. 2019, A\&A, 627, A40

\bibitem[{Escala {et~al.}(2020)Escala, Gilbert, Kirby, Wojno, Cunningham, \&
  Guhathakurta}]{Escala2020a}
Escala, I., Gilbert, K.~M., Kirby, E.~N., {et~al.} 2020, ApJ, 889, 177

\bibitem[{Escala {et~al.}(2019)Escala, Kirby, Gilbert, Cunningham, \&
  Wojno}]{Escala2019}
Escala, I., Kirby, E.~N., Gilbert, K.~M., Cunningham, E.~C., \& Wojno, J. 2019,
  ApJ, 878, 42

\bibitem[{Forbes(2020)}]{Forbes2020}
Forbes, D.~A. 2020, MNRAS, 493, 847

\bibitem[{Forte {et~al.}(1981)Forte, Strom, \& Strom}]{Forte1981}
Forte, J.~C., Strom, S.~E., \& Strom, K.~M. 1981, ApJ, 245, L9

\bibitem[{Frebel(2010)}]{Frebel2010}
Frebel, A. 2010, Astronomische Nachrichten, 331, 474

\bibitem[{Freedman \& Madore(1988)}]{Freedman1988}
Freedman, W.~L. \& Madore, B.~F. 1988, ApJ, 332, L63

\bibitem[{Freeman \& Bland-Hawthorn(2002)}]{Freeman2002}
Freeman, K. \& Bland-Hawthorn, J. 2002, ARA\&A, 40, 487

\bibitem[{Fulbright(2002)}]{Fulbright2002}
Fulbright, J.~P. 2002, AJ, 123, 404

\bibitem[{Fulbright {et~al.}(2006)Fulbright, McWilliam, \&
  Rich}]{Fulbright2006}
Fulbright, J.~P., McWilliam, A., \& Rich, R.~M. 2006, ApJ, 636, 821

\bibitem[{Furenlid(1988)}]{Furenlid1988}
Furenlid, I. 1988, in IAU Symp. 132, The Impact of Very High S/N Spectroscopy
  on Stellar Physics, ed. G.~{Cayrel de Strobel} \& M.~Spite (Dordrecht: Kluwer
  Academic Publishers), 435

\bibitem[{Gallagher {et~al.}(2020)Gallagher, Bergemann, Collet, Plez,
  Leenaarts, Carlsson, Yakovleva, \& Belyaev}]{Gallagher2020}
Gallagher, A.~J., Bergemann, M., Collet, R., {et~al.} 2020, A\&A, 634, A55

\bibitem[{Galleti {et~al.}(2004)Galleti, Federici, Bellazzini, Pecci, \&
  Macrina}]{Galleti2004}
Galleti, S., Federici, L., Bellazzini, M., Pecci, F.~F., \& Macrina, S. 2004,
  A\&A, 416, 917

\bibitem[{Gehren {et~al.}(2004)Gehren, Liang, Shi, Zhang, \& Zhao}]{Gehren2004}
Gehren, T., Liang, Y.~C., Shi, J.~R., Zhang, H.~W., \& Zhao, G. 2004, A\&A,
  413, 1045

\bibitem[{Gehren {et~al.}(2006)Gehren, Shi, Zhang, Zhao, \& Korn}]{Gehren2006}
Gehren, T., Shi, J.~R., Zhang, H.~W., Zhao, G., \& Korn, A.~J. 2006, A\&A, 451,
  1065

\bibitem[{Genel {et~al.}(2014)Genel, Vogelsberger, Springel, Sijacki, Nelson,
  Snyder, Rodriguez-Gomez, Torrey, Hernquist, Genel, Vogelsberger, Springel,
  Sijacki, Nelson, Snyder, Rodriguez-Gomez, Torrey, \& Hernquist}]{Genel2014}
Genel, S., Vogelsberger, M., Springel, V., {et~al.} 2014, MNRAS, 445, 175

\bibitem[{Gieren {et~al.}(2013)Gieren, G{\'{o}}rski, Pietrzy{\'{n}}ski,
  Konorski, Suchomska, Graczyk, Pilecki, Bresolin, Kudritzki, Storm,
  Karczmarek, Gallenne, Calder{\'{o}}n, \& Geisler}]{Gieren2013}
Gieren, W., G{\'{o}}rski, M., Pietrzy{\'{n}}ski, G., {et~al.} 2013, ApJ, 773,
  69

\bibitem[{Gilbert {et~al.}(2020)Gilbert, Wojno, Kirby, Escala, Beaton,
  Guhathakurta, \& Majewski}]{Gilbert2020}
Gilbert, K.~M., Wojno, J., Kirby, E.~N., {et~al.} 2020, AJ, 160, 41

\bibitem[{Gilmore \& Wyse(1998)}]{Gilmore1998}
Gilmore, G. \& Wyse, R. F.~G. 1998, AJ, 116, 748

\bibitem[{Gorgas {et~al.}(1997)Gorgas, Pedraz, Guzman, Cardiel, \&
  Gonzalez}]{Gorgas1997}
Gorgas, J., Pedraz, S., Guzman, R., Cardiel, N., \& Gonzalez, J.~J. 1997, ApJ,
  481, L19

\bibitem[{Gratton {et~al.}(2019)Gratton, Bragaglia, Carretta, D'Orazi,
  Lucatello, \& Sollima}]{Gratton2019}
Gratton, R., Bragaglia, A., Carretta, E., {et~al.} 2019, A\&A, 27, 8

\bibitem[{Gratton {et~al.}(2004)Gratton, Sneden, \& Carretta}]{Gratton2004}
Gratton, R., Sneden, C., \& Carretta, E. 2004, ARA\&A, 42, 385

\bibitem[{Gratton {et~al.}(2012)Gratton, Carretta, \& Bragaglia}]{Gratton2012}
Gratton, R.~G., Carretta, E., \& Bragaglia, A. 2012, A\&AR, 20, 1

\bibitem[{Gratton {et~al.}(2003)Gratton, Carretta, Claudi, Lucatello, \&
  Barbieri}]{Gratton2003a}
Gratton, R.~G., Carretta, E., Claudi, R., Lucatello, S., \& Barbieri, M. 2003,
  A\&A, 404, 187

\bibitem[{Gratton {et~al.}(1999)Gratton, Carretta, Eriksson, \&
  Gustafsson}]{Gratton1999}
Gratton, R.~G., Carretta, E., Eriksson, K., \& Gustafsson, B. 1999, A\&A, 350,
  955

\bibitem[{Gratton \& Sneden(1987)}]{Gratton1987}
Gratton, R.~G. \& Sneden, C. 1987, A{\&}A, 178, 179

\bibitem[{Gratton {et~al.}(2000)Gratton, Sneden, Carretta, \&
  Bragaglia}]{Gratton2000}
Gratton, R.~G., Sneden, C., Carretta, E., \& Bragaglia, A. 2000, A\&A, 354, 169

\bibitem[{Graves \& Schiavon(2008)}]{Graves2008}
Graves, G.~J. \& Schiavon, R.~P. 2008, ApJS, 177, 446

\bibitem[{Greener {et~al.}(2021)Greener, Merrifield, Arag{\'{o}}n-Salamanca,
  Peterken, Andrews, \& Lane}]{Greener2021}
Greener, M.~J., Merrifield, M., Arag{\'{o}}n-Salamanca, A., {et~al.} 2021,
  MNRAS, 502, L95

\bibitem[{Grevesse \& Sauval(1998)}]{Grevesse1998}
Grevesse, N. \& Sauval, A.~J. 1998, Space Science Reviews, 85, 161

\bibitem[{Grumer \& Barklem(2020)}]{Grumer2020}
Grumer, J. \& Barklem, P.~S. 2020, A{\&}A, 637, A28

\bibitem[{Gustafsson {et~al.}(2008)Gustafsson, Edvardsson, Eriksson,
  J{\o}rgensen, Nordlund, \& Plez}]{Gustafsson2008}
Gustafsson, B., Edvardsson, B., Eriksson, K., {et~al.} 2008, A\&A, 486, 951

\bibitem[{Hansen {et~al.}(2018)Hansen, El-Souri, Monaco, Villanova, Bonifacio,
  Caffau, \& Sbordone}]{Hansen2018}
Hansen, C.~J., El-Souri, M., Monaco, L., {et~al.} 2018, ApJ, 855, 83

\bibitem[{Hardy(2002)}]{Hardy2002}
Hardy, E. 2002, in IAU Symp. 207, Extragalactic Star Clusters, ed. D.~Geisler,
  E.~K. Grebel, \& D.~Minniti (San Francisco: Astronomical Society of the
  Pacific), 62

\bibitem[{Harris(1996)}]{Harris1996}
Harris, W.~E. 1996, AJ, 112, 1487

\bibitem[{Harris {et~al.}(2007)Harris, Harris, Layden, \& Wehner}]{Harris2007}
Harris, W.~E., Harris, G. L.~H., Layden, A.~C., \& Wehner, E. M.~H. 2007, ApJ,
  666, 903

\bibitem[{Hasselquist {et~al.}(2017)Hasselquist, Shetrone, Smith, Holtzman,
  McWilliam, Fern{\'{a}}ndez-Trincado, Beers, Majewski, Nidever, Tang, Tissera,
  Alvar, Prieto, Almeida, Anguiano, Battaglia, Carigi, Inglada, Frinchaboy,
  Garc{\'{i}}a-Hern{\'{a}}ndez, Geisler, Minniti, Placco, Schultheis, Sobeck,
  \& Villanova}]{Hasselquist2017}
Hasselquist, S., Shetrone, M., Smith, V., {et~al.} 2017, ApJ, 845, 162

\bibitem[{Helmi {et~al.}(2018)Helmi, Babusiaux, Koppelman, Massari, Veljanoski,
  \& Brown}]{Helmi2018}
Helmi, A., Babusiaux, C., Koppelman, H.~H., {et~al.} 2018, Nature, 563, 85

\bibitem[{Hendricks {et~al.}(2016)Hendricks, Boeche, Johnson, Frank, Koch,
  Mateo, \& Bailey}]{Hendricks2016}
Hendricks, B., Boeche, C., Johnson, C.~I., {et~al.} 2016, A\&A, 585, A86

\bibitem[{Hendricks {et~al.}(2014)Hendricks, Koch, Walker, Johnson,
  Pe{\~{n}}arrubia, \& Gilmore}]{Hendricks2014}
Hendricks, B., Koch, A., Walker, M., {et~al.} 2014, A\&A, 572, A82

\bibitem[{Herbig(1975)}]{Herbig1975}
Herbig, G.~H. 1975, ApJ, 196, 129

\bibitem[{Hernandez {et~al.}(2018)Hernandez, Larsen, Trager, Kaper, \&
  Groot}]{Hernandez2018}
Hernandez, S., Larsen, S., Trager, S., Kaper, L., \& Groot, P. 2018, MNRAS,
  476, 5189

\bibitem[{Herschel(1847)}]{Herschel1847}
Herschel, J. F.~W. 1847, {Results of astronomical observations made during the
  years 1834, 5, 6, 7, 8, at the Cape of Good Hope; being the completion of a
  telescopic survey of the whole surface of the visible heavens, commenced in
  1825} (London, Smith, Elder and co.)

\bibitem[{Hidalgo {et~al.}(2018)Hidalgo, Pietrinferni, Cassisi, Salaris,
  Mucciarelli, Savino, Aparicio, Aguirre, \& Verma}]{Hidalgo2018}
Hidalgo, S.~L., Pietrinferni, A., Cassisi, S., {et~al.} 2018, ApJ, 856, 125

\bibitem[{Hiltner(1960)}]{Hiltner1960}
Hiltner, W.~A. 1960, ApJ, 131, 163

\bibitem[{Hodge(1961)}]{Hodge1961}
Hodge, P.~W. 1961, AJ, 66, 83

\bibitem[{Hodge {et~al.}(1999)Hodge, Dolphin, Smith, \& Mateo}]{Hodge1999}
Hodge, P.~W., Dolphin, A.~E., Smith, T.~R., \& Mateo, M. 1999, ApJ, 521, 577

\bibitem[{Hopkins(2018)}]{Hopkins2018}
Hopkins, A.~M. 2018, PASA, 35, e039

\bibitem[{Horta {et~al.}(2020)Horta, Schiavon, Mackereth, Beers,
  Fern{\'{a}}ndez-Trincado, Frinchaboy, Garc{\'{i}}a-Hern{\'{a}}ndez, Geisler,
  Hasselquist, J{\"{o}}nsson, Lane, Majewski, M{\'{e}}sz{\'{a}}ros, Bidin,
  Nataf, Roman-Lopes, Nitschelm, Vargas-Gonz{\'{a}}lez, \&
  Zasowski}]{Horta2020}
Horta, D., Schiavon, R.~P., Mackereth, J.~T., {et~al.} 2020, MNRAS, 493, 3363

\bibitem[{Hubble(1932)}]{Hubble1932}
Hubble, E. 1932, ApJ, 76, 44

\bibitem[{Hubble(1925)}]{Hubble1925}
Hubble, E.~P. 1925, ApJ, 62, 409

\bibitem[{Humason {et~al.}(1956)Humason, Mayall, \& Sandage}]{Humason1956}
Humason, M.~L., Mayall, N.~U., \& Sandage, A.~R. 1956, AJ, 61, 97

\bibitem[{Huxor {et~al.}(2013)Huxor, Ferguson, Veljanoski, Mackey, \&
  Tanvir}]{Huxor2013}
Huxor, A., Ferguson, A., Veljanoski, J., Mackey, D., \& Tanvir, N. 2013, MNRAS,
  429, 1039

\bibitem[{Huxor {et~al.}(2009)Huxor, Ferguson, Barker, Tanvir, Irwin, Chapman,
  Ibata, \& Lewis}]{Huxor2009}
Huxor, A., Ferguson, A. M.~N., Barker, M.~K., {et~al.} 2009, ApJL, 698, L77

\bibitem[{Huxor {et~al.}(2014)Huxor, Mackey, Ferguson, Irwin, Martin, Tanvir,
  Veljanoski, McConnachie, Fishlock, Ibata, \& Lewis}]{Huxor2014}
Huxor, A.~P., Mackey, A.~D., Ferguson, A. M.~N., {et~al.} 2014, MNRAS, 442,
  2165

\bibitem[{Hwang {et~al.}(2011)Hwang, Lee, Lee, Park, Park, Kim, \&
  Park}]{Hwang2011}
Hwang, N., Lee, M.~G., Lee, J.~C., {et~al.} 2011, ApJ, 738, 58

\bibitem[{Ibata {et~al.}(1994)Ibata, Gilmore, \& Irwin}]{Ibata1994}
Ibata, R.~A., Gilmore, G., \& Irwin, M.~J. 1994, Nature, 370, 194

\bibitem[{Ibata {et~al.}(2014)Ibata, Lewis, McConnachie, Martin, Irwin,
  Ferguson, Babul, Bernard, Chapman, Collins, Fardal, Mackey, Navarro,
  Pe{\~{n}}arrubia, Rich, Tanvir, \& Widrow}]{Ibata2014}
Ibata, R.~A., Lewis, G.~F., McConnachie, A.~W., {et~al.} 2014, ApJ, 780, 128

\bibitem[{Igenbergs {et~al.}(2008)Igenbergs, Schweinzer, Bray, Bridi, \&
  Aumayr}]{Igenbergs2008}
Igenbergs, K., Schweinzer, J., Bray, I., Bridi, D., \& Aumayr, F. 2008, ADNDT,
  94, 981

\bibitem[{Irwin(1981)}]{Irwin1981}
Irwin, A.~W. 1981, ApJS, 45, 621

\bibitem[{Ishigaki {et~al.}(2013)Ishigaki, Aoki, \& Chiba}]{Ishigaki2013}
Ishigaki, M.~N., Aoki, W., \& Chiba, M. 2013, ApJ, 771, 67

\bibitem[{Jefferies(1968)}]{Jefferies1968}
Jefferies, J.~T. 1968, {Spectral line formation} (Waltham, Massachusetts:
  Blaisdell Publishing Company)

\bibitem[{Ji {et~al.}(2016)Ji, Frebel, Chiti, \& Simon}]{Ji2016}
Ji, A.~P., Frebel, A., Chiti, A., \& Simon, J.~D. 2016, Nature, 531, 610

\bibitem[{Jofr{\'{e}} {et~al.}(2019)Jofr{\'{e}}, Heiter, \&
  Soubiran}]{Jofre2019}
Jofr{\'{e}}, P., Heiter, U., \& Soubiran, C. 2019, ARA{\&}A, 57, 571

\bibitem[{Johnson {et~al.}(2012)Johnson, Rich, Kobayashi, \&
  Fulbright}]{Johnson2012b}
Johnson, C.~I., Rich, R.~M., Kobayashi, C., \& Fulbright, J.~P. 2012, ApJ, 749,
  175

\bibitem[{Keenan \& McNeil(1989)}]{Keenan1989}
Keenan, P.~C. \& McNeil, R.~C. 1989, ApJS, 71, 245

\bibitem[{Kinman(1959)}]{Kinman1959}
Kinman, T.~D. 1959, MNRAS, 119, 538

\bibitem[{Kirby {et~al.}(2011)Kirby, Lanfranchi, Simon, Cohen, \&
  Guhathakurta}]{Kirby2011a}
Kirby, E.~N., Lanfranchi, G.~A., Simon, J.~D., Cohen, J.~G., \& Guhathakurta,
  P. 2011, ApJ, 727, 78

\bibitem[{Kirby {et~al.}(2019)Kirby, Xie, Guo, de~los Reyes, Bergemann,
  Kovalev, Shen, Piro, \& McWilliam}]{Kirby2019}
Kirby, E.~N., Xie, J.~L., Guo, R., {et~al.} 2019, ApJ, 881, 45

\bibitem[{Kobayashi {et~al.}(2019)Kobayashi, Haynes, \&
  Vincenzo}]{Kobayashi2019}
Kobayashi, C., Haynes, C.~J., \& Vincenzo, F. 2019, in IAU Symp. 343, Why
  Galaxies Care About AGB Stars: A Continuing Challenge through Cosmic Time
  (Cambridge University Press), 247--257

\bibitem[{Kobayashi {et~al.}(2020)Kobayashi, Karakas, \&
  Lugaro}]{Kobayashi2020}
Kobayashi, C., Karakas, A.~I., \& Lugaro, M. 2020, ApJ, 900, 179

\bibitem[{Koch \& McWilliam(2008)}]{Koch2008}
Koch, A. \& McWilliam, A. 2008, AJ, 135, 1551

\bibitem[{Koch-Hansen {et~al.}(2021)Koch-Hansen, Hansen, \&
  McWilliam}]{Koch-Hansen2021}
Koch-Hansen, A.~J., Hansen, C.~J., \& McWilliam, A. 2021, A\&A, 653, A2

\bibitem[{Kondo {et~al.}(2019)Kondo, Fukue, Matsunaga, Ikeda, Taniguchi,
  Kobayashi, Sameshima, Hamano, Arai, Kawakita, Yasui, Izumi, Mizumoto, Otsubo,
  Takenaka, Watase, Asano, Yoshikawa, \& Tsujimoto}]{Kondo2019}
Kondo, S., Fukue, K., Matsunaga, N., {et~al.} 2019, ApJ, 875, 129

\bibitem[{Korotin {et~al.}(2018)Korotin, Andrievsky, \& Zhukova}]{Korotin2018}
Korotin, S.~A., Andrievsky, S.~M., \& Zhukova, A.~V. 2018, MNRAS, 480, 965

\bibitem[{Kovalev {et~al.}(2019)Kovalev, Bergemann, Ting, \& Rix}]{Kovalev2019}
Kovalev, M., Bergemann, M., Ting, Y.~S., \& Rix, H.~W. 2019, A\&A, 628, 54

\bibitem[{Kramida {et~al.}(2013)Kramida, Ralchenko, \& Reader}]{NIST}
Kramida, A., Ralchenko, Y., \& Reader, J. 2013, {NIST Atomic Spectra Database}

\bibitem[{Kriek {et~al.}(2019)Kriek, Price, Conroy, Suess, Mowla, Pasha,
  Bezanson, van Dokkum, \& Barro}]{Kriek2019}
Kriek, M., Price, S.~H., Conroy, C., {et~al.} 2019, ApJL, 880, L31

\bibitem[{Kron \& Mayall(1960)}]{Kron1960}
Kron, G.~E. \& Mayall, N.~U. 1960, AJ, 65, 581

\bibitem[{Kroupa(2001)}]{Kroupa2001}
Kroupa, P. 2001, MNRAS, 322, 231

\bibitem[{Kruijssen {et~al.}(2020)Kruijssen, Pfeffer, Chevance, Bonaca,
  Trujillo-Gomez, Bastian, Reina-Campos, Crain, \& Hughes}]{Kruijssen2020}
Kruijssen, J. M.~D., Pfeffer, J.~L., Chevance, M., {et~al.} 2020, MNRAS, 498,
  2472

\bibitem[{Kuntschner(2000)}]{Kuntschner2000}
Kuntschner, H. 2000, MNRAS, 315, 184

\bibitem[{Kupka {et~al.}(1999)Kupka, Piskunov, Ryabchikova, Stempels, \&
  Weiss}]{Kupka1999}
Kupka, F., Piskunov, N., Ryabchikova, T.~A., Stempels, H.~C., \& Weiss, W.~W.
  1999, A\&AS, 138, 119

\bibitem[{Kurucz(1970)}]{Kurucz1970}
Kurucz, R.~L. 1970, {Atlas: a Computer Program for Calculating Model Stellar
  Atmospheres,
  http://kurucz.harvard.edu/papers/sao309/saospecialreport309.pdf}, Tech. rep.,
  Smithsonian Astrophysical Observatory

\bibitem[{Kurucz(2005)}]{Kurucz2005}
Kurucz, R.~L. 2005, Memorie della Societ{\`{a}} Astronomica Italiana
  Supplement, 8, 14

\bibitem[{Kurucz \& Avrett(1981)}]{Kurucz1981}
Kurucz, R.~L. \& Avrett, E.~H. 1981, {Solar Spectrum Synthesis. I. A Sample
  Atlas from 224 to 300 nm,
  http://kurucz.harvard.edu/papers/sao391/saosr391.pdf}, Tech. rep.,
  Smithsonian Astrophysical Observatory

\bibitem[{Kurucz {et~al.}(1984)Kurucz, Furenlid, Brault, \&
  Testerman}]{Kurucz1984}
Kurucz, R.~L., Furenlid, I., Brault, J., \& Testerman, L. 1984, {Solar flux
  atlas from 296 to 1300 nm} (Sunspot, New Mexico: National Solar Observatory)

\bibitem[{{La Barbera} {et~al.}(2017){La Barbera}, Vazdekis, Ferreras,
  Pasquali, Prieto, Rock, Aguado, \& Peletier}]{LaBarbera2017}
{La Barbera}, F., Vazdekis, A., Ferreras, I., {et~al.} 2017, MNRAS, 464, 3597

\bibitem[{Lach {et~al.}(2020)Lach, R{\"{o}}pke, Seitenzahl, Cot{\'{e}}, Gronow,
  \& Ruiter}]{Lach2020}
Lach, F., R{\"{o}}pke, F.~K., Seitenzahl, I.~R., {et~al.} 2020, A\&A, 644, A118

\bibitem[{Lamers {et~al.}(2017)Lamers, Kruijssen, Bastian, Rejkuba, Hilker, \&
  Kissler-Patig}]{Lamers2017}
Lamers, H. J. G. L.~M., Kruijssen, J. M.~D., Bastian, N., {et~al.} 2017, A\&A,
  606, A85

\bibitem[{Lapenna {et~al.}(2014)Lapenna, Mucciarelli, Lanzoni, Ferraro,
  Dalessandro, Origlia, \& Massari}]{Lapenna2014}
Lapenna, E., Mucciarelli, A., Lanzoni, B., {et~al.} 2014, ApJ, 797, 124

\bibitem[{Lardo {et~al.}(2012)Lardo, Pancino, Mucciarelli, \&
  Milone}]{Lardo2012}
Lardo, C., Pancino, E., Mucciarelli, A., \& Milone, A.~P. 2012, A{\&}A, 548,
  A107

\bibitem[{Larsen(2020)}]{ispy3}
Larsen, S. 2020, {ISPy3: Integrated-light Spectroscopy for Python3,
  https://github.com/soerenslarsen/ISPy3, DOI 10.5281/zenodo.4036092}

\bibitem[{Larsen {et~al.}(2002{\natexlab{a}})Larsen, Brodie, Beasley, \&
  Forbes}]{Larsen2002e}
Larsen, S., Brodie, J., Beasley, M., \& Forbes, D. 2002{\natexlab{a}}, AJ, 124

\bibitem[{Larsen {et~al.}(2014)Larsen, Brodie, Forbes, \& Strader}]{Larsen2014}
Larsen, S.~S., Brodie, J.~P., Forbes, D.~A., \& Strader, J. 2014, A\&A, 565,
  A98

\bibitem[{Larsen {et~al.}(2002{\natexlab{b}})Larsen, Brodie, Sarajedini, \&
  Huchra}]{Larsen2002b}
Larsen, S.~S., Brodie, J.~P., Sarajedini, A., \& Huchra, J.~P.
  2002{\natexlab{b}}, AJ, 124, 2615

\bibitem[{Larsen {et~al.}(2012)Larsen, Brodie, \& Strader}]{Larsen2012a}
Larsen, S.~S., Brodie, J.~P., \& Strader, J. 2012, A\&A, 546, A53

\bibitem[{Larsen {et~al.}(2017)Larsen, Brodie, \& Strader}]{Larsen2017}
Larsen, S.~S., Brodie, J.~P., \& Strader, J. 2017, A\&A, 601, A96

\bibitem[{Larsen {et~al.}(2018{\natexlab{a}})Larsen, Brodie, Wasserman, \&
  Strader}]{Larsen2018}
Larsen, S.~S., Brodie, J.~P., Wasserman, A., \& Strader, J. 2018{\natexlab{a}},
  A\&A, 613, A56

\bibitem[{Larsen {et~al.}(2018{\natexlab{b}})Larsen, Pugliese, \&
  Brodie}]{Larsen2018a}
Larsen, S.~S., Pugliese, G., \& Brodie, J.~P. 2018{\natexlab{b}}, A\&A, 617,
  A119

\bibitem[{Larsen {et~al.}(2021)Larsen, Romanowsky, \& Brodie}]{Larsen2021}
Larsen, S.~S., Romanowsky, A.~J., \& Brodie, J.~P. 2021, A\&A, 651, A102

\bibitem[{Larsen {et~al.}(2020)Larsen, Romanowsky, Brodie, \&
  Wasserman}]{Larsen2020}
Larsen, S.~S., Romanowsky, A.~J., Brodie, J.~P., \& Wasserman, A. 2020,
  Science, 370, 970

\bibitem[{Laverick {et~al.}(2019)Laverick, Lobel, Royer, Merle, Martayan, van
  Hoof, {Van der Swaelmen}, David, Hensberge, \& Thienpont}]{Laverick2019}
Laverick, M., Lobel, A., Royer, P., {et~al.} 2019, A\&A, 624, A60

\bibitem[{Leaman {et~al.}(2013)Leaman, Venn, Brooks, Battaglia, Cole, Ibata,
  Irwin, McConnachie, Mendel, Starkenburg, \& Tolstoy}]{Leaman2013}
Leaman, R., Venn, K.~A., Brooks, A.~M., {et~al.} 2013, ApJ, 767, 131

\bibitem[{Lemasle {et~al.}(2014)Lemasle, de~Boer, Hill, Tolstoy, Irwin,
  Jablonka, Venn, Battaglia, Starkenburg, Shetrone, Letarte, Francois, Helmi,
  Primas, Kaufer, \& Szeifert}]{Lemasle2014}
Lemasle, B., de~Boer, T., Hill, V., {et~al.} 2014, A\&A, 572, A88

\bibitem[{Letarte {et~al.}(2006)Letarte, Hill, Jablonka, Tolstoy,
  Fran{\c{c}}ois, \& Meylan}]{Letarte2006}
Letarte, B., Hill, V., Jablonka, P., {et~al.} 2006, A\&A, 453, 547

\bibitem[{Letarte {et~al.}(2010)Letarte, Hill, Tolstoy, Jablonka, Shetrone,
  Venn, Spite, Irwin, Battaglia, Helmi, Primas, Fran{\c{c}}ois, Kaufer,
  Szeifert, Arimoto, \& Sadakane}]{Letarte2010}
Letarte, B., Hill, V., Tolstoy, E., {et~al.} 2010, A\&A, 523, A17

\bibitem[{Limongi \& Chieffi(2018)}]{Limongi2018}
Limongi, M. \& Chieffi, A. 2018, ApJS, 237, 13

\bibitem[{Lind {et~al.}(2011)Lind, Asplund, Barklem, \& Belyaev}]{Lind2011}
Lind, K., Asplund, M., Barklem, P.~S., \& Belyaev, A.~K. 2011, A\&A, 528, A103

\bibitem[{Liu {et~al.}(2019)Liu, Romero-Romero, Garand, Lantis, Minamisono, \&
  Stracener}]{Liu2019}
Liu, Y., Romero-Romero, E., Garand, D., {et~al.} 2019, Spectrochimica Acta Part
  B: Atomic Spectroscopy, 158, 105640

\bibitem[{Lodders(2003)}]{Lodders2003}
Lodders, K. 2003, ApJ, 591, 1220

\bibitem[{Luck \& Bond(1981)}]{Luck1981}
Luck, R.~E. \& Bond, H.~E. 1981, ApJ, 244, 919

\bibitem[{Mackereth {et~al.}(2019)Mackereth, Schiavon, Pfeffer, Hayes, Bovy,
  Anguiano, {Allende Prieto}, Hasselquist, Holtzman, Johnson, Majewski,
  O'Connell, Shetrone, Tissera, \& Fern{\'{a}}ndez-Trincado}]{Mackereth2019}
Mackereth, J.~T., Schiavon, R.~P., Pfeffer, J., {et~al.} 2019, MNRAS, 482, 3426

\bibitem[{Mackey {et~al.}(2019)Mackey, Ferguson, Huxor, Veljanoski, Lewis,
  McConnachie, Martin, Ibata, Irwin, C{\^{o}}t{\'{e}}, Collins, Tanvir, \&
  Bate}]{Mackey2019a}
Mackey, A.~D., Ferguson, A.~M., Huxor, A.~P., {et~al.} 2019, MNRAS, 484, 1756

\bibitem[{Mackey \& Gilmore(2003)}]{Mackey2003a}
Mackey, A.~D. \& Gilmore, G.~F. 2003, MNRAS, 340, 175

\bibitem[{Magic {et~al.}(2013)Magic, Collet, Asplund, Trampedach, Hayek,
  Chiavassa, Stein, \& Nordlund}]{Magic2013}
Magic, Z., Collet, R., Asplund, M., {et~al.} 2013, A{\&}A, 557, A26

\bibitem[{Malhan {et~al.}(2018)Malhan, Ibata, \& Martin}]{Malhan2018}
Malhan, K., Ibata, R.~A., \& Martin, N.~F. 2018, MNRAS, 481, 3442

\bibitem[{Marino {et~al.}(2021)Marino, Milone, Renzini, Yong, Asplund, {Da
  Costa}, Jerjen, Cordoni, Carlos, Dondoglio, Lagioia, Jang, Tailo, Marino,
  Milone, Renzini, Yong, Asplund, {Da Costa}, Jerjen, Cordoni, Carlos,
  Dondoglio, Lagioia, Jang, \& Tailo}]{Marino2021}
Marino, A.~F., Milone, A.~P., Renzini, A., {et~al.} 2021, arXiv,
  arXiv:2106.15978

\bibitem[{Marshall {et~al.}(2019)Marshall, Hansen, Simon, Li, Bernstein, Kuehn,
  Pace, DePoy, Palmese, Pieres, Strigari, Drlica-Wagner, Bechtol, Lidman,
  Nagasawa, Bertin, Brooks, Buckley-Geer, Burke, {Carnero Rosell}, {Carrasco
  Kind}, Carretero, Cunha, D'Andrea, da~Costa, {De Vicente}, Desai, Doel,
  Eifler, Flaugher, Fosalba, Frieman, Garc{\'{i}}a-Bellido, Gaztanaga, Gerdes,
  Gruendl, Gschwend, Gutierrez, Hartley, Hollowood, Honscheid, Hoyle, James,
  Kuropatkin, Maia, Menanteau, Miller, Miquel, Plazas, Sanchez, Santiago,
  Scarpine, Schubnell, Serrano, Sevilla-Noarbe, Smith, Soares-Santos, Suchyta,
  Swanson, Tarle, Wester, \& Collaboration}]{Marshall2019}
Marshall, J.~L., Hansen, T., Simon, J.~D., {et~al.} 2019, ApJ, 882, 177

\bibitem[{Martell {et~al.}(2008)Martell, Smith, \& Briley}]{Martell2008}
Martell, S.~L., Smith, G.~H., \& Briley, M.~M. 2008, AJ, 136, 2522

\bibitem[{Martocchia {et~al.}(2020)Martocchia, Dalessandro, Salaris, Larsen, \&
  Rejkuba}]{Martocchia2020}
Martocchia, S., Dalessandro, E., Salaris, M., Larsen, S., \& Rejkuba, M. 2020,
  MNRAS, 495, 4518

\bibitem[{Martocchia {et~al.}(2021)Martocchia, Lardo, Rejkuba, Kamann, Bastian,
  Larsen, Cabrera-Ziri, Chantereau, Dalessandro, Kacharov, \&
  Salaris}]{Martocchia2021}
Martocchia, S., Lardo, C., Rejkuba, M., {et~al.} 2021, MNRAS, 505, 5389

\bibitem[{Mashonkina {et~al.}(2017{\natexlab{a}})Mashonkina, Jablonka,
  Pakhomov, Sitnova, \& North}]{Mashonkina2017a}
Mashonkina, L., Jablonka, P., Pakhomov, Y., Sitnova, T., \& North, P.
  2017{\natexlab{a}}, A{\&}A, 604, A129

\bibitem[{Mashonkina {et~al.}(2017{\natexlab{b}})Mashonkina, Jablonka, Sitnova,
  Pakhomov, \& North}]{Mashonkina2017}
Mashonkina, L., Jablonka, P., Sitnova, T., Pakhomov, Y., \& North, P.
  2017{\natexlab{b}}, A\&A, 608, 89

\bibitem[{Mashonkina {et~al.}(2007)Mashonkina, Korn, \&
  Przybilla}]{Mashonkina2007}
Mashonkina, L., Korn, A.~J., \& Przybilla, N. 2007, A\&A, 461, 261

\bibitem[{Mashonkina {et~al.}(2019)Mashonkina, Neretina, Sitnova, \&
  Pakhomov}]{Mashonkina2019}
Mashonkina, L.~I., Neretina, M.~D., Sitnova, T.~M., \& Pakhomov, Y.~V. 2019,
  Astronomy Reports, 63, 726

\bibitem[{Masseron(2006)}]{Masseron2006}
Masseron, T. 2006, {PhD thesis}, Tech. rep., Observatoire de Paris

\bibitem[{Masseron {et~al.}(2014)Masseron, Plez, {Van Eck}, Colin, Daoutidis,
  Godefroid, Coheur, Bernath, Jorissen, \& Christlieb}]{Masseron2014}
Masseron, T., Plez, B., {Van Eck}, S., {et~al.} 2014, A\&A, 571, A47

\bibitem[{Matsuno {et~al.}(2021{\natexlab{a}})Matsuno, Aoki, Casagrande,
  Ishigaki, Shi, Takata, Xiang, Yong, Li, Suda, Xing, \& Zhao}]{Matsuno2021}
Matsuno, T., Aoki, W., Casagrande, L., {et~al.} 2021{\natexlab{a}}, ApJ, 912,
  72

\bibitem[{Matsuno {et~al.}(2021{\natexlab{b}})Matsuno, Hirai, Tarumi,
  Hotokezaka, Tanaka, \& Helmi}]{Matsuno2021a}
Matsuno, T., Hirai, Y., Tarumi, Y., {et~al.} 2021{\natexlab{b}}, A{\&}A, 650,
  A110

\bibitem[{Matteucci \& Brocato(1990)}]{Matteucci1990}
Matteucci, F. \& Brocato, E. 1990, ApJ, 365, 539

\bibitem[{Matteucci \& Greggio(1986)}]{Matteucci1986}
Matteucci, F. \& Greggio, L. 1986, A\&A, 154, 279

\bibitem[{Mayall(1946)}]{Mayall1946}
Mayall, N.~U. 1946, ApJ, 104, 290

\bibitem[{McConnachie {et~al.}(2018)McConnachie, Ibata, Martin, Ferguson,
  Collins, Gwyn, Irwin, Lewis, Mackey, Davidge, Arias, Conn, C{\^{o}}t{\'{e}},
  Crnojevic, Huxor, Penarrubia, Spengler, Tanvir, Valls-Gabaud, Babul, Barmby,
  Bate, Bernard, Chapman, Dotter, Harris, McMonigal, Navarro, Puzia, Rich,
  Thomas, \& Widrow}]{McConnachie2018}
McConnachie, A.~W., Ibata, R., Martin, N., {et~al.} 2018, ApJ, 868, 55

\bibitem[{McWilliam(1997)}]{McWilliam1997}
McWilliam, A. 1997, ARA\&A, 35, 503

\bibitem[{McWilliam(1998)}]{McWilliam1998}
McWilliam, A. 1998, AJ, 115, 1640

\bibitem[{McWilliam \& Bernstein(2008)}]{McWilliam2008}
McWilliam, A. \& Bernstein, R.~A. 2008, ApJ, 684, 326

\bibitem[{McWilliam {et~al.}(2003)McWilliam, Rich, \&
  Smecker-Hane}]{McWilliam2003}
McWilliam, A., Rich, R.~M., \& Smecker-Hane, T.~A. 2003, ApJL, 592, L21

\bibitem[{McWilliam {et~al.}(2013)McWilliam, Wallerstein, \&
  Mottini}]{McWilliam2013}
McWilliam, A., Wallerstein, G., \& Mottini, M. 2013, ApJ, 778, 149

\bibitem[{Melnick \& D'Odorico(1978)}]{Melnick1978}
Melnick, J. \& D'Odorico, S. 1978, A\&AS, 34, 249

\bibitem[{Mihalas(1970)}]{Mihalas1970}
Mihalas, D. 1970, {Stellar atmospheres} (San Francisco: Freeman {\&} Co.)

\bibitem[{Minelli {et~al.}(2021)Minelli, Mucciarelli, Massari, Bellazzini,
  Romano, \& Ferraro}]{Minelli2021}
Minelli, A., Mucciarelli, A., Massari, D., {et~al.} 2021, ApJL, 918, L32

\bibitem[{Mishenina {et~al.}(2017)Mishenina, Pignatari, Cot'e, Thielemann,
  Soubiran, Basak, Gorbaneva, Korotin, Kovtyukh, Wehmeyer, Bisterzo, Travaglio,
  Gibson, Jordan, Paul, Ritter, \& Herwig}]{Mishenina2017}
Mishenina, T., Pignatari, M., Cot'e, B., {et~al.} 2017, MNRAS, 469, 4378

\bibitem[{Molero {et~al.}(2021)Molero, Romano, Reichert, Matteucci, Arcones,
  Cescutti, Simonetti, Hansen, Lanfranchi, Molero, Romano, Reichert, Matteucci,
  Arcones, Cescutti, Simonetti, Hansen, \& Lanfranchi}]{Molero2021}
Molero, M., Romano, D., Reichert, M., {et~al.} 2021, MNRAS, 505, 2913

\bibitem[{Moltzer(2020)}]{Moltzer2020}
Moltzer, C. A.~S. 2020, Bachelor's thesis, Radboud University

\bibitem[{Morgan(1956)}]{Morgan1956}
Morgan, W.~W. 1956, PASP, 68, 509

\bibitem[{Navarro \& White(1994)}]{Navarro1994}
Navarro, J.~F. \& White, S. D.~M. 1994, MNRAS, 267, 401

\bibitem[{Nissen {et~al.}(1994)Nissen, Gustafsson, Edvardsson, \&
  Gilmore}]{Nissen1994}
Nissen, P.~E., Gustafsson, B., Edvardsson, B., \& Gilmore, G. 1994, A{\&}A,
  285, 440

\bibitem[{Nissen \& Schuster(1997)}]{Nissen1997}
Nissen, P.~E. \& Schuster, W.~J. 1997, A\&A, 326, 751

\bibitem[{Nissen \& Schuster(2010)}]{Nissen2010}
Nissen, P.~E. \& Schuster, W.~J. 2010, A\&A, 511, L10

\bibitem[{Pace {et~al.}(2021)Pace, Walker, Koposov, Caldwell, Mateo, Olszewski,
  Bailey, \& Wang}]{Pace2021}
Pace, A.~B., Walker, M.~G., Koposov, S.~E., {et~al.} 2021, arXiv:2105.00064
  [\eprint[arXiv]{2105.00064}]

\bibitem[{Pagel \& Tautvaisiene(1997)}]{Pagel1997a}
Pagel, B. E.~J. \& Tautvaisiene, G. 1997, MNRAS, 288, 108

\bibitem[{Palla(2021)}]{Palla2021}
Palla, M. 2021, MNRAS, 503, 3216

\bibitem[{Pancino {et~al.}(2017)Pancino, Romano, Tang, Tautvai{\v{s}}ienė,
  Casey, Gruyters, Geisler, {San Roman}, Randich, Alfaro, Bragaglia, Flaccomio,
  Korn, Recio-Blanco, Smiljanic, Carraro, Bayo, Costado, Damiani, Jofr{\'{e}},
  Lardo, de~Laverny, Monaco, Morbidelli, Sbordone, Sousa, \&
  Villanova}]{Pancino2017}
Pancino, E., Romano, D., Tang, B., {et~al.} 2017, A{\&}A, 601, A112

\bibitem[{Parikh {et~al.}(2019)Parikh, Thomas, Maraston, Westfall, Lian,
  Fraser-McKelvie, Andrews, Drory, \& Meneses-Goytia}]{Parikh2019}
Parikh, T., Thomas, D., Maraston, C., {et~al.} 2019, MNRAS, 483, 3420

\bibitem[{Pavlenko {et~al.}(2012)Pavlenko, Jenkins, Jones, Ivanyuk, \&
  Pinfield}]{Pavlenko2012}
Pavlenko, Y.~V., Jenkins, J.~S., Jones, H. R.~A., Ivanyuk, O., \& Pinfield,
  D.~J. 2012, MNRAS, 422, 542

\bibitem[{Pedregosa {et~al.}(2011)Pedregosa, Varoquaux, Gramfort, Michel,
  Thirion, Grisel, Blondel, Prettenhofer, Weiss, Dubourg, Vanderplas, Passos,
  Cournapeau, Brucher, Perrot, \& Duchesnay}]{Pedregosa2011}
Pedregosa, F., Varoquaux, G., Gramfort, A., {et~al.} 2011, Journal of Machine
  Learning Research, 12, 2825

\bibitem[{Peterken {et~al.}(2020)Peterken, Merrifield, Arag{\'{o}}n-Salamanca,
  Fraser-McKelvie, Avila-Reese, Riffel, Knapen, Drory, Peterken, Merrifield,
  Arag{\'{o}}n-Salamanca, Fraser-McKelvie, Avila-Reese, Riffel, Knapen, \&
  Drory}]{Peterken2020}
Peterken, T., Merrifield, M., Arag{\'{o}}n-Salamanca, A., {et~al.} 2020, MNRAS,
  495, 3387

\bibitem[{Pietrinferni {et~al.}(2021)Pietrinferni, Hidalgo, Cassisi, Salaris,
  Savino, Mucciarelli, Verma, Aguirre, Aparicio, Ferguson, Pietrinferni,
  Hidalgo, Cassisi, Salaris, Savino, Mucciarelli, Verma, {Silva Aguirre},
  Aparicio, \& Ferguson}]{Pietrinferni2021}
Pietrinferni, A., Hidalgo, S.~L., Cassisi, S., {et~al.} 2021, ApJ, 908, 102

\bibitem[{Pilachowski {et~al.}(1980)Pilachowski, Leep, \&
  Wallerstein}]{Pilachowski1980}
Pilachowski, C.~A., Leep, E.~M., \& Wallerstein, G. 1980, ApJ, 236, 508

\bibitem[{Pilachowski {et~al.}(1996)Pilachowski, Sneden, \&
  Kraft}]{Pilachowski1996}
Pilachowski, C.~A., Sneden, C., \& Kraft, R.~P. 1996, AJ, 111, 1689

\bibitem[{Pillepich {et~al.}(2015)Pillepich, Madau, \& Mayer}]{Pillepich2015}
Pillepich, A., Madau, P., \& Mayer, L. 2015, ApJ, 799, 184

\bibitem[{Piskunov {et~al.}(1995)Piskunov, Kupka, Ryabchikova, Weiss, \&
  Jeffery}]{Piskunov1995}
Piskunov, N.~E., Kupka, F., Ryabchikova, T.~A., Weiss, W.~W., \& Jeffery, C.~S.
  1995, A\&A, 112, 525

\bibitem[{Plez(1998)}]{Plez1998}
Plez, B. 1998, A\&A, 337, 495

\bibitem[{Plez(2012)}]{Plez2012}
Plez, B. 2012, Astrophysics Source Code Library, 1205.004

\bibitem[{Prantzos {et~al.}(2018)Prantzos, Abia, Limongi, Chieffi, \&
  Cristallo}]{Prantzos2018}
Prantzos, N., Abia, C., Limongi, M., Chieffi, A., \& Cristallo, S. 2018, MNRAS,
  476, 3432

\bibitem[{Pritzl {et~al.}(2005)Pritzl, Venn, \& Irwin}]{Pritzl2005}
Pritzl, B.~J., Venn, K.~A., \& Irwin, M. 2005, AJ, 130, 2140

\bibitem[{Puzia {et~al.}(2005)Puzia, Kissler-Patig, Thomas, Maraston, Saglia,
  Bender, Goudfrooij, \& Hempel}]{Puzia2005}
Puzia, T.~H., Kissler-Patig, M., Thomas, D., {et~al.} 2005, A\&A, 439, 997

\bibitem[{Puzia {et~al.}(2006)Puzia, Kissler‐Patig, \&
  Goudfrooij}]{Puzia2006}
Puzia, T.~H., Kissler‐Patig, M., \& Goudfrooij, P. 2006, ApJ, 648, 383

\bibitem[{Radburn-Smith {et~al.}(2011)Radburn-Smith, de~Jong, Seth, Bailin,
  Bell, Brown, Bullock, Courteau, Dalcanton, Ferguson, Goudfrooij, Holfeltz,
  Holwerda, Purcell, Sick, Streich, Vlajic, \& Zucker}]{Radburn-Smith2011}
Radburn-Smith, D.~J., de~Jong, R.~S., Seth, A.~C., {et~al.} 2011, ApJS, 195, 18

\bibitem[{Ram{\'{i}}rez \& {Allende Prieto}(2011)}]{Ramirez2011}
Ram{\'{i}}rez, I. \& {Allende Prieto}, C. 2011, ApJ, 743, 135

\bibitem[{Recchi {et~al.}(2014)Recchi, Calura, Gibson, \& Kroupa}]{Recchi2014}
Recchi, S., Calura, F., Gibson, B.~K., \& Kroupa, P. 2014, MNRAS, 437, 994

\bibitem[{Reggiani {et~al.}(2021)Reggiani, Schlaufman, Casey, Simon, \&
  Ji}]{Reggiani2021}
Reggiani, H., Schlaufman, K.~C., Casey, A.~R., Simon, J.~D., \& Ji, A.~P. 2021,
  AJ, 162, 229

\bibitem[{Reichert {et~al.}(2020)Reichert, Hansen, Hanke,
  Sk{\'{u}}lad{\'{o}}ttir, Arcones, \& Grebel}]{Reichert2020}
Reichert, M., Hansen, C.~J., Hanke, M., {et~al.} 2020, A{\&}A, 641, A127

\bibitem[{Renn{\'{o}} {et~al.}(2020)Renn{\'{o}}, Barbuy, Moura, \&
  Trevisan}]{Renno2020}
Renn{\'{o}}, C., Barbuy, B., Moura, T.~C., \& Trevisan, M. 2020, MNRAS, 498,
  5834

\bibitem[{Roederer(2011)}]{Roederer2011b}
Roederer, I.~U. 2011, ApJ, 732, L17

\bibitem[{Roederer {et~al.}(2016)Roederer, Mateo, {Bailey, John I.}, Song,
  Bell, Crane, Loebman, Nidever, Olszewski, Shectman, Thompson, Valluri, \&
  Walker}]{Roederer2016}
Roederer, I.~U., Mateo, M., {Bailey, John I.}, I., {et~al.} 2016, AJ, 151, 82

\bibitem[{Roederer {et~al.}(2014)Roederer, Preston, Thompson, Shectman, Sneden,
  Burley, \& Kelson}]{Roederer2014}
Roederer, I.~U., Preston, G.~W., Thompson, I.~B., {et~al.} 2014, AJ, 147, 136

\bibitem[{Roediger {et~al.}(2014)Roediger, Courteau, Graves, \&
  Schiavon}]{Roediger2014}
Roediger, J.~C., Courteau, S., Graves, G., \& Schiavon, R.~P. 2014, ApJS, 210,
  10

\bibitem[{Romano \& Matteucci(2007)}]{Romano2007}
Romano, D. \& Matteucci, F. 2007, MNRAS, 378, L59

\bibitem[{Romanowsky {et~al.}(2012)Romanowsky, Strader, Brodie, Mihos, Spitler,
  Forbes, Foster, \& Arnold}]{Romanowsky2012}
Romanowsky, A.~J., Strader, J., Brodie, J.~P., {et~al.} 2012, ApJ, 748, 29

\bibitem[{Saglia {et~al.}(2009)Saglia, Fabricius, Bender, Montalto, Lee,
  Riffeser, Seitz, Morganti, Gerhard, \& Hopp}]{Saglia2009}
Saglia, R.~P., Fabricius, M., Bender, R., {et~al.} 2009, A{\&}A, 509, A61

\bibitem[{Saglia {et~al.}(2018)Saglia, Opitsch, Fabricius, Bender, Bla{\~{n}}a,
  \& Gerhard}]{Saglia2018}
Saglia, R.~P., Opitsch, M., Fabricius, M.~H., {et~al.} 2018, A\&A, 618, A156

\bibitem[{Sakari {et~al.}(2013)Sakari, Shetrone, Venn, McWilliam, \&
  Dotter}]{Sakari2013}
Sakari, C.~M., Shetrone, M., Venn, K., McWilliam, A., \& Dotter, A. 2013,
  MNRAS, 434, 358

\bibitem[{Sakari {et~al.}(2016)Sakari, Shetrone, Schiavon, Bizyaev, Prieto,
  Beers, Caldwell, Garcia-Hernandez, Lucatello, Majewski, O'Connell, Pan, \&
  Strader}]{Sakari2016}
Sakari, C.~M., Shetrone, M.~D., Schiavon, R.~P., {et~al.} 2016, ApJ, 829, 116

\bibitem[{Sakari {et~al.}(2014)Sakari, Venn, Shetrone, Dotter, \&
  Mackey}]{Sakari2014}
Sakari, C.~M., Venn, K., Shetrone, M., Dotter, A., \& Mackey, D. 2014, MNRAS,
  443, 2285

\bibitem[{Sakari {et~al.}(2015)Sakari, Venn, Mackey, Shetrone, Dotter,
  Ferguson, \& Huxor}]{Sakari2015}
Sakari, C.~M., Venn, K.~A., Mackey, D., {et~al.} 2015, MNRAS, 448, 1314

\bibitem[{Salgado {et~al.}(2019)Salgado, {Da Costa}, Norris, \&
  Yong}]{Salgado2019}
Salgado, C., {Da Costa}, G.~S., Norris, J.~E., \& Yong, D. 2019, MNRAS, 484,
  3093

\bibitem[{Salpeter(1955)}]{Salpeter1955}
Salpeter, E.~E. 1955, ApJ, 121, 161

\bibitem[{Sandage \& Carlson(1985)}]{Sandage1985}
Sandage, A. \& Carlson, G. 1985, AJ, 90, 1464

\bibitem[{Sanders {et~al.}(2021)Sanders, Belokurov, \& Man}]{Sanders2021}
Sanders, J.~L., Belokurov, V., \& Man, K. T.~F. 2021, MNRAS, 506, 4321

\bibitem[{Sarajedini {et~al.}(2007)Sarajedini, Bedin, Chaboyer, Dotter, Siegel,
  Anderson, Aparicio, King, Majewski, Mar{\'{i}}n-Franch, Piotto, Reid, \&
  Rosenberg}]{Sarajedini2007}
Sarajedini, A., Bedin, L.~R., Chaboyer, B., {et~al.} 2007, AJ, 133, 1658

\bibitem[{Sarajedini {et~al.}(1998)Sarajedini, Geisler, Harding, \&
  Schommer}]{Sarajedini1998}
Sarajedini, A., Geisler, D., Harding, P., \& Schommer, R. 1998, ApJ, 508, L37

\bibitem[{Sarajedini {et~al.}(2000)Sarajedini, Geisler, Schommer, \&
  Harding}]{Sarajedini2000}
Sarajedini, A., Geisler, D., Schommer, R., \& Harding, P. 2000, AJ, 120, 2437

\bibitem[{Sargent {et~al.}(1977)Sargent, Kowal, Hartwick, \& van~den
  Bergh}]{Sargent1977}
Sargent, W. L.~W., Kowal, C.~T., Hartwick, F. D.~A., \& van~den Bergh, S. 1977,
  AJ, 82, 947

\bibitem[{Sbordone {et~al.}(2007)Sbordone, Bonifacio, Buonanno, Marconi,
  Monaco, \& Zaggia}]{Sbordone2007}
Sbordone, L., Bonifacio, P., Buonanno, R., {et~al.} 2007, A\&A, 465, 815

\bibitem[{Sbordone {et~al.}(2004)Sbordone, Bonifacio, Castelli, \&
  Kurucz}]{Sbordone2004}
Sbordone, L., Bonifacio, P., Castelli, F., \& Kurucz, R.~L. 2004, Memorie della
  Societ{\`{a}} Astronomica Italiana Supplement, 5, 93

\bibitem[{Schaye {et~al.}(2015)Schaye, Crain, Bower, Furlong, Schaller, Theuns,
  {Dalla Vecchia}, Frenk, McCarthy, Helly, Jenkins, Rosas-Guevara, White, Baes,
  Booth, Camps, Navarro, Qu, Rahmati, Sawala, Thomas, \& Trayford}]{Schaye2015}
Schaye, J., Crain, R.~A., Bower, R.~G., {et~al.} 2015, MNRAS, 446, 521

\bibitem[{Schiavon {et~al.}(2013)Schiavon, Caldwell, Conroy, Graves, Strader,
  MacArthur, Courteau, \& Harding}]{Schiavon2013}
Schiavon, R.~P., Caldwell, N., Conroy, C., {et~al.} 2013, ApJ, 776, L7

\bibitem[{Schwenke(1998)}]{Schwenke1998}
Schwenke, D.~W. 1998, Faraday Discussions, 109, 321

\bibitem[{Searle \& Zinn(1978)}]{Searle1978}
Searle, L. \& Zinn, R. 1978, ApJ, 225, 357

\bibitem[{Semenova {et~al.}(2020)Semenova, Bergemann, Deal, Serenelli, Hansen,
  Gallagher, Bayo, Bensby, Bragaglia, Carraro, Morbidelli, Pancino, \&
  Smiljanic}]{Semenova2020}
Semenova, E., Bergemann, M., Deal, M., {et~al.} 2020, A\&A, 643, A164

\bibitem[{{\c S}en {et~al.}(2018){\c S}en, Peletier, Boselli, den Brok,
  Falc{\'{o}}n-Barroso, Hensler, Janz, Laurikainen, Lisker, Mentz, Paudel,
  Salo, Sybilska, Toloba, van~de Ven, Vazdekis, \& Yesilyaprak}]{Sen2018}
{\c S}en, S., Peletier, R.~F., Boselli, A., {et~al.} 2018, MNRAS, 475, 3453

\bibitem[{Shapley(1939)}]{Shapley1939}
Shapley, H. 1939, Proceedings of the National Academy of Sciences, 25, 565

\bibitem[{Sharina \& Davoust(2009)}]{Sharina2009}
Sharina, M. \& Davoust, E. 2009, A\&A, 497, 65

\bibitem[{Sharina {et~al.}(2006)Sharina, Afanasiev, \& Puzia}]{Sharina2006}
Sharina, M.~E., Afanasiev, V.~L., \& Puzia, T.~H. 2006, MNRAS, 372, 1259

\bibitem[{Shetrone {et~al.}(2015)Shetrone, Bizyaev, Lawler, Prieto, Johnson,
  Smith, Cunha, Holtzman, P{\'{e}}rez, M{\'{e}}sz{\'{a}}ros, Sobeck, Zamora,
  Garc{\'{i}}a-Hern{\'{a}}ndez, Souto, Chojnowski, Koesterke, Majewski, \&
  Zasowski}]{Shetrone2015}
Shetrone, M., Bizyaev, D., Lawler, J.~E., {et~al.} 2015, ApJS, 221, 24

\bibitem[{Shetrone {et~al.}(2001)Shetrone, C{\^{o}}t{\'{e}}, \&
  Sargent}]{Shetrone2001}
Shetrone, M.~D., C{\^{o}}t{\'{e}}, P., \& Sargent, W. L.~W. 2001, ApJ, 548, 592

\bibitem[{Shetrone \& Keane(2000)}]{Shetrone2000}
Shetrone, M.~D. \& Keane, M.~J. 2000, AJ, 119, 840

\bibitem[{Shi {et~al.}(2018)Shi, Yan, Zhou, \& Zhao}]{Shi2018}
Shi, J.~R., Yan, H.~L., Zhou, Z.~M., \& Zhao, G. 2018, ApJ, 862, 71

\bibitem[{Sk{\'{u}}lad{\'{o}}ttir \& Salvadori(2020)}]{Skuladottir2020}
Sk{\'{u}}lad{\'{o}}ttir, {\'{A}}. \& Salvadori, S. 2020, A\&A, 634, L2

\bibitem[{Smith \& Raggett(1981)}]{Smith1981}
Smith, G. \& Raggett, D.~J. 1981, Journal of Physics B: Atomic and Molecular
  Physics, 14, 4015

\bibitem[{Sneden {et~al.}(1991)Sneden, Gratton, \& Crocker}]{Sneden1991a}
Sneden, C., Gratton, R.~G., \& Crocker, D.~A. 1991, A\&A, 246, 354

\bibitem[{Sneden {et~al.}(1997)Sneden, Kraft, Shetrone, Smith, Langer, \&
  Prosser}]{Sneden1997}
Sneden, C., Kraft, R.~P., Shetrone, M.~D., {et~al.} 1997, AJ, 114, 1964

\bibitem[{Sneden {et~al.}(1979)Sneden, Lambert, \& Whitaker}]{Sneden1979}
Sneden, C., Lambert, D.~L., \& Whitaker, R.~W. 1979, ApJ, 234, 964

\bibitem[{Sollima \& Baumgardt(2017)}]{Sollima2017}
Sollima, A. \& Baumgardt, H. 2017, MNRAS, 471, 3668

\bibitem[{Stanek \& Garnavich(1998)}]{Stanek1998}
Stanek, K.~Z. \& Garnavich, P.~M. 1998, ApJ, 503, L131

\bibitem[{Strader {et~al.}(2005)Strader, Brodie, Cenarro, Beasley, \&
  Forbes}]{Strader2005}
Strader, J., Brodie, J.~P., Cenarro, A.~J., Beasley, M.~A., \& Forbes, D.~A.
  2005, AJ, 130, 1315

\bibitem[{Strader {et~al.}(2003)Strader, Brodie, Forbes, Beasley, \&
  Huchra}]{Strader2003}
Strader, J., Brodie, J.~P., Forbes, D.~A., Beasley, M.~A., \& Huchra, J.~P.
  2003, AJ, 125, 1291

\bibitem[{Strader {et~al.}(2011)Strader, Caldwell, \& Seth}]{Strader2011}
Strader, J., Caldwell, N., \& Seth, A.~C. 2011, AJ, 142, 8

\bibitem[{Strader {et~al.}(2009)Strader, Smith, Larsen, Brodie, \&
  Huchra}]{Strader2009}
Strader, J., Smith, G.~H., Larsen, S., Brodie, J.~P., \& Huchra, J.~P. 2009,
  AJ, 138, 547

\bibitem[{Suda {et~al.}(2008)Suda, Katsuta, Yamada, Suwa, Ishizuka, Komiya,
  Sorai, Aikawa, \& Fujimoto}]{Suda2008}
Suda, T., Katsuta, Y., Yamada, S., {et~al.} 2008, Publications of the
  Astronomical Society of Japan, 60, 1159

\bibitem[{Takeda {et~al.}(2003)Takeda, Zhao, Takada-Hidai, Chen, Saito, \&
  Zhang}]{Takeda2003}
Takeda, Y., Zhao, G., Takada-Hidai, M., {et~al.} 2003, Chinese Journal of
  Astronomy {\&} Astrophysics, 3, 316

\bibitem[{Tammann \& Sandage(1968)}]{Tammann1968}
Tammann, G.~A. \& Sandage, A. 1968, ApJ, 151, 825

\bibitem[{Tautvai{\v{s}}ienė {et~al.}(2021)Tautvai{\v{s}}ienė, {Viscasillas
  V{\'{a}}zquez}, Mikolaitis, Stonkutė, Minkevi{\v{c}}iūtė, Drazdauskas, \&
  Bagdonas}]{Tautvaisiene2021}
Tautvai{\v{s}}ienė, G., {Viscasillas V{\'{a}}zquez}, C., Mikolaitis, {\v{S}}.,
  {et~al.} 2021, A\&A, 649, A126

\bibitem[{Thomas {et~al.}(2005)Thomas, Maraston, Bender, \&
  de~Oliveira}]{Thomas2005}
Thomas, D., Maraston, C., Bender, R., \& de~Oliveira, C.~M. 2005, ApJ, 621, 673

\bibitem[{Thygesen {et~al.}(2014)Thygesen, Sbordone, Andrievsky, Korotin, Yong,
  Zaggia, Ludwig, Collet, Asplund, Ventura, D'Antona, Mel{\'{e}}ndez, \&
  D'Ercole}]{Thygesen2014}
Thygesen, A.~O., Sbordone, L., Andrievsky, S., {et~al.} 2014, A\&A, 572, A108

\bibitem[{Tinsley(1979)}]{Tinsley1979}
Tinsley, B.~M. 1979, ApJ, 229, 1046

\bibitem[{Tolstoy {et~al.}(2009)Tolstoy, Hill, \& Tosi}]{Tolstoy2009}
Tolstoy, E., Hill, V., \& Tosi, M. 2009, ARA\&A, 47, 371

\bibitem[{Tolstoy {et~al.}(2003)Tolstoy, Venn, Shetrone, Primas, Hill, Kaufer,
  \& Szeifert}]{Tolstoy2003}
Tolstoy, E., Venn, K.~A., Shetrone, M., {et~al.} 2003, AJ, 125, 707

\bibitem[{Trager {et~al.}(2000)Trager, Faber, Worthey, \&
  Gonz{\'{a}}lez}]{Trager2000}
Trager, S.~C., Faber, S.~M., Worthey, G., \& Gonz{\'{a}}lez, J.~J. 2000, AJ,
  120, 165

\bibitem[{Tsujimoto {et~al.}(1997)Tsujimoto, Yoshii, Nomoto, Matteucci,
  Thielemann, Hashimoto, Tsujimoto, Yoshii, Nomoto, Matteucci, Thielemann, \&
  Hashimoto}]{Tsujimoto1997}
Tsujimoto, T., Yoshii, Y., Nomoto, K., {et~al.} 1997, ApJ, 483, 228

\bibitem[{Valenti \& Fischer(2005)}]{Valenti2005}
Valenti, J.~A. \& Fischer, D.~A. 2005, ApJS, 159, 141

\bibitem[{{Van der Swaelmen} {et~al.}(2016){Van der Swaelmen}, Barbuy, Hill,
  Zoccali, Minniti, Ortolani, \& Gomez}]{VanderSwaelmen2016}
{Van der Swaelmen}, M., Barbuy, B., Hill, V., {et~al.} 2016, A\&A, 586, A1

\bibitem[{{Van der Swaelmen} {et~al.}(2013){Van der Swaelmen}, Hill, Primas, \&
  Cole}]{vanderSwaelmen2013}
{Van der Swaelmen}, M., Hill, V., Primas, F., \& Cole, A.~A. 2013, A\&A, 560,
  A44

\bibitem[{Vanderbeke {et~al.}(2013)Vanderbeke, West, {De Propris}, Peng,
  Blakeslee, Jord{\'{a}}n, C{\^{o}}t{\'{e}}, Gregg, Ferrarese, Takamiya, Baes,
  Vanderbeke, West, {De Propris}, Peng, Blakeslee, Jord{\'{a}}n,
  C{\^{o}}t{\'{e}}, Gregg, Ferrarese, Takamiya, \& Baes}]{Vanderbeke2013a}
Vanderbeke, J., West, M.~J., {De Propris}, R., {et~al.} 2013, MNRAS, 437, 1734

\bibitem[{Vargas {et~al.}(2014)Vargas, Geha, \& Tollerud}]{Vargas2014}
Vargas, L.~C., Geha, M.~C., \& Tollerud, E.~J. 2014, ApJ, 790, 73

\bibitem[{Vazdekis {et~al.}(2015)Vazdekis, Coelho, Cassisi, Ricciardelli,
  Falc{\'{o}}n-Barroso, S{\'{a}}nchez-Bl{\'{a}}zquez, {La Barbera}, Beasley, \&
  Pietrinferni}]{Vazdekis2015}
Vazdekis, A., Coelho, P., Cassisi, S., {et~al.} 2015, MNRAS, 449, 1177

\bibitem[{Veljanoski {et~al.}(2013)Veljanoski, Ferguson, Huxor, Mackey,
  Fishlock, Irwin, Tanvir, Chapman, Ibata, Lewis, \&
  McConnachie}]{Veljanoski2013}
Veljanoski, J., Ferguson, A. M.~N., Huxor, A.~P., {et~al.} 2013, MNRAS, 435,
  3654

\bibitem[{Veljanoski {et~al.}(2015)Veljanoski, Ferguson, Mackey, Huxor, Hurley,
  Bernard, Cote, Irwin, Martin, Burgett, Chambers, Flewelling, Kudritzki, \&
  Waters}]{Veljanoski2015}
Veljanoski, J., Ferguson, A. M.~N., Mackey, A.~D., {et~al.} 2015, MNRAS, 452,
  320

\bibitem[{Venn {et~al.}(2004)Venn, Irwin, Shetrone, Tout, Hill, \&
  Tolstoy}]{Venn2004}
Venn, K.~A., Irwin, M., Shetrone, M.~D., {et~al.} 2004, AJ, 128, 1177

\bibitem[{Vincenzo {et~al.}(2015)Vincenzo, Matteucci, Recchi, Calura,
  McWilliam, \& Lanfranchi}]{Vincenzo2015}
Vincenzo, F., Matteucci, F., Recchi, S., {et~al.} 2015, MNRAS, 449, 1327

\bibitem[{Vogt {et~al.}(1994)Vogt, Allen, Bigelow, Bresee, Brown, Cantrall,
  Conrad, Couture, Delaney, Epps, Hilyard, Hilyard, Horn, Jern, Kanto, Keane,
  Kibrick, Lewis, Osborne, Pardeilhan, Pfister, Ricketts, Robinson, Stover,
  Tucker, Ward, \& Wei}]{Vogt1994}
Vogt, S.~S., Allen, S.~L., Bigelow, B.~C., {et~al.} 1994, in Proc. SPIE, ed.
  D.~L. Crawford \& E.~R. Craine, Vol. 2198, 362

\bibitem[{Wallace {et~al.}(2000)Wallace, Hinkle, Valenti, \&
  Harmer}]{Wallace2000}
Wallace, L., Hinkle, K., Valenti, J., \& Harmer, D. 2000, {Visible and near
  infrared atlas of the Arcturus spectrum, 3727-9300 AA} (San Francisco:
  Astronomical Society of the Pacific), 375

\bibitem[{Webbink(1985)}]{Webbink1985}
Webbink, R.~F. 1985, in IAU Symposium 113: Dynamics of star clusters, ed.
  J.~Goodman \& P.~Hut (Dordrecht: Reidel Publishing Co.), 541--577

\bibitem[{Weidner \& Kroupa(2005)}]{Weidner2005}
Weidner, C. \& Kroupa, P. 2005, ApJ, 625, 754

\bibitem[{Wiese {et~al.}(1969)Wiese, Smith, \& Miles}]{Wiese1969}
Wiese, W.~L., Smith, M.~W., \& Miles, B.~M. 1969, {Atomic transition
  probabilities. Vol. 2: Sodium through Calcium. A critical data compilation}
  (Washington, D.C.: National Bureau of Standards)

\bibitem[{Woodley {et~al.}(2010)Woodley, Harris, Puzia, G{\'{o}}mez, Harris, \&
  Geisler}]{Woodley2010}
Woodley, K.~A., Harris, W.~E., Puzia, T.~H., {et~al.} 2010, ApJ, 708, 1335

\bibitem[{Woody \& Schlaufman(2021)}]{Woody2021}
Woody, T. \& Schlaufman, K.~C. 2021, AJ, 162, 42

\bibitem[{Woosley \& Weaver(1995)}]{Woosley1995}
Woosley, S.~E. \& Weaver, T.~A. 1995, ApJS, 101, 181

\bibitem[{Worley {et~al.}(2009)Worley, Cottrell, Freeman, \& {Wylie-de
  Boer}}]{Worley2009}
Worley, C.~C., Cottrell, P.~L., Freeman, K.~C., \& {Wylie-de Boer}, E.~C. 2009,
  MNRAS, 400, 1039

\bibitem[{Worley {et~al.}(2013)Worley, Hill, Sobeck, \& Carretta}]{Worley2013}
Worley, C.~C., Hill, V., Sobeck, J., \& Carretta, E. 2013, A\&A, 553, A47

\bibitem[{Worthey {et~al.}(1992)Worthey, Faber, \& Gonzalez}]{Worthey1992}
Worthey, G., Faber, S.~M., \& Gonzalez, J.~J. 1992, ApJ, 398, 69

\bibitem[{Worthey {et~al.}(2014)Worthey, Tang, \& Serven}]{Worthey2014}
Worthey, G., Tang, B., \& Serven, J. 2014, ApJ, 783, 20

\bibitem[{Wyder {et~al.}(2000)Wyder, Hodge, \& Zucker}]{Wyder2000}
Wyder, T.~K., Hodge, P.~W., \& Zucker, D.~B. 2000, PASP, 112, 1162

\bibitem[{Wyse(1998)}]{Wyse1998}
Wyse, R. F.~G. 1998, ASPC, 142, 89

\bibitem[{Wyse \& Gilmore(1992)}]{Wyse1992}
Wyse, R. F.~G. \& Gilmore, G. 1992, AJ, 104, 144

\bibitem[{Yan {et~al.}(2015)Yan, Shi, \& Zhao}]{Yan2015}
Yan, H.~L., Shi, J.~R., \& Zhao, G. 2015, ApJ, 802, 36

\bibitem[{Yong {et~al.}(2005)Yong, Carney, \& de~Almeida}]{Yong2005}
Yong, D., Carney, B.~W., \& de~Almeida, M. L.~T. 2005, AJ, 130, 597

\bibitem[{Zevin {et~al.}(2019)Zevin, Kremer, Siegel, Coughlin, Tsang, Berry, \&
  Kalogera}]{Zevin2019}
Zevin, M., Kremer, K., Siegel, D.~M., {et~al.} 2019, ApJ, 886, 4

\bibitem[{Zhao {et~al.}(2016)Zhao, Mashonkina, Yan, Alexeeva, Kobayashi,
  Pakhomov, Shi, Sitnova, Tan, Zhang, Zhang, Zhou, Bolte, Chen, Li, Liu, \&
  Zhai}]{Zhao2016}
Zhao, G., Mashonkina, L., Yan, H.~L., {et~al.} 2016, ApJ, 833, 225

\end{thebibliography}

\onecolumn

\begin{appendix}

\section{Observations}

\begin{table*}[h!]
\caption{Observations}
\label{tab:obs}
\centering
{\small
\begin{tabular}{llllllccc}
\hline\hline
Cluster & Instrument & Date (UT) &  Range (\AA ) & T$_\mathrm{exp}$ (s) & S/N  & $V$ & \multicolumn{2}{c}{RA, Dec (J2000.0)} \\ 
\hline
NGC 104 & UVES & 22 Jul 2015$^{1}$ & 4150 - 6200 & $2\times1500$ & 546  & 3.95$^8$ & 00:24:05.67 & $-$72:04:52.6$^8$ \\ 
  & UVES & 6 Aug 2019$^{2}$ & 5660 - 9460 & $2\times1500$ & 583$^\star$    \\
NGC 362 & UVES & 22 Jul 2015$^{1}$ & 4150 - 6200 & $1\times1500$ & 408 &  6.40$^8$ & 01:03:14.26 & $-$70:50:55.6$^8$ \\
  & UVES & 6 Aug 2019$^{2}$ & 5660 - 9460 & $2\times1500$ & 596$^\star$  \\ 
NGC 6254 & UVES & 22-23 Jul 2015$^{1}$ & 4150 - 6200 & $8\times1800$ & 373  & 6.60$^8$ & 16:57:09.05 & $-$04:06:01.1$^8$ \\
  & UVES & 6-7 Aug 2019$^{2}$ & 5660 - 9460 & $8\times1800$ & 466$^\star$  \\
NGC 6388 & UVES & 22 Jul 2015$^{1}$  & 4150 - 6200 & $2\times1200$ & 561  & 6.72$^8$ & 17:36:17.23 & $-$44:44:07.8$^8$ \\
  & UVES & 6 Aug 2019$^{2}$ & 5660 - 9460 & $2\times1200$ & 635$^\star$  \\
NGC 6752 & UVES & 23 Jul 2015$^{1}$  & 4150 - 6200 & $4\times1800$ & 617  & 5.40$^8$ & 19:10:52.11 & $-$59:59:04.4$^8$ \\
  & UVES & 7 Aug 2019$^{2}$ & 5660 - 9460 & $4\times1800$ & 677$^\star$  \\
NGC 7078 & UVES & 22 Jul 2015$^{1}$  & 4150 - 6200 & $2\times1800$ & 580  & 6.20$^8$ & 21:29:58.33 & $+$12:10:01.2$^8$ \\
  & UVES & 6 Aug 2019$^{2}$ & 5660 - 9460 & $2\times1800$ & 563$^\star$  \\
NGC 7099 & UVES & 23 Jul 2015$^{1}$  & 4150 - 6200 & $4\times1800$ & 440  & 7.19$^8$ & 21:40:22.12 & $-$23:10:47.5$^8$ \\
  & UVES & 7 Aug 2019$^{2}$ & 5660 - 9460 & $4\times1800$ & 380$^\star$  \\
M31 006-058 & HIRES & 2-3 Oct 2007$^{2}$ & 3900 - 8350 & $10\times1800$ & 329  & 15.50$^9$ & 00:40:26.49 & $+$41:27:26.7$^9$ \\
M31 012-064 & HIRES & 19 Oct 2007$^{2}$ & 3550 - 6300 & $4\times1800$ & 190  & 15.09$^9$ & 00:40:32.47 & $+$41:21:44.2$^9$ \\
M31 019-072 & HIRES & 18 Oct 2007$^{2}$ & 3550 - 6300 & $4\times1800$ & 285  & 14.93$^9$ & 00:40:52.53 & $+$41:18:53.4$^9$ \\
M31 058-119 & HIRES & 1 Oct 2007$^{2}$ & 3900 - 8350 & $9\times1800$ & 394  & 14.97$^9$ & 00:41:53.01 & $+$40:47:09.7$^9$ \\
M31 082-144 & HIRES & 19 Oct 2007$^{2}$ & 3550 - 6300 & $4\times1800$ & 108  & 15.54$^9$ & 00:42:15.84 & $+$41:01:14.3$^9$ \\
M31 163-217 & HIRES & 3 Oct 2007$^{2}$ & 3900 - 8350 & $7\times1800$ & 386  & 15.05$^9$ & 00:43:17.64 & $+$41:27:44.9$^9$ \\
M31 171-222 & HIRES & 3 Oct 2007$^{2}$ & 3900 - 8350 & $5\times1800$ & 305  & 15.22$^9$ & 00:43:25.61 & $+$41:15:37.1$^9$ \\
M31 174-226 & HIRES & 18 Oct 2007$^{2}$ & 3550 - 6300 & $4\times1800$ & 139  & 15.47$^9$ & 00:43:30.30 & $+$41:38:56.2$^9$ \\
M31 225-280 & HIRES & 1 Oct 2007$^{2}$ & 3900 - 8350 & $7\times1800$ & 618  & 14.16$^9$ & 00:44:29.56 & $+$41:21:35.3$^9$ \\
M31 338-076 & HIRES & 18 Oct 2007$^{2}$ & 3550 - 6300 & $4\times1800$ & 316  & 14.25$^9$ & 00:40:58.87 & $+$40:35:47.8$^9$ \\
M31 358-219 & HIRES & 2 Oct 2007$^{2}$ & 3900 - 8350 & $8\times1800$ & 300  & 15.22$^9$ & 00:43:17.86 & $+$39:49:13.2$^9$ \\
M31 EXT8    & HIRES & 25 Oct 2019$^{3}$ & 3850 - 8170 & $2\times1200$ & 216  & 15.60$^9$ & 00:53:14.53 & $+$41:33:24.5$^9$ \\
M33 H38 & HIRES & 26 Oct 1998$^{4}$ & 3730 - 6170 & $7\times1800$ & 44  & 17.25$^{10}$ & 01:33:52.12 & $+$30:29:03.6$^{19}$ \\
M33 M9 & HIRES & 26 Oct 1998$^{4}$ & 3730 - 6170 & $9\times1800$ & 81  & 17.12$^{10}$ & 01:34:30.22 & $+$30:38:12.7$^{19}$ \\
M33 R12 & HIRES & 25 Oct 1998$^{4}$ & 3730 - 6170 & $7\times1800$ & 94  & 16.38$^{10}$ & 01:34:08.01 & $+$30:38:38.0$^{19}$ \\
M33 U49 & HIRES & 25 Oct 1998$^{4}$ & 3730 - 6170 & $8\times1800$ & 57  & 16.25$^{10}$ & 01:33:45.01 & $+$30:47:46.7$^{19}$ \\
M33 R14 & HIRES & 01 Nov 2018$^{2}$ & 3900 - 8160 & $4\times1800$ & 142  & 16.48$^{10}$ & 01:34:02.44 & $+$30:40:40.6$^{19}$ \\
M33 U77 & HIRES & 25 Oct 2019$^{2}$ & 3850 - 8170 & $6\times1800$ & 91  & 17.19$^{10}$ & 01:33:28.68 & $+$30:41:34.9$^{19}$ \\
M33 CBF28 & HIRES & 25 Oct 2019$^{2}$ & 3850 - 8170 & $4\times1800$ & 216  & 16.37$^{11}$ & 01:34:01.91 & $+$30:39:45.9$^{11}$ \\
M33 HM33B & HIRES & 25 Oct 2019$^{2}$ & 3850 - 8170 & $6\times1800$ & 32  & 17.76$^{12}$ & 01:36:02.12 & $+$29:57:49.4$^{12}$ \\
N147 Hodge II & HIRES & 25 Sep 2016$^{5}$ & 3930 - 8170 & $5\times1800$ & 39  & 18.06$^{13}$ & 00:33:13.6 & $+$48:28:48.7$^{13}$ \\
N147 Hodge III & HIRES & 05 Oct 2015$^{5}$ & 3610 - 8170 & $4\times1800$ & 96  & 16.58$^{13}$ & 00:33:15.2 & $+$48:27:23.1$^{13}$ \\
N147 PA1 & HIRES & 05 Oct 2015$^{5}$ & 3970 - 8170 & $1800+865$ & 92  & 16.96$^{13}$ & 00:32:35.3 & $+$48:19:48.0$^{13}$ \\
           & & 25 Sep 2016$^{5}$ & & $2\times1800$ \\
N147 PA2 & HIRES & 25 Sep 2016$^{5}$ & 3930 - 8170 & $5\times1800$ & 112  & 17.37$^{13}$ & 00:33:43.3 & $+$48:38:45.0$^{13}$ \\
N147 SD7 & HIRES & 05 Oct 2015$^{5}$ & 3610 - 8170 & $4\times1800$ & 103  & 17.00$^{13}$ & 00:32:22.2 & $+$48:31:27.0$^{13}$ \\
N185 FJJ-III & HIRES & 25 Sep 2017$^{2}$ & 3850 - 8170 & $4\times1800$ & 79  & 16.58$^{13}$ & 00:39:03.8 & $+$48:19:57.5$^{13}$ \\
N185 FJJ-V & HIRES & 25 Sep 2017$^{2}$ & 3850 - 8170 & $4\times1800$ & 100  & 16.71$^{13}$ & 00:39:13.4 & $+$48:23:04.9$^{13}$ \\
N185 FJJ-VIII & HIRES & 26 Sep 2017$^{2}$ & 3850 - 8170 & $5\times1800$ & 91  & 17.57$^{13}$ & 00:39:23.7 & $+$48:18:45.1$^{13}$ \\
N205 Hubble I & HIRES & 25 Sep 2017$^{2}$ & 3850 - 8170 & $4\times1800$ & 163  & 16.9$^{14}$ & 00:40:30.70 & $+$41:36:55.7$^{20}$ \\
N205 Hubble II & HIRES & 25 Sep 2017$^{2}$ & 3850 - 8170 & $4\times1800$ & 183  & 16.7$^{14}$ & 00:40:31.88 & $+$41:39:17.0$^{20}$ \\
N6822 Hubble VII & HIRES & 25 Sep 2017$^{2}$ & 3850 - 8170 & $2\times1800$ & 168  & 15.79$^{15}$ & 19:44:55.8 & $-$14:48:56.2$^{15}$\\
N6822 SC6 & HIRES & 25 Sep 2016$^{5}$ & 3930 - 8170 & $2\times1800$ & 142  & 15.97$^{15}$ & 19:45:37.0 & $-$14:41:10.8$^{15}$ \\
N6822 SC7 & HIRES & 05 Oct 2015$^{5}$ & 3610 - 8170 & $1800+1139$ & 181  & 15.42$^{15}$ & 19:46:00.7 & $-$14:32:35.0$^{15}$ \\
WLM-GC & UVES & 29 Jul -  & 4170 - 6200 & $12\times1475$ & 205  & 16.06$^{16}$ & 00:01:49.54 & $-$15:27:31.0$^{20}$ \\
 & & 02 Aug 2006$^{6}$ \\
Fornax 3 & UVES &  19 Nov 2006$^{7}$ & 4170 - 6210 & $4\times2400$ & 264  & 12.61$^{17}$ & 02:39:52.5 & $-$34:16:08$^{21}$ \\
Fornax 4 & UVES &  19-20 Nov 2006$^{7}$ & 4170 - 6210 & $4\times2400$ & 192  & 13.57$^{17}$ & 02:40:09.0 & $-$34:32:24$^{21}$ \\
Fornax 5 & UVES &  20 Nov 2006$^{7}$ & 4170 - 6210 & $4\times2400$ & 144  & 13.42$^{17}$ & 02:42:21.15 & $-$34:06:04.7$^{21}$ \\
NGC~2403 F46 & HIRES & 25 Oct 2019$^{2}$ & 3850 - 8170 & $2\times1800+1200$ & 121  & 17.8$^{18}$ & 07:36:29.17 & $+$65:40:33.5$^{20}$ \\
\hline
\end{tabular}
\tablefoot{The S/N values are given per \AA\ at 5000~\AA, except for entries marked with a star ($^\star$) for which they are given at 6000~\AA .}
\tablebib{
Spectroscopy: (1)~\citet{Larsen2017}; (2)~this work; (3)~\citet{Larsen2020}; 
(4)~\citet{Larsen2002b}; (5)~\cite{Larsen2018}; (6)~\citet{Larsen2014}; (7)~\citet{Larsen2012a};
Integrated magnitudes and coordinates: 
(8)~\citet[2010 revision]{Harris1996}; 
(9)~Revised Bologna Catalogue v5.0 \citep{Galleti2004};
(10)~\citet{Sarajedini1998};
(11)~\citet{Chandar1999};
(12)~\citet{Huxor2009};
(13)~\citet{Veljanoski2013};
(14)~\citet{Sharina2006}; 
(15)~\citet{Veljanoski2015}; 
(16)~\citet{Sandage1985};
(17)~\citet{Webbink1985};
(18)~Sect.~\ref{sec:individual_remarks}; 
(19)~\citet{Beasley2015};
(20)~Gaia EDR3 \citep{Brown2021a} via ESASky; 
(21)~\citet{Mackey2003a}.
}
}
\end{table*}

\clearpage

\section{Spectral windows}
\label{app:specwin}

\begin{table}[!h]
\caption{Windows used for spectral fitting.}
\label{tab:windows}
\begin{adjustbox}{height=0.45\textheight}
\centering
\small
\begin{tabular}{l l l l}
\hline\hline
 Elem. & Range (\AA) & Elem. & Range (\AA) \\
\hline
Fe & 4573.0--4600.0 & Sc & 4739.0--4758.0 \\
Fe & 4600.0--4618.0 & Sc & 5026.0--5036.0 \\
Fe & 4631.0--4660.0 & Sc & 5521.0--5531.0 \\
Fe & 4671.0--4686.0 & Sc & 5638.0--5690.0 \\
Fe & 4705.0--4714.0 & Sc & 6206.0--6216.0 \\
Fe & 4724.0--4750.0 & Ti & 4500.0--4519.5 \\
Fe & 4866.0--4883.0 & Ti & 4551.0--4570.0 \\
Fe & 4886.0--4896.0 & Ti & 4586.5--4596.0 \\
Fe & 4897.0--4915.0 & Ti & 4638.0--4660.0 \\
Fe & 4915.0--4929.0 & Ti & 4680.0--4698.0 \\
Fe & 4936.0--4944.0 & Ti & 4802.0--4821.0 \\
Fe & 4944.0--4953.0 & Ti & 4975.0--5000.0 \\
Fe & 4952.0--4962.0 & Ti & 5000.0--5030.0 \\
Fe & 4963.0--4976.0 & Ti & 5060.0--5075.0 \\
Fe & 4975.0--4998.0 & Ti & 5331.0--5341.0 \\
Fe & 5008.0--5017.0 & Ti & 5376.0--5386.0 \\
Fe & 5045.0--5064.0 & Ti & 5510.0--5520.0 \\
Fe & 5066.0--5115.0 & Ti & 5860.0--5875.0 \\
Fe & 5118.0--5150.0 & Ti & 5912.0--5922.0 \\
Fe & 5250.0--5259.0 & Cr & 4535.0--4550.0 \\
Fe & 5271.0--5289.0 & Cr & 4611.0--4631.0 \\
Fe & 5300.0--5345.0 & Cr & 4646.0--4657.0 \\
Fe & 5358.0--5375.0 & Cr & 4703.0--4723.0 \\
Fe & 5378.0--5400.0 & Cr & 4751.0--4761.0 \\
Fe & 5400.0--5420.0 & Cr & 4796.0--4806.0 \\
Fe & 5420.0--5460.0 & Cr & 4824.0--4834.0 \\
Fe & 5460.0--5475.5 & Cr & 4866.0--4876.0 \\
Fe & 5494.0--5510.0 & Cr & 4931.0--4947.0 \\
Fe & 5529.0--5539.0 & Cr & 5063.0--5096.0 \\
Fe & 5566.5--5590.0 & Cr & 5117.0--5127.0 \\
Fe & 5610.0--5630.0 & Cr & 5270.0--5281.0 \\
Fe & 5682.0--5714.0 & Cr & 5292.0--5302.0 \\
Fe & 5858.5--5865.0 & Cr & 5341.0--5353.0 \\
Fe & 5970.0--5980.0 & Cr & 5407.0--5413.0 \\
Fe & 6001.0--6019.0 & Cr & 5783.0--5793.0 \\
Fe & 6021.0--6029.5 & Cr & 6325.0--6335.0 \\
Fe & 6053.0--6082.0 & Cr & 6973.0--6983.0 \\
Fe & 6131.0--6140.0 & Mn & 4750.0--4790.0 \\
Fe & 6144.0--6160.0 & Mn & 6010.0--6030.0 \\
Fe & 6170.0--6185.0 & Ni & 4600.0--4610.0 \\
Na & 5677.0--5695.0 & Ni & 4644.0--4654.0 \\
Na & 6149.0--6166.0 & Ni & 4681.0--4691.0 \\
Mg & 4347.0--4357.0 & Ni & 4709.0--4719.0 \\
Mg & 4565.0--4576.0 & Ni & 4824.0--4835.0 \\
Mg & 4700.0--4707.0 & Ni & 4899.0--4909.0 \\
Mg & 5523.0--5531.5 & Ni & 4931.0--4942.0 \\
Mg & 5705.0--5715.0 & Ni & 4975.0--4985.0 \\
Si & 5661.0--5671.0 & Ni & 5075.0--5089.0 \\
Si & 5685.0--5695.0 & Ni & 5098.0--5108.0 \\
Si & 5767.0--5777.0 & Ni & 5141.0--5151.0 \\
Si & 6150.0--6160.0 & Ni & 5472.0--5482.0 \\
Si & 6232.0--6250.0 & Ni & 5707.0--5717.0 \\
Si & 7400.0--7427.0 & Ni & 6103.0--6113.0 \\
Ca & 4420.0--4440.0 & Ni & 6172.0--6182.0 \\
Ca & 4451.0--4461.0 & Cu & 5101.0--5112.0 \\
Ca & 4573.0--4590.0 & Cu & 5777.0--5787.0 (DIB) \\
Ca & 5255.0--5268.0 & Zn & 4717.0--4727.0 \\
Ca & 5347.0--5357.0 & Zn & 4805.0--4815.0 \\
Ca & 5507.0--5517.0 & Zr & 6124.0--6147.0 \\
Ca & 5576.0--5602.0 & Ba & 4551.0--4560.0 \\
Ca & 5852.0--5862.0 & Ba & 4929.0--4939.0 \\
Ca & 6098.0--6127.0 & Ba & 5849.0--5859.0 \\
Ca & 6151.0--6174.0 & Ba & 6135.0--6145.0 \\
& & Ba & 6492.0--6502.0 \\
& & Eu & 4431.0--4441.0 \\
& & Eu & 6640.0--6650.0 \\
\hline
\end{tabular}
\end{adjustbox}
\tablefoot{The Cu window at 5777~\AA - 5787~\AA\ is affected by a diffuse interstellar absorption band. 
}
\end{table}

\clearpage

\section{NLTE abundance measurements}

Tables~\ref{tab:results1}-\ref{tab:results3} list the average abundance measurements from our analysis, based on DSEP isochrones and empirical horizontal branches. The measurements in these Tables include \ac{nlte} corrections on the abundance measurements when available. 

\begin{table*}[!h]
\caption{Results for Fe, Na, Mg, Si, Ca, and Ti.}
\label{tab:results1}
\begin{adjustbox}{width=1\textwidth}
\centering
{\small
\begin{tabular}{l cccccccccccccccccccccccc}
\hline\hline
Cluster       & [Fe/H] & $\sigma_{\langle\mathrm{Fe}\rangle}$ & $S_\mathrm{Fe}$ & N      & [Na/Fe]  & $\sigma_{\langle\mathrm{Na}\rangle}$ & $S_\mathrm{Na}$ & N    & [Mg/Fe]  & $\sigma_{\langle\mathrm{Mg}\rangle}$ & $S_\mathrm{Mg}$ & N    & [Si/Fe]  & $\sigma_{\langle\mathrm{Si}\rangle}$ & $S_\mathrm{Si}$ & N    & [Ca/Fe]  & $\sigma_{\langle\mathrm{Ca}\rangle}$ & $S_\mathrm{Ca}$ & N    & [Ti/Fe]  & $\sigma_{\langle\mathrm{Ti}\rangle}$ & $S_\mathrm{Ti}$ & N    \\ \hline
NGC 0104      & $-0.735$ & 0.008 & 0.019 & 39 & $+0.237$ & 0.038 & 0.065 & 2 & $+0.404$ & 0.027 & 0.058 & 4 & $+0.376$ & 0.023 & 0.044 & 6 & $+0.238$ & 0.018 & 0.030 & 9 & $+0.412$ & 0.015 & 0.026 & 14 \\
NGC 0362      & $-1.073$ & 0.009 & 0.019 & 39 & $-0.207$ & 0.043 & 0.005 & 2 & $+0.169$ & 0.030 & 0.027 & 4 & $+0.155$ & 0.026 & 0.051 & 6 & $+0.140$ & 0.019 & 0.028 & 9 & $+0.417$ & 0.016 & 0.033 & 14 \\
NGC 6254      & $-1.485$ & 0.009 & 0.018 & 39 & $-0.204$ & 0.048 & 0.009 & 2 & $+0.295$ & 0.031 & 0.050 & 4 & $+0.279$ & 0.032 & 0.051 & 6 & $+0.260$ & 0.020 & 0.048 & 9 & $+0.422$ & 0.017 & 0.034 & 14 \\
NGC 6388      & $-0.537$ & 0.008 & 0.025 & 39 & $+0.202$ & 0.038 & 0.090 & 2 & $+0.113$ & 0.028 & 0.073 & 4 & $+0.218$ & 0.023 & 0.060 & 6 & $+0.040$ & 0.018 & 0.047 & 9 & $+0.262$ & 0.016 & 0.027 & 14 \\
NGC 6752      & $-1.704$ & 0.009 & 0.014 & 39 & $+0.064$ & 0.043 & 0.037 & 2 & $+0.312$ & 0.030 & 0.074 & 4 & $+0.408$ & 0.028 & 0.016 & 6 & $+0.326$ & 0.018 & 0.022 & 9 & $+0.361$ & 0.016 & 0.028 & 14 \\
NGC 7078      & $-2.293$ & 0.010 & 0.020 & 38 & $-0.057$ & 0.091 & 0.002 & 2 & $+0.147$ & 0.035 & 0.031 & 4 & $+0.471$ & 0.065 & 0.115 & 3 & $+0.262$ & 0.022 & 0.031 & 9 & $+0.431$ & 0.021 & 0.046 & 12 \\
NGC 7099      & $-2.243$ & 0.010 & 0.020 & 39 & $-0.035$ & 0.082 & 0.025 & 2 & $+0.295$ & 0.035 & 0.039 & 4 & $+0.456$ & 0.054 & 0.084 & 5 & $+0.260$ & 0.022 & 0.023 & 9 & $+0.372$ & 0.020 & 0.020 & 14 \\
N147 HII      & $-1.441$ & 0.029 & 0.059 & 37 & $+0.250$ & 0.119 & 0.048 & 2 & $+0.201$ & 0.165 & 0.120 & 4 & $+0.679$ & 0.115 & 0.062 & 2 & $+0.253$ & 0.062 & 0.050 & 8 & $+0.661$ & 0.134 & 0.080 & 6 \\
N147 HIII     & $-2.363$ & 0.024 & 0.041 & 34 & $+0.408$ & 0.146 & \ldots & 1 & $+0.069$ & 0.120 & 0.123 & 3 & $+0.152$ & 0.274 & \ldots & 1 & $+0.299$ & 0.052 & 0.060 & 6 & $+0.370$ & 0.084 & 0.067 & 8 \\
N147 PA-1     & $-2.216$ & 0.023 & 0.042 & 34 & $+0.281$ & 0.218 & \ldots & 1 & $+0.209$ & 0.110 & 0.101 & 4 & $+0.574$ & 0.363 & 0.159 & 2 & $+0.178$ & 0.051 & 0.039 & 7 & $+0.379$ & 0.089 & 0.076 & 8 \\
N147 PA-2     & $-1.919$ & 0.017 & 0.026 & 38 & $+0.110$ & 0.127 & 0.019 & 2 & $+0.337$ & 0.084 & 0.061 & 4 & $+0.150$ & 0.127 & 0.232 & 2 & $+0.332$ & 0.036 & 0.042 & 9 & $+0.426$ & 0.052 & 0.064 & 11 \\
N147 SD7      & $-1.887$ & 0.017 & 0.037 & 37 & $-0.184$ & 0.179 & 0.316 & 2 & $+0.368$ & 0.095 & 0.156 & 3 & $+0.549$ & 0.066 & 0.105 & 4 & $+0.198$ & 0.035 & 0.061 & 9 & $+0.435$ & 0.054 & 0.051 & 10 \\
N185 FJJ-III  & $-1.754$ & 0.020 & 0.031 & 36 & $+0.323$ & 0.114 & \ldots & 1 & $+0.472$ & 0.076 & 0.147 & 5 & $+0.419$ & 0.099 & 0.142 & 6 & $+0.342$ & 0.045 & 0.067 & 9 & $+0.443$ & 0.059 & 0.082 & 11 \\
N185 FJJ-V    & $-1.780$ & 0.017 & 0.031 & 36 & $-0.114$ & 0.162 & \ldots & 1 & $+0.186$ & 0.070 & 0.123 & 5 & $+0.329$ & 0.091 & 0.157 & 5 & $+0.289$ & 0.035 & 0.040 & 9 & $+0.412$ & 0.052 & 0.077 & 11 \\
N185 FJJ-VIII & $-1.749$ & 0.017 & 0.026 & 37 & $+0.026$ & 0.123 & 0.010 & 2 & $-0.049$ & 0.077 & 0.080 & 5 & $+0.286$ & 0.088 & 0.166 & 4 & $+0.317$ & 0.036 & 0.043 & 9 & $+0.387$ & 0.056 & 0.070 & 11 \\
N205 HubbleI  & $-1.410$ & 0.012 & 0.025 & 37 & $-0.124$ & 0.073 & 0.072 & 2 & $+0.417$ & 0.041 & 0.060 & 5 & $+0.301$ & 0.048 & 0.025 & 6 & $+0.236$ & 0.026 & 0.031 & 9 & $+0.386$ & 0.028 & 0.048 & 12 \\
N205 HubbleII & $-1.348$ & 0.011 & 0.023 & 37 & $+0.014$ & 0.063 & 0.012 & 2 & $+0.318$ & 0.040 & 0.054 & 5 & $+0.315$ & 0.046 & 0.083 & 6 & $+0.218$ & 0.024 & 0.043 & 9 & $+0.326$ & 0.027 & 0.030 & 12 \\
N6822 SC6     & $-1.689$ & 0.015 & 0.025 & 37 & $+0.015$ & 0.104 & 0.031 & 2 & $+0.255$ & 0.058 & 0.076 & 5 & $+0.043$ & 0.111 & 0.144 & 2 & $+0.206$ & 0.034 & 0.044 & 9 & $+0.282$ & 0.046 & 0.063 & 12 \\
N6822 SC7     & $-1.130$ & 0.011 & 0.019 & 36 & $-0.663$ & 0.084 & 0.041 & 2 & $-0.235$ & 0.050 & 0.027 & 4 & $-0.010$ & 0.045 & 0.085 & 6 & $-0.047$ & 0.025 & 0.033 & 9 & $-0.006$ & 0.031 & 0.066 & 12 \\
N6822 HVII    & $-1.666$ & 0.014 & 0.035 & 36 & $+0.049$ & 0.079 & 0.014 & 2 & $-0.031$ & 0.059 & 0.059 & 5 & $+0.283$ & 0.050 & 0.114 & 6 & $+0.073$ & 0.030 & 0.040 & 9 & $+0.264$ & 0.038 & 0.075 & 12 \\
M33 H38       & $-1.090$ & 0.023 & 0.051 & 30 & $-0.262$ & 0.153 & \ldots & 1 & $+0.144$ & 0.136 & 0.208 & 2 & \ldots & \ldots & \ldots & \ldots & $+0.324$ & 0.076 & 0.146 & 5 & $+0.393$ & 0.058 & 0.086 & 11 \\
M33 M9        & $-1.634$ & 0.019 & 0.027 & 30 & $-0.014$ & 0.123 & 0.054 & 2 & $-0.090$ & 0.129 & 0.051 & 2 & $+0.489$ & 0.256 & 0.006 & 2 & $+0.220$ & 0.051 & 0.067 & 7 & $+0.472$ & 0.046 & 0.056 & 13 \\
M33 R12       & $-0.861$ & 0.014 & 0.023 & 30 & $-0.151$ & 0.064 & 0.084 & 2 & $+0.161$ & 0.082 & 0.096 & 2 & $+0.196$ & 0.101 & 0.140 & 3 & $+0.202$ & 0.037 & 0.050 & 7 & $+0.346$ & 0.031 & 0.043 & 14 \\
M33 U49       & $-1.333$ & 0.024 & 0.036 & 30 & $-0.098$ & 0.141 & 0.068 & 2 & $+0.304$ & 0.183 & 0.054 & 2 & $+0.453$ & 0.209 & 0.040 & 2 & $+0.149$ & 0.069 & 0.132 & 6 & $+0.526$ & 0.058 & 0.060 & 14 \\
M33 R14       & $-1.020$ & 0.013 & 0.021 & 37 & $+0.004$ & 0.063 & 0.001 & 2 & $+0.162$ & 0.046 & 0.031 & 5 & $+0.350$ & 0.043 & 0.041 & 6 & $+0.209$ & 0.028 & 0.039 & 9 & $+0.230$ & 0.036 & 0.048 & 12 \\
M33 U77       & $-1.797$ & 0.020 & 0.038 & 35 & $-0.285$ & 0.232 & \ldots & 1 & $+0.147$ & 0.093 & 0.187 & 4 & $+0.329$ & 0.129 & 0.056 & 2 & $+0.400$ & 0.043 & 0.070 & 8 & $+0.446$ & 0.061 & 0.049 & 12 \\
M33 CBF28     & $-1.128$ & 0.011 & 0.019 & 35 & $+0.046$ & 0.051 & 0.142 & 2 & $+0.209$ & 0.034 & 0.065 & 5 & $+0.181$ & 0.037 & 0.045 & 6 & $+0.201$ & 0.023 & 0.015 & 9 & $+0.312$ & 0.024 & 0.039 & 12 \\
M33 HM33B     & $-1.218$ & 0.032 & 0.068 & 33 & $-0.087$ & 0.199 & 0.479 & 2 & $-0.267$ & 0.197 & 0.234 & 4 & $+0.365$ & 0.188 & 0.061 & 3 & $-0.000$ & 0.083 & 0.162 & 9 & $+0.300$ & 0.109 & 0.136 & 11 \\
WLM GC        & $-1.849$ & 0.014 & 0.029 & 37 & $+0.021$ & 0.151 & 0.009 & 2 & $+0.038$ & 0.064 & 0.073 & 4 & $+0.185$ & 0.305 & \ldots & 1 & $+0.255$ & 0.034 & 0.056 & 9 & $+0.407$ & 0.032 & 0.056 & 14 \\
Fornax 3      & $-2.284$ & 0.013 & 0.021 & 39 & $-0.073$ & 0.202 & \ldots & 1 & $-0.071$ & 0.057 & 0.099 & 4 & $+0.642$ & 0.219 & 0.161 & 2 & $+0.154$ & 0.034 & 0.056 & 8 & $+0.290$ & 0.030 & 0.042 & 13 \\
Fornax 4      & $-1.237$ & 0.010 & 0.019 & 39 & $-0.390$ & 0.074 & 0.003 & 2 & $-0.069$ & 0.044 & 0.043 & 4 & $+0.065$ & 0.075 & 0.122 & 4 & $+0.033$ & 0.025 & 0.027 & 9 & $+0.162$ & 0.023 & 0.025 & 13 \\
Fornax 5      & $-2.058$ & 0.016 & 0.027 & 39 & $+0.319$ & 0.138 & \ldots & 1 & $+0.048$ & 0.078 & 0.080 & 4 & $+0.669$ & 0.620 & \ldots & 1 & $+0.208$ & 0.040 & 0.055 & 9 & $+0.309$ & 0.040 & 0.079 & 11 \\
M31 006-058   & $-0.527$ & 0.009 & 0.018 & 37 & $+0.300$ & 0.039 & 0.044 & 2 & $+0.341$ & 0.026 & 0.035 & 5 & $+0.311$ & 0.026 & 0.029 & 6 & $+0.216$ & 0.020 & 0.026 & 8 & $+0.369$ & 0.017 & 0.035 & 13 \\
M31 012-064   & $-1.708$ & 0.016 & 0.021 & 38 & $+0.111$ & 0.101 & 0.090 & 2 & $+0.016$ & 0.078 & 0.086 & 4 & $+0.688$ & 0.082 & 0.051 & 4 & $+0.283$ & 0.039 & 0.089 & 8 & $+0.289$ & 0.039 & 0.066 & 12 \\
M31 019-072   & $-0.693$ & 0.010 & 0.025 & 38 & $+0.299$ & 0.045 & 0.040 & 2 & $+0.246$ & 0.034 & 0.091 & 4 & $+0.213$ & 0.039 & 0.055 & 5 & $+0.206$ & 0.023 & 0.043 & 8 & $+0.432$ & 0.019 & 0.048 & 14 \\
M31 058-119   & $-0.985$ & 0.009 & 0.023 & 37 & $-0.013$ & 0.045 & 0.079 & 2 & $+0.191$ & 0.028 & 0.125 & 5 & $+0.329$ & 0.029 & 0.026 & 6 & $+0.172$ & 0.021 & 0.022 & 8 & $+0.319$ & 0.019 & 0.038 & 13 \\
M31 082-114   & $-0.689$ & 0.016 & 0.037 & 36 & $+0.339$ & 0.079 & 0.229 & 2 & $+0.386$ & 0.065 & 0.146 & 4 & $+0.387$ & 0.068 & 0.069 & 5 & $+0.239$ & 0.037 & 0.030 & 8 & $+0.406$ & 0.047 & 0.079 & 12 \\
M31 163-217   & $-0.132$ & 0.009 & 0.028 & 37 & $+0.518$ & 0.039 & 0.073 & 2 & $+0.224$ & 0.025 & 0.047 & 5 & $+0.274$ & 0.025 & 0.080 & 6 & $+0.083$ & 0.019 & 0.058 & 9 & $+0.329$ & 0.017 & 0.040 & 13 \\
M31 171-222   & $-0.217$ & 0.009 & 0.024 & 37 & $+0.441$ & 0.039 & 0.036 & 2 & $+0.290$ & 0.027 & 0.109 & 5 & $+0.235$ & 0.027 & 0.063 & 6 & $+0.083$ & 0.020 & 0.034 & 8 & $+0.295$ & 0.018 & 0.036 & 13 \\
M31 174-226   & $-1.012$ & 0.013 & 0.024 & 38 & $+0.091$ & 0.079 & 0.091 & 2 & $+0.214$ & 0.054 & 0.033 & 4 & $+0.319$ & 0.073 & 0.042 & 5 & $+0.254$ & 0.033 & 0.030 & 8 & $+0.396$ & 0.030 & 0.050 & 12 \\
M31 225-280   & $-0.342$ & 0.009 & 0.026 & 34 & $+0.399$ & 0.038 & 0.159 & 2 & $+0.242$ & 0.024 & 0.117 & 5 & $+0.348$ & 0.024 & 0.075 & 6 & $+0.118$ & 0.018 & 0.060 & 9 & $+0.442$ & 0.016 & 0.039 & 13 \\
M31 338-076   & $-1.065$ & 0.010 & 0.023 & 38 & $+0.025$ & 0.059 & 0.178 & 2 & $+0.244$ & 0.038 & 0.149 & 4 & $+0.246$ & 0.050 & 0.049 & 5 & $+0.226$ & 0.025 & 0.033 & 8 & $+0.362$ & 0.021 & 0.043 & 14 \\
M31 358-219   & $-2.199$ & 0.013 & 0.023 & 36 & $+0.202$ & 0.086 & 0.121 & 2 & $+0.165$ & 0.044 & 0.074 & 5 & $+0.369$ & 0.082 & 0.105 & 5 & $+0.300$ & 0.027 & 0.034 & 8 & $+0.307$ & 0.030 & 0.043 & 12 \\
M31 EXT8      & $-2.808$ & 0.024 & 0.043 & 28 & \ldots & \ldots & \ldots & \ldots & $-0.344$ & 0.220 & 0.015 & 2 & $+0.547$ & 0.316 & \ldots & 1 & $+0.269$ & 0.055 & 0.075 & 8 & $+0.335$ & 0.082 & 0.088 & 10 \\
N2403 F46     & $-1.707$ & 0.019 & 0.027 & 34 & $-0.172$ & 0.153 & \ldots & 1 & $-0.228$ & 0.099 & 0.079 & 4 & $+0.237$ & 0.113 & 0.171 & 4 & $+0.215$ & 0.042 & 0.066 & 8 & $+0.303$ & 0.062 & 0.083 & 10 \\
\hline
\end{tabular}
}
\end{adjustbox}
\tablefoot{
The listed abundances include NLTE corrections for Fe, Na, Mg, Ca, and Ti. The columns $\sigma_{\langle\mathrm{X}\rangle}$ list the uncertainties on the mean abundances of elements $X$ from propagation of the measurement errors, while the columns $S_X$ give the standard errors on the means estimated from the dispersions of the measurements. The columns labelled N list the number of individual measurements. 
}
\end{table*}

\begin{table*}
\caption{Results for Sc, Cr, Mn, and Ni.}
\label{tab:results2}
\begin{adjustbox}{width=1\textwidth}
\centering
{\small
\begin{tabular}{l cccccccccccccccc}
\hline\hline
Cluster       & [Sc/Fe] & $\sigma_{\langle\mathrm{Sc}\rangle}$ & $S_\mathrm{Sc}$ & N    & [Cr/Fe] & $\sigma_{\langle\mathrm{Cr}\rangle}$ & $S_\mathrm{Cr}$ & N     & [Mn/Fe] & $\sigma_{\langle\mathrm{Mn}\rangle}$ & $S_\mathrm{Mn}$ & N    & [Ni/Fe] & $\sigma_{\langle\mathrm{Ni}\rangle}$ & $S_\mathrm{Ni}$ & N    \\ \hline
NGC 0104      & $+0.197$ & 0.028 & 0.067 & 5 & $-0.029$ & 0.014 & 0.033 & 17 & $-0.188$ & 0.037 & 0.041 & 2 & $+0.060$ & 0.015 & 0.053 & 14 \\
NGC 0362      & $+0.133$ & 0.031 & 0.050 & 5 & $-0.025$ & 0.015 & 0.037 & 17 & $-0.255$ & 0.039 & 0.110 & 2 & $-0.087$ & 0.017 & 0.051 & 14 \\
NGC 6254      & $+0.238$ & 0.036 & 0.056 & 4 & $-0.124$ & 0.019 & 0.043 & 17 & $-0.338$ & 0.042 & 0.050 & 2 & $+0.023$ & 0.020 & 0.046 & 14 \\
NGC 6388      & $+0.145$ & 0.028 & 0.075 & 5 & $-0.063$ & 0.014 & 0.038 & 17 & $-0.147$ & 0.038 & 0.006 & 2 & $-0.015$ & 0.016 & 0.074 & 14 \\
NGC 6752      & $+0.130$ & 0.034 & 0.060 & 4 & $-0.111$ & 0.017 & 0.025 & 17 & $-0.248$ & 0.040 & 0.012 & 2 & $+0.072$ & 0.017 & 0.041 & 14 \\
NGC 7078      & $+0.119$ & 0.041 & 0.014 & 3 & $-0.267$ & 0.030 & 0.041 & 12 & $-0.139$ & 0.057 & 0.182 & 2 & $+0.115$ & 0.029 & 0.066 & 13 \\
NGC 7099      & $+0.129$ & 0.041 & 0.066 & 4 & $-0.176$ & 0.027 & 0.046 & 13 & $-0.203$ & 0.056 & 0.032 & 2 & $+0.097$ & 0.028 & 0.059 & 14 \\
N147 HII      & $+0.318$ & 0.128 & 0.220 & 2 & $+0.026$ & 0.088 & 0.110 & 12 & $-0.033$ & 0.147 & 0.307 & 2 & $+0.191$ & 0.106 & 0.148 & 11 \\
N147 HIII     & $-0.190$ & 0.199 & 0.097 & 2 & $+0.130$ & 0.116 & 0.231 & 6 & $+0.333$ & 0.125 & 0.304 & 2 & $-0.341$ & 0.118 & 0.259 & 5 \\
N147 PA-1     & $-0.406$ & 0.319 & 0.107 & 2 & $-0.026$ & 0.076 & 0.101 & 8 & $+0.118$ & 0.125 & \ldots & 1 & $+0.233$ & 0.175 & 0.039 & 3 \\
N147 PA-2     & $+0.021$ & 0.083 & 0.133 & 2 & $+0.019$ & 0.055 & 0.091 & 10 & $-0.126$ & 0.125 & 0.087 & 2 & $+0.106$ & 0.072 & 0.094 & 10 \\
N147 SD7      & $-0.183$ & 0.112 & 0.098 & 2 & $-0.022$ & 0.056 & 0.083 & 14 & $-0.316$ & 0.111 & 0.130 & 2 & $+0.172$ & 0.059 & 0.094 & 12 \\
N185 FJJ-III  & $+0.096$ & 0.127 & 0.126 & 2 & $+0.221$ & 0.053 & 0.114 & 12 & $-0.081$ & 0.135 & \ldots & 1 & $+0.259$ & 0.063 & 0.093 & 12 \\
N185 FJJ-V    & $-0.093$ & 0.097 & 0.173 & 2 & $-0.147$ & 0.062 & 0.092 & 11 & $-0.004$ & 0.096 & 0.121 & 2 & $+0.197$ & 0.055 & 0.092 & 12 \\
N185 FJJ-VIII & $+0.177$ & 0.092 & 0.012 & 2 & $-0.054$ & 0.049 & 0.063 & 15 & $-0.060$ & 0.078 & 0.101 & 2 & $+0.130$ & 0.055 & 0.091 & 13 \\
N205 HubbleI  & $+0.278$ & 0.052 & 0.045 & 4 & $-0.038$ & 0.028 & 0.047 & 17 & $-0.200$ & 0.052 & 0.078 & 2 & $-0.020$ & 0.032 & 0.057 & 14 \\
N205 HubbleII & $+0.334$ & 0.058 & 0.148 & 2 & $-0.066$ & 0.024 & 0.042 & 17 & $-0.247$ & 0.050 & 0.045 & 2 & $+0.072$ & 0.027 & 0.065 & 14 \\
N6822 SC6     & $+0.188$ & 0.084 & 0.062 & 2 & $-0.066$ & 0.043 & 0.051 & 15 & $-0.334$ & 0.085 & 0.056 & 2 & $-0.053$ & 0.053 & 0.063 & 13 \\
N6822 SC7     & $-0.292$ & 0.061 & 0.035 & 3 & $-0.128$ & 0.026 & 0.036 & 17 & $-0.301$ & 0.051 & 0.038 & 2 & $-0.217$ & 0.030 & 0.056 & 14 \\
N6822 HVII    & $-0.065$ & 0.072 & 0.164 & 3 & $+0.087$ & 0.032 & 0.085 & 17 & $-0.027$ & 0.059 & 0.225 & 2 & $+0.034$ & 0.043 & 0.100 & 14 \\
M33 H38       & $+0.022$ & 0.107 & 0.104 & 3 & $+0.626$ & 0.158 & 0.161 & 3 & $-0.303$ & 0.113 & 0.257 & 2 & $+0.118$ & 0.073 & 0.100 & 12 \\
M33 M9        & $+0.202$ & 0.087 & 0.049 & 2 & $-0.298$ & 0.126 & 0.393 & 4 & $-0.276$ & 0.106 & 0.078 & 2 & $-0.014$ & 0.063 & 0.058 & 13 \\
M33 R12       & $+0.090$ & 0.063 & 0.098 & 3 & $-0.411$ & 0.111 & 0.148 & 3 & $-0.132$ & 0.061 & 0.154 & 2 & $-0.007$ & 0.036 & 0.069 & 13 \\
M33 U49       & $+0.352$ & 0.095 & 0.026 & 3 & $+0.135$ & 0.190 & 0.507 & 3 & $+0.014$ & 0.101 & 0.148 & 2 & $+0.067$ & 0.078 & 0.112 & 10 \\
M33 R14       & $+0.119$ & 0.061 & 0.036 & 3 & $-0.056$ & 0.030 & 0.050 & 17 & $-0.151$ & 0.057 & 0.023 & 2 & $+0.003$ & 0.035 & 0.069 & 14 \\
M33 U77       & $+0.328$ & 0.105 & 0.128 & 2 & $+0.049$ & 0.056 & 0.089 & 13 & $-0.165$ & 0.113 & 0.064 & 2 & $-0.001$ & 0.068 & 0.064 & 12 \\
M33 CBF28     & $+0.050$ & 0.049 & 0.061 & 4 & $-0.084$ & 0.022 & 0.026 & 18 & $-0.257$ & 0.047 & 0.053 & 2 & $-0.062$ & 0.024 & 0.049 & 14 \\
M33 HM33B     & $-0.227$ & 0.193 & 0.051 & 2 & $+0.210$ & 0.079 & 0.117 & 16 & $-0.137$ & 0.226 & \ldots & 1 & $-0.197$ & 0.105 & 0.092 & 12 \\
WLM GC        & $+0.159$ & 0.073 & 0.089 & 3 & $+0.019$ & 0.046 & 0.071 & 10 & $-0.262$ & 0.079 & 0.111 & 2 & $+0.131$ & 0.044 & 0.095 & 13 \\
Fornax 3      & $+0.109$ & 0.066 & 0.120 & 3 & $-0.216$ & 0.048 & 0.046 & 12 & $-0.258$ & 0.100 & 0.256 & 2 & $-0.003$ & 0.051 & 0.079 & 12 \\
Fornax 4      & $-0.134$ & 0.049 & 0.062 & 3 & $-0.108$ & 0.025 & 0.034 & 15 & $-0.275$ & 0.049 & 0.053 & 2 & $-0.163$ & 0.023 & 0.051 & 15 \\
Fornax 5      & $-0.119$ & 0.112 & 0.226 & 3 & $-0.004$ & 0.057 & 0.087 & 11 & $-0.286$ & 0.114 & \ldots & 1 & $+0.151$ & 0.052 & 0.093 & 13 \\
M31 006-058   & $+0.224$ & 0.032 & 0.077 & 5 & $-0.012$ & 0.017 & 0.031 & 15 & $-0.157$ & 0.039 & 0.074 & 2 & $+0.014$ & 0.017 & 0.059 & 15 \\
M31 012-064   & $+0.344$ & 0.075 & 0.147 & 3 & $-0.071$ & 0.043 & 0.062 & 13 & $-0.336$ & 0.086 & 0.177 & 2 & $+0.031$ & 0.051 & 0.076 & 14 \\
M31 019-072   & $+0.190$ & 0.040 & 0.036 & 5 & $-0.037$ & 0.019 & 0.037 & 16 & $-0.182$ & 0.043 & 0.066 & 2 & $+0.054$ & 0.020 & 0.064 & 15 \\
M31 058-119   & $+0.138$ & 0.037 & 0.090 & 5 & $-0.078$ & 0.021 & 0.048 & 15 & $-0.251$ & 0.042 & 0.098 & 2 & $+0.016$ & 0.019 & 0.061 & 15 \\
M31 082-114   & $+0.327$ & 0.091 & 0.118 & 2 & $-0.068$ & 0.044 & 0.066 & 15 & $-0.326$ & 0.092 & 0.111 & 2 & $+0.245$ & 0.044 & 0.079 & 15 \\
M31 163-217   & $+0.136$ & 0.031 & 0.076 & 5 & $-0.057$ & 0.017 & 0.039 & 15 & $-0.030$ & 0.040 & 0.028 & 2 & $+0.108$ & 0.016 & 0.074 & 15 \\
M31 171-222   & $+0.219$ & 0.033 & 0.082 & 5 & $-0.028$ & 0.017 & 0.041 & 15 & $-0.015$ & 0.041 & 0.049 & 2 & $+0.096$ & 0.017 & 0.068 & 15 \\
M31 174-226   & $+0.235$ & 0.069 & 0.056 & 3 & $-0.033$ & 0.031 & 0.041 & 16 & $-0.267$ & 0.063 & 0.151 & 2 & $+0.059$ & 0.037 & 0.074 & 14 \\
M31 225-280   & $+0.163$ & 0.030 & 0.125 & 5 & $-0.004$ & 0.016 & 0.053 & 15 & $-0.103$ & 0.039 & 0.004 & 2 & $+0.112$ & 0.016 & 0.085 & 15 \\
M31 338-076   & $+0.077$ & 0.046 & 0.095 & 5 & $-0.075$ & 0.022 & 0.037 & 16 & $-0.280$ & 0.047 & 0.046 & 2 & $-0.008$ & 0.023 & 0.057 & 15 \\
M31 358-219   & $-0.029$ & 0.060 & 0.021 & 3 & $-0.163$ & 0.049 & 0.053 & 9 & $-0.237$ & 0.083 & 0.067 & 2 & $+0.047$ & 0.042 & 0.073 & 12 \\
M31 EXT8      & $+0.442$ & 0.118 & 0.148 & 2 & $-0.234$ & 0.163 & 0.149 & 5 & $+1.155$ & 0.167 & \ldots & 1 & $+0.462$ & 0.089 & 0.218 & 8 \\
N2403 F46     & $-0.063$ & 0.103 & 0.133 & 2 & $-0.041$ & 0.057 & 0.084 & 13 & $-0.313$ & 0.109 & 0.076 & 2 & $+0.210$ & 0.057 & 0.112 & 12 \\
\hline
\end{tabular}
}
\end{adjustbox}
\tablefoot{
The listed abundances include NLTE corrections for Mn and Ni. See notes to Table~\ref{tab:results1} for further explanations. 
}
\end{table*}

\begin{table*}
\caption{Results for Cu, Zn, Zr, Ba, and Eu.}
\label{tab:results3}
\begin{adjustbox}{width=1\textwidth}
\centering
{\small
\begin{tabular}{l cccccccccccccccccccc}
\hline\hline
Cluster       & [Cu/Fe] & $\sigma_{\langle\mathrm{Cu}\rangle}$ & $S_\mathrm{Cu}$ & N    & [Zn/Fe] & $\sigma_{\langle\mathrm{Zn}\rangle}$ & $S_\mathrm{Zn}$ & N    & [Zr/Fe] & $\sigma_{\langle\mathrm{Zr}\rangle}$ & $S_\mathrm{Zr}$ & N     & [Ba/Fe] & $\sigma_{\langle\mathrm{Ba}\rangle}$ & $S_\mathrm{Ba}$ & N    & [Eu/Fe] & $\sigma_{\langle\mathrm{Eu}\rangle}$ & $S_\mathrm{Eu}$ & N    \\ \hline
NGC 0104      & $-0.049$ & 0.060 & \ldots & 1 & $+0.113$ & 0.044 & 0.046 & 2 & $+0.224$ & 0.065 & \ldots & 1 & $+0.133$ & 0.025 & 0.070 & 5 & $+0.225$ & 0.050 & 0.006 & 2 \\
NGC 0362      & $-0.330$ & 0.067 & \ldots & 1 & $-0.098$ & 0.048 & 0.066 & 2 & $+0.396$ & 0.076 & \ldots & 1 & $+0.303$ & 0.025 & 0.053 & 5 & $+0.625$ & 0.047 & 0.021 & 2 \\
NGC 6254      & $-0.588$ & 0.076 & \ldots & 1 & $-0.005$ & 0.056 & 0.008 & 2 & $-0.130$ & 0.150 & \ldots & 1 & $+0.328$ & 0.028 & 0.049 & 5 & $+0.188$ & 0.090 & \ldots & 1 \\
NGC 6388      & $-0.102$ & 0.060 & \ldots & 1 & $-0.159$ & 0.050 & 0.076 & 2 & $+0.294$ & 0.063 & \ldots & 1 & $+0.120$ & 0.024 & 0.066 & 5 & $-0.055$ & 0.071 & \ldots & 1 \\
NGC 6752      & $-0.441$ & 0.065 & \ldots & 1 & $+0.098$ & 0.045 & 0.029 & 2 & $+0.295$ & 0.134 & \ldots & 1 & $+0.153$ & 0.027 & 0.077 & 5 & $+0.390$ & 0.054 & 0.093 & 2 \\
NGC 7078      & \ldots & \ldots & \ldots & \ldots & $-0.189$ & 0.103 & 0.060 & 2 & \ldots & \ldots & \ldots & \ldots & $+0.259$ & 0.030 & 0.036 & 5 & $+0.531$ & 0.109 & 0.082 & 2 \\
NGC 7099      & $-0.437$ & 0.114 & \ldots & 1 & $-0.080$ & 0.077 & 0.092 & 2 & \ldots & \ldots & \ldots & \ldots & $-0.052$ & 0.032 & 0.145 & 5 & $+0.332$ & 0.110 & \ldots & 1 \\
N147 HII      & $-0.425$ & 0.362 & \ldots & 1 & $+0.655$ & 0.731 & \ldots & 1 & $+0.452$ & 0.458 & \ldots & 1 & $-0.224$ & 0.176 & 0.308 & 3 & \ldots & \ldots & \ldots & \ldots \\
N147 HIII     & $+0.386$ & 0.228 & \ldots & 1 & $+0.179$ & 0.392 & \ldots & 1 & \ldots & \ldots & \ldots & \ldots & $-0.615$ & 0.107 & 0.092 & 5 & \ldots & \ldots & \ldots & \ldots \\
N147 PA-1     & $-0.203$ & 0.292 & \ldots & 1 & $+0.026$ & 0.405 & \ldots & 1 & \ldots & \ldots & \ldots & \ldots & $+0.048$ & 0.086 & 0.083 & 5 & \ldots & \ldots & \ldots & \ldots \\
N147 PA-2     & $-0.840$ & 0.350 & \ldots & 1 & $+0.307$ & 0.205 & 0.322 & 3 & \ldots & \ldots & \ldots & \ldots & $+0.209$ & 0.062 & 0.033 & 5 & \ldots & \ldots & \ldots & \ldots \\
N147 SD7      & $-0.503$ & 0.312 & \ldots & 1 & $+0.038$ & 0.200 & 0.329 & 2 & \ldots & \ldots & \ldots & \ldots & $+0.181$ & 0.057 & 0.157 & 5 & \ldots & \ldots & \ldots & \ldots \\
N185 FJJ-III  & $-0.792$ & 0.243 & \ldots & 1 & $+0.141$ & 0.263 & 0.179 & 2 & \ldots & \ldots & \ldots & \ldots & $+0.255$ & 0.075 & 0.101 & 5 & $+0.395$ & 0.564 & \ldots & 1 \\
N185 FJJ-V    & $-0.710$ & 0.317 & \ldots & 1 & $-0.024$ & 0.255 & \ldots & 1 & \ldots & \ldots & \ldots & \ldots & $+0.248$ & 0.069 & 0.145 & 5 & \ldots & \ldots & \ldots & \ldots \\
N185 FJJ-VIII & $-0.720$ & 0.282 & \ldots & 1 & $+0.494$ & 0.153 & 0.037 & 2 & \ldots & \ldots & \ldots & \ldots & $+0.211$ & 0.059 & 0.170 & 5 & $+0.253$ & 0.420 & \ldots & 1 \\
N205 HubbleI  & $-0.472$ & 0.136 & \ldots & 1 & $-0.199$ & 0.117 & 0.104 & 3 & $+0.548$ & 0.148 & \ldots & 1 & $+0.206$ & 0.041 & 0.063 & 5 & $+0.741$ & 0.187 & \ldots & 1 \\
N205 HubbleII & $-0.530$ & 0.093 & \ldots & 1 & $-0.051$ & 0.078 & 0.182 & 3 & $-0.291$ & 0.241 & \ldots & 1 & $+0.102$ & 0.036 & 0.076 & 5 & $+0.112$ & 0.220 & \ldots & 1 \\
N6822 SC6     & $-0.612$ & 0.177 & \ldots & 1 & $+0.129$ & 0.157 & 0.143 & 2 & \ldots & \ldots & \ldots & \ldots & $+0.194$ & 0.054 & 0.034 & 5 & $+0.554$ & 0.246 & \ldots & 1 \\
N6822 SC7     & \ldots & \ldots & \ldots & \ldots & $-0.405$ & 0.098 & 0.059 & 3 & $-0.470$ & 0.333 & \ldots & 1 & $+0.109$ & 0.035 & 0.030 & 5 & $+0.145$ & 0.152 & \ldots & 1 \\
N6822 HVII    & $-0.313$ & 0.167 & \ldots & 1 & $-0.359$ & 0.173 & 0.104 & 2 & \ldots & \ldots & \ldots & \ldots & $+0.353$ & 0.041 & 0.082 & 5 & $-0.023$ & 0.329 & \ldots & 1 \\
M33 H38       & $-0.037$ & 0.335 & \ldots & 1 & $-0.573$ & 0.429 & \ldots & 1 & \ldots & \ldots & \ldots & \ldots & $+0.475$ & 0.062 & 0.110 & 4 & \ldots & \ldots & \ldots & \ldots \\
M33 M9        & $-0.235$ & 0.180 & \ldots & 1 & $+0.309$ & 0.220 & 0.069 & 2 & \ldots & \ldots & \ldots & \ldots & $+0.504$ & 0.065 & 0.089 & 4 & \ldots & \ldots & \ldots & \ldots \\
M33 R12       & $-0.441$ & 0.128 & \ldots & 1 & $+0.061$ & 0.143 & 0.115 & 3 & \ldots & \ldots & \ldots & \ldots & $+0.250$ & 0.055 & 0.082 & 4 & $+0.483$ & 0.205 & \ldots & 1 \\
M33 U49       & $-0.598$ & 0.297 & \ldots & 1 & $+0.587$ & 0.304 & 0.081 & 2 & \ldots & \ldots & \ldots & \ldots & $+0.461$ & 0.091 & 0.235 & 4 & \ldots & \ldots & \ldots & \ldots \\
M33 R14       & $-0.565$ & 0.132 & \ldots & 1 & $-0.134$ & 0.136 & 0.196 & 3 & $+0.197$ & 0.176 & \ldots & 1 & $+0.242$ & 0.042 & 0.084 & 5 & \ldots & \ldots & \ldots & \ldots \\
M33 U77       & $-0.599$ & 0.250 & \ldots & 1 & $+0.216$ & 0.217 & 0.327 & 3 & \ldots & \ldots & \ldots & \ldots & $+0.297$ & 0.075 & 0.078 & 5 & \ldots & \ldots & \ldots & \ldots \\
M33 CBF28     & $-0.347$ & 0.088 & \ldots & 1 & $-0.177$ & 0.074 & 0.118 & 3 & $+0.605$ & 0.103 & \ldots & 1 & $+0.168$ & 0.033 & 0.025 & 5 & $+0.657$ & 0.106 & \ldots & 1 \\
M33 HM33B     & $-0.012$ & 0.279 & \ldots & 1 & $-0.619$ & 1.038 & \ldots & 1 & $+0.884$ & 0.482 & \ldots & 1 & $+0.714$ & 0.149 & 0.008 & 2 & $+0.660$ & 0.499 & \ldots & 1 \\
WLM GC        & $-1.113$ & 0.524 & \ldots & 1 & $+0.055$ & 0.134 & 0.228 & 2 & \ldots & \ldots & \ldots & \ldots & $-0.110$ & 0.064 & 0.053 & 4 & $+0.080$ & 0.278 & \ldots & 1 \\
Fornax 3      & \ldots & \ldots & \ldots & \ldots & $-0.090$ & 0.161 & \ldots & 1 & \ldots & \ldots & \ldots & \ldots & $+0.354$ & 0.044 & 0.061 & 4 & \ldots & \ldots & \ldots & \ldots \\
Fornax 4      & $-0.885$ & 0.106 & \ldots & 1 & $-0.287$ & 0.088 & 0.091 & 2 & $+0.055$ & 0.279 & \ldots & 1 & $-0.003$ & 0.042 & 0.009 & 4 & $+0.194$ & 0.120 & \ldots & 1 \\
Fornax 5      & $-0.543$ & 0.293 & \ldots & 1 & $-0.027$ & 0.175 & 0.019 & 2 & \ldots & \ldots & \ldots & \ldots & $-0.183$ & 0.072 & 0.064 & 4 & $+0.314$ & 0.323 & \ldots & 1 \\
M31 006-058   & $-0.066$ & 0.070 & \ldots & 1 & $+0.002$ & 0.053 & 0.104 & 3 & $+0.378$ & 0.070 & \ldots & 1 & $+0.067$ & 0.027 & 0.100 & 5 & $-0.023$ & 0.092 & \ldots & 1 \\
M31 012-064   & $-0.730$ & 0.235 & \ldots & 1 & $-0.256$ & 0.218 & 0.079 & 2 & \ldots & \ldots & \ldots & \ldots & $+0.290$ & 0.060 & 0.081 & 4 & \ldots & \ldots & \ldots & \ldots \\
M31 019-072   & $-0.277$ & 0.085 & \ldots & 1 & $-0.008$ & 0.069 & 0.066 & 3 & $+0.302$ & 0.111 & \ldots & 1 & $-0.021$ & 0.037 & 0.072 & 4 & \ldots & \ldots & \ldots & \ldots \\
M31 058-119   & $-0.407$ & 0.080 & \ldots & 1 & $-0.267$ & 0.062 & 0.117 & 3 & $-0.015$ & 0.109 & \ldots & 1 & $+0.316$ & 0.029 & 0.028 & 4 & $+0.340$ & 0.079 & 0.062 & 2 \\
M31 082-114   & $-0.318$ & 0.193 & \ldots & 1 & $+0.330$ & 0.269 & 0.308 & 3 & $+0.429$ & 0.230 & \ldots & 1 & $+0.164$ & 0.085 & 0.232 & 4 & \ldots & \ldots & \ldots & \ldots \\
M31 163-217   & $+0.127$ & 0.073 & \ldots & 1 & $-0.045$ & 0.055 & 0.104 & 3 & $+0.060$ & 0.075 & \ldots & 1 & $-0.109$ & 0.029 & 0.119 & 5 & $-0.127$ & 0.092 & \ldots & 1 \\
M31 171-222   & $+0.096$ & 0.078 & \ldots & 1 & $-0.064$ & 0.060 & 0.142 & 3 & $+0.243$ & 0.076 & \ldots & 1 & $+0.062$ & 0.028 & 0.145 & 5 & $-0.216$ & 0.112 & \ldots & 1 \\
M31 174-226   & $-0.349$ & 0.166 & \ldots & 1 & $-0.266$ & 0.149 & 0.160 & 3 & $+0.488$ & 0.196 & \ldots & 1 & $+0.375$ & 0.048 & 0.051 & 4 & \ldots & \ldots & \ldots & \ldots \\
M31 225-280   & $-0.257$ & 0.065 & \ldots & 1 & $-0.180$ & 0.049 & 0.038 & 3 & $+0.098$ & 0.077 & \ldots & 1 & $+0.136$ & 0.025 & 0.135 & 5 & $-0.028$ & 0.072 & 0.048 & 2 \\
M31 338-076   & $-0.395$ & 0.086 & \ldots & 1 & $+0.063$ & 0.070 & 0.104 & 3 & $+0.082$ & 0.165 & \ldots & 1 & $+0.363$ & 0.034 & 0.062 & 4 & \ldots & \ldots & \ldots & \ldots \\
M31 358-219   & $-0.816$ & 0.387 & \ldots & 1 & $+0.019$ & 0.121 & 0.028 & 2 & \ldots & \ldots & \ldots & \ldots & $+0.011$ & 0.041 & 0.048 & 5 & $+0.222$ & 0.303 & \ldots & 1 \\
M31 EXT8      & $+0.068$ & 0.475 & \ldots & 1 & \ldots & \ldots & \ldots & \ldots & \ldots & \ldots & \ldots & \ldots & $+0.348$ & 0.069 & 0.028 & 4 & \ldots & \ldots & \ldots & \ldots \\
N2403 F46     & \ldots & \ldots & \ldots & \ldots & $+0.185$ & 0.222 & 0.282 & 2 & \ldots & \ldots & \ldots & \ldots & $+0.152$ & 0.066 & 0.143 & 5 & $+0.423$ & 0.340 & \ldots & 1 \\
\hline
\end{tabular}
}
\end{adjustbox}
\tablefoot{
The listed abundances include NLTE corrections for Ba. See notes to Table~\ref{tab:results1} for further explanations.
}
\end{table*}

\clearpage

\section{LTE abundance measurements}
\label{app:lte}

Tables~\ref{tab:results1v4lte}-\ref{tab:results3v4lte} list the average \ac{lte} abundance measurements from our  analysis, based on DSEP isochrones and empirical horizontal branches.

\begin{table*}[!h]
\caption{LTE results for Fe, Na, Mg, Si, Ca, and Ti.}
\label{tab:results1v4lte}
\begin{adjustbox}{width=1\textwidth}
\centering
{\small
\begin{tabular}{l cccccccccccccccccccccccc}
\hline\hline
Cluster       & [Fe/H] & $\sigma_{\langle\mathrm{Fe}\rangle}$ & $S_\mathrm{Fe}$ & N      & [Na/Fe]  & $\sigma_{\langle\mathrm{Na}\rangle}$ & $S_\mathrm{Na}$ & N    & [Mg/Fe]  & $\sigma_{\langle\mathrm{Mg}\rangle}$ & $S_\mathrm{Mg}$ & N    & [Si/Fe]  & $\sigma_{\langle\mathrm{Si}\rangle}$ & $S_\mathrm{Si}$ & N    & [Ca/Fe]  & $\sigma_{\langle\mathrm{Ca}\rangle}$ & $S_\mathrm{Ca}$ & N    & [Ti/Fe]  & $\sigma_{\langle\mathrm{Ti}\rangle}$ & $S_\mathrm{Ti}$ & N    \\ \hline
NGC 0104      & $-0.748$ & 0.008 & 0.019 & 39 & $+0.393$ & 0.038 & 0.070 & 2 & $+0.404$ & 0.024 & 0.053 & 5 & $+0.389$ & 0.023 & 0.044 & 6 & $+0.296$ & 0.017 & 0.029 & 10 & $+0.337$ & 0.015 & 0.031 & 14 \\
NGC 0362      & $-1.092$ & 0.009 & 0.019 & 39 & $-0.078$ & 0.043 & 0.013 & 2 & $+0.161$ & 0.027 & 0.048 & 5 & $+0.174$ & 0.026 & 0.051 & 6 & $+0.208$ & 0.018 & 0.025 & 10 & $+0.326$ & 0.016 & 0.034 & 14 \\
NGC 6254      & $-1.507$ & 0.009 & 0.018 & 39 & $-0.088$ & 0.048 & 0.001 & 2 & $+0.318$ & 0.029 & 0.048 & 5 & $+0.302$ & 0.032 & 0.051 & 6 & $+0.313$ & 0.019 & 0.043 & 10 & $+0.313$ & 0.017 & 0.047 & 14 \\
NGC 6388      & $-0.548$ & 0.008 & 0.025 & 39 & $+0.360$ & 0.038 & 0.088 & 2 & $+0.092$ & 0.026 & 0.082 & 5 & $+0.229$ & 0.023 & 0.060 & 6 & $+0.080$ & 0.017 & 0.045 & 10 & $+0.185$ & 0.015 & 0.033 & 14 \\
NGC 6752      & $-1.734$ & 0.009 & 0.014 & 39 & $+0.193$ & 0.043 & 0.027 & 2 & $+0.362$ & 0.027 & 0.060 & 5 & $+0.438$ & 0.028 & 0.016 & 6 & $+0.378$ & 0.017 & 0.018 & 10 & $+0.252$ & 0.016 & 0.042 & 14 \\
NGC 7078      & $-2.343$ & 0.010 & 0.020 & 38 & $+0.074$ & 0.091 & 0.002 & 2 & $+0.193$ & 0.031 & 0.028 & 5 & $+0.521$ & 0.066 & 0.115 & 3 & $+0.307$ & 0.021 & 0.023 & 10 & $+0.322$ & 0.021 & 0.060 & 12 \\
NGC 7099      & $-2.293$ & 0.010 & 0.020 & 39 & $+0.097$ & 0.082 & 0.024 & 2 & $+0.273$ & 0.032 & 0.090 & 5 & $+0.506$ & 0.054 & 0.084 & 5 & $+0.294$ & 0.020 & 0.020 & 10 & $+0.257$ & 0.020 & 0.029 & 14 \\
N147 HII      & $-1.460$ & 0.029 & 0.059 & 37 & $+0.393$ & 0.119 & 0.032 & 2 & $+0.243$ & 0.165 & 0.123 & 4 & $+0.698$ & 0.115 & 0.062 & 2 & $+0.284$ & 0.062 & 0.066 & 9 & $+0.585$ & 0.134 & 0.103 & 6 \\
N147 HIII     & $-2.411$ & 0.024 & 0.041 & 31 & $+0.554$ & 0.146 & \ldots & 1 & $+0.131$ & 0.104 & 0.084 & 4 & $+0.200$ & 0.273 & \ldots & 1 & $+0.348$ & 0.049 & 0.046 & 8 & $+0.292$ & 0.084 & 0.056 & 8 \\
N147 PA-1     & $-2.258$ & 0.023 & 0.043 & 33 & $+0.418$ & 0.218 & \ldots & 1 & $+0.232$ & 0.099 & 0.084 & 5 & $+0.616$ & 0.362 & 0.159 & 2 & $+0.219$ & 0.048 & 0.034 & 9 & $+0.316$ & 0.089 & 0.086 & 8 \\
N147 PA-2     & $-1.948$ & 0.017 & 0.028 & 36 & $+0.238$ & 0.127 & 0.021 & 2 & $+0.353$ & 0.077 & 0.062 & 5 & $+0.178$ & 0.126 & 0.232 & 2 & $+0.378$ & 0.034 & 0.040 & 11 & $+0.347$ & 0.052 & 0.075 & 11 \\
N147 SD7      & $-1.916$ & 0.017 & 0.037 & 37 & $-0.072$ & 0.179 & 0.312 & 2 & $+0.316$ & 0.085 & 0.167 & 4 & $+0.579$ & 0.066 & 0.105 & 4 & $+0.231$ & 0.034 & 0.054 & 11 & $+0.335$ & 0.054 & 0.047 & 10 \\
N185 FJJ-III  & $-1.780$ & 0.020 & 0.031 & 36 & $+0.469$ & 0.113 & \ldots & 1 & $+0.497$ & 0.073 & 0.135 & 6 & $+0.445$ & 0.099 & 0.142 & 6 & $+0.368$ & 0.043 & 0.066 & 11 & $+0.356$ & 0.059 & 0.072 & 11 \\
N185 FJJ-V    & $-1.806$ & 0.017 & 0.031 & 36 & $+0.005$ & 0.162 & \ldots & 1 & $+0.209$ & 0.067 & 0.111 & 6 & $+0.355$ & 0.091 & 0.157 & 5 & $+0.316$ & 0.033 & 0.040 & 11 & $+0.328$ & 0.052 & 0.084 & 11 \\
N185 FJJ-VIII & $-1.775$ & 0.017 & 0.027 & 37 & $+0.151$ & 0.123 & 0.004 & 2 & $-0.053$ & 0.074 & 0.111 & 6 & $+0.313$ & 0.088 & 0.166 & 4 & $+0.332$ & 0.035 & 0.053 & 11 & $+0.287$ & 0.056 & 0.062 & 11 \\
N205 HubbleI  & $-1.428$ & 0.012 & 0.025 & 37 & $-0.001$ & 0.073 & 0.082 & 2 & $+0.440$ & 0.039 & 0.056 & 6 & $+0.319$ & 0.048 & 0.025 & 6 & $+0.254$ & 0.025 & 0.037 & 11 & $+0.277$ & 0.028 & 0.055 & 12 \\
N205 HubbleII & $-1.364$ & 0.011 & 0.022 & 37 & $+0.142$ & 0.063 & 0.004 & 2 & $+0.297$ & 0.038 & 0.088 & 6 & $+0.331$ & 0.046 & 0.083 & 6 & $+0.279$ & 0.023 & 0.039 & 11 & $+0.221$ & 0.027 & 0.036 & 12 \\
N6822 SC6     & $-1.716$ & 0.015 & 0.024 & 37 & $+0.146$ & 0.104 & 0.035 & 2 & $+0.254$ & 0.055 & 0.089 & 6 & $+0.069$ & 0.111 & 0.144 & 2 & $+0.256$ & 0.032 & 0.039 & 11 & $+0.184$ & 0.046 & 0.060 & 12 \\
N6822 SC7     & $-1.148$ & 0.011 & 0.020 & 36 & $-0.523$ & 0.084 & 0.051 & 2 & $-0.267$ & 0.046 & 0.081 & 5 & $+0.010$ & 0.045 & 0.085 & 6 & $+0.003$ & 0.023 & 0.028 & 11 & $-0.125$ & 0.031 & 0.071 & 12 \\
N6822 HVII    & $-1.693$ & 0.014 & 0.034 & 36 & $+0.180$ & 0.080 & 0.023 & 2 & $-0.063$ & 0.057 & 0.131 & 6 & $+0.311$ & 0.050 & 0.114 & 6 & $+0.058$ & 0.029 & 0.055 & 11 & $+0.156$ & 0.038 & 0.064 & 12 \\
M33 H38       & $-1.108$ & 0.023 & 0.051 & 30 & $-0.123$ & 0.153 & \ldots & 1 & $+0.168$ & 0.136 & 0.215 & 2 & \ldots & \ldots & \ldots & \ldots & $+0.384$ & 0.074 & 0.142 & 7 & $+0.305$ & 0.058 & 0.073 & 11 \\
M33 M9        & $-1.659$ & 0.019 & 0.028 & 30 & $+0.115$ & 0.124 & 0.058 & 2 & $-0.102$ & 0.112 & 0.071 & 3 & $+0.514$ & 0.256 & 0.006 & 2 & $+0.258$ & 0.049 & 0.055 & 9 & $+0.362$ & 0.046 & 0.055 & 13 \\
M33 R12       & $-0.875$ & 0.014 & 0.023 & 30 & $-0.010$ & 0.064 & 0.100 & 2 & $+0.246$ & 0.075 & 0.116 & 3 & $+0.209$ & 0.101 & 0.140 & 3 & $+0.258$ & 0.035 & 0.048 & 9 & $+0.248$ & 0.031 & 0.041 & 14 \\
M33 U49       & $-1.348$ & 0.024 & 0.036 & 30 & $+0.025$ & 0.141 & 0.054 & 2 & $+0.204$ & 0.164 & 0.191 & 3 & $+0.468$ & 0.209 & 0.040 & 2 & $+0.202$ & 0.067 & 0.124 & 8 & $+0.430$ & 0.058 & 0.070 & 14 \\
M33 R14       & $-1.037$ & 0.013 & 0.020 & 37 & $+0.146$ & 0.063 & 0.015 & 2 & $+0.179$ & 0.045 & 0.049 & 6 & $+0.367$ & 0.043 & 0.041 & 6 & $+0.237$ & 0.027 & 0.052 & 11 & $+0.129$ & 0.036 & 0.040 & 12 \\
M33 U77       & $-1.824$ & 0.020 & 0.039 & 35 & $-0.174$ & 0.232 & \ldots & 1 & $+0.235$ & 0.088 & 0.161 & 5 & $+0.356$ & 0.128 & 0.056 & 2 & $+0.448$ & 0.042 & 0.068 & 9 & $+0.355$ & 0.061 & 0.067 & 12 \\
M33 CBF28     & $-1.147$ & 0.011 & 0.019 & 35 & $+0.188$ & 0.051 & 0.159 & 2 & $+0.208$ & 0.032 & 0.079 & 6 & $+0.200$ & 0.037 & 0.045 & 6 & $+0.263$ & 0.022 & 0.016 & 10 & $+0.210$ & 0.024 & 0.034 & 12 \\
M33 HM33B     & $-1.235$ & 0.032 & 0.067 & 33 & $+0.039$ & 0.200 & 0.462 & 2 & $-0.221$ & 0.183 & 0.185 & 5 & $+0.382$ & 0.188 & 0.061 & 3 & $+0.025$ & 0.081 & 0.152 & 10 & $+0.202$ & 0.109 & 0.115 & 11 \\
WLM GC        & $-1.879$ & 0.014 & 0.028 & 37 & $+0.147$ & 0.151 & 0.006 & 2 & $+0.071$ & 0.060 & 0.057 & 5 & $+0.215$ & 0.305 & \ldots & 1 & $+0.258$ & 0.032 & 0.060 & 10 & $+0.284$ & 0.032 & 0.061 & 14 \\
Fornax F3     & $-2.331$ & 0.013 & 0.021 & 39 & $+0.056$ & 0.202 & \ldots & 1 & $+0.014$ & 0.050 & 0.081 & 5 & $+0.689$ & 0.219 & 0.161 & 2 & $+0.187$ & 0.031 & 0.054 & 9 & $+0.166$ & 0.030 & 0.056 & 13 \\
Fornax F4     & $-1.256$ & 0.010 & 0.019 & 39 & $-0.282$ & 0.074 & 0.010 & 2 & $-0.076$ & 0.041 & 0.065 & 5 & $+0.084$ & 0.075 & 0.122 & 4 & $+0.061$ & 0.023 & 0.031 & 10 & $+0.026$ & 0.023 & 0.030 & 13 \\
Fornax F5     & $-2.098$ & 0.016 & 0.027 & 39 & $+0.463$ & 0.138 & \ldots & 1 & $+0.178$ & 0.070 & 0.098 & 5 & $+0.709$ & 0.620 & \ldots & 1 & $+0.211$ & 0.038 & 0.064 & 10 & $+0.178$ & 0.040 & 0.082 & 11 \\
M31 006-058   & $-0.538$ & 0.009 & 0.018 & 37 & $+0.455$ & 0.039 & 0.038 & 2 & $+0.341$ & 0.024 & 0.039 & 6 & $+0.322$ & 0.026 & 0.029 & 6 & $+0.248$ & 0.019 & 0.034 & 9 & $+0.294$ & 0.017 & 0.037 & 13 \\
M31 012-064   & $-1.734$ & 0.016 & 0.021 & 38 & $+0.241$ & 0.101 & 0.081 & 2 & $+0.047$ & 0.067 & 0.067 & 5 & $+0.714$ & 0.082 & 0.050 & 4 & $+0.331$ & 0.038 & 0.090 & 8 & $+0.176$ & 0.039 & 0.077 & 12 \\
M31 019-072   & $-0.705$ & 0.010 & 0.025 & 38 & $+0.458$ & 0.045 & 0.041 & 2 & $+0.211$ & 0.031 & 0.100 & 5 & $+0.225$ & 0.039 & 0.055 & 5 & $+0.268$ & 0.023 & 0.048 & 8 & $+0.360$ & 0.019 & 0.049 & 14 \\
M31 058-119   & $-1.001$ & 0.009 & 0.023 & 37 & $+0.130$ & 0.044 & 0.095 & 2 & $+0.183$ & 0.027 & 0.122 & 6 & $+0.345$ & 0.029 & 0.026 & 6 & $+0.230$ & 0.020 & 0.022 & 9 & $+0.209$ & 0.019 & 0.041 & 13 \\
M31 082-144   & $-0.700$ & 0.016 & 0.037 & 36 & $+0.496$ & 0.079 & 0.228 & 2 & $+0.382$ & 0.064 & 0.141 & 5 & $+0.399$ & 0.068 & 0.069 & 5 & $+0.287$ & 0.037 & 0.023 & 8 & $+0.321$ & 0.047 & 0.078 & 12 \\
M31 163-217   & $-0.145$ & 0.009 & 0.028 & 37 & $+0.646$ & 0.039 & 0.045 & 2 & $+0.223$ & 0.023 & 0.048 & 6 & $+0.287$ & 0.025 & 0.080 & 6 & $+0.110$ & 0.018 & 0.056 & 10 & $+0.267$ & 0.017 & 0.043 & 13 \\
M31 171-222   & $-0.228$ & 0.009 & 0.024 & 37 & $+0.577$ & 0.039 & 0.013 & 2 & $+0.266$ & 0.025 & 0.105 & 6 & $+0.246$ & 0.027 & 0.063 & 6 & $+0.108$ & 0.019 & 0.038 & 9 & $+0.225$ & 0.018 & 0.040 & 13 \\
M31 174-226   & $-1.029$ & 0.013 & 0.024 & 38 & $+0.241$ & 0.079 & 0.106 & 2 & $+0.187$ & 0.050 & 0.078 & 5 & $+0.336$ & 0.073 & 0.042 & 5 & $+0.319$ & 0.033 & 0.034 & 8 & $+0.297$ & 0.030 & 0.050 & 12 \\
M31 225-280   & $-0.351$ & 0.009 & 0.033 & 34 & $+0.543$ & 0.038 & 0.141 & 2 & $+0.266$ & 0.022 & 0.097 & 6 & $+0.357$ & 0.024 & 0.075 & 6 & $+0.151$ & 0.017 & 0.055 & 10 & $+0.381$ & 0.016 & 0.042 & 13 \\
M31 338-076   & $-1.083$ & 0.010 & 0.023 & 38 & $+0.172$ & 0.058 & 0.194 & 2 & $+0.213$ & 0.035 & 0.143 & 5 & $+0.264$ & 0.050 & 0.049 & 5 & $+0.293$ & 0.025 & 0.035 & 8 & $+0.264$ & 0.021 & 0.052 & 14 \\
M31 358-219   & $-2.245$ & 0.013 & 0.023 & 36 & $+0.341$ & 0.086 & 0.120 & 2 & $+0.180$ & 0.040 & 0.075 & 6 & $+0.415$ & 0.081 & 0.105 & 5 & $+0.338$ & 0.026 & 0.027 & 9 & $+0.184$ & 0.030 & 0.053 & 12 \\
M31 EXT8      & $-2.882$ & 0.024 & 0.045 & 24 & \ldots & \ldots & \ldots & \ldots & $-0.271$ & 0.220 & 0.015 & 2 & $+0.620$ & 0.316 & \ldots & 1 & $+0.371$ & 0.050 & 0.067 & 9 & $+0.244$ & 0.082 & 0.084 & 10 \\
N2403 F46     & $-1.732$ & 0.019 & 0.028 & 34 & $-0.054$ & 0.153 & \ldots & 1 & $-0.188$ & 0.090 & 0.064 & 5 & $+0.261$ & 0.113 & 0.171 & 4 & $+0.290$ & 0.040 & 0.069 & 9 & $+0.208$ & 0.062 & 0.085 & 10 \\
\hline
\end{tabular}
}
\end{adjustbox}
\tablefoot{
See notes to Table~\ref{tab:results1} for explanations of the columns.
}
\end{table*}

\begin{table*}
\caption{LTE results for Sc, Cr, Mn, and Ni.}
\label{tab:results2v4lte}
\begin{adjustbox}{width=1\textwidth}
\centering
{\small
\begin{tabular}{l cccccccccccccccc}
\hline\hline
Cluster       & [Sc/Fe] & $\sigma_{\langle\mathrm{Sc}\rangle}$ & $S_\mathrm{Sc}$ & N    & [Cr/Fe] & $\sigma_{\langle\mathrm{Cr}\rangle}$ & $S_\mathrm{Cr}$ & N     & [Mn/Fe] & $\sigma_{\langle\mathrm{Mn}\rangle}$ & $S_\mathrm{Mn}$ & N    & [Ni/Fe] & $\sigma_{\langle\mathrm{Ni}\rangle}$ & $S_\mathrm{Ni}$ & N    \\ \hline
NGC 0104      & $+0.209$ & 0.028 & 0.067 & 5 & $-0.017$ & 0.014 & 0.033 & 17 & $-0.256$ & 0.037 & 0.043 & 2 & $+0.028$ & 0.015 & 0.053 & 14 \\
NGC 0362      & $+0.153$ & 0.031 & 0.050 & 5 & $-0.006$ & 0.015 & 0.036 & 17 & $-0.355$ & 0.039 & 0.110 & 2 & $-0.170$ & 0.017 & 0.053 & 14 \\
NGC 6254      & $+0.261$ & 0.036 & 0.056 & 4 & $-0.102$ & 0.019 & 0.044 & 17 & $-0.495$ & 0.042 & 0.059 & 2 & $-0.107$ & 0.020 & 0.045 & 14 \\
NGC 6388      & $+0.156$ & 0.028 & 0.074 & 5 & $-0.052$ & 0.014 & 0.038 & 17 & $-0.193$ & 0.038 & 0.006 & 2 & $-0.035$ & 0.016 & 0.075 & 14 \\
NGC 6752      & $+0.160$ & 0.034 & 0.060 & 4 & $-0.081$ & 0.017 & 0.025 & 17 & $-0.411$ & 0.040 & 0.001 & 2 & $-0.081$ & 0.017 & 0.037 & 14 \\
NGC 7078      & $+0.168$ & 0.041 & 0.014 & 3 & $-0.218$ & 0.030 & 0.041 & 12 & $-0.352$ & 0.057 & 0.168 & 2 & $-0.090$ & 0.029 & 0.060 & 13 \\
NGC 7099      & $+0.178$ & 0.041 & 0.066 & 4 & $-0.127$ & 0.027 & 0.046 & 13 & $-0.412$ & 0.056 & 0.021 & 2 & $-0.107$ & 0.028 & 0.052 & 14 \\
N147 HII      & $+0.337$ & 0.128 & 0.220 & 2 & $+0.046$ & 0.088 & 0.110 & 12 & $-0.143$ & 0.147 & 0.300 & 2 & $+0.101$ & 0.106 & 0.146 & 11 \\
N147 HIII     & $-0.141$ & 0.199 & 0.097 & 2 & $+0.178$ & 0.116 & 0.231 & 6 & $+0.170$ & 0.125 & 0.294 & 2 & $-0.569$ & 0.118 & 0.282 & 5 \\
N147 PA-1     & $-0.364$ & 0.319 & 0.107 & 2 & $+0.016$ & 0.077 & 0.101 & 8 & $-0.048$ & 0.125 & \ldots & 1 & $+0.025$ & 0.175 & 0.036 & 3 \\
N147 PA-2     & $+0.050$ & 0.083 & 0.133 & 2 & $+0.048$ & 0.055 & 0.092 & 10 & $-0.296$ & 0.125 & 0.092 & 2 & $-0.074$ & 0.072 & 0.081 & 10 \\
N147 SD7      & $-0.153$ & 0.112 & 0.098 & 2 & $+0.008$ & 0.056 & 0.083 & 14 & $-0.515$ & 0.111 & 0.119 & 2 & $+0.006$ & 0.059 & 0.088 & 12 \\
N185 FJJ-III  & $+0.123$ & 0.127 & 0.126 & 2 & $+0.246$ & 0.053 & 0.114 & 12 & $-0.220$ & 0.135 & \ldots & 1 & $+0.139$ & 0.063 & 0.085 & 12 \\
N185 FJJ-V    & $-0.068$ & 0.098 & 0.173 & 2 & $-0.120$ & 0.062 & 0.092 & 11 & $-0.136$ & 0.096 & 0.126 & 2 & $+0.056$ & 0.055 & 0.086 & 12 \\
N185 FJJ-VIII & $+0.203$ & 0.092 & 0.012 & 2 & $-0.028$ & 0.049 & 0.062 & 15 & $-0.203$ & 0.078 & 0.091 & 2 & $-0.029$ & 0.055 & 0.084 & 13 \\
N205 HubbleI  & $+0.295$ & 0.052 & 0.045 & 4 & $-0.021$ & 0.028 & 0.047 & 17 & $-0.329$ & 0.052 & 0.070 & 2 & $-0.146$ & 0.032 & 0.057 & 14 \\
N205 HubbleII & $+0.350$ & 0.057 & 0.148 & 2 & $-0.049$ & 0.024 & 0.042 & 17 & $-0.377$ & 0.051 & 0.039 & 2 & $-0.025$ & 0.027 & 0.066 & 14 \\
N6822 SC6     & $+0.215$ & 0.084 & 0.061 & 2 & $-0.039$ & 0.043 & 0.051 & 15 & $-0.505$ & 0.085 & 0.063 & 2 & $-0.222$ & 0.053 & 0.060 & 13 \\
N6822 SC7     & $-0.273$ & 0.061 & 0.035 & 3 & $-0.109$ & 0.025 & 0.036 & 17 & $-0.413$ & 0.051 & 0.037 & 2 & $-0.318$ & 0.030 & 0.056 & 14 \\
N6822 HVII    & $-0.038$ & 0.072 & 0.164 & 3 & $+0.113$ & 0.032 & 0.085 & 17 & $-0.157$ & 0.059 & 0.215 & 2 & $-0.109$ & 0.043 & 0.096 & 14 \\
M33 H38       & $+0.040$ & 0.107 & 0.104 & 3 & $+0.643$ & 0.158 & 0.161 & 3 & $-0.412$ & 0.113 & 0.258 & 2 & $+0.049$ & 0.073 & 0.098 & 12 \\
M33 M9        & $+0.227$ & 0.087 & 0.049 & 2 & $-0.273$ & 0.126 & 0.393 & 4 & $-0.438$ & 0.106 & 0.085 & 2 & $-0.172$ & 0.063 & 0.056 & 13 \\
M33 R12       & $+0.103$ & 0.063 & 0.098 & 3 & $-0.397$ & 0.111 & 0.148 & 3 & $-0.207$ & 0.061 & 0.158 & 2 & $-0.063$ & 0.036 & 0.066 & 13 \\
M33 U49       & $+0.368$ & 0.095 & 0.026 & 3 & $+0.150$ & 0.189 & 0.507 & 3 & $-0.083$ & 0.101 & 0.151 & 2 & $-0.030$ & 0.078 & 0.115 & 10 \\
M33 R14       & $+0.136$ & 0.061 & 0.036 & 3 & $-0.039$ & 0.030 & 0.050 & 17 & $-0.238$ & 0.057 & 0.025 & 2 & $-0.055$ & 0.035 & 0.070 & 14 \\
M33 U77       & $+0.355$ & 0.105 & 0.128 & 2 & $+0.075$ & 0.056 & 0.089 & 13 & $-0.328$ & 0.113 & 0.055 & 2 & $-0.186$ & 0.068 & 0.060 & 12 \\
M33 CBF28     & $+0.069$ & 0.049 & 0.061 & 4 & $-0.065$ & 0.022 & 0.026 & 18 & $-0.363$ & 0.047 & 0.052 & 2 & $-0.151$ & 0.024 & 0.048 & 14 \\
M33 HM33B     & $-0.210$ & 0.193 & 0.051 & 2 & $+0.227$ & 0.079 & 0.117 & 16 & $-0.238$ & 0.226 & \ldots & 1 & $-0.333$ & 0.105 & 0.087 & 12 \\
WLM GC        & $+0.189$ & 0.073 & 0.089 & 3 & $+0.049$ & 0.046 & 0.071 & 10 & $-0.444$ & 0.078 & 0.103 & 2 & $-0.043$ & 0.044 & 0.099 & 13 \\
Fornax F3     & $+0.156$ & 0.067 & 0.120 & 3 & $-0.168$ & 0.048 & 0.046 & 12 & $-0.484$ & 0.100 & 0.242 & 2 & $-0.229$ & 0.051 & 0.068 & 12 \\
Fornax F4     & $-0.116$ & 0.049 & 0.062 & 3 & $-0.090$ & 0.025 & 0.035 & 15 & $-0.397$ & 0.049 & 0.049 & 2 & $-0.292$ & 0.023 & 0.052 & 15 \\
Fornax F5     & $-0.077$ & 0.112 & 0.226 & 3 & $+0.035$ & 0.057 & 0.087 & 11 & $-0.484$ & 0.114 & \ldots & 1 & $-0.048$ & 0.052 & 0.088 & 13 \\
M31 006-058   & $+0.234$ & 0.032 & 0.077 & 5 & $-0.001$ & 0.017 & 0.031 & 15 & $-0.204$ & 0.039 & 0.073 & 2 & $-0.006$ & 0.017 & 0.060 & 15 \\
M31 012-064   & $+0.370$ & 0.075 & 0.147 & 3 & $-0.045$ & 0.043 & 0.062 & 13 & $-0.516$ & 0.086 & 0.167 & 2 & $-0.128$ & 0.051 & 0.079 & 14 \\
M31 019-072   & $+0.202$ & 0.040 & 0.037 & 5 & $-0.025$ & 0.019 & 0.037 & 16 & $-0.245$ & 0.044 & 0.067 & 2 & $+0.026$ & 0.020 & 0.065 & 15 \\
M31 058-119   & $+0.154$ & 0.037 & 0.090 & 5 & $-0.062$ & 0.021 & 0.048 & 15 & $-0.346$ & 0.042 & 0.099 & 2 & $-0.047$ & 0.019 & 0.062 & 15 \\
M31 082-144   & $+0.339$ & 0.090 & 0.118 & 2 & $-0.057$ & 0.044 & 0.065 & 15 & $-0.405$ & 0.092 & 0.114 & 2 & $+0.242$ & 0.044 & 0.082 & 15 \\
M31 163-217   & $+0.149$ & 0.031 & 0.075 & 5 & $-0.044$ & 0.017 & 0.039 & 15 & $-0.039$ & 0.040 & 0.027 & 2 & $+0.121$ & 0.016 & 0.074 & 15 \\
M31 171-222   & $+0.230$ & 0.032 & 0.082 & 5 & $-0.018$ & 0.017 & 0.041 & 15 & $-0.029$ & 0.041 & 0.048 & 2 & $+0.103$ & 0.017 & 0.068 & 15 \\
M31 174-226   & $+0.251$ & 0.068 & 0.056 & 3 & $-0.017$ & 0.031 & 0.042 & 16 & $-0.365$ & 0.063 & 0.153 & 2 & $-0.014$ & 0.037 & 0.077 & 14 \\
M31 225-280   & $+0.173$ & 0.030 & 0.126 & 5 & $+0.005$ & 0.016 & 0.053 & 15 & $-0.134$ & 0.039 & 0.005 & 2 & $+0.111$ & 0.016 & 0.085 & 15 \\
M31 338-076   & $+0.095$ & 0.046 & 0.095 & 5 & $-0.057$ & 0.022 & 0.037 & 16 & $-0.383$ & 0.047 & 0.046 & 2 & $-0.081$ & 0.023 & 0.058 & 15 \\
M31 358-219   & $+0.017$ & 0.060 & 0.021 & 3 & $-0.116$ & 0.049 & 0.053 & 9 & $-0.446$ & 0.084 & 0.057 & 2 & $-0.155$ & 0.042 & 0.070 & 12 \\
M31 EXT8      & $+0.515$ & 0.118 & 0.148 & 2 & $-0.160$ & 0.163 & 0.149 & 5 & $+0.890$ & 0.167 & \ldots & 1 & $+0.277$ & 0.089 & 0.206 & 8 \\
N2403 F46     & $-0.038$ & 0.103 & 0.133 & 2 & $-0.015$ & 0.057 & 0.084 & 13 & $-0.489$ & 0.110 & 0.066 & 2 & $+0.083$ & 0.057 & 0.111 & 12 \\
\hline
\end{tabular}
}
\end{adjustbox}
\tablefoot{
See notes to Table~\ref{tab:results1} for explanations of the columns.
}
\end{table*}

\begin{table*}
\caption{LTE results for Cu, Zn, Zr, Ba, and Eu.}
\label{tab:results3v4lte}
\begin{adjustbox}{width=1\textwidth}
\centering
{\small
\begin{tabular}{l cccccccccccccccccccc}
\hline\hline
Cluster       & [Cu/Fe] & $\sigma_{\langle\mathrm{Cu}\rangle}$ & $S_\mathrm{Cu}$ & N    & [Zn/Fe] & $\sigma_{\langle\mathrm{Zn}\rangle}$ & $S_\mathrm{Zn}$ & N    & [Zr/Fe] & $\sigma_{\langle\mathrm{Zr}\rangle}$ & $S_\mathrm{Zr}$ & N     & [Ba/Fe] & $\sigma_{\langle\mathrm{Ba}\rangle}$ & $S_\mathrm{Ba}$ & N    & [Eu/Fe] & $\sigma_{\langle\mathrm{Eu}\rangle}$ & $S_\mathrm{Eu}$ & N    \\ \hline
NGC 0104      & $-0.036$ & 0.060 & \ldots & 1 & $+0.126$ & 0.044 & 0.046 & 2 & $+0.237$ & 0.064 & \ldots & 1 & $+0.205$ & 0.025 & 0.082 & 5 & $+0.238$ & 0.050 & 0.006 & 2 \\
NGC 0362      & $-0.311$ & 0.067 & \ldots & 1 & $-0.079$ & 0.048 & 0.066 & 2 & $+0.415$ & 0.076 & \ldots & 1 & $+0.395$ & 0.025 & 0.069 & 5 & $+0.644$ & 0.047 & 0.021 & 2 \\
NGC 6254      & $-0.565$ & 0.075 & \ldots & 1 & $+0.018$ & 0.056 & 0.008 & 2 & $-0.107$ & 0.150 & \ldots & 1 & $+0.437$ & 0.028 & 0.072 & 5 & $+0.211$ & 0.090 & \ldots & 1 \\
NGC 6388      & $-0.090$ & 0.060 & \ldots & 1 & $-0.148$ & 0.051 & 0.076 & 2 & $+0.306$ & 0.063 & \ldots & 1 & $+0.177$ & 0.024 & 0.079 & 5 & $-0.043$ & 0.071 & \ldots & 1 \\
NGC 6752      & $-0.411$ & 0.065 & \ldots & 1 & $+0.128$ & 0.045 & 0.029 & 2 & $+0.325$ & 0.134 & \ldots & 1 & $+0.282$ & 0.027 & 0.094 & 5 & $+0.420$ & 0.054 & 0.093 & 2 \\
NGC 7078      & \ldots & \ldots & \ldots & \ldots & $-0.139$ & 0.102 & 0.060 & 2 & \ldots & \ldots & \ldots & \ldots & $+0.420$ & 0.030 & 0.057 & 5 & $+0.580$ & 0.109 & 0.082 & 2 \\
NGC 7099      & $-0.388$ & 0.114 & \ldots & 1 & $-0.031$ & 0.077 & 0.092 & 2 & \ldots & \ldots & \ldots & \ldots & $+0.084$ & 0.032 & 0.159 & 5 & $+0.381$ & 0.110 & \ldots & 1 \\
N147 HII      & $-0.407$ & 0.362 & \ldots & 1 & $+0.673$ & 0.731 & \ldots & 1 & $+0.470$ & 0.457 & \ldots & 1 & $-0.125$ & 0.177 & 0.287 & 3 & \ldots & \ldots & \ldots & \ldots \\
N147 HIII     & $+0.434$ & 0.228 & \ldots & 1 & $+0.227$ & 0.392 & \ldots & 1 & \ldots & \ldots & \ldots & \ldots & $-0.501$ & 0.107 & 0.092 & 5 & \ldots & \ldots & \ldots & \ldots \\
N147 PA-1     & $-0.162$ & 0.293 & \ldots & 1 & $+0.067$ & 0.405 & \ldots & 1 & \ldots & \ldots & \ldots & \ldots & $+0.215$ & 0.086 & 0.107 & 5 & \ldots & \ldots & \ldots & \ldots \\
N147 PA-2     & $-0.811$ & 0.349 & \ldots & 1 & $+0.336$ & 0.205 & 0.322 & 3 & \ldots & \ldots & \ldots & \ldots & $+0.358$ & 0.062 & 0.033 & 5 & \ldots & \ldots & \ldots & \ldots \\
N147 SD7      & $-0.473$ & 0.312 & \ldots & 1 & $+0.069$ & 0.200 & 0.329 & 2 & \ldots & \ldots & \ldots & \ldots & $+0.323$ & 0.057 & 0.157 & 5 & \ldots & \ldots & \ldots & \ldots \\
N185 FJJ-III  & $-0.766$ & 0.243 & \ldots & 1 & $+0.167$ & 0.263 & 0.179 & 2 & \ldots & \ldots & \ldots & \ldots & $+0.381$ & 0.075 & 0.100 & 5 & $+0.421$ & 0.564 & \ldots & 1 \\
N185 FJJ-V    & $-0.684$ & 0.317 & \ldots & 1 & $+0.002$ & 0.255 & \ldots & 1 & \ldots & \ldots & \ldots & \ldots & $+0.395$ & 0.069 & 0.160 & 5 & \ldots & \ldots & \ldots & \ldots \\
N185 FJJ-VIII & $-0.694$ & 0.282 & \ldots & 1 & $+0.520$ & 0.153 & 0.037 & 2 & \ldots & \ldots & \ldots & \ldots & $+0.361$ & 0.059 & 0.180 & 5 & $+0.279$ & 0.420 & \ldots & 1 \\
N205 HubbleI  & $-0.455$ & 0.135 & \ldots & 1 & $-0.182$ & 0.117 & 0.104 & 3 & $+0.565$ & 0.147 & \ldots & 1 & $+0.315$ & 0.041 & 0.074 & 5 & $+0.758$ & 0.187 & \ldots & 1 \\
N205 HubbleII & $-0.514$ & 0.093 & \ldots & 1 & $-0.036$ & 0.078 & 0.182 & 3 & $-0.275$ & 0.242 & \ldots & 1 & $+0.208$ & 0.036 & 0.096 & 5 & $+0.128$ & 0.220 & \ldots & 1 \\
N6822 SC6     & $-0.585$ & 0.178 & \ldots & 1 & $+0.155$ & 0.157 & 0.143 & 2 & \ldots & \ldots & \ldots & \ldots & $+0.322$ & 0.054 & 0.047 & 5 & $+0.581$ & 0.247 & \ldots & 1 \\
N6822 SC7     & \ldots & \ldots & \ldots & \ldots & $-0.386$ & 0.098 & 0.059 & 3 & $-0.452$ & 0.333 & \ldots & 1 & $+0.213$ & 0.035 & 0.052 & 5 & $+0.163$ & 0.152 & \ldots & 1 \\
N6822 HVII    & $-0.286$ & 0.166 & \ldots & 1 & $-0.333$ & 0.173 & 0.104 & 2 & \ldots & \ldots & \ldots & \ldots & $+0.482$ & 0.041 & 0.102 & 5 & $+0.004$ & 0.328 & \ldots & 1 \\
M33 H38       & $-0.019$ & 0.335 & \ldots & 1 & $-0.555$ & 0.429 & \ldots & 1 & \ldots & \ldots & \ldots & \ldots & $+0.518$ & 0.062 & 0.098 & 4 & \ldots & \ldots & \ldots & \ldots \\
M33 M9        & $-0.210$ & 0.180 & \ldots & 1 & $+0.334$ & 0.220 & 0.069 & 2 & \ldots & \ldots & \ldots & \ldots & $+0.609$ & 0.065 & 0.081 & 4 & \ldots & \ldots & \ldots & \ldots \\
M33 R12       & $-0.427$ & 0.129 & \ldots & 1 & $+0.075$ & 0.143 & 0.115 & 3 & \ldots & \ldots & \ldots & \ldots & $+0.324$ & 0.055 & 0.095 & 4 & $+0.497$ & 0.205 & \ldots & 1 \\
M33 U49       & $-0.582$ & 0.298 & \ldots & 1 & $+0.602$ & 0.304 & 0.081 & 2 & \ldots & \ldots & \ldots & \ldots & $+0.552$ & 0.091 & 0.246 & 4 & \ldots & \ldots & \ldots & \ldots \\
M33 R14       & $-0.548$ & 0.132 & \ldots & 1 & $-0.118$ & 0.136 & 0.196 & 3 & $+0.214$ & 0.177 & \ldots & 1 & $+0.345$ & 0.042 & 0.103 & 5 & \ldots & \ldots & \ldots & \ldots \\
M33 U77       & $-0.572$ & 0.250 & \ldots & 1 & $+0.243$ & 0.217 & 0.327 & 3 & \ldots & \ldots & \ldots & \ldots & $+0.438$ & 0.075 & 0.083 & 5 & \ldots & \ldots & \ldots & \ldots \\
M33 CBF28     & $-0.328$ & 0.088 & \ldots & 1 & $-0.158$ & 0.074 & 0.118 & 3 & $+0.624$ & 0.103 & \ldots & 1 & $+0.268$ & 0.033 & 0.035 & 5 & $+0.676$ & 0.106 & \ldots & 1 \\
M33 HM33B     & $+0.005$ & 0.279 & \ldots & 1 & $-0.602$ & 1.038 & \ldots & 1 & $+0.901$ & 0.482 & \ldots & 1 & $+0.764$ & 0.149 & 0.034 & 2 & $+0.677$ & 0.499 & \ldots & 1 \\
WLM GC        & $-1.083$ & 0.525 & \ldots & 1 & $+0.085$ & 0.134 & 0.227 & 2 & \ldots & \ldots & \ldots & \ldots & $-0.011$ & 0.064 & 0.067 & 4 & $+0.110$ & 0.278 & \ldots & 1 \\
Fornax F3     & \ldots & \ldots & \ldots & \ldots & $-0.043$ & 0.161 & \ldots & 1 & \ldots & \ldots & \ldots & \ldots & $+0.494$ & 0.044 & 0.061 & 4 & \ldots & \ldots & \ldots & \ldots \\
Fornax F4     & $-0.867$ & 0.106 & \ldots & 1 & $-0.269$ & 0.088 & 0.091 & 2 & $+0.073$ & 0.279 & \ldots & 1 & $+0.080$ & 0.042 & 0.015 & 4 & $+0.212$ & 0.120 & \ldots & 1 \\
Fornax F5     & $-0.503$ & 0.294 & \ldots & 1 & $+0.013$ & 0.175 & 0.019 & 2 & \ldots & \ldots & \ldots & \ldots & $-0.088$ & 0.072 & 0.049 & 4 & $+0.354$ & 0.322 & \ldots & 1 \\
M31 006-058   & $-0.056$ & 0.070 & \ldots & 1 & $+0.014$ & 0.053 & 0.104 & 3 & $+0.388$ & 0.070 & \ldots & 1 & $+0.126$ & 0.027 & 0.111 & 5 & $-0.013$ & 0.092 & \ldots & 1 \\
M31 012-064   & $-0.705$ & 0.236 & \ldots & 1 & $-0.230$ & 0.218 & 0.079 & 2 & \ldots & \ldots & \ldots & \ldots & $+0.390$ & 0.060 & 0.093 & 4 & \ldots & \ldots & \ldots & \ldots \\
M31 019-072   & $-0.265$ & 0.085 & \ldots & 1 & $+0.004$ & 0.068 & 0.066 & 3 & $+0.314$ & 0.110 & \ldots & 1 & $+0.036$ & 0.037 & 0.079 & 4 & \ldots & \ldots & \ldots & \ldots \\
M31 058-119   & $-0.391$ & 0.080 & \ldots & 1 & $-0.251$ & 0.062 & 0.117 & 3 & $+0.001$ & 0.109 & \ldots & 1 & $+0.397$ & 0.029 & 0.053 & 4 & $+0.356$ & 0.078 & 0.061 & 2 \\
M31 082-144   & $-0.307$ & 0.193 & \ldots & 1 & $+0.341$ & 0.269 & 0.308 & 3 & $+0.440$ & 0.230 & \ldots & 1 & $+0.231$ & 0.085 & 0.240 & 4 & \ldots & \ldots & \ldots & \ldots \\
M31 163-217   & $+0.140$ & 0.073 & \ldots & 1 & $-0.034$ & 0.055 & 0.104 & 3 & $+0.073$ & 0.075 & \ldots & 1 & $-0.064$ & 0.029 & 0.131 & 5 & $-0.114$ & 0.092 & \ldots & 1 \\
M31 171-222   & $+0.107$ & 0.078 & \ldots & 1 & $-0.052$ & 0.060 & 0.142 & 3 & $+0.254$ & 0.075 & \ldots & 1 & $+0.102$ & 0.028 & 0.157 & 5 & $-0.205$ & 0.112 & \ldots & 1 \\
M31 174-226   & $-0.332$ & 0.166 & \ldots & 1 & $-0.250$ & 0.149 & 0.160 & 3 & $+0.505$ & 0.196 & \ldots & 1 & $+0.436$ & 0.048 & 0.058 & 4 & \ldots & \ldots & \ldots & \ldots \\
M31 225-280   & $-0.248$ & 0.064 & \ldots & 1 & $-0.171$ & 0.049 & 0.038 & 3 & $+0.107$ & 0.077 & \ldots & 1 & $+0.176$ & 0.025 & 0.143 & 5 & $-0.020$ & 0.072 & 0.048 & 2 \\
M31 338-076   & $-0.377$ & 0.086 & \ldots & 1 & $+0.082$ & 0.070 & 0.104 & 3 & $+0.100$ & 0.164 & \ldots & 1 & $+0.432$ & 0.034 & 0.077 & 4 & \ldots & \ldots & \ldots & \ldots \\
M31 358-219   & $-0.770$ & 0.387 & \ldots & 1 & $+0.065$ & 0.121 & 0.028 & 2 & \ldots & \ldots & \ldots & \ldots & $+0.152$ & 0.041 & 0.059 & 5 & $+0.268$ & 0.304 & \ldots & 1 \\
M31 EXT8      & $+0.141$ & 0.475 & \ldots & 1 & \ldots & \ldots & \ldots & \ldots & \ldots & \ldots & \ldots & \ldots & $+0.517$ & 0.069 & 0.046 & 4 & \ldots & \ldots & \ldots & \ldots \\
N2403 F46     & \ldots & \ldots & \ldots & \ldots & $+0.211$ & 0.222 & 0.282 & 2 & \ldots & \ldots & \ldots & \ldots & $+0.283$ & 0.066 & 0.148 & 5 & $+0.448$ & 0.340 & \ldots & 1 \\
\hline
\end{tabular}
}
\end{adjustbox}
\tablefoot{
See notes to Table~\ref{tab:results1} for explanations of the columns.
}
\end{table*}

\clearpage

\section{NLTE abundance measurements (BaSTI isochrones)}
\label{app:BaSTIfits}

Tables~\ref{tab:results1v5}-\ref{tab:results3v5} list the average abundance measurements based on BaSTI isochrones. The measurements in these Tables include \ac{nlte} corrections on the abundance measurements when available.

\begin{table*}[!h]
\caption{Results for Fe, Na, Mg, Si, Ca, and Ti.}
\label{tab:results1v5}
\begin{adjustbox}{width=1\textwidth}
\centering
{\small
\begin{tabular}{l cccccccccccccccccccccccc}
\hline\hline
Cluster       & [Fe/H] & $\sigma_{\langle\mathrm{Fe}\rangle}$ & $S_\mathrm{Fe}$ & N      & [Na/Fe]  & $\sigma_{\langle\mathrm{Na}\rangle}$ & $S_\mathrm{Na}$ & N    & [Mg/Fe]  & $\sigma_{\langle\mathrm{Mg}\rangle}$ & $S_\mathrm{Mg}$ & N    & [Si/Fe]  & $\sigma_{\langle\mathrm{Si}\rangle}$ & $S_\mathrm{Si}$ & N    & [Ca/Fe]  & $\sigma_{\langle\mathrm{Ca}\rangle}$ & $S_\mathrm{Ca}$ & N    & [Ti/Fe]  & $\sigma_{\langle\mathrm{Ti}\rangle}$ & $S_\mathrm{Ti}$ & N    \\ \hline
NGC 0104      & $-0.724$ & 0.008 & 0.019 & 39 & $+0.237$ & 0.038 & 0.075 & 2 & $+0.392$ & 0.027 & 0.054 & 4 & $+0.368$ & 0.023 & 0.042 & 6 & $+0.239$ & 0.018 & 0.033 & 9 & $+0.421$ & 0.015 & 0.029 & 14 \\
NGC 0362      & $-1.097$ & 0.009 & 0.019 & 39 & $-0.194$ & 0.043 & 0.004 & 2 & $+0.186$ & 0.030 & 0.019 & 4 & $+0.138$ & 0.026 & 0.052 & 6 & $+0.165$ & 0.019 & 0.032 & 9 & $+0.426$ & 0.016 & 0.033 & 14 \\
NGC 6254      & $-1.477$ & 0.009 & 0.018 & 39 & $-0.186$ & 0.048 & 0.011 & 2 & $+0.284$ & 0.031 & 0.037 & 4 & $+0.250$ & 0.032 & 0.052 & 6 & $+0.270$ & 0.020 & 0.047 & 9 & $+0.421$ & 0.017 & 0.029 & 14 \\
NGC 6388      & $-0.573$ & 0.008 & 0.026 & 39 & $+0.207$ & 0.038 & 0.090 & 2 & $+0.085$ & 0.028 & 0.072 & 4 & $+0.234$ & 0.024 & 0.067 & 6 & $+0.042$ & 0.018 & 0.052 & 9 & $+0.270$ & 0.016 & 0.032 & 14 \\
NGC 6752      & $-1.661$ & 0.009 & 0.013 & 39 & $+0.052$ & 0.042 & 0.036 & 2 & $+0.293$ & 0.030 & 0.050 & 4 & $+0.383$ & 0.028 & 0.016 & 6 & $+0.319$ & 0.018 & 0.018 & 9 & $+0.359$ & 0.016 & 0.023 & 14 \\
NGC 7078      & $-2.232$ & 0.010 & 0.021 & 38 & $-0.106$ & 0.093 & \ldots & 1 & $+0.141$ & 0.035 & 0.036 & 4 & $+0.432$ & 0.065 & 0.105 & 4 & $+0.249$ & 0.022 & 0.029 & 9 & $+0.427$ & 0.021 & 0.046 & 12 \\
NGC 7099      & $-2.226$ & 0.010 & 0.020 & 39 & $-0.034$ & 0.082 & 0.009 & 2 & $+0.278$ & 0.035 & 0.046 & 4 & $+0.449$ & 0.054 & 0.098 & 4 & $+0.259$ & 0.022 & 0.025 & 9 & $+0.359$ & 0.020 & 0.021 & 14 \\
N147 HII      & $-1.437$ & 0.028 & 0.059 & 37 & $+0.252$ & 0.116 & 0.057 & 2 & $+0.199$ & 0.163 & 0.131 & 4 & $+0.647$ & 0.111 & 0.066 & 2 & $+0.259$ & 0.060 & 0.052 & 8 & $+0.665$ & 0.119 & 0.064 & 7 \\
N147 HIII     & $-2.309$ & 0.024 & 0.041 & 34 & $+0.378$ & 0.145 & \ldots & 1 & $+0.054$ & 0.122 & 0.130 & 3 & $+0.103$ & 0.275 & \ldots & 1 & $+0.324$ & 0.050 & 0.077 & 7 & $+0.359$ & 0.086 & 0.069 & 8 \\
N147 PA-1     & $-2.158$ & 0.023 & 0.043 & 34 & $+0.242$ & 0.225 & \ldots & 1 & $+0.193$ & 0.112 & 0.101 & 4 & $+0.560$ & 0.353 & 0.188 & 2 & $+0.147$ & 0.052 & 0.036 & 8 & $+0.357$ & 0.091 & 0.078 & 8 \\
N147 PA-2     & $-1.855$ & 0.017 & 0.028 & 38 & $+0.089$ & 0.127 & 0.021 & 2 & $+0.314$ & 0.082 & 0.055 & 4 & $+0.110$ & 0.127 & 0.232 & 2 & $+0.327$ & 0.036 & 0.041 & 9 & $+0.405$ & 0.052 & 0.058 & 11 \\
N147 SD7      & $-1.824$ & 0.017 & 0.037 & 37 & $-0.213$ & 0.179 & 0.311 & 2 & $+0.307$ & 0.091 & 0.155 & 3 & $+0.508$ & 0.066 & 0.107 & 4 & $+0.197$ & 0.035 & 0.062 & 9 & $+0.434$ & 0.054 & 0.059 & 10 \\
N185 FJJ-III  & $-1.697$ & 0.020 & 0.032 & 36 & $+0.309$ & 0.114 & \ldots & 1 & $+0.451$ & 0.076 & 0.131 & 5 & $+0.379$ & 0.100 & 0.145 & 6 & $+0.336$ & 0.045 & 0.067 & 9 & $+0.429$ & 0.059 & 0.084 & 11 \\
N185 FJJ-V    & $-1.723$ & 0.017 & 0.033 & 36 & $-0.142$ & 0.162 & \ldots & 1 & $+0.154$ & 0.070 & 0.127 & 5 & $+0.299$ & 0.091 & 0.160 & 5 & $+0.283$ & 0.035 & 0.044 & 9 & $+0.400$ & 0.053 & 0.074 & 11 \\
N185 FJJ-VIII & $-1.694$ & 0.017 & 0.026 & 37 & $-0.002$ & 0.123 & 0.009 & 2 & $-0.075$ & 0.075 & 0.092 & 5 & $+0.253$ & 0.088 & 0.166 & 4 & $+0.311$ & 0.036 & 0.042 & 9 & $+0.378$ & 0.056 & 0.073 & 11 \\
N205 HubbleI  & $-1.420$ & 0.012 & 0.024 & 37 & $-0.091$ & 0.072 & 0.073 & 2 & $+0.419$ & 0.040 & 0.060 & 5 & $+0.286$ & 0.048 & 0.026 & 6 & $+0.259$ & 0.026 & 0.032 & 9 & $+0.386$ & 0.028 & 0.053 & 12 \\
N205 HubbleII & $-1.319$ & 0.011 & 0.022 & 37 & $+0.008$ & 0.063 & 0.010 & 2 & $+0.329$ & 0.040 & 0.056 & 5 & $+0.287$ & 0.046 & 0.085 & 6 & $+0.233$ & 0.025 & 0.047 & 9 & $+0.339$ & 0.027 & 0.030 & 12 \\
N6822 SC6     & $-1.618$ & 0.015 & 0.025 & 37 & $+0.004$ & 0.105 & 0.032 & 2 & $+0.228$ & 0.058 & 0.060 & 5 & $-0.007$ & 0.111 & 0.150 & 2 & $+0.199$ & 0.034 & 0.045 & 9 & $+0.276$ & 0.046 & 0.067 & 12 \\
N6822 SC7     & $-1.139$ & 0.011 & 0.019 & 36 & $-0.651$ & 0.085 & 0.043 & 2 & $-0.227$ & 0.050 & 0.023 & 4 & $-0.016$ & 0.045 & 0.087 & 6 & $-0.037$ & 0.025 & 0.034 & 9 & $-0.004$ & 0.031 & 0.069 & 12 \\
N6822 HVII    & $-1.614$ & 0.014 & 0.035 & 36 & $+0.030$ & 0.080 & 0.015 & 2 & $-0.041$ & 0.059 & 0.067 & 5 & $+0.259$ & 0.050 & 0.118 & 6 & $+0.068$ & 0.030 & 0.040 & 9 & $+0.263$ & 0.038 & 0.075 & 12 \\
M33 H38       & $-1.126$ & 0.023 & 0.051 & 30 & $-0.207$ & 0.116 & 0.051 & 2 & $+0.109$ & 0.133 & 0.288 & 2 & \ldots & \ldots & \ldots & \ldots & $+0.346$ & 0.077 & 0.153 & 5 & $+0.420$ & 0.057 & 0.083 & 12 \\
M33 M9        & $-1.604$ & 0.019 & 0.028 & 30 & $-0.023$ & 0.123 & 0.046 & 2 & $-0.116$ & 0.125 & 0.003 & 2 & $+0.465$ & 0.254 & 0.006 & 2 & $+0.204$ & 0.050 & 0.066 & 7 & $+0.471$ & 0.045 & 0.058 & 13 \\
M33 R12       & $-0.826$ & 0.014 & 0.023 & 30 & $-0.167$ & 0.064 & 0.082 & 2 & $+0.157$ & 0.081 & 0.090 & 2 & $+0.154$ & 0.101 & 0.138 & 3 & $+0.219$ & 0.038 & 0.052 & 7 & $+0.348$ & 0.031 & 0.043 & 14 \\
M33 U49       & $-1.324$ & 0.025 & 0.037 & 30 & $-0.097$ & 0.141 & 0.069 & 2 & $+0.318$ & 0.185 & 0.050 & 2 & $+0.433$ & 0.205 & 0.056 & 2 & $+0.160$ & 0.070 & 0.134 & 6 & $+0.538$ & 0.058 & 0.064 & 14 \\
M33 R14       & $-1.052$ & 0.013 & 0.021 & 37 & $+0.017$ & 0.063 & 0.000 & 2 & $+0.184$ & 0.046 & 0.035 & 5 & $+0.342$ & 0.043 & 0.045 & 6 & $+0.235$ & 0.029 & 0.041 & 9 & $+0.238$ & 0.036 & 0.045 & 12 \\
M33 U77       & $-1.741$ & 0.020 & 0.038 & 35 & $-0.286$ & 0.235 & \ldots & 1 & $+0.115$ & 0.093 & 0.174 & 4 & $+0.287$ & 0.128 & 0.055 & 2 & $+0.393$ & 0.043 & 0.072 & 8 & $+0.434$ & 0.061 & 0.047 & 12 \\
M33 CBF28     & $-1.145$ & 0.011 & 0.019 & 35 & $+0.065$ & 0.051 & 0.135 & 2 & $+0.224$ & 0.034 & 0.072 & 5 & $+0.166$ & 0.037 & 0.045 & 6 & $+0.221$ & 0.023 & 0.017 & 9 & $+0.325$ & 0.024 & 0.038 & 12 \\
M33 HM33B     & $-1.202$ & 0.032 & 0.068 & 33 & $-0.084$ & 0.200 & 0.482 & 2 & $-0.258$ & 0.201 & 0.237 & 4 & $+0.326$ & 0.189 & 0.068 & 3 & $+0.018$ & 0.084 & 0.167 & 9 & $+0.325$ & 0.109 & 0.135 & 11 \\
WLM GC        & $-1.874$ & 0.014 & 0.028 & 37 & $+0.025$ & 0.154 & 0.006 & 2 & $+0.043$ & 0.063 & 0.077 & 4 & $+0.176$ & 0.302 & \ldots & 1 & $+0.260$ & 0.034 & 0.054 & 9 & $+0.392$ & 0.032 & 0.055 & 14 \\
Fornax 3      & $-2.219$ & 0.013 & 0.023 & 39 & $-0.121$ & 0.203 & \ldots & 1 & $-0.085$ & 0.057 & 0.097 & 4 & $+0.622$ & 0.218 & 0.192 & 2 & $+0.137$ & 0.034 & 0.059 & 8 & $+0.281$ & 0.030 & 0.042 & 13 \\
Fornax 4      & $-1.249$ & 0.010 & 0.019 & 39 & $-0.396$ & 0.073 & 0.012 & 2 & $-0.069$ & 0.045 & 0.045 & 4 & $+0.064$ & 0.075 & 0.120 & 4 & $+0.039$ & 0.025 & 0.027 & 9 & $+0.162$ & 0.023 & 0.025 & 13 \\
Fornax 5      & $-2.006$ & 0.016 & 0.027 & 39 & $+0.304$ & 0.137 & \ldots & 1 & $+0.025$ & 0.078 & 0.090 & 4 & $+0.632$ & 0.626 & \ldots & 1 & $+0.203$ & 0.041 & 0.055 & 9 & $+0.325$ & 0.040 & 0.078 & 11 \\
M31 006-058   & $-0.515$ & 0.009 & 0.019 & 37 & $+0.283$ & 0.040 & 0.050 & 2 & $+0.306$ & 0.025 & 0.036 & 5 & $+0.320$ & 0.026 & 0.029 & 6 & $+0.210$ & 0.020 & 0.030 & 8 & $+0.369$ & 0.017 & 0.037 & 13 \\
M31 012-064   & $-1.651$ & 0.016 & 0.021 & 38 & $+0.079$ & 0.102 & 0.089 & 2 & $-0.010$ & 0.077 & 0.099 & 4 & $+0.656$ & 0.081 & 0.051 & 4 & $+0.280$ & 0.038 & 0.089 & 8 & $+0.290$ & 0.039 & 0.063 & 12 \\
M31 019-072   & $-0.668$ & 0.010 & 0.026 & 38 & $+0.295$ & 0.045 & 0.046 & 2 & $+0.246$ & 0.034 & 0.088 & 4 & $+0.201$ & 0.038 & 0.056 & 5 & $+0.219$ & 0.023 & 0.045 & 8 & $+0.446$ & 0.019 & 0.049 & 14 \\
M31 058-119   & $-0.986$ & 0.010 & 0.022 & 37 & $-0.011$ & 0.045 & 0.076 & 2 & $+0.207$ & 0.028 & 0.125 & 5 & $+0.299$ & 0.029 & 0.031 & 6 & $+0.194$ & 0.021 & 0.024 & 8 & $+0.330$ & 0.019 & 0.036 & 13 \\
M31 082-114   & $-0.694$ & 0.017 & 0.029 & 35 & $+0.378$ & 0.080 & 0.243 & 2 & $+0.418$ & 0.064 & 0.139 & 4 & $+0.406$ & 0.068 & 0.063 & 5 & $+0.288$ & 0.037 & 0.033 & 8 & $+0.433$ & 0.049 & 0.083 & 11 \\
M31 163-217   & $-0.203$ & 0.009 & 0.029 & 37 & $+0.498$ & 0.038 & 0.088 & 2 & $+0.191$ & 0.025 & 0.068 & 5 & $+0.337$ & 0.025 & 0.072 & 6 & $+0.064$ & 0.019 & 0.062 & 9 & $+0.309$ & 0.017 & 0.043 & 13 \\
M31 171-222   & $-0.282$ & 0.009 & 0.024 & 37 & $+0.453$ & 0.039 & 0.043 & 2 & $+0.290$ & 0.026 & 0.111 & 5 & $+0.290$ & 0.027 & 0.060 & 6 & $+0.096$ & 0.020 & 0.037 & 8 & $+0.282$ & 0.018 & 0.039 & 13 \\
M31 174-226   & $-1.014$ & 0.013 & 0.024 & 38 & $+0.090$ & 0.080 & 0.093 & 2 & $+0.234$ & 0.054 & 0.031 & 4 & $+0.294$ & 0.073 & 0.043 & 5 & $+0.280$ & 0.033 & 0.034 & 8 & $+0.409$ & 0.030 & 0.049 & 12 \\
M31 225-280   & $-0.389$ & 0.009 & 0.026 & 34 & $+0.412$ & 0.038 & 0.147 & 2 & $+0.234$ & 0.024 & 0.122 & 5 & $+0.369$ & 0.024 & 0.071 & 6 & $+0.120$ & 0.018 & 0.066 & 9 & $+0.428$ & 0.016 & 0.044 & 13 \\
M31 338-076   & $-1.058$ & 0.010 & 0.023 & 38 & $+0.020$ & 0.059 & 0.181 & 2 & $+0.255$ & 0.038 & 0.144 & 4 & $+0.199$ & 0.051 & 0.047 & 5 & $+0.251$ & 0.025 & 0.036 & 8 & $+0.374$ & 0.021 & 0.043 & 14 \\
M31 358-219   & $-2.123$ & 0.013 & 0.024 & 36 & $+0.159$ & 0.087 & 0.121 & 2 & $+0.147$ & 0.044 & 0.078 & 5 & $+0.330$ & 0.081 & 0.116 & 5 & $+0.287$ & 0.027 & 0.032 & 8 & $+0.302$ & 0.030 & 0.042 & 12 \\
M31 EXT8      & $-2.808$ & 0.024 & 0.043 & 28 & \ldots & \ldots & \ldots & \ldots & $-0.344$ & 0.220 & 0.015 & 2 & $+0.547$ & 0.316 & \ldots & 1 & $+0.269$ & 0.055 & 0.075 & 8 & $+0.335$ & 0.082 & 0.088 & 10 \\
N2403 F46     & $-1.634$ & 0.019 & 0.027 & 34 & $-0.205$ & 0.155 & \ldots & 1 & $-0.249$ & 0.098 & 0.058 & 4 & $+0.208$ & 0.111 & 0.177 & 4 & $+0.208$ & 0.041 & 0.062 & 8 & $+0.293$ & 0.062 & 0.088 & 10 \\
\hline
\end{tabular}
}
\end{adjustbox}
\tablefoot{
The listed abundances include NLTE corrections for Fe, Na, Mg, Ca, and Ti. See notes to Table~\ref{tab:results1} for explanations of the columns.
}
\end{table*}

\begin{table*}
\caption{Results for Sc, Cr, Mn, and Ni.}
\label{tab:results2v5}
\begin{adjustbox}{width=1\textwidth}
\centering
{\small
\begin{tabular}{l cccccccccccccccc}
\hline\hline
Cluster       & [Sc/Fe] & $\sigma_{\langle\mathrm{Sc}\rangle}$ & $S_\mathrm{Sc}$ & N    & [Cr/Fe] & $\sigma_{\langle\mathrm{Cr}\rangle}$ & $S_\mathrm{Cr}$ & N     & [Mn/Fe] & $\sigma_{\langle\mathrm{Mn}\rangle}$ & $S_\mathrm{Mn}$ & N    & [Ni/Fe] & $\sigma_{\langle\mathrm{Ni}\rangle}$ & $S_\mathrm{Ni}$ & N    \\ \hline
NGC 0104      & $+0.207$ & 0.028 & 0.085 & 5 & $-0.036$ & 0.014 & 0.035 & 17 & $-0.205$ & 0.037 & 0.028 & 2 & $+0.068$ & 0.015 & 0.053 & 14 \\
NGC 0362      & $+0.135$ & 0.031 & 0.044 & 5 & $-0.018$ & 0.015 & 0.036 & 17 & $-0.265$ & 0.039 & 0.111 & 2 & $-0.100$ & 0.017 & 0.050 & 14 \\
NGC 6254      & $+0.194$ & 0.035 & 0.038 & 4 & $-0.096$ & 0.019 & 0.042 & 17 & $-0.305$ & 0.042 & 0.034 & 2 & $+0.023$ & 0.019 & 0.050 & 14 \\
NGC 6388      & $+0.155$ & 0.028 & 0.094 & 5 & $-0.055$ & 0.014 & 0.040 & 17 & $-0.152$ & 0.038 & 0.008 & 2 & $+0.004$ & 0.016 & 0.074 & 14 \\
NGC 6752      & $+0.106$ & 0.034 & 0.052 & 4 & $-0.101$ & 0.017 & 0.023 & 17 & $-0.241$ & 0.041 & 0.004 & 2 & $+0.068$ & 0.017 & 0.040 & 14 \\
NGC 7078      & $+0.086$ & 0.042 & 0.015 & 3 & $-0.277$ & 0.030 & 0.038 & 12 & $-0.148$ & 0.056 & 0.181 & 2 & $+0.104$ & 0.029 & 0.063 & 13 \\
NGC 7099      & $+0.095$ & 0.041 & 0.073 & 4 & $-0.168$ & 0.027 & 0.047 & 13 & $-0.194$ & 0.056 & 0.037 & 2 & $+0.099$ & 0.028 & 0.061 & 14 \\
N147 HII      & $+0.270$ & 0.127 & 0.228 & 2 & $+0.058$ & 0.085 & 0.108 & 12 & $-0.012$ & 0.145 & 0.318 & 2 & $+0.171$ & 0.102 & 0.145 & 11 \\
N147 HIII     & $-0.233$ & 0.200 & 0.096 & 2 & $+0.118$ & 0.119 & 0.260 & 5 & $+0.306$ & 0.125 & 0.300 & 2 & $-0.342$ & 0.123 & 0.258 & 5 \\
N147 PA-1     & $-0.436$ & 0.309 & 0.113 & 2 & $-0.025$ & 0.079 & 0.096 & 9 & $+0.101$ & 0.128 & \ldots & 1 & $+0.207$ & 0.173 & 0.045 & 3 \\
N147 PA-2     & $-0.009$ & 0.083 & 0.134 & 2 & $+0.054$ & 0.056 & 0.100 & 9 & $-0.129$ & 0.125 & 0.088 & 2 & $+0.102$ & 0.072 & 0.094 & 10 \\
N147 SD7      & $-0.219$ & 0.113 & 0.102 & 2 & $-0.007$ & 0.056 & 0.080 & 14 & $-0.306$ & 0.111 & 0.132 & 2 & $+0.167$ & 0.059 & 0.093 & 12 \\
N185 FJJ-III  & $+0.070$ & 0.127 & 0.126 & 2 & $+0.235$ & 0.053 & 0.108 & 12 & $-0.070$ & 0.136 & \ldots & 1 & $+0.251$ & 0.063 & 0.094 & 12 \\
N185 FJJ-V    & $-0.112$ & 0.097 & 0.170 & 2 & $-0.114$ & 0.062 & 0.093 & 11 & $+0.007$ & 0.097 & 0.127 & 2 & $+0.184$ & 0.055 & 0.090 & 13 \\
N185 FJJ-VIII & $+0.162$ & 0.092 & 0.004 & 2 & $-0.037$ & 0.050 & 0.058 & 15 & $-0.058$ & 0.079 & 0.092 & 2 & $+0.129$ & 0.056 & 0.093 & 13 \\
N205 HubbleI  & $+0.244$ & 0.052 & 0.034 & 4 & $-0.005$ & 0.027 & 0.043 & 17 & $-0.170$ & 0.051 & 0.094 & 2 & $-0.011$ & 0.032 & 0.060 & 14 \\
N205 HubbleII & $+0.315$ & 0.058 & 0.162 & 2 & $-0.057$ & 0.025 & 0.042 & 17 & $-0.253$ & 0.051 & 0.040 & 2 & $+0.062$ & 0.028 & 0.064 & 14 \\
N6822 SC6     & $+0.153$ & 0.084 & 0.066 & 2 & $-0.050$ & 0.043 & 0.047 & 15 & $-0.330$ & 0.085 & 0.054 & 2 & $-0.067$ & 0.053 & 0.063 & 13 \\
N6822 SC7     & $-0.308$ & 0.061 & 0.031 & 3 & $-0.124$ & 0.026 & 0.037 & 17 & $-0.308$ & 0.051 & 0.037 & 2 & $-0.224$ & 0.031 & 0.054 & 14 \\
N6822 HVII    & $-0.088$ & 0.073 & 0.166 & 3 & $+0.091$ & 0.033 & 0.083 & 17 & $-0.034$ & 0.059 & 0.218 & 2 & $+0.025$ & 0.043 & 0.099 & 14 \\
M33 H38       & $+0.024$ & 0.108 & 0.113 & 3 & $+0.620$ & 0.157 & 0.156 & 3 & $-0.303$ & 0.111 & 0.259 & 2 & $+0.120$ & 0.074 & 0.099 & 12 \\
M33 M9        & $+0.177$ & 0.086 & 0.069 & 2 & $-0.266$ & 0.127 & 0.392 & 4 & $-0.266$ & 0.106 & 0.079 & 2 & $-0.021$ & 0.062 & 0.057 & 13 \\
M33 R12       & $+0.077$ & 0.063 & 0.099 & 3 & $-0.434$ & 0.114 & 0.150 & 3 & $-0.141$ & 0.062 & 0.142 & 2 & $-0.016$ & 0.036 & 0.069 & 13 \\
M33 U49       & $+0.347$ & 0.097 & 0.028 & 3 & $+0.118$ & 0.189 & 0.498 & 3 & $+0.015$ & 0.100 & 0.138 & 2 & $+0.056$ & 0.078 & 0.113 & 10 \\
M33 R14       & $+0.124$ & 0.061 & 0.031 & 3 & $-0.047$ & 0.031 & 0.052 & 17 & $-0.161$ & 0.057 & 0.022 & 2 & $-0.018$ & 0.035 & 0.064 & 14 \\
M33 U77       & $+0.303$ & 0.104 & 0.131 & 2 & $+0.057$ & 0.057 & 0.086 & 13 & $-0.163$ & 0.115 & 0.063 & 2 & $-0.015$ & 0.068 & 0.063 & 12 \\
M33 CBF28     & $+0.053$ & 0.049 & 0.050 & 4 & $-0.075$ & 0.022 & 0.028 & 18 & $-0.264$ & 0.047 & 0.060 & 2 & $-0.072$ & 0.024 & 0.048 & 14 \\
M33 HM33B     & $-0.251$ & 0.196 & 0.048 & 2 & $+0.226$ & 0.079 & 0.117 & 16 & $-0.126$ & 0.228 & \ldots & 1 & $-0.122$ & 0.112 & 0.045 & 11 \\
WLM GC        & $+0.154$ & 0.073 & 0.088 & 3 & $+0.028$ & 0.047 & 0.072 & 10 & $-0.260$ & 0.079 & 0.108 & 2 & $+0.127$ & 0.044 & 0.093 & 13 \\
Fornax 3      & $+0.070$ & 0.068 & 0.118 & 3 & $-0.217$ & 0.049 & 0.048 & 12 & $-0.277$ & 0.099 & 0.250 & 2 & $-0.016$ & 0.051 & 0.077 & 12 \\
Fornax 4      & $-0.136$ & 0.049 & 0.068 & 3 & $-0.109$ & 0.026 & 0.036 & 15 & $-0.287$ & 0.049 & 0.043 & 2 & $-0.172$ & 0.023 & 0.049 & 15 \\
Fornax 5      & $-0.133$ & 0.113 & 0.228 & 3 & $-0.008$ & 0.058 & 0.083 & 11 & $-0.294$ & 0.116 & \ldots & 1 & $+0.143$ & 0.053 & 0.091 & 13 \\
M31 006-058   & $+0.244$ & 0.032 & 0.101 & 5 & $-0.027$ & 0.017 & 0.032 & 15 & $-0.166$ & 0.040 & 0.060 & 2 & $+0.026$ & 0.017 & 0.059 & 15 \\
M31 012-064   & $+0.322$ & 0.074 & 0.135 & 3 & $-0.066$ & 0.044 & 0.064 & 13 & $-0.333$ & 0.087 & 0.174 & 2 & $+0.019$ & 0.052 & 0.077 & 14 \\
M31 019-072   & $+0.193$ & 0.040 & 0.047 & 5 & $-0.033$ & 0.019 & 0.038 & 16 & $-0.196$ & 0.044 & 0.053 & 2 & $+0.052$ & 0.020 & 0.064 & 15 \\
M31 058-119   & $+0.130$ & 0.037 & 0.089 & 5 & $-0.075$ & 0.021 & 0.048 & 15 & $-0.261$ & 0.042 & 0.094 & 2 & $+0.002$ & 0.019 & 0.059 & 15 \\
M31 082-114   & $+0.384$ & 0.092 & 0.147 & 2 & $-0.032$ & 0.045 & 0.071 & 15 & $-0.314$ & 0.097 & 0.098 & 2 & $+0.287$ & 0.045 & 0.080 & 15 \\
M31 163-217   & $+0.137$ & 0.030 & 0.098 & 5 & $-0.082$ & 0.017 & 0.040 & 15 & $-0.061$ & 0.040 & 0.018 & 2 & $+0.114$ & 0.016 & 0.075 & 15 \\
M31 171-222   & $+0.225$ & 0.032 & 0.104 & 5 & $-0.040$ & 0.017 & 0.041 & 15 & $-0.021$ & 0.041 & 0.052 & 2 & $+0.107$ & 0.017 & 0.067 & 15 \\
M31 174-226   & $+0.224$ & 0.070 & 0.065 & 3 & $-0.025$ & 0.032 & 0.044 & 16 & $-0.277$ & 0.063 & 0.149 & 2 & $+0.052$ & 0.037 & 0.074 & 15 \\
M31 225-280   & $+0.155$ & 0.030 & 0.132 & 5 & $-0.023$ & 0.016 & 0.056 & 15 & $-0.130$ & 0.039 & 0.007 & 2 & $+0.107$ & 0.016 & 0.084 & 15 \\
M31 338-076   & $+0.062$ & 0.047 & 0.088 & 5 & $-0.064$ & 0.022 & 0.038 & 16 & $-0.281$ & 0.047 & 0.044 & 2 & $-0.023$ & 0.023 & 0.055 & 15 \\
M31 358-219   & $-0.071$ & 0.061 & 0.025 & 3 & $-0.158$ & 0.050 & 0.054 & 9 & $-0.247$ & 0.084 & 0.067 & 2 & $+0.039$ & 0.042 & 0.071 & 12 \\
M31 EXT8      & $+0.442$ & 0.118 & 0.148 & 2 & $-0.234$ & 0.163 & 0.149 & 5 & $+1.155$ & 0.167 & \ldots & 1 & $+0.462$ & 0.089 & 0.218 & 8 \\
N2403 F46     & $-0.095$ & 0.103 & 0.134 & 2 & $-0.016$ & 0.056 & 0.083 & 13 & $-0.308$ & 0.110 & 0.073 & 2 & $+0.199$ & 0.057 & 0.115 & 12 \\
\hline
\end{tabular}
}
\end{adjustbox}
\tablefoot{
The listed abundances include NLTE corrections for Mn and Ni. See notes to Table~\ref{tab:results1} for explanations of the columns.
}
\end{table*}

\begin{table*}
\caption{Results for Cu, Zn, Zr, Ba, and Eu.}
\label{tab:results3v5}
\begin{adjustbox}{width=1\textwidth}
\centering
{\small
\begin{tabular}{l cccccccccccccccccccc}
\hline\hline
Cluster       & [Cu/Fe] & $\sigma_{\langle\mathrm{Cu}\rangle}$ & $S_\mathrm{Cu}$ & N    & [Zn/Fe] & $\sigma_{\langle\mathrm{Zn}\rangle}$ & $S_\mathrm{Zn}$ & N    & [Zr/Fe] & $\sigma_{\langle\mathrm{Zr}\rangle}$ & $S_\mathrm{Zr}$ & N     & [Ba/Fe] & $\sigma_{\langle\mathrm{Ba}\rangle}$ & $S_\mathrm{Ba}$ & N    & [Eu/Fe] & $\sigma_{\langle\mathrm{Eu}\rangle}$ & $S_\mathrm{Eu}$ & N    \\ \hline
NGC 0104      & $-0.028$ & 0.061 & \ldots & 1 & $+0.165$ & 0.045 & 0.044 & 2 & $+0.165$ & 0.063 & \ldots & 1 & $+0.159$ & 0.025 & 0.071 & 5 & $+0.258$ & 0.050 & 0.029 & 2 \\
NGC 0362      & $-0.372$ & 0.067 & \ldots & 1 & $-0.108$ & 0.049 & 0.073 & 2 & $+0.443$ & 0.074 & \ldots & 1 & $+0.300$ & 0.026 & 0.052 & 5 & $+0.595$ & 0.047 & 0.019 & 2 \\
NGC 6254      & $-0.582$ & 0.076 & \ldots & 1 & $-0.052$ & 0.054 & 0.007 & 2 & $-0.030$ & 0.155 & \ldots & 1 & $+0.261$ & 0.028 & 0.071 & 5 & $+0.081$ & 0.094 & \ldots & 1 \\
NGC 6388      & $-0.089$ & 0.060 & \ldots & 1 & $-0.129$ & 0.051 & 0.074 & 2 & $+0.245$ & 0.063 & \ldots & 1 & $+0.128$ & 0.024 & 0.073 & 5 & $-0.059$ & 0.072 & \ldots & 1 \\
NGC 6752      & $-0.434$ & 0.067 & \ldots & 1 & $+0.080$ & 0.045 & 0.027 & 2 & $+0.282$ & 0.136 & \ldots & 1 & $+0.119$ & 0.026 & 0.084 & 5 & $+0.365$ & 0.055 & 0.097 & 2 \\
NGC 7078      & \ldots & \ldots & \ldots & \ldots & $-0.222$ & 0.103 & 0.059 & 2 & \ldots & \ldots & \ldots & \ldots & $+0.271$ & 0.030 & 0.030 & 5 & $+0.531$ & 0.105 & 0.058 & 2 \\
NGC 7099      & $-0.426$ & 0.114 & \ldots & 1 & $-0.109$ & 0.076 & 0.091 & 2 & \ldots & \ldots & \ldots & \ldots & $-0.083$ & 0.031 & 0.145 & 5 & $+0.283$ & 0.112 & \ldots & 1 \\
N147 HII      & $-0.405$ & 0.363 & \ldots & 1 & $+0.535$ & 0.657 & \ldots & 1 & $+0.555$ & 0.453 & \ldots & 1 & $-0.282$ & 0.172 & 0.305 & 3 & \ldots & \ldots & \ldots & \ldots \\
N147 HIII     & $+0.363$ & 0.231 & \ldots & 1 & $+0.147$ & 0.399 & \ldots & 1 & \ldots & \ldots & \ldots & \ldots & $-0.626$ & 0.111 & 0.094 & 5 & \ldots & \ldots & \ldots & \ldots \\
N147 PA-1     & $-0.221$ & 0.297 & \ldots & 1 & $-0.013$ & 0.414 & \ldots & 1 & \ldots & \ldots & \ldots & \ldots & $+0.042$ & 0.086 & 0.082 & 5 & \ldots & \ldots & \ldots & \ldots \\
N147 PA-2     & $-0.849$ & 0.359 & \ldots & 1 & $+0.281$ & 0.202 & 0.319 & 3 & \ldots & \ldots & \ldots & \ldots & $+0.193$ & 0.061 & 0.036 & 5 & \ldots & \ldots & \ldots & \ldots \\
N147 SD7      & $-0.511$ & 0.315 & \ldots & 1 & $+0.008$ & 0.200 & 0.332 & 2 & \ldots & \ldots & \ldots & \ldots & $+0.158$ & 0.056 & 0.163 & 5 & \ldots & \ldots & \ldots & \ldots \\
N185 FJJ-III  & $-0.799$ & 0.251 & \ldots & 1 & $+0.130$ & 0.258 & 0.177 & 2 & \ldots & \ldots & \ldots & \ldots & $+0.214$ & 0.075 & 0.099 & 5 & $+0.382$ & 0.589 & \ldots & 1 \\
N185 FJJ-V    & $-0.720$ & 0.328 & \ldots & 1 & $-0.041$ & 0.252 & \ldots & 1 & \ldots & \ldots & \ldots & \ldots & $+0.226$ & 0.068 & 0.145 & 5 & \ldots & \ldots & \ldots & \ldots \\
N185 FJJ-VIII & $-0.728$ & 0.292 & \ldots & 1 & $+0.464$ & 0.146 & 0.041 & 2 & \ldots & \ldots & \ldots & \ldots & $+0.189$ & 0.058 & 0.174 & 5 & $+0.235$ & 0.441 & \ldots & 1 \\
N205 HubbleI  & $-0.447$ & 0.137 & \ldots & 1 & $-0.241$ & 0.112 & 0.099 & 3 & $+0.650$ & 0.151 & \ldots & 1 & $+0.153$ & 0.041 & 0.070 & 5 & $+0.656$ & 0.179 & \ldots & 1 \\
N205 HubbleII & $-0.548$ & 0.095 & \ldots & 1 & $-0.072$ & 0.079 & 0.184 & 3 & $-0.223$ & 0.244 & \ldots & 1 & $+0.119$ & 0.036 & 0.074 & 5 & $+0.128$ & 0.214 & \ldots & 1 \\
N6822 SC6     & $-0.614$ & 0.182 & \ldots & 1 & $+0.092$ & 0.154 & 0.139 & 2 & \ldots & \ldots & \ldots & \ldots & $+0.161$ & 0.054 & 0.038 & 5 & $+0.527$ & 0.252 & \ldots & 1 \\
N6822 SC7     & \ldots & \ldots & \ldots & \ldots & $-0.401$ & 0.099 & 0.059 & 3 & $-0.455$ & 0.331 & \ldots & 1 & $+0.115$ & 0.035 & 0.029 & 5 & $+0.137$ & 0.151 & \ldots & 1 \\
N6822 HVII    & $-0.316$ & 0.170 & \ldots & 1 & $-0.371$ & 0.173 & 0.108 & 2 & \ldots & \ldots & \ldots & \ldots & $+0.350$ & 0.041 & 0.080 & 5 & $-0.003$ & 0.311 & \ldots & 1 \\
M33 H38       & $-0.083$ & 0.339 & \ldots & 1 & $-0.580$ & 0.435 & \ldots & 1 & \ldots & \ldots & \ldots & \ldots & $+0.473$ & 0.061 & 0.103 & 4 & \ldots & \ldots & \ldots & \ldots \\
M33 M9        & $-0.216$ & 0.186 & \ldots & 1 & $+0.266$ & 0.208 & 0.058 & 2 & \ldots & \ldots & \ldots & \ldots & $+0.466$ & 0.066 & 0.092 & 4 & \ldots & \ldots & \ldots & \ldots \\
M33 R12       & $-0.443$ & 0.135 & \ldots & 1 & $+0.081$ & 0.147 & 0.110 & 3 & \ldots & \ldots & \ldots & \ldots & $+0.264$ & 0.054 & 0.084 & 4 & $+0.493$ & 0.213 & \ldots & 1 \\
M33 U49       & $-0.620$ & 0.297 & \ldots & 1 & $+0.585$ & 0.314 & 0.089 & 2 & \ldots & \ldots & \ldots & \ldots & $+0.478$ & 0.091 & 0.234 & 4 & \ldots & \ldots & \ldots & \ldots \\
M33 R14       & $-0.613$ & 0.129 & \ldots & 1 & $-0.139$ & 0.140 & 0.201 & 3 & $+0.278$ & 0.167 & \ldots & 1 & $+0.249$ & 0.042 & 0.084 & 5 & \ldots & \ldots & \ldots & \ldots \\
M33 U77       & $-0.609$ & 0.258 & \ldots & 1 & $+0.211$ & 0.213 & 0.321 & 3 & \ldots & \ldots & \ldots & \ldots & $+0.270$ & 0.074 & 0.084 & 5 & \ldots & \ldots & \ldots & \ldots \\
M33 CBF28     & $-0.384$ & 0.087 & \ldots & 1 & $-0.189$ & 0.076 & 0.124 & 3 & $+0.672$ & 0.098 & \ldots & 1 & $+0.181$ & 0.033 & 0.025 & 5 & $+0.629$ & 0.106 & \ldots & 1 \\
M33 HM33B     & $-0.024$ & 0.282 & \ldots & 1 & $-0.651$ & 1.015 & \ldots & 1 & $+0.890$ & 0.482 & \ldots & 1 & $+0.715$ & 0.149 & 0.006 & 2 & $+0.654$ & 0.493 & \ldots & 1 \\
WLM GC        & \ldots & \ldots & \ldots & \ldots & $+0.072$ & 0.134 & 0.230 & 2 & \ldots & \ldots & \ldots & \ldots & $-0.137$ & 0.061 & 0.052 & 4 & $+0.054$ & 0.284 & \ldots & 1 \\
Fornax 3      & \ldots & \ldots & \ldots & \ldots & $-0.126$ & 0.162 & \ldots & 1 & \ldots & \ldots & \ldots & \ldots & $+0.371$ & 0.044 & 0.065 & 4 & \ldots & \ldots & \ldots & \ldots \\
Fornax 4      & $-0.921$ & 0.107 & \ldots & 1 & $-0.269$ & 0.089 & 0.092 & 2 & $+0.051$ & 0.267 & \ldots & 1 & $+0.009$ & 0.042 & 0.011 & 4 & $+0.200$ & 0.123 & \ldots & 1 \\
Fornax 5      & $-0.569$ & 0.301 & \ldots & 1 & $-0.047$ & 0.177 & 0.022 & 2 & \ldots & \ldots & \ldots & \ldots & $-0.180$ & 0.071 & 0.061 & 4 & $+0.291$ & 0.325 & \ldots & 1 \\
M31 006-058   & $-0.039$ & 0.071 & \ldots & 1 & $+0.075$ & 0.054 & 0.106 & 3 & $+0.282$ & 0.070 & \ldots & 1 & $+0.104$ & 0.027 & 0.105 & 5 & $+0.013$ & 0.094 & \ldots & 1 \\
M31 012-064   & $-0.751$ & 0.247 & \ldots & 1 & $-0.269$ & 0.216 & 0.072 & 2 & \ldots & \ldots & \ldots & \ldots & $+0.255$ & 0.060 & 0.095 & 4 & \ldots & \ldots & \ldots & \ldots \\
M31 019-072   & $-0.264$ & 0.088 & \ldots & 1 & $+0.034$ & 0.070 & 0.071 & 3 & $+0.227$ & 0.109 & \ldots & 1 & $-0.003$ & 0.036 & 0.069 & 4 & \ldots & \ldots & \ldots & \ldots \\
M31 058-119   & $-0.418$ & 0.080 & \ldots & 1 & $-0.279$ & 0.064 & 0.123 & 3 & $+0.053$ & 0.106 & \ldots & 1 & $+0.320$ & 0.029 & 0.027 & 4 & $+0.320$ & 0.078 & 0.040 & 2 \\
M31 082-114   & $-0.303$ & 0.203 & \ldots & 1 & $+0.441$ & 0.274 & 0.277 & 3 & $+0.325$ & 0.231 & \ldots & 1 & $+0.235$ & 0.082 & 0.233 & 4 & \ldots & \ldots & \ldots & \ldots \\
M31 163-217   & $+0.122$ & 0.073 & \ldots & 1 & $+0.004$ & 0.055 & 0.086 & 3 & $-0.063$ & 0.074 & \ldots & 1 & $-0.110$ & 0.028 & 0.116 & 5 & $-0.180$ & 0.097 & \ldots & 1 \\
M31 171-222   & $+0.099$ & 0.078 & \ldots & 1 & $+0.003$ & 0.060 & 0.131 & 3 & $+0.138$ & 0.075 & \ldots & 1 & $+0.068$ & 0.028 & 0.144 & 5 & $-0.222$ & 0.114 & \ldots & 1 \\
M31 174-226   & $-0.373$ & 0.165 & \ldots & 1 & $-0.279$ & 0.152 & 0.170 & 3 & $+0.531$ & 0.187 & \ldots & 1 & $+0.382$ & 0.048 & 0.050 & 4 & \ldots & \ldots & \ldots & \ldots \\
M31 225-280   & $-0.261$ & 0.065 & \ldots & 1 & $-0.143$ & 0.049 & 0.026 & 3 & $-0.030$ & 0.075 & \ldots & 1 & $+0.105$ & 0.026 & 0.123 & 5 & $+0.027$ & 0.067 & 0.006 & 2 \\
M31 338-076   & $-0.399$ & 0.086 & \ldots & 1 & $+0.058$ & 0.072 & 0.110 & 3 & $+0.159$ & 0.157 & \ldots & 1 & $+0.365$ & 0.034 & 0.062 & 4 & \ldots & \ldots & \ldots & \ldots \\
M31 358-219   & $-0.839$ & 0.416 & \ldots & 1 & $-0.019$ & 0.122 & 0.030 & 2 & \ldots & \ldots & \ldots & \ldots & $+0.017$ & 0.041 & 0.044 & 5 & $+0.242$ & 0.271 & \ldots & 1 \\
M31 EXT8      & $+0.068$ & 0.475 & \ldots & 1 & \ldots & \ldots & \ldots & \ldots & \ldots & \ldots & \ldots & \ldots & $+0.348$ & 0.069 & 0.028 & 4 & \ldots & \ldots & \ldots & \ldots \\
N2403 F46     & \ldots & \ldots & \ldots & \ldots & $+0.155$ & 0.218 & 0.279 & 2 & \ldots & \ldots & \ldots & \ldots & $+0.114$ & 0.066 & 0.142 & 5 & $+0.391$ & 0.349 & \ldots & 1 \\
\hline
\end{tabular}
}
\end{adjustbox}
\tablefoot{
The listed abundances include NLTE corrections for Ba. See notes to Table~\ref{tab:results1} for explanations of the columns.
}
\end{table*}

\clearpage

\section{Individual abundance measurements}
\label{app:iam}

This appendix gives the individual abundance measurements per spectral window for each cluster, as well as the Sun and Arcturus. Two examples are shown here. For each measurement, the Table columns give the wavelength range, the \ac{lte} abundance ([X/H]) obtained from the spectral modelling, the \ac{nlte} correction ($\Delta_\mathrm{NLTE}$), and the formal uncertainty on the measurement ($\sigma_i$). 
For the \ac{nlte} corrections, a value of $+99.990$ indicates that no \ac{nlte} correction was computed for the corresponding spectral window. 

\begin{table}
\caption{Abundances for NGC~104.}
{\tiny
\begin{tabular}{lccc} \\ \hline\hline
Wavelengths & $\mathrm{[X/H]}$ (LTE) & $\Delta_\mathrm{NLTE}$ & $\sigma_i$ \\ \hline
\mbox{[Fe/H]} \\
4573.0--4600.0 & $-0.901$ & $+0.026$ & 0.010 \\
4600.0--4618.0 & $-0.789$ & $+0.031$ & 0.015 \\
4631.0--4660.0 & $-0.838$ & $+0.021$ & 0.012 \\
4671.0--4686.0 & $-0.892$ & $+0.017$ & 0.014 \\
4705.0--4714.0 & $-0.576$ & $+0.014$ & 0.018 \\
4724.0--4750.0 & $-0.721$ & $+0.014$ & 0.011 \\
4866.0--4883.0 & $-0.727$ & $+0.010$ & 0.009 \\
4886.0--4896.0 & $-0.709$ & $+0.011$ & 0.010 \\
4897.0--4915.0 & $-0.917$ & $+0.012$ & 0.013 \\
4915.0--4929.0 & $-0.611$ & $+0.012$ & 0.008 \\
4936.0--4944.0 & $-0.600$ & $+0.017$ & 0.014 \\
4944.0--4953.0 & $-0.615$ & $-0.000$ & 0.022 \\
4952.0--4962.0 & $-0.707$ & $+0.013$ & 0.008 \\
4963.0--4976.0 & $-0.772$ & $+0.009$ & 0.014 \\
4975.0--4998.0 & $-0.905$ & $+0.009$ & 0.011 \\
5008.0--5017.0 & $-0.813$ & $+0.017$ & 0.013 \\
5045.0--5064.0 & $-0.890$ & $+0.020$ & 0.013 \\
5066.0--5115.0 & $-0.667$ & $+0.018$ & 0.007 \\
5118.0--5150.0 & $-0.839$ & $+0.009$ & 0.005 \\
5250.0--5259.0 & $-0.415$ & $+0.022$ & 0.019 \\
5271.0--5289.0 & $-0.803$ & $+0.012$ & 0.013 \\
5300.0--5345.0 & $-0.941$ & $+0.017$ & 0.006 \\
5358.0--5375.0 & $-0.750$ & $+0.013$ & 0.011 \\
5378.0--5400.0 & $-0.799$ & $+0.012$ & 0.010 \\
5400.0--5420.0 & $-0.806$ & $+0.006$ & 0.009 \\
5420.0--5460.0 & $-0.841$ & $+0.015$ & 0.007 \\
5460.0--5475.5 & $-0.815$ & $+0.003$ & 0.016 \\
5494.0--5510.0 & $-0.544$ & $+0.024$ & 0.017 \\
5529.0--5539.0 & $-0.800$ & $+0.038$ & 0.019 \\
5566.5--5590.0 & $-0.712$ & $-0.003$ & 0.011 \\
5610.0--5630.0 & $-0.669$ & $+0.002$ & 0.011 \\
5682.0--5714.0 & $-0.772$ & $+0.012$ & 0.013 \\
5858.5--5865.0 & $-0.660$ & $+0.006$ & 0.039 \\
5970.0--5980.0 & $-0.815$ & $+0.003$ & 0.025 \\
6001.0--6030.0 & $-0.628$ & $-0.010$ & 0.015 \\
6053.0--6082.0 & $-0.811$ & $-0.006$ & 0.014 \\
6131.0--6140.0 & $-0.722$ & $+0.004$ & 0.013 \\
6144.0--6160.0 & $-0.665$ & $+0.021$ & 0.018 \\
6170.0--6185.0 & $-0.599$ & $+0.017$ & 0.020 \\
\mbox{[Na/H]} \\
5677.0--5695.0 & $-0.285$ & $-0.148$ & 0.019\\
6149.0--6166.0 & $-0.425$ & $-0.138$ & 0.019\\
\mbox{[Mg/H]} \\
4347.0--4357.0 & $-0.449$ & $-0.005$ & 0.027\\
4565.0--4576.0 & $-0.457$ & $+0.000$ & 0.034\\
4700.0--4707.0 & $-0.337$ & $-0.008$ & 0.013\\
5523.0--5531.5 & $-0.352$ & $-0.016$ & 0.013\\
5705.0--5715.0 & $-0.157$ & $-0.015$ & 0.021\\
\mbox{[Si/H]} \\
5661.0--5671.0 & $-0.343$ & $+99.990$ & 0.032\\
5685.0--5695.0 & $-0.349$ & $+99.990$ & 0.032\\
5767.0--5777.0 & $-0.270$ & $+99.990$ & 0.032\\
6150.0--6160.0 & $-0.340$ & $+99.990$ & 0.024\\
6232.0--6250.0 & $-0.555$ & $+99.990$ & 0.018\\
7400.0--7427.0 & $-0.275$ & $+99.990$ & 0.018\\
\mbox{[Ca/H]} \\
4420.0--4440.0 & $-0.511$ & $-0.017$ & 0.012\\
4451.0--4461.0 & $-0.369$ & $-0.017$ & 0.014\\
4573.0--4590.0 & $-0.489$ & $-0.047$ & 0.019\\
5256.0--5268.0 & $-0.385$ & $-0.053$ & 0.024\\
5347.0--5357.0 & $-0.466$ & $-0.090$ & 0.034\\
5507.0--5517.0 & $-0.577$ & $-0.087$ & 0.033\\
5576.0--5602.0 & $-0.479$ & $-0.043$ & 0.014\\
5852.0--5862.0 & $-0.248$ & $-0.112$ & 0.021\\
6098.0--6127.0 & $-0.503$ & $-0.022$ & 0.011\\
6151.0--6174.0 & $-0.503$ & $-0.022$ & 0.009\\
\hline
\end{tabular}
}
\end{table}

\addtocounter{table}{-1}
\begin{table}
\caption{continued. Abundances for NGC~104.}
{\tiny
\begin{tabular}{lccc} \\ \hline\hline
Wavelengths & $\mathrm{[X/H]}$ (LTE) & $\Delta_\mathrm{NLTE}$ & $\sigma_i$ \\ \hline
\mbox{[Sc/H]} \\
4739.0--4758.0 & $-0.772$ & $+99.990$ & 0.055\\
5026.0--5036.0 & $-0.412$ & $+99.990$ & 0.038\\
5521.0--5531.0 & $-0.415$ & $+99.990$ & 0.036\\
5638.0--5690.0 & $-0.513$ & $+99.990$ & 0.013\\
6206.0--6216.0 & $-0.688$ & $+99.990$ & 0.047\\
\mbox{[Ti/H]} \\
4500.0--4519.5 & $-0.507$ & $+0.050$ & 0.017\\
4551.0--4570.0 & $-0.417$ & $+0.054$ & 0.015\\
4586.5--4596.0 & $-0.140$ & $-0.056$ & 0.046\\
4638.0--4660.0 & $-0.530$ & $+0.138$ & 0.015\\
4680.0--4698.0 & $-0.339$ & $+0.158$ & 0.025\\
4802.0--4821.0 & $-0.494$ & $+0.093$ & 0.021\\
4975.0--5000.0 & $-0.510$ & $+0.097$ & 0.012\\
5000.0--5030.0 & $-0.392$ & $+0.119$ & 0.011\\
5060.0--5075.0 & $-0.377$ & $+0.159$ & 0.018\\
5331.0--5341.0 & $-0.308$ & $-0.012$ & 0.037\\
5376.0--5386.0 & $-0.203$ & $-0.002$ & 0.040\\
5510.0--5520.0 & $-0.433$ & $+0.133$ & 0.021\\
5860.0--5875.0 & $-0.385$ & $+0.111$ & 0.038\\
5912.0--5922.0 & $-0.621$ & $+0.111$ & 0.063\\
\mbox{[Cr/H]} \\
4537.0--4550.0 & $-0.822$ & $+99.990$ & 0.022\\
4612.0--4631.0 & $-0.885$ & $+99.990$ & 0.018\\
4646.0--4657.0 & $-0.760$ & $+99.990$ & 0.025\\
4703.0--4723.0 & $-0.884$ & $+99.990$ & 0.028\\
4751.0--4761.0 & $-0.574$ & $+99.990$ & 0.034\\
4796.0--4806.0 & $-0.802$ & $+99.990$ & 0.043\\
4824.0--4834.0 & $-0.719$ & $+99.990$ & 0.025\\
4866.0--4876.0 & $-0.761$ & $+99.990$ & 0.044\\
4931.0--4947.0 & $-0.836$ & $+99.990$ & 0.030\\
5063.0--5096.0 & $-0.779$ & $+99.990$ & 0.026\\
5270.0--5281.0 & $-0.790$ & $+99.990$ & 0.021\\
5292.0--5304.0 & $-0.846$ & $+99.990$ & 0.014\\
5341.0--5353.0 & $-0.790$ & $+99.990$ & 0.021\\
5407.0--5413.0 & $-0.746$ & $+99.990$ & 0.028\\
5779.0--5793.0 & $-0.960$ & $+99.990$ & 0.025\\
6325.0--6335.0 & $-0.525$ & $+99.990$ & 0.045\\
6973.0--6983.0 & $-0.406$ & $+99.990$ & 0.024\\
\mbox{[Mn/H]} \\
4750.0--4790.0 & $-1.045$ & $+0.082$ & 0.014\\
6010.0--6030.0 & $-0.959$ & $+0.078$ & 0.019\\
\mbox{[Ni/H]} \\
4600.0--4610.0 & $-0.519$ & $+0.062$ & 0.028\\
4644.0--4654.0 & $-0.661$ & $+0.060$ & 0.040\\
4681.0--4691.0 & $-0.920$ & $+99.990$ & 0.036\\
4709.0--4719.0 & $-0.662$ & $+0.040$ & 0.026\\
4824.0--4835.0 & $-1.130$ & $+0.068$ & 0.028\\
4899.0--4909.0 & $-0.799$ & $+0.063$ & 0.037\\
4931.0--4942.0 & $-0.690$ & $+99.990$ & 0.025\\
4975.0--4985.0 & $-0.523$ & $+0.047$ & 0.018\\
5098.0--5108.0 & $-0.603$ & $+0.053$ & 0.021\\
5141.0--5151.0 & $-0.830$ & $+0.067$ & 0.030\\
5472.0--5482.0 & $-0.998$ & $-0.001$ & 0.018\\
5707.0--5717.0 & $-0.496$ & $+0.008$ & 0.028\\
6103.0--6113.0 & $-0.504$ & $+0.017$ & 0.025\\
6172.0--6182.0 & $-0.780$ & $+0.064$ & 0.019\\
\mbox{[Cu/H]} \\
5101.0--5112.0 & $-0.784$ & $+99.990$ & 0.034\\
\mbox{[Zn/H]} \\
4717.0--4727.0 & $-0.574$ & $+99.990$ & 0.040\\
4805.0--4815.0 & $-0.666$ & $+99.990$ & 0.036\\
\mbox{[Zr/H]} \\
6124.0--6147.0 & $-0.511$ & $+99.990$ & 0.041\\
\mbox{[Ba/H]} \\
4551.0--4560.0 & $-0.592$ & $-0.008$ & 0.013\\
4929.0--4939.0 & $-0.736$ & $-0.035$ & 0.028\\
5849.0--5859.0 & $-0.462$ & $-0.067$ & 0.035\\
6135.0--6145.0 & $-0.658$ & $-0.076$ & 0.025\\
6492.0--6502.0 & $-0.270$ & $-0.116$ & 0.021\\
\mbox{[Eu/H]} \\
4431.0--4441.0 & $-0.506$ & $+99.990$ & 0.032\\
6640.0--6650.0 & $-0.519$ & $+99.990$ & 0.080\\
\hline
\end{tabular}
}
\end{table}

\begin{table}
\caption{Abundances for NGC~7078.}
{\tiny
\begin{tabular}{lccc} \\ \hline\hline
Wavelengths & $\mathrm{[X/H]}$ (LTE) & $\Delta_\mathrm{NLTE}$ & $\sigma_i$ \\ \hline
\mbox{[Fe/H]} \\
4573.0--4600.0 & $-2.374$ & $+0.055$ & 0.034 \\
4600.0--4618.0 & $-2.443$ & $+0.042$ & 0.040 \\
4631.0--4660.0 & $-2.333$ & $+0.097$ & 0.026 \\
4671.0--4686.0 & $-2.599$ & $+0.081$ & 0.075 \\
4705.0--4714.0 & $-2.410$ & $+0.113$ & 0.056 \\
4724.0--4750.0 & $-2.409$ & $+0.071$ & 0.036 \\
4866.0--4883.0 & $-2.430$ & $+0.029$ & 0.029 \\
4886.0--4896.0 & $-2.282$ & $+0.026$ & 0.028 \\
4897.0--4915.0 & $-2.538$ & $+0.046$ & 0.037 \\
4915.0--4929.0 & $-2.236$ & $+0.027$ & 0.024 \\
4936.0--4944.0 & $-2.264$ & $+0.041$ & 0.032 \\
4944.0--4953.0 & $-2.341$ & $+0.036$ & 0.079 \\
4952.0--4962.0 & $-2.240$ & $+0.012$ & 0.021 \\
4963.0--4976.0 & $-2.462$ & $+0.066$ & 0.055 \\
4975.0--4998.0 & $-2.413$ & $+0.067$ & 0.025 \\
5008.0--5017.0 & $-2.159$ & $+0.068$ & 0.031 \\
5045.0--5064.0 & $-2.237$ & $+0.041$ & 0.033 \\
5066.0--5115.0 & $-2.236$ & $+0.049$ & 0.013 \\
5118.0--5150.0 & $-2.391$ & $+0.049$ & 0.017 \\
5250.0--5259.0 & $-2.079$ & $+0.033$ & 0.041 \\
5271.0--5289.0 & $-2.377$ & $+0.031$ & 0.024 \\
5300.0--5345.0 & $-2.321$ & $+0.040$ & 0.015 \\
5358.0--5375.0 & $-2.337$ & $+0.060$ & 0.025 \\
5378.0--5400.0 & $-2.310$ & $+0.044$ & 0.025 \\
5400.0--5420.0 & $-2.430$ & $+0.056$ & 0.022 \\
5420.0--5460.0 & $-2.391$ & $+0.048$ & 0.015 \\
5460.0--5475.5 & $-2.613$ & $+0.106$ & 0.059 \\
5494.0--5510.0 & $-2.027$ & $+0.034$ & 0.028 \\
5529.0--5539.0 & $-2.228$ & $+0.209$ & 0.085 \\
5566.5--5590.0 & $-2.447$ & $+0.030$ & 0.025 \\
5610.0--5630.0 & $-2.464$ & $+0.020$ & 0.026 \\
5682.0--5714.0 & $-2.327$ & $+0.036$ & 0.042 \\
5858.5--5865.0 & $-2.451$ & $+0.110$ & 0.114 \\
5970.0--5980.0 & \multicolumn{2}{c}{$\ldots$} \\
6001.0--6030.0 & $-2.700$ & $+0.109$ & 0.075 \\
6053.0--6082.0 & $-2.393$ & $+0.014$ & 0.043 \\
6131.0--6140.0 & $-2.280$ & $+0.013$ & 0.029 \\
6144.0--6160.0 & $-2.168$ & $+0.122$ & 0.079 \\
6170.0--6185.0 & $-2.329$ & $+0.031$ & 0.074 \\
\mbox{[Na/H]} \\
5677.0--5695.0 & $-2.268$ & $-0.082$ & 0.077\\
6149.0--6166.0 & $-2.284$ & $-0.079$ & 0.777\\
\mbox{[Mg/H]} \\
4347.0--4357.0 & $-2.210$ & $-0.006$ & 0.050\\
4565.0--4576.0 & $-2.082$ & $+0.003$ & 0.057\\
4700.0--4707.0 & $-2.208$ & $-0.011$ & 0.034\\
5523.0--5531.5 & $-2.104$ & $-0.020$ & 0.032\\
5705.0--5715.0 & $-2.117$ & $-0.008$ & 0.086\\
\mbox{[Si/H]} \\
5661.0--5671.0 & $-1.503$ & $+99.990$ & 0.170\\
5685.0--5695.0 & $-1.616$ & $+99.990$ & 0.171\\
5767.0--5777.0 & \multicolumn{2}{c}{$\ldots$} \\
6150.0--6160.0 & \multicolumn{2}{c}{$\ldots$} \\
6232.0--6250.0 & \multicolumn{2}{c}{$\ldots$} \\
7400.0--7427.0 & $-1.920$ & $+99.990$ & 0.058\\
\mbox{[Ca/H]} \\
4420.0--4440.0 & $-2.007$ & $+0.005$ & 0.029\\
4451.0--4461.0 & $-2.103$ & $-0.006$ & 0.040\\
4573.0--4590.0 & $-2.099$ & $+0.048$ & 0.056\\
5256.0--5268.0 & $-1.917$ & $+0.069$ & 0.037\\
5347.0--5357.0 & $-2.060$ & $+0.034$ & 0.079\\
5507.0--5517.0 & $-2.204$ & $+0.035$ & 0.135\\
5576.0--5602.0 & $-2.044$ & $+0.026$ & 0.022\\
5852.0--5862.0 & $-2.082$ & $+0.015$ & 0.043\\
6098.0--6127.0 & $-1.946$ & $-0.030$ & 0.025\\
6151.0--6174.0 & $-2.087$ & $-0.042$ & 0.022\\
\hline
\end{tabular}
}
\end{table}

\addtocounter{table}{-1}
\begin{table}
\caption{continued. Abundances for NGC~7078.}
{\tiny
\begin{tabular}{lccc} \\ \hline\hline
Wavelengths & $\mathrm{[X/H]}$ (LTE) & $\Delta_\mathrm{NLTE}$ & $\sigma_i$ \\ \hline
\mbox{[Sc/H]} \\
4739.0--4758.0 & \multicolumn{2}{c}{$\ldots$} \\
5026.0--5036.0 & $-2.154$ & $+99.990$ & 0.065\\
5521.0--5531.0 & $-2.204$ & $+99.990$ & 0.057\\
5638.0--5690.0 & $-2.165$ & $+99.990$ & 0.037\\
6206.0--6216.0 & \multicolumn{2}{c}{$\ldots$} \\
\mbox{[Ti/H]} \\
4500.0--4519.5 & $-1.980$ & $+0.112$ & 0.035\\
4551.0--4570.0 & $-1.950$ & $+0.101$ & 0.032\\
4586.5--4596.0 & $-1.579$ & $+0.050$ & 0.059\\
4638.0--4660.0 & $-2.145$ & $+0.175$ & 0.058\\
4680.0--4698.0 & $-2.032$ & $+0.332$ & 0.065\\
4802.0--4821.0 & $-2.039$ & $+0.148$ & 0.093\\
4975.0--5000.0 & $-2.213$ & $+0.157$ & 0.033\\
5000.0--5030.0 & $-2.273$ & $+0.215$ & 0.026\\
5060.0--5075.0 & $-2.185$ & $+0.325$ & 0.058\\
5331.0--5341.0 & $-1.796$ & $+0.050$ & 0.059\\
5376.0--5386.0 & $-1.767$ & $+0.080$ & 0.063\\
5510.0--5520.0 & $-1.967$ & $+0.169$ & 0.079\\
5860.0--5875.0 & \multicolumn{2}{c}{$\ldots$} \\
5912.0--5922.0 & \multicolumn{2}{c}{$\ldots$} \\
\mbox{[Cr/H]} \\
4537.0--4550.0 & $-2.593$ & $+99.990$ & 0.128\\
4612.0--4631.0 & $-2.534$ & $+99.990$ & 0.059\\
4646.0--4657.0 & $-2.670$ & $+99.990$ & 0.059\\
4703.0--4723.0 & $-2.244$ & $+99.990$ & 0.197\\
4751.0--4761.0 & \multicolumn{2}{c}{$\ldots$} \\
4796.0--4806.0 & \multicolumn{2}{c}{$\ldots$} \\
4824.0--4834.0 & $-2.417$ & $+99.990$ & 0.113\\
4866.0--4876.0 & $-2.272$ & $+99.990$ & 0.286\\
4931.0--4947.0 & $-2.409$ & $+99.990$ & 0.247\\
5063.0--5096.0 & \multicolumn{2}{c}{$\ldots$} \\
5270.0--5281.0 & $-2.330$ & $+99.990$ & 0.122\\
5292.0--5304.0 & $-2.651$ & $+99.990$ & 0.054\\
5341.0--5353.0 & $-2.692$ & $+99.990$ & 0.052\\
5407.0--5413.0 & $-2.500$ & $+99.990$ & 0.060\\
5779.0--5793.0 & $-2.125$ & $+99.990$ & 0.185\\
6325.0--6335.0 & \multicolumn{2}{c}{$\ldots$} \\
6973.0--6983.0 & \multicolumn{2}{c}{$\ldots$} \\
\mbox{[Mn/H]} \\
4750.0--4790.0 & $-2.801$ & $+0.253$ & 0.045\\
6010.0--6030.0 & $-2.430$ & $+0.283$ & 0.093\\
\mbox{[Ni/H]} \\
4600.0--4610.0 & $-2.726$ & $+0.330$ & 0.214\\
4644.0--4654.0 & $-2.430$ & $+0.270$ & 0.102\\
4681.0--4691.0 & \multicolumn{2}{c}{$\ldots$} \\
4709.0--4719.0 & $-2.381$ & $+0.136$ & 0.067\\
4824.0--4835.0 & $-2.361$ & $+0.249$ & 0.098\\
4899.0--4909.0 & $-2.808$ & $+0.254$ & 0.176\\
4931.0--4942.0 & $-2.353$ & $+99.990$ & 0.145\\
4975.0--4985.0 & $-2.294$ & $+0.267$ & 0.071\\
5098.0--5108.0 & $-2.212$ & $+0.272$ & 0.080\\
5141.0--5151.0 & $-2.446$ & $+0.271$ & 0.085\\
5472.0--5482.0 & $-2.802$ & $+0.208$ & 0.052\\
5707.0--5717.0 & $-2.387$ & $+0.342$ & 0.082\\
6103.0--6113.0 & $-2.417$ & $+0.328$ & 0.107\\
6172.0--6182.0 & $-2.062$ & $+0.290$ & 0.113\\
\mbox{[Cu/H]} \\
5101.0--5112.0 & \multicolumn{2}{c}{$\ldots$} \\
\mbox{[Zn/H]} \\
4717.0--4727.0 & $-2.587$ & $+99.990$ & 0.202\\
4805.0--4815.0 & $-2.448$ & $+99.990$ & 0.107\\
\mbox{[Zr/H]} \\
6124.0--6147.0 & \multicolumn{2}{c}{$\ldots$} \\
\mbox{[Ba/H]} \\
4551.0--4560.0 & $-2.053$ & $-0.041$ & 0.040\\
4929.0--4939.0 & $-2.006$ & $-0.091$ & 0.033\\
5849.0--5859.0 & $-1.892$ & $-0.075$ & 0.061\\
6135.0--6145.0 & $-1.889$ & $-0.169$ & 0.042\\
6492.0--6502.0 & $-1.732$ & $-0.185$ & 0.044\\
\mbox{[Eu/H]} \\
4431.0--4441.0 & $-1.803$ & $+99.990$ & 0.111\\
6640.0--6650.0 & $-1.595$ & $+99.990$ & 0.242\\
\hline
\end{tabular}
}
\end{table}

\end{appendix}

\end{document}